 \documentclass[preprint2]{aastex}

\shorttitle{1-mm line survey of EGOs}
\shortauthors{He, Takahashi \and Chen}

\begin{document}


\title{A 1-mm spectral line survey toward GLIMPSE
  Extended Green Objects (EGOs).  \footnote{The observation
    project was funded by Academia Sinica, Institute of Astronomy and
    Astrophysics, Taipei}}


\author{J.H. He\altaffilmark{1},
        S. Takahashi\altaffilmark{2}, and 
        X. Chen\altaffilmark{3}}

  \email{jinhuahe@ynao.ac.cn, satoko\_t@asiaa.sinica.edu.tw, chenxi@shao.ac.cn}

\altaffiltext{1}{Key Laboratory for the Structure and Evolution of Celestial Objects, 
  Yunnan Astronomical Observatory/National Astronomical Observatory, 
  Chinese Academy of Sciences, 
  P.O. Box 110, Kunming, 650011, Yunnan Province, PR China }

\altaffiltext{2}{Academia Sinica, Institute of Astronomy and Astrophysics,
  P.O. Box 23-141, Taipei 10617}

\altaffiltext{3}{Key Laboratory for Research in Galaxies and Cosmology,
  Key Laboratory for Research in Galaxies and Cosmology,
  Shanghai Astronomical Observatory, Chinese Academy of Sciences, 80
  Nandan Road, Shanghai 200030, PR China }


%\altaffiltext{1}{Visiting Astronomer, Cerro Tololo Inter-American Observatory.
%CTIO is operated by AURA, Inc.\ under contract to the National Science
%\altaffiltext{2}{Society of Fellows, Harvard University.}
%\altaffiltext{3}{present address: Center for Astrophysics,
%    60 Garden Street, Cambridge, MA 02138}
%\altaffiltext{4}{Visiting Programmer, Space Telescope Science Institute}
%\altaffiltext{5}{Patron, Alonso's Bar and Grill}


\begin{abstract}
A northern subsample of 89 {\it Spitzer GLIMPSE} 
extended green objects (EGOs), the candidate massive young stellar objects, are surveyed for
molecular lines in two 1-GHz ranges: 251.5-252.5 and 260.188-261.188\,GHz.
A comprehensive catalog of observed
molecular line data and spectral plots are presented. 
Eight molecular species are undoubtedly detected:
H$^{13}$CO$^+$, SiO, SO, CH$_3$OH, CH$_3$OCH$_3$, CH$_3$CH$_2$CN,
HCOOCH$_3$, and HN$^{13}$C. H$^{13}$CO$^+$\,3-2 line is detected in 70 EGOs 
among which 37 ones also show SiO\,6-5 line, demonstrating their 
association to dense gas and supporting the outflow
interpretation of the extended $4.5\,\mu$m excess emission.  
Our major dense gas and outflow tracers (H$^{13}$CO$^+$, SiO, SO and CH$_3$OH) are combined with 
our previous survey of $^{13}$CO, $^{12}$CO and C$^{18}$O\,1-0 
toward the same sample of EGOs for a multi-line multi-cloud analysis of line width and luminosity correlations. 
Good log-linear correlations are found among all considered line luminosities, which requires a universal similarity 
of density and thermal structures and probably of shock properties among all EGO clouds to explain. It also 
requires that the shocks should be produced within the natal clouds of the EGOs. Diverse degrees of correlation are 
found among the line widths. However, both the line width and luminosity correlations tend to 
progressively worsen across larger cloud subcomponent size-scales, depicting the increase of randomness across cloud subcomponent sizes. 
Moreover, the line width correlations
among the three isotopic CO\,1-0 lines show data scatter as linear functions 
of the line width itself, indicating that the velocity randomness also increases with whole-cloud sizes and 
has some regularity behind. 
\end{abstract}





\section{Introduction}

Formation and evolution of massive young stars is still a matter of
debate. As reviewed by \citet{zinn07}, the high
luminosity, high protostar temperature, and much shorter formation
time scales make the formation of massive stars distinct from that
of low mass stars. Particularly, many observed massive protostars
reside in proto-clusters, which further complicates
their formation processes by introducing possible interplay
between neighboring forming massive stars and their lower mass siblings. 

The paucity of known massive star forming regions is one of the
major obstacles in observational studies.
Thus, the recently uncovered new sample of over 300 massive young stellar
object (MYSO) candidates --
Extended Green Objects (EGOs) -- that are 
identified from the {\it Spitzer} Galactic Legacy Infrared Mid-Plane Survey
Extraordinaire ({\it GLIMPSE}) at 3.6, 4.5, 5.8, 8.0, and {\it MIPSGAL} at 24\,$\mu$m, 
have significantly extended the working sample for massive star formation studies
\citep{cyga08}. The EGOs are so named because they show excess
emission in extended structures in the Infrared Array Camera
({\it IRAC}) $4.5\,\mu$m band images that are
conventionally coded as green in the {\it IRAC} false color images.

EGOs are candidate birth places of MYSOs because, as shown by
\citet{cyga08}, many of them are associated with Class II CH$_3$OH
masers and/or infrared dark clouds (IRDCs). Both the Class II CH$_3$OH
masers \citep{crag05,elli06} and the IRDCs
\citep{simo06a,simo06b,rath06,rath07} are signposts of massive star
formation sites. A more detailed interferometric thermal millimeter
line mapping of two EGOs by \citet{cyga11} has lent further support to EGOs
being MYSOs or massive protoclusters.

EGOs very possibly possess prominent outflows, because the characteristic extended
green structures in them are possibly shock originated \citep{nori04,smit05,debu10}. Studies have confirmed 
that it is the $4.5\,\mu$m excess instead of the $4.5\,\mu$m total flux that traces 
the outflow shocks \citep[e.g.,][]{simp12,lee12}. The outflow nature of the
EGO samples is further confirmed by the high
detection rate of SiO\,5-4 line in a selected subsample of 10 EGOs
\citep{cyga09}. Furthermore, the interferometric mapping of two EGOs by
\citet{cyga11} has also revealed their outflow nature.

These EGOs are possibly still in early cloud collapse stage, because
their {\it Spitzer} [3.6]-[5.8] and [8.0]-[24] colors mimic
that of the youngest massive star formation models that are still
in early cloud collapse phase \citep{cyga08}.  
Furthermore, infall tracers such as HCO$^{+}\,1-0$ have been
surveyed by \citet{chen10} toward a subsample of 88 EGOs with a
single dish telescope in the
northern hemisphere. They found more blue skewed HCO$^{+}\,1-0$ line 
profiles than red skewed ones, which statistically supports the
existence of infall motions in these EGOs. The early stage of star formation 
in EGOs is also confirmed by the non-detection of continuum emission 
at 3.6 or 1.3\,cm (before the onset of UC\,H\,II region) by 
\citet{cyga11b}. 

The existing methanol (CH$_3$OH) maser data have provided rich
information to diagnose the physical conditions in EGOs. 
Many EGOs associate with both Class\,I and II methanol masers.
On the basis of methanol maser surveys in
literature, \citet{chen09} found that about two third of the observed EGOs are
spatially coincident with known Class\,I methanol masers at 44 and/or
95\,GHz. The high detection rate of Class\,I methanol masers has been 
confirmed by the 55 per cent detection rates of the 95\,GHz methanol maser in a recent 
Mopra survey of almost all EGOs in the \citet{cyga08} catalog by the same group \citep{chen11}.
Thus, more than half of the EGOs should be undoubtedly associated with 
active outflow shocks.

The VLA mapping of both classes of methanol masers by
\citet{cyga09} illustrated that the two classes of masers actually
occur in different spatial components of the objects: Class\,II
6.7\,GHz maser spots concentrate on the peak of the $24\,\mu$m
emission sources, tracing
the warm star forming cores, whilst the Class\,I 44\,GHz masers usually scatter
around the extended $4.5\,\mu$m emission structures, tracing the interface
between the proposed outflows and interstellar medium. The coexistence
of both classes of methanol masers in some EGOs may complicate the
interpretation of methanol thermal lines observed by single dishes,
because part of the detected emission could be mainly excited by strong
IR radiation in the hot star-forming cores while the rest part could 
be contributed by collisional excitation in the hot and dense shock regions around the outflows.

Up to now, these new objects are still lack of large scale survey of thermal molecular lines 
beside our previous survey of the three isotopic CO\,1-0 and HCO$^+$\,1-0 lines \citep{chen10}. 
To further explore the nature of the clouds around EGOs (EGO clouds hereafter), we have performed a survey of
outflow tracers, SiO\,6-5 and CH$_3$OH\,J$_3$-J$_2$\,A line series,
toward a sample of 89 EGOs mainly in the northern sky, nearly the same sample 
as in \citet{chen10}. 
Our observations also 
simultaneously cover some rotational transitions of H$^{13}$CO$^+$, SO, and other 
interesting complex species such as CH$_3$OCH$_3$, CH$_3$CH$_2$CN 
and HCOOCH$_3$. 

In this paper, we mainly present the
spectral plots, line parameters and discuss the observed line width and luminosity correlations
among the considered molecules,
while more detailed analysis of the results will be presented in future papers. 
Sects.~\ref{sample} and \ref{obs} of this paper present target selection and
observations. Some notes on line identification are given in
Sect.~\ref{lineiden}. We
present the results in Sect.~\ref{results} and also discuss the
dense gas, outflows, distance and line width and luminosity correlations in Sect.~\ref{discuss}.
In the end, the main points are summarized in Sect.~\ref{summary}.

\section[]{Sample selection}
\label{sample}

We have selected 89 EGOs with DEC $>-38\,\deg$ from the over 300 EGOs in \citet{cyga08} 
as our working sample. The infrared positions from the same paper are used. 
It is nearly identical to the sample of the isotopic CO\,1-0 observations by
\citet{chen10}. During the object selection, we found five pairs of EGOs that are
so close to each other that our telescope beam (29$\arcsec$) can hardly get them
resolved. Thus, only one source in each pair was selected and
the observed lines should be deemed as contributed by both EGOs. The
five dropped sources were \object[EGO G24.11-0.18]{G24.11-0.18}, \object[EGO
G43.04-0.45(b)]{G43.04-0.45(b)}, \object[EGO G54.45+1.02]{G54.45+1.02},
\object[EGO G54.11-0.05]{G54.11-0.05} \footnote{Later we recognized that this EGO is actually far enough away from
G54.11-0.04 so that it should be included in our sample. We
realized this only after the observations were done.} and \object[EGO
G58.78+0.65]{G58.78+0.65}. In addition, two EGOs, \object[EGO G19.01-0.03 O-N]{G19.01-0.03\,O-N}
and \object[EGO G19.01-0.03 O-S]{G19.01-0.03\,O-S}, were also excluded
from our sample, because they were covered by the beam toward their
associated point source, G19.01-0.03.

\section[]{Observations and data reduction}
\label{obs}

The observations were done with the Arizona Radio Observatory 10-m
Submillimeter Telescope (AROSMT) in several nights during two epochs:
2009 April 24 to May 11 and 2010 March 29 to April 27. The ALMA band-6
sideband-separating receiver was used with two Filter Banks to record
the upper and lower sidebands (USB and LSB) separately. The observed
frequency ranges were 251.5-252.5\,GHz (USB) and 260.188-261.188\,GHz
(LSB). The LRS velocities of most of our EGOs from the C$^{18}$O\,1-0 observations of \citet{chen10} 
had been used for tuning. The spectral resolution was
1\,MHz in the 1\,GHz bandwidth in both sidebands.
We tested several off
positions that were selected on the basis of the CO\,1-0 survey result
of \citet{dame01} at locations several arc minutes to degrees away from some targets but found
no line emission. However, our test runs in position-switch mode were found to suffer from strong 
ripples in baseline. Thus, we adopted beam switch
mode with a 2-arcmin throw at 1\,Hz for all our targets to improve the 
spectral baseline. This was
viable perhaps because our observed lines are mostly high density
tracers. This strategy was justified by the fact that no obvious
absorption like features were seen in the final
data. Although the existence of weak emission at the 
off-positions can not be excluded, their effects to 
our data should be hopefully small.

The telescope
pointing and focus was checked roughly every two
hours with a planet. The representative beam width is $29\arcsec$ (at
260\,GHz), which is corresponding to a linear resolution of roughly
0.56\,pc at a representative distance of 4\,kpc. 
Main beam efficiencies of 0.65 and 0.55 were applied
to the LSB and USB, respectively, to obtain the main beam
temperature $T_{\rm mb}$ from the antenna temperature $T_{\rm A}^{\rm *}$. 
Line flux can be computed from $T_{\rm mb}$ with the nominal conversion factor of 46.5\,Jy/K (at
260\,GHz) for the ARO SMT. The typical T$_{\rm sys}$ during most of the
observations were $\sim\,280$\,K in the LSB and $\sim\,311$\,K in the USB.

Sideband leakage can produce false line features to make trouble
  to line identification. Fortunately, it can be checked in our data by comparing the USB
and LSB data which are obtained simultaneously and separately.
We did not find leakage in most of our data,
except for the cases of \object[EGO G14.33-0.64]{G14.33-0.64} and
\object[EGO G24.33+0.14]{G24.33+0.14} of which the strong SO
line in the LSB leaked into the USB as a small artificial line feature. 
The rareness of the leakage is due to 
the image rejection ratio of about
10 to 20\,dB during most of our observations.

Gildas/CLASS software\footnote{\url{http://www.iram.fr/IRAMFR/GILDAS}} was used to reduce the data and analyze the line
profiles. A linear baseline was removed from almost all spectra, except the 
LSB spectrum of \object[EGO G11.92-0.61]{G11.92-0.61} for which a cubic baseline was
subtracted to correct for the gently curvy baseline. With an on+off exposure time of 10 minutes, 
the average baseline RMS (in mainbeam temperature) are $\sim\,25$\,mK in the LSB
and $\sim\,33$\,mK in the USB at a resolution of 1\,km\,s$^{-1}$. The rms of individual objects 
are listed in Table~\ref{tab2} (together with their LRS velocity and distance that will be discussed in later sections).

\section{Line identification}
\label{lineiden}

Systemic velocity ($V_{\rm LSR}$) is needed for line
identification. Although some EGOs already have their velocity known
from the C$^{18}$O observations of \citet{chen10}, we still use the
strongest lines in our results, H$^{13}$CO$^+$ 3-2 in the USB,
SO\,$^{3}\Sigma\,5_6-4_5$ and the CH$_3$OH\,J$_3$-J$_2$\,A line series
(with blended lines excluded) in the LSB, to determine
the systemic velocity for each target independently.  A single Gaussian profile is fit
to each line to determine observed line frequencies and uncertainties. (Note that 
some lines with broad line wings will be fit with double Gauss profiles to determined 
their line parameters in later sections.)
The velocities
determined from all lines are averaged together with the weight $T_{\rm mb}/\sigma_{\rm
  V}$, where $\sigma_{\rm V}$ is the statistic error of
the velocity and $T_{\rm mb}$ is the line peak temperature. With this
weighting scheme, the
strongest line, H$^{13}$CO$^+$ 3-2, actually always plays the major role in
determining $V_{\rm LSR}$. The uncertainty of the systemic velocity is
also computed during the weighted average.

Line identification turns out to be a hard task because of the limited
frequency coverage. In the first step, we recognize all line-like features that are
stronger than $3\sigma$ level in terms of integrated line
intensity in the spectra data. Then, the obtained line frequencies
(corrected for the object velocity) are compared with the line frequency lists,
SPLATALOGUE~\footnote{splatalogue at \url{http://www.splatalogue.net/}},
CDMS~\footnote{CMDS at \url{http://www.astro.uni-koeln.de/cdms/catalog}}\citep{muel01,muel05}
and
JPL~\footnote{JPL line list at \url{http://spec.jpl.nasa.gov/ftp/pub/catalog/catform.html}}\citep{pick98},
to identify the carriers of the lines. For simple molecules of which only one transition is observed in our 
observed frequency ranges, their identification is usually tentative (e.g., H$_2$CCO and OC$^{34}$S), unless they are 
well known abundant species (e.g., H$^{13}$CO$^+$, SiO, CH$_3$OH and SO). On the opposite side, line-rich spectra of some
complex molecules (some are the so called `weeds' lines) are covered by our observations and used as
fingerprints for these species. We plot the expected relative line strengths at various
representative LTE temperatures in optically thin cases to help the
line identification by eyes. 

\section{Results}
\label{results}

%(prepared by hand)
\begin{table*}
\centering
\begin{minipage}{155mm}
\footnotesize\rm
\caption{\label{tab2}Parameters of the EGOs: baseline RMS level at a
  resolution of 1\,km\,s$^{-1}$, $V_{\rm sys}$
    (Sect.~\ref{lineiden}), and kinematic distances (Sect.~\ref{distance}).}
\begin{tabular*}{\textwidth}{l@{ }l@{ }l@{ }l@{ }l@{ }r@{ }l@{ }r@{ }r@{ }l}
\hline
              &            &           & RMS-LSB & RMS-USB & $V_{\rm sys}(\sigma)$ &                        &\multicolumn{2}{l}{D$_{\rm kin}$}  &                          \\
Object        & RA(2000)   & DEC(2000) & (mK)    & (mK)    & (km\,s$^{-1}$)        & NoteV\tablenotemark{a} & \multicolumn{2}{l}{(kpc)}         & NoteD$\tablenotemark{b}$ \\
\hline
\object[EGO G10.29-0.13]{G10.29$-$0.13}    & 18$^h$08$^m$49.3$^s$ & -20$^d$05$^m$57$^s$ & 24 & 30   &  13.32(17)                           &                       &  $ 2.11^{+0.72}_{-0.92}$ & $                     $ & IRDC      \\ 
\object[EGO G10.34-0.14]{G10.34$-$0.14}    & 18$^h$09$^m$00.0$^s$ & -20$^d$03$^m$35$^s$ & 23 & 34   &  12.32(08)                           &                       &  $ 1.99^{+0.74}_{-0.95}$ & $                     $ & IRDC      \\
\object[EGO G11.11-0.11]{G11.11$-$0.11}    & 18$^h$10$^m$28.3$^s$ & -19$^d$22$^m$31$^s$ & 38 & 50   &  29.79(10)                           &                       &  $ 3.45^{+0.45}_{-0.53}$ & $                     $ & IRDC      \\
\object[EGO G11.92-0.61]{G11.92$-$0.61}    & 18$^h$13$^m$58.1$^s$ & -18$^d$54$^m$17$^s$ & 23 & 33   &  35.88(06)                           &                       &  $ 3.72^{+0.39}_{-0.46}$ & $                     $ & IRDC      \\
\object[EGO G12.02-0.21]{G12.02$-$0.21}    & 18$^h$12$^m$40.4$^s$ & -18$^d$37$^m$11$^s$ & 26 & 34   &  -3.43(30)                           & weak                  &  $16.92^{+1.57}_{-1.17}$ & $                     $ & Outer     \\
\object[EGO G12.20-0.03]{G12.20$-$0.03}    & 18$^h$12$^m$23.6$^s$ & -18$^d$22$^m$54$^s$ & 39 & 51   &  51.15(14)                           &                       &  $ 4.47^{+0.30}_{-0.33}$ & $11.95^{+0.33}_{-0.30}$ & N/F       \\
\object[EGO G12.42+0.50]{G12.42$+$0.50}    & 18$^h$10$^m$51.1$^s$ & -17$^d$55$^m$50$^s$ & 18 & 28   &  17.76(01)                           &                       &  $ 2.30^{+0.59}_{-0.71}$ & $                     $ & Z1/2      \\
\object[EGO G12.68-0.18]{G12.68$-$0.18}    & 18$^h$13$^m$54.7$^s$ & -18$^d$01$^m$47$^s$ & 40 & 52   &  55.54(10)                           & new                   &  $ 4.59^{+0.28}_{-0.31}$ & $11.80^{+0.31}_{-0.28}$ & N/F       \\
\object[EGO G12.91-0.03]{G12.91$-$0.03}    & 18$^h$13$^m$48.2$^s$ & -17$^d$45$^m$39$^s$ & 27 & 33   &  56.46(08)                           &                       &  $ 4.60^{+0.27}_{-0.31}$ & $                     $ & IRDC      \\
\object[EGO G12.91-0.26]{G12.91$-$0.26}    & 18$^h$14$^m$39.5$^s$ & -17$^d$52$^m$00$^s$ & 24 & 33   &  37.40(03)                           &                       &  $ 3.67^{+0.38}_{-0.44}$ & $12.71^{+0.44}_{-0.38}$ & N/F       \\
\object[EGO G14.33-0.64]{G14.33$-$0.64}    & 18$^h$18$^m$54.4$^s$ & -16$^d$47$^m$46$^s$ & 26 & 35   &  22.16(02)                           &                       &  $ 2.47^{+0.51}_{-0.59}$ & $                     $ & IRDC      \\
\object[EGO G14.63-0.58]{G14.63$-$0.58}    & 18$^h$19$^m$15.4$^s$ & -16$^d$30$^m$07$^s$ & 25 & 30   &  18.32(04)                           &                       &  $ 2.13^{+0.55}_{-0.65}$ & $                     $ & IRDC      \\
\object[EGO G16.58-0.08]{G16.58$-$0.08}    & 18$^h$21$^m$15.0$^s$ & -14$^d$33$^m$02$^s$ & 21 & 29   &  39.74(72)                           & weak                  &  $ 3.38^{+0.36}_{-0.40}$ & $                     $ & IRDC      \\
\object[EGO G16.59-0.05]{G16.59$-$0.05}    & 18$^h$21$^m$09.1$^s$ & -14$^d$31$^m$48$^s$ & 24 & 33   &  59.54(05)                           &                       &  $ 4.31^{+0.27}_{-0.30}$ & $                     $ & IRDC      \\
\object[EGO G16.61-0.24]{G16.61$-$0.24}    & 18$^h$21$^m$52.7$^s$ & -14$^d$35$^m$51$^s$ & 26 & 32   &  42.87(22)                           & new                   &  $ 3.54^{+0.34}_{-0.38}$ & $                     $ & IRDC      \\
\object[EGO G17.96+0.08]{G17.96$+$0.08}    & 18$^h$23$^m$21.0$^s$ & -13$^d$15$^m$11$^s$ & 24 & 32   &  22.30(26)                           & weak                  &  $ 2.18^{+0.47}_{-0.54}$ & $                     $ & HISA      \\
\object[EGO G18.67+0.03]{G18.67$+$0.03}    & 18$^h$24$^m$53.7$^s$ & -12$^d$39$^m$20$^s$ & 22 & 29   &  80.04(28)                           &                       &  $11.01^{+0.25}_{-0.23}$ & $                     $ & HISA      \\
\object[EGO G18.89-0.47]{G18.89$-$0.47}    & 18$^h$27$^m$07.9$^s$ & -12$^d$41$^m$36$^s$ & 23 & 28   &  66.21(07)                           &                       &  $ 4.38^{+0.26}_{-0.28}$ & $                     $ & IRDC      \\
\object[EGO G19.01-0.03]{G19.01$-$0.03}    & 18$^h$25$^m$44.8$^s$ & -12$^d$22$^m$46$^s$ & 15 & 22   &  59.78(04)                           &                       &  $ 4.11^{+0.28}_{-0.30}$ & $                     $ & IRDC      \\
\object[EGO G19.36-0.03]{G19.36$-$0.03}    & 18$^h$26$^m$25.8$^s$ & -12$^d$03$^m$57$^s$ & 22 & 28   &  26.69(05)                           &                       &  $ 2.38^{+0.43}_{-0.48}$ & $                     $ & IRDC      \\
\object[EGO G19.61-0.12]{G19.61$-$0.12}    & 18$^h$27$^m$13.6$^s$ & -11$^d$53$^m$20$^s$ & 26 & 34   &  57.03(20)                           &                       &  $ 3.95^{+0.29}_{-0.31}$ & $11.87^{+0.31}_{-0.29}$ & N/F       \\
\object[EGO G19.61-0.14]{G19.61$-$0.14}    & 18$^h$27$^m$16.8$^s$ & -11$^d$53$^m$51$^s$ & 19 & 23   &  56.98(19)                           & new                   &  $ 3.95^{+0.29}_{-0.31}$ & $11.87^{+0.31}_{-0.29}$ & N/F       \\
\object[EGO G19.88-0.53]{G19.88$-$0.53}    & 18$^h$29$^m$14.7$^s$ & -11$^d$50$^m$23$^s$ & 21 & 29   &  43.67(02)                           &                       &  $ 3.31^{+0.34}_{-0.37}$ & $                     $ & HISA      \\
\object[EGO G20.24+0.07]{G20.24$+$0.07}    & 18$^h$27$^m$44.6$^s$ & -11$^d$14$^m$54$^s$ & 22 & 29   &  70.03(32)                           & weak                  &  $ 4.44^{+0.26}_{-0.28}$ & $11.32^{+0.28}_{-0.26}$ & N/F       \\
\object[EGO G21.24+0.19]{G21.24$+$0.19}    & 18$^h$29$^m$10.2$^s$ & -10$^d$18$^m$11$^s$ & 22 & 28   &  25.40(09)                           &                       &  $ 2.19^{+0.43}_{-0.47}$ & $                     $ & HISA      \\
\object[EGO G22.04+0.22]{G22.04$+$0.22}    & 18$^h$30$^m$34.7$^s$ & -09$^d$34$^m$47$^s$ & 21 & 30   &  51.16(05)                           &                       &  $ 3.54^{+0.31}_{-0.34}$ & $                     $ & HISA      \\
\object[EGO G23.01-0.41]{G23.01$-$0.41}    & 18$^h$34$^m$40.2$^s$ & -09$^d$00$^m$38$^s$ & 18 & 28   &  77.45(05)                           &                       &  $ 4.58^{+0.26}_{-0.27}$ & $                     $ & HISA      \\
\object[EGO G23.82+0.38]{G23.82$+$0.38}    & 18$^h$33$^m$19.5$^s$ & -07$^d$55$^m$37$^s$ & 21 & 28   &  76.41(24)                           &                       &  $ 4.52^{+0.27}_{-0.28}$ & $10.85^{+0.28}_{-0.27}$ & N/F       \\
\object[EGO G23.96-0.11]{G23.96$-$0.11}    & 18$^h$35$^m$22.3$^s$ & -08$^d$01$^m$28$^s$ & 21 & 30   &  72.10(10)                           & new                   &  $ 4.34^{+0.27}_{-0.29}$ & $                     $ & 21cm      \\
\object[EGO G24.00-0.10]{G24.00$-$0.10}    & 18$^h$35$^m$23.5$^s$ & -07$^d$59$^m$32$^s$ & 21 & 30   &  70.61(16)                           &                       &  $ 4.28^{+0.28}_{-0.29}$ & $                     $ & 21cm      \\
\object[EGO G24.11-0.17]{G24.11$-$0.17}    & 18$^h$35$^m$52.6$^s$ & -07$^d$55$^m$17$^s$ & 22 & 32   &  80.48(20)                           &                       &  $ 4.66^{+0.27}_{-0.27}$ & $                     $ & IRDC      \\
\object[EGO G24.17-0.02]{G24.17$-$0.02}    & 18$^h$35$^m$25.0$^s$ & -07$^d$48$^m$15$^s$ & 21 & 30   &                                      &                       &  $                     $ & $                     $ &           \\
\object[EGO G24.33+0.14]{G24.33$+$0.14}    & 18$^h$35$^m$08.1$^s$ & -07$^d$35$^m$04$^s$ & 22 & 30   & 113.79(07)                           & new                   &  $ 5.93^{+0.30}_{-0.28}$ & $ 9.37^{+0.28}_{-0.30}$ & N/F       \\
\object[EGO G24.63+0.15]{G24.63$+$0.15}    & 18$^h$35$^m$40.1$^s$ & -07$^d$18$^m$35$^s$ & 23 & 30   &  53.16(16)                           &                       &  $ 3.51^{+0.31}_{-0.33}$ & $                     $ & HISA      \\
\object[EGO G24.94+0.07]{G24.94$+$0.07}    & 18$^h$36$^m$31.5$^s$ & -07$^d$04$^m$16$^s$ & 24 & 32   &  41.25(16)                           &                       &  $ 2.92^{+0.35}_{-0.37}$ & $12.32^{+0.37}_{-0.35}$ & N/F       \\
\object[EGO G25.27-0.43]{G25.27$-$0.43}    & 18$^h$38$^m$57.0$^s$ & -07$^d$00$^m$48$^s$ & 26 & 35   &  59.14(03)                           & chen10                &  $ 3.75^{+0.30}_{-0.32}$ & $                     $ & HISA      \\
\object[EGO G25.38-0.15]{G25.38$-$0.15}    & 18$^h$38$^m$08.1$^s$ & -06$^d$46$^m$53$^s$ & 20 & 28   &  95.12(07)                           &                       &  $ 5.21^{+0.28}_{-0.27}$ & $                     $ & IRDC      \\
\object[EGO G27.97-0.47]{G27.97$-$0.47}    & 18$^h$44$^m$03.6$^s$ & -04$^d$38$^m$02$^s$ & 25 & 34   &                                      &                       &  $                     $ & $                     $ &           \\
\object[EGO G28.28-0.36]{G28.28$-$0.36}    & 18$^h$44$^m$13.2$^s$ & -04$^d$18$^m$04$^s$ & 32 & 45   &  49.07(24)                           &                       &  $ 3.20^{+0.33}_{-0.35}$ & $11.60^{+0.35}_{-0.33}$ & N/F       \\
\object[EGO G28.83-0.25]{G28.83$-$0.25}    & 18$^h$44$^m$51.3$^s$ & -03$^d$45$^m$48$^s$ & 32 & 44   &  87.31(12)                           &                       &  $ 4.92^{+0.32}_{-0.31}$ & $                     $ & HISA      \\
\object[EGO G28.85-0.23]{G28.85$-$0.23}    & 18$^h$44$^m$47.5$^s$ & -03$^d$44$^m$15$^s$ & 24 & 31   &  96.68(23)                           & weak,new                 &  $ 5.35^{+0.35}_{-0.32}$ & $ 9.37^{+0.32}_{-0.35}$ & N/F       \\
\object[EGO G29.84-0.47]{G29.84$-$0.47}    & 18$^h$47$^m$28.8$^s$ & -02$^d$58$^m$03$^s$ & 32 & 42   &                                      &                       &  $                     $ & $                     $ &           \\
\object[EGO G29.89-0.77]{G29.89$-$0.77}    & 18$^h$48$^m$37.7$^s$ & -03$^d$03$^m$44$^s$ & 33 & 45   &  83.85(02)                           & chen10                &  $ 4.78^{+0.33}_{-0.32}$ & $                     $ & IRDC      \\
\object[EGO G29.91-0.81]{G29.91$-$0.81}    & 18$^h$48$^m$47.6$^s$ & -03$^d$03$^m$31$^s$ & 33 & 44   &  83.90(02)                           & chen10                &  $ 4.78^{+0.33}_{-0.32}$ & $                     $ & Z1/2      \\
\object[EGO G29.96-0.79]{G29.96$-$0.79}    & 18$^h$48$^m$50.0$^s$ & -03$^d$00$^m$21$^s$ & 33 & 40   &  85.12(03)                           & chen10                &  $ 4.84^{+0.33}_{-0.32}$ & $                     $ & IRDC      \\
\object[EGO G34.26+0.15]{G34.26$+$0.15}    & 18$^h$53$^m$16.4$^s$ & +01$^d$15$^m$07$^s$ & 23 & 30   &  58.88(03)                           &                       &  $ 3.64^{+0.36}_{-0.36}$ & $                     $ & 21cm      \\
\object[EGO G34.28+0.18]{G34.28$+$0.18}    & 18$^h$53$^m$15.0$^s$ & +01$^d$17$^m$11$^s$ & 27 & 36   &  55.99(09)                           &                       &  $ 3.49^{+0.36}_{-0.36}$ & $                     $ & IRDC      \\
\object[EGO G34.39+0.22]{G34.39$+$0.22}    & 18$^h$53$^m$19.0$^s$ & +01$^d$24$^m$08$^s$ & 24 & 31   &  57.46(05)                           &                       &  $ 3.57^{+0.36}_{-0.36}$ & $                     $ & IRDC      \\
\object[EGO G34.41+0.24]{G34.41$+$0.24}    & 18$^h$53$^m$17.9$^s$ & +01$^d$25$^m$25$^s$ & 25 & 36   &  57.86(04)                           &                       &  $ 3.59^{+0.36}_{-0.36}$ & $                     $ & IRDC      \\
\object[EGO G35.03+0.35]{G35.03$+$0.35}    & 18$^h$54$^m$00.5$^s$ & +02$^d$01$^m$18$^s$ & 27 & 36   &  53.11(05)                           &                       &  $ 3.35^{+0.37}_{-0.37}$ & $10.41^{+0.37}_{-0.37}$ & N/F       \\
\object[EGO G35.04-0.47]{G35.04$-$0.47}    & 18$^h$56$^m$58.1$^s$ & +01$^d$39$^m$37$^s$ & 28 & 38   &  50.73(12)                           &                       &  $ 3.23^{+0.37}_{-0.37}$ & $                     $ & IRDC      \\
\hline
\end{tabular*}
\end{minipage}
\end{table*}
\begin{table*}
\addtocounter{table}{-1}
\centering
\begin{minipage}{155mm}
\footnotesize\rm
\caption{(continued)}
\begin{tabular*}{\textwidth}{l@{ }l@{ }l@{ }l@{ }l@{ }r@{ }l@{ }r@{ }r@{ }l}
\hline
              &            &           & RMS-LSB & RMS-USB & $V_{\rm sys}(\sigma)$ &                        &\multicolumn{2}{l}{D$_{\rm kin}$}  &                          \\
Object        & RA(2000)   & DEC(2000) & (mK)    & (mK)    & (km\,s$^{-1}$)        & NoteV\tablenotemark{a} & \multicolumn{2}{l}{(kpc)}         & NoteD$\tablenotemark{b}$ \\
\hline
\object[EGO G35.13-0.74]{G35.13$-$0.74}    & 18$^h$58$^m$06.4$^s$ & +01$^d$37$^m$01$^s$ & 28 & 38   &  34.20(04)                           &                       &  $ 2.34^{+0.38}_{-0.40}$ & $                     $ & HISA      \\
\object[EGO G35.15+0.80]{G35.15$+$0.80}    & 18$^h$52$^m$36.6$^s$ & +02$^d$20$^m$26$^s$ & 28 & 40   &  74.89(18)                           &                       &  $ 4.53^{+0.42}_{-0.39}$ & $                     $ & Z1/2      \\
\object[EGO G35.20-0.74]{G35.20$-$0.74}    & 18$^h$58$^m$12.9$^s$ & +01$^d$40$^m$33$^s$ & 28 & 39   &  33.49(03)                           &                       &  $ 2.30^{+0.38}_{-0.40}$ & $                     $ & Z1/2      \\
\object[EGO G35.68-0.18]{G35.68$-$0.18}    & 18$^h$57$^m$05.0$^s$ & +02$^d$22$^m$00$^s$ & 28 & 37   &  27.76(10)                           &                       &  $ 1.97^{+0.40}_{-0.41}$ & $                     $ & IRDC      \\
\object[EGO G35.79-0.17]{G35.79$-$0.17}    & 18$^h$57$^m$16.7$^s$ & +02$^d$27$^m$56$^s$ & 28 & 35   &  61.54(13)                           & new                   &  $ 3.81^{+0.39}_{-0.38}$ & $                     $ & IRDC      \\
\object[EGO G35.83-0.20]{G35.83$-$0.20}    & 18$^h$57$^m$26.9$^s$ & +02$^d$29$^m$00$^s$ & 23 & 30   &  28.38(04)                           & chen10                &  $ 2.01^{+0.40}_{-0.41}$ & $                     $ & HISA      \\
\object[EGO G36.01-0.20]{G36.01$-$0.20}    & 18$^h$57$^m$45.9$^s$ & +02$^d$39$^m$05$^s$ & 27 & 34   &  87.25(04)                           & chen10                &  $ 5.44^{+0.52}_{-0.52}$ & $ 8.15^{+0.52}_{-0.52}$ & N/F       \\
\object[EGO G37.48-0.10]{G37.48$-$0.10}    & 19$^h$00$^m$07.0$^s$ & +03$^d$59$^m$53$^s$ & 29 & 39   &  58.93(20)                           &                       &  $ 3.72^{+0.41}_{-0.39}$ & $ 9.61^{+0.39}_{-0.41}$ & N/F       \\
\object[EGO G37.55+0.20]{G37.55$+$0.20}    & 18$^h$59$^m$07.5$^s$ & +04$^d$12$^m$31$^s$ & 28 & 38   &                                      &                       &  $                     $ & $                     $ &           \\
\object[EGO G39.10+0.49]{G39.10$+$0.49}    & 19$^h$00$^m$58.1$^s$ & +05$^d$42$^m$44$^s$ & 28 & 40   &  23.10(23)                           &                       &  $11.34^{+0.43}_{-0.42}$ & $                     $ & HISA      \\
\object[EGO G39.39-0.14]{G39.39$-$0.14}    & 19$^h$03$^m$45.3$^s$ & +05$^d$40$^m$43$^s$ & 31 & 42   &  65.64(15)                           &                       &  $ 4.24^{+0.51}_{-0.45}$ & $ 8.74^{+0.45}_{-0.51}$ & N/F       \\
\object[EGO G40.28-0.22]{G40.28$-$0.22}    & 19$^h$05$^m$41.3$^s$ & +06$^d$26$^m$13$^s$ & 24 & 32   &  73.52(04)                           &                       &  $ 4.97^{+0.58}_{-0.58}$ & $ 7.85^{+0.58}_{-0.58}$ & N/F       \\
\object[EGO G40.28-0.27]{G40.28$-$0.27}    & 19$^h$05$^m$51.5$^s$ & +06$^d$24$^m$39$^s$ & 26 & 36   &  71.26(42)                           & weak                  &  $ 4.76^{+0.54}_{-0.54}$ & $                     $ & Z1/2      \\
\object[EGO G40.60-0.72]{G40.60$-$0.72}    & 19$^h$08$^m$03.3$^s$ & +06$^d$29$^m$15$^s$ & 20 & 27   &  64.43(88)                           & weak                  &  $ 4.27^{+0.56}_{-0.48}$ & $ 8.49^{+0.48}_{-0.56}$ & N/F       \\
\object[EGO G43.04-0.45(a)]{G43.04$-$0.45} & 19$^h$11$^m$38.9$^s$ & +08$^d$46$^m$39$^s$ & 27 & 34   &  57.50(12)                           &                       &  $ 3.99^{+0.61}_{-0.51}$ & $ 8.29^{+0.51}_{-0.61}$ & N/F       \\
\object[EGO G44.01-0.03]{G44.01$-$0.03}    & 19$^h$11$^m$57.2$^s$ & +09$^d$50$^m$05$^s$ & 27 & 35   &  64.48(05)                           & chen10                &  $ 4.80^{+0.70}_{-0.70}$ & $ 7.28^{+0.70}_{-0.70}$ & N/F       \\
\object[EGO G45.47+0.05]{G45.47$+$0.05}    & 19$^h$14$^m$25.6$^s$ & +11$^d$09$^m$28$^s$ & 27 & 35   &  62.81(05)                           &                       &  $ 5.89^{+1.74}_{-1.74}$ & $                     $ & Ave       \\
\object[EGO G45.47+0.13]{G45.47$+$0.13}    & 19$^h$14$^m$07.3$^s$ & +11$^d$12$^m$16$^s$ & 27 & 34   &  61.61(08)                           &                       &  $ 6.95^{+0.79}_{-0.79}$ & $                     $ & HII       \\
\object[EGO G45.50+0.12]{G45.50$+$0.12}    & 19$^h$14$^m$13.0$^s$ & +11$^d$13$^m$30$^s$ & 18 & 25   &  61.19(05)                           & chen10                &  $ 7.00^{+0.77}_{-0.77}$ & $                     $ & HII       \\
\object[EGO G45.80-0.36]{G45.80$-$0.36}    & 19$^h$16$^m$31.1$^s$ & +11$^d$16$^m$11$^s$ & 19 & 26   &  59.18(14)                           & new                   &  $ 4.59^{+0.73}_{-0.73}$ & $ 7.12^{+0.73}_{-0.73}$ & N/F       \\
\object[EGO G48.66-0.30]{G48.66$-$0.30}    & 19$^h$21$^m$48.0$^s$ & +13$^d$49$^m$21$^s$ & 26 & 36   &  33.60(03)                           & chen10                &  $ 8.48^{+0.56}_{-0.51}$ & $                     $ & HISA      \\
\object[EGO G49.07-0.33]{G49.07$-$0.33}    & 19$^h$22$^m$41.9$^s$ & +14$^d$10$^m$12$^s$ & 19 & 26   &  60.85(13)                           & new                   &  $ 5.50^{+1.57}_{-1.57}$ & $                     $ & Tan       \\
\object[EGO G49.27-0.32]{G49.27$-$0.32}    & 19$^h$23$^m$02.2$^s$ & +14$^d$20$^m$52$^s$ & 20 & 26   &  66.04(07)                           & chen10                &  $ 5.48^{+1.58}_{-1.58}$ & $                     $ & Tan       \\
\object[EGO G49.27-0.34]{G49.27$-$0.34}    & 19$^h$23$^m$06.7$^s$ & +14$^d$20$^m$13$^s$ & 28 & 36   &  67.76(05)                           &                       &  $ 5.48^{+1.58}_{-1.58}$ & $                     $ & Tan       \\
\object[EGO G49.42+0.33]{G49.42$+$0.33}    & 19$^h$20$^m$59.1$^s$ & +14$^d$46$^m$53$^s$ & 19 & 26   & -20.76(30)                           & weak                  &  $12.20^{+0.56}_{-0.53}$ & $                     $ & Outer     \\
\object[EGO G49.91+0.37]{G49.91$+$0.37}    & 19$^h$21$^m$47.5$^s$ & +15$^d$14$^m$26$^s$ & 20 & 27   &   8.33(26)                           & new                   &  $ 0.84^{+0.49}_{-0.49}$ & $ 9.98^{+0.49}_{-0.49}$ & N/F       \\
\object[EGO G50.36-0.42]{G50.36$-$0.42}    & 19$^h$25$^m$32.8$^s$ & +15$^d$15$^m$38$^s$ & 20 & 25   &  38.87(23)                           & weak,new                 &  $ 3.18^{+0.61}_{-0.61}$ & $                     $ & HISA      \\
\object[EGO G53.92-0.07]{G53.92$-$0.07}    & 19$^h$31$^m$23.0$^s$ & +18$^d$33$^m$00$^s$ & 20 & 25   &  42.90(09)                           &                       &  $ 4.95^{+1.68}_{-1.68}$ & $                     $ & Tan       \\
\object[EGO G54.11-0.04]{G54.11$-$0.04}    & 19$^h$31$^m$40.0$^s$ & +18$^d$43$^m$53$^s$ & 20 & 27   &                                      &                       &  $                     $ & $                     $ &           \\
\object[EGO G54.11-0.08]{G54.11$-$0.08}    & 19$^h$31$^m$48.8$^s$ & +18$^d$42$^m$57$^s$ & 20 & 26   &  38.41(13)                           &                       &  $ 3.81^{+0.93}_{-0.93}$ & $ 6.04^{+0.93}_{-0.93}$ & N/F       \\
\object[EGO G54.45+1.01]{G54.45$+$1.01}    & 19$^h$28$^m$25.7$^s$ & +19$^d$32$^m$20$^s$ & 26 & 35   &  34.72(35)                           & weak                  &  $ 3.32^{+0.79}_{-0.79}$ & $                     $ & Z1/2      \\
\object[EGO G56.13+0.22]{G56.13$+$0.22}    & 19$^h$34$^m$51.5$^s$ & +20$^d$37$^m$28$^s$ & 29 & 36   &                                      &                       &  $                     $ & $                     $ &           \\
\object[EGO G57.61+0.02]{G57.61$+$0.02}    & 19$^h$38$^m$40.8$^s$ & +21$^d$49$^m$35$^s$ & 27 & 35   &  37.01(14)                           & new                   &  $ 4.50^{+1.76}_{-1.76}$ & $                     $ & Tan       \\
\object[EGO G58.09-0.34]{G58.09$-$0.34}    & 19$^h$41$^m$03.9$^s$ & +22$^d$03$^m$39$^s$ & 28 & 39   &                                      &                       &  $                     $ & $                     $ &           \\
\object[EGO G58.78+0.64]{G58.78$+$0.64}    & 19$^h$38$^m$49.6$^s$ & +23$^d$08$^m$40$^s$ & 25 & 35   &  32.55(13)                           &                       &  $ 4.35^{+1.78}_{-1.78}$ & $                     $ & Tan       \\
\object[EGO G58.79+0.63]{G58.79$+$0.63}    & 19$^h$38$^m$55.3$^s$ & +23$^d$09$^m$04$^s$ & 26 & 37   &                                      &                       &  $                     $ & $                     $ &           \\
\object[EGO G59.79+0.63]{G59.79$+$0.63}    & 19$^h$41$^m$03.1$^s$ & +24$^d$01$^m$15$^s$ & 26 & 36   &  33.97(12)                           &                       &  $ 4.23^{+1.80}_{-1.80}$ & $                     $ & Tan       \\
\object[EGO G62.70-0.51]{G62.70$-$0.51}    & 19$^h$51$^m$51.1$^s$ & +25$^d$57$^m$40$^s$ & 26 & 34   &                                      &                       &  $                     $ & $                     $ &           \\
\hline
\end{tabular*}
{\it A machine-readable format of this table is also available through ADS.}
\tablenotetext{a}{
{\bf new} -- new velocity not presented in \citet{chen10};\\
{\bf chen10} -- velocity from \citet{chen10}, non-detection of H$^{13}$CO$^+$\,3-2 in this work;\\
{\bf weak} -- the detection level of H$^{13}$CO$^+$\,3-2 is lower than $5\,\sigma$ in terms of integrated line area in our observations. Thus the velocity should be used with caution.}
\tablenotetext{b}{
{\bf Tan} -- No near/far kinematic distance ambiguity at the tangential point; \\
{\bf Outer} -- No near/far kinematic distance ambiguity outside of the solar circle; \\
{\bf IRDC} -- Near kinematic distance determined by the association to IRDCs; \\
{\bf Z1/2} -- Near distance is preferred if the perpendicular distance is smaller than the scale hight of 60\,pc of the Galaxy 
              but larger than or nearly equal to two times of the scale hight;\\
{\bf HISA} -- H\,I self absorption method \citep{roma09};\\
{\bf 21cm} -- The 21cm continuum absorption line method \citep{roma09};\\
{\bf HII } -- H\,I absorption line method for H\,II regions \citep{kolp03};\\
{\bf Ave } -- Average of near and far kinematic distances, with the near/far difference included in uncertainty; \\
{\bf N/F} -- Both the near and far kinematic distances are given. }
\end{minipage}
\end{table*}

%(prepared by hand)
\begin{table*}
\centering
\begin{minipage}{160mm}
\footnotesize\rm
\caption{\label{tab1}Detected lines toward the EGOs. Note that the string `(E+A)' after the
  transition CH$_3$OCH$_3$,$10_{2,9}-9_{1,8}$ means it is a blending of four transitions: EA, AE, EE, and AA.}
\begin{tabular*}{\textwidth}{ll@{}c|ll@{}c}
\hline
$\nu_{\rm Catalog}$ & Transition                                &       &$\nu_{\rm Catalog}$  & Transition           &                          \\
(MHz)           & (Lower Sideband)                              & Note\tablenotemark{a}  & (MHz)  & (Lower Sideband) & Note\tablenotemark{a}    \\
\hline                                                                                                                                      
251501.40980    &       CH$_3$CH$_2$CN\,$28_{5,24}-27_{5,23}$   &       &  260191.98200   &           H$_2$CCO\,$13_{1,13}-12_{1,12}$ &         \\ 
251517.26200    &               CH$_3$OH\,$8_{3,5}-8_{2,6}-+$   &       &  260221.65800   &     CH$_3$CH$_2$CN\,$29_{6,24}-28_{6,23}$ &         \\
251527.30230    &                  c-HCCCH\,$6_{2,5}-5_{1,4}$   &       &  260229.15800   &     CH$_3$CH$_2$CN\,$29_{6,23}-28_{6,22}$ &         \\ 
251583.62730    &      CH$_3$OCH$_3$\,$10_{2,9}-9_{1,8}$(E+A)   &       &  260244.49800   &        HCOOCH$_3$\,$21_{3,18}-20_{3,17}$E &         \\ 
251607.12090    &       CH$_3$CH$_2$CN\,$28_{5,23}-27_{5,22}$   &       &  260255.08000   &        HCOOCH$_3$\,$21_{3,18}-20_{3,17}$A & b       \\ 
251641.66700    &               CH$_3$OH\,$7_{3,4}-7_{2,5}-+$   &       &  260255.33900   &                       H$^{13}$CO$^+$\,3-2 & b       \\ 
251668.84660    &       CH$_3$CH$_2$CN\,$28_{4,25}-27_{4,24}$   &       &  260329.31200   &   CH$_3$OCH$_3$\,$19_{5,15}-19_{4,16}$EE  &         \\ 
251738.52000    &               CH$_3$OH\,$6_{3,3}-6_{2,4}-+$   &       &  260365.00000   &                                  U260365  &         \\ 
251811.88200    &               CH$_3$OH\,$5_{3,2}-5_{2,3}-+$   &       &  260384.26800   &        HCOOCH$_3$\,$21_{8,13}-20_{8,12}$E &         \\ 
251825.77000    &                SO\,$^3\Sigma$\,$6_{5}-5_{4}$  &       &  260403.39210   &    CH$_3$OCH$_3$\,$16_{5,11}-16_{4,12}$EE & b       \\ 
251866.57900    &               CH$_3$OH\,$4_{3,1}-4_{2,2}-+$   &       &  260404.02600   &        HCOOCH$_3$\,$21_{8,14}-20_{8,13}$E & b       \\ 
251890.90100    &               CH$_3$OH\,$5_{3,3}-5_{2,4}+-$   &       &  260518.12200   &                                  SiO,6-5  &         \\ 
251895.72800    &               CH$_3$OH\,$6_{3,4}-6_{2,5}+-$   & p     &  260664.76700   &     CH$_3$CH$_2$CN\,$29_{5,26}-28_{5,25}$ & b       \\ 
251900.49500    &               CH$_3$OH\,$4_{3,2}-4_{2,3}+-$   & p     &  260664.77000   &     CH$_3$CH$_2$CN\,$29_{4,26}-28_{4,25}$ & b       \\ 
251905.81200    &               CH$_3$OH\,$3_{3,0}-3_{2,1}-+$   & p     &  260727.86540   &    CH$_3$OCH$_3$\,$18_{5,14}-18_{4,15}$EE &         \\ 
251917.04200    &               CH$_3$OH\,$3_{3,1}-3_{2,2}+-$   & p     &  260758.38210   &        CH$_3$OCH$_3$\,$6_{3,3}-5_{2,4}$EE &         \\ 
251923.63100    &               CH$_3$OH\,$7_{3,5}-7_{2,6}+-$   & p     &  260796.74435   &                                  U260797  &         \\ 
251984.70200    &               CH$_3$OH\,$8_{3,6}-8_{2,7}+-$   & p     &  260864.87361   &                                  U260865  &         \\ 
252090.38000    &               CH$_3$OH\,$9_{3,7}-9_{2,8}+-$   &       &  260991.80790   &                         OC$^{34}$S\,22-21 &         \\ 
252252.85000    &             CH$_3$OH\,$10_{3,8}-10_{2,9}+-$   &       &  261147.89810   &   CH$_3$OCH$_3$\,$17_{5,13}-17_{4,14}$EE  & b       \\ 
252477.92270    &                                    U252478    &       &  261148.90400   &        HCOOCH$_3$\,$21_{5,17}-20_{5,16}$E & b       \\ 
252485.64900    &            CH$_3$OH\,$11_{3,9}-11_{2,10}+-$   &       &  261165.45600   &        HCOOCH$_3$\,$21_{5,17}-20_{5,16}$A &         \\ 
252557.33758    &                                    U252557    &       &  261263.31010   &                           HN$^{13}$C\,3-2 &         \\ 
\hline
\end{tabular*}
\tablenotetext{a}{{\bf p} = usually partially blended with neigbouring
  lines, depending on the line width in each object; {\bf b} = totally blended (undistinguishable).}
\end{minipage}
\end{table*}

\subsection{Detected lines}
\label{lines}

All the detected lines are listed in Table~\ref{tab1} which includes
46 entries. We confidently identified eight molecular species in our EGOs: 
H$^{13}$CO$^+$ (70 objects), 
SiO (37 objects), 
SO (43 objects), 
CH$_3$OH (38 objects),
CH$_3$OCH$_3$ (8 objects), 
CH$_3$CH$_2$CN (5 objects), 
HCOOCH$_3$ (6 objects), and 
HN$^{13}$C (1 objects), and
possibly three additional species: 
H$_2$CCO (9 objects), 
c-HCCCH (7 objects) and
OC$^{34}$S (1 objects). 
There are also five weak unidentified lines (U-lines).
All the lines in Table~\ref{tab1} are grouped for the two sidebands and sorted in increasing frequency order, 
for the ease of comparison with the spectral plots presented in Fig.~\ref{figapp1}.

\subsection{Line parameters and spectral plots}
\label{paraplot}

The detailed line parameters of each detected lines are presented in
Table~A\ref{tabapp1}. The columns are 
(1) object name, 
(2) transition, 
(3-4) observed frequency $\nu$ and uncertainty $\sigma_\nu$ (MHz), 
(5-6) full width at half maximum (FWHM) of the line and uncertainty $\sigma_{\rm FWHM}$ (km\,s$^{-1}$), 
(7-8) mainbeam temperature $T_{\rm MB}$ and uncertainty (mK), 
(9-10) line area $I_{\rm int}$ and uncertainty $\sigma_I$ (K\,km/s), and 
(11) signal to noise ratio (${\rm SNR} =I_{\rm int}/\sigma_I$). 
Most of the line parameters determined by fitting a single Gaussian profile using GILDAS/CLASS
software, while the integrated line area ($I_{\rm int}$) is usually obtained by directly integrating the observed line
profiles. However, there are several special cases: 
1) Some lines that show obvious broad line wings are fit with
two-Gaussian profiles forced to the same line center position. The two
overlapped components are listed as separate records in
Table~A\ref{tabapp1}; 
2) Blended lines, such as the methanol lines in the LSB, 
are fit with multiple Gaussian profiles at fixed catalog frequencies.
The integrated line areas ($I_{\rm int}$) of these lines are computed from the fitting; 
3) Some blended lines that are not 
differentiable ($\Delta\nu < 1$\,MHz) are fit with a
single Gaussian profile and the same line parameters are assigned to both
blended lines. Baseline RMS is used to estimate upper limit for undetected lines. 

The spectral plots of both sidebands are presented in
Fig.~A\ref{figapp1} source by source. The molecular species is marked for each detected line. The
identity of each transition in the plots can be recognized by referring
to the line list in Table~\ref{tab1} in which lines are sorted in
increasing frequency order for the LSB and USB separately.

\subsection{Individual species}
\label{individual}

The detected species can be divided
into two groups: surely detected species -- those well known abundant
species or those with characteristic multiple lines in our frequency
ranges; tentatively detected species -- those with only one
transitions detectable in our frequency ranges and unidentified lines.

\subsubsection{Surely detected species}
\underline{\bf H$^{13}$CO$^+$\,3-2 (260255\,MHz)} is detected
       in 79 per cent (70 out of 89) of the EGOs. The main beam
       temperature ranges from 67\,mK (3.12\,Jy) to 3.084\,K
       (143.4\,Jy). Five EGOs show broad H$^{13}$CO$^+$\,3-2 line
       wings that could be
       caused by high velocity outflows. We fit these broad lines with double
       Gaussian profiles that are forced to the same line center and
       the second component (usually the broader one) is marked by `s' in the last column of
       Table~A\ref{tabapp1}. The
       widths of the broad component range from 5.5-31.6\,km\,s$^{-1}$. The widths of the narrow component of these EGOs
       and that of the other narrow-line EGOs, range from
       1.5-5.7\,km\,s$^{-1}$, with equal mean and median width of 3.4\,km\,s$^{-1}$.

\underline{\bf SiO\,6-5 (260518\,MHz)} is detected in about 42 per cent
      (37 out of 89) of our observed EGO samples. Excluding those H$^{13}$CO$^+$\,3-2 non-detections 
      (due to limited sensitivity), a higher detection rate of $\sim 53$ per cent among 
      the 70 H$^{13}$CO$^+$\,3-2 detected EGOs is found and it may be closer to the true occurrence frequency 
      of outflows among EGOs. The line 
      strength ranges from 40-448\,mK. Most of them have broad
      line widths (FWHM$>$5\,km\,s$^{-1}$), supporting that the SiO
      line might be emitted from outflow regions. For 11 EGOs that show obvious
      line wings, we fit the SiO
      line profiles with a double Gaussian profile. The obtained line
      width (FWHM) of the broad components has a median value of
      $\sim 37$\,km\,s$^{-1}$, indicating the existence of fast
      outflows. The FWHM of the narrow component ranges from
      2.1-9.2\,km\,s$^{-1}$. 
      The diversity of the SiO line profiles will be discussed
      in detail in a separate paper.

\underline{\bf CH$_3$OH line series (251517-252486\,MHz)} 
      are detected in 43 per cent (38 out of 89) of our EGOs.
      They appear only in the lower
      sideband, all belonging to the A-type J$_3$-J$_2$
      inter-K ladder transitions. They are composed of
      two series corresponding to the $+$ and $-$ parities of the
      upper levels,
      respectively. Several lines are partially blended in the center of our LSB. 
      The fitted line strength ranges from 25-489\,mK. The fitted FWHM
      ranges from 1.8-25.2\,km\,s$^{-1}$, with a mean of
      5.7\,km\,s$^{-1}$. Three EGOs, G10.34-0.14, G12.91-0.26 and G23.01-0.41, show clear evidence of broad line
      wings. With a double Gaussian profile fitting, the FWHM of the broad
      components are found to range from 13.6-23.52\,km\,s$^{-1}$, possibly
      indicating fast outflows.

\underline{\bf SO\,$^3\Sigma$,$6_{5}$-$5_{4}$ (251826\,MHz)} is detected
      in 43 EGOs, with a detection rate of $\sim 48$ per cent among all 89 EGOs
      or $\sim 61$ per cent among the 70 H$^{13}$CO$^+$
      emitters. The line strength ranges from 43-2097\,mK. Eight EGOs show broad SO line wings. Double
      Gaussian profiles are fit to them. The width
      of the broad SO line component of
      the eight EGOs ranges from 11.0-21.1\,km\,s$^{-1}$ which possibly
      originate from fast outflows. The central narrow component of the
      eight EGOs and the other 35 detected objects have their FWHM line width
      in the 2.2-14.2\,km\,s$^{-1}$ range, with a mean of
      5.8\,km\,s$^{-1}$. 

\underline{\bf CH$_3$OCH$_3$ line series (251583-251971,
        26} \underline{\bf 1329-261148\,MHz)} All strong transitions of the EE symmetry
      species of dimethyl
      ether with its lower energy level up to 300\,cm$^{-1}$ are definitely detected at
      least in three EGOs (all associated with well known H\,II regions):
      \object[EGO G012.91-0.26]{G12.91$-$0.26}, 
      \object[EGO G14.33-0.64]{G14.33$-$0.64}, and
      \object[EGO G34.41+0.24]{G34.41$+$0.24}. We do not detect the lines of the other three
      symmetry species: AA, AE, and EA, except the case of the line at
      251583.62730\,MHz that could be a mixture of the AA, AE, EA, and EE
      transitions. However, because no other lines of
      the AA, AE, and EA states are seen, we assign this line for the EE
      species only (see in Table~\ref{tab1}). Weaker CH$_3$OCH$_3$\,EE
      lines are possibly also 
      detected in another six objects. The line strength ranges from
      48-203\,mK. Their FWHM ranges from
      1.9-10.6\,km\,s$^{-1}$, with a mean width of 6.9\,km\,s$^{-1}$.

\underline{\bf HCOOCH$_3$ line series (260244-261166\,M} \underline{\bf Hz)} Similarly, many lines of methyl
      formate fall in the USB of our observations. Several strongest lines
      whose lower energy levels are lower than $\sim 120$\,K are clearly
      detected in the USB at least in two EGOs (both associated with well known objects):
      \object[EGO G012.91-0.26]{G12.91-0.26}, and
      \object[EGO G34.41+0.24]{G34.41+0.24}, and possibly in another
      three EGOs: 
      \object[EGO G35.20-0.74]{G35.20-0.74},
      \object[EGO G12.68-0.18]{G12.68-0.18}, and
      \object[EGO G23.01-0.41]{G23.01-0.41}. For the case of another object,
      \object[EGO G24.33+0.14]{G24.33+0.14}, we take it
      as non-detection, due to
      the blending of the lines with that of CH$_3$OCH$_3$. The detection
      of a single line, HCOOCH$_3$ $30_{3,28}-30_{2,29}$, in
      \object[EGO G14.33-0.64]{G14.33$-$0.64} is also doubtful, unless it is a maser
      line, because the other transitions that are expected to be equally strong are not
      detected. The line strength ranges from 55-186\,mK, the FWHM 
      from 1.9-12.3\,km\,s$^{-1}$, with a mean width
      of 6.9 \,km\,s$^{-1}$. 

\underline{\bf CH$_3$CH$_2$CN line series (251503-251669, 2} \underline{\bf 60221-260665\,MHz)} Many lines of ethyl cyanide are
      covered by our observed frequency ranges. Several strongest
      transitions are clearly detected at least in two EGOs,
      \object[EGO G012.91-0.26]{G12.91-0.26} and
      \object[EGO G34.41+0.24]{G34.41+0.24}, and possibly in another
      three EGOs,
      \object[EGO G12.68-0.18]{G12.68-0.18},
      \object[EGO G23.01-0.41]{G23.01-0.41}, and
      \object[EGO G24.33+0.14]{G24.33+0.14}. The line strengths are in
      the 44-243\,mK range. Their FWHM ranges from
      1.5-9.3\,km\,s$^{-1}$, with a mean width of 6.1\,km\,s$^{-1}$. 

\underline{\bf HN$^{13}$C\,3-2 (261263\,MHz)} is surely detected by
      chance in only one source,
      \object[EGO G24.33+0.14]{G24.33+0.14}. This object has a very large
      $V_{\rm sys}\approx 114\,$km\,s$^{-1}$, so that the HN$^{13}$C\,3-2 line comes
      into our frequency coverage by large Doppler shift when the
      object was initially
      observed by assuming zero source velocity. The frequency of this line is
      not covered by any other of our observations. The
      strength of this line is stronger than one half of the
      H$^{13}$CO$^+$\,6-5 line and thus is in agreement with an
      abundant species like HN$^{13}$C.

\subsubsection{Tentatively detected species}

\underline{\bf H$_2$CCO (260192\,MHz)} Only a group of the $13_{\rm Ka,Kc}-12_{\rm Ka,Kc}$ transitions
      of ethenone fall in our frequency range, among which only the
      $13_{1,13}-12_{1,12}$ transition is strong enough to be detectable by our
      observations. This line is clearly detected in 9 of our observed
      EGOs. Further observations of more transitions will hopefully confirm the
      detection of this molecule in these star forming regions.

\underline{\bf c-HCCCH (251527\,MHz)} Cyclopropenylidene is an asymmetric top with ring structure. Some
      transitions fall in our covered frequency ranges. However, only
      the ortho c-HCCCH transition $6_{2,5}$-$5_{1,4}$ is possibly seen in our data. This line is clearly detected in
      seven EGOs: 
      \object[EGO G12.42+0.50]{G12.42+0.50}, 
      \object[EGO G012.91-0.26]{G12.91-0.26},
      \object[EGO G16.59-0.05]{G16.59-0.05},
      \object[EGO G19.88-0.53]{G19.88-0.53},
      \object[EGO G34.41+0.24]{G34.41+0.24},
      \object[EGO G35.20-0.74]{G35.20-0.74}, and
      \object[EGO G49.27-0.34]{G49.27-0.34}. We note that the line looks relative very strong in
      \object[EGO G49.27-0.34]{G49.27-0.34}, say, it reaches about one half of the
      strength of the SO\,$^{3}\Sigma\,5_6-4_5$ line and is stronger than all
      CH$_3$OH lines which are almost not detected in this
      object. We caution that, because only one line is detected for
      this species, other transitions need to be observed to get its identity confirmed.

\underline{\bf OC$^{34}$S\,22-21 (260992\,MHz)} The only transition
      of this species covered in our frequency ranges is possibly detected in
      one target, 
      \object[EGO G23.01-0.41]{G23.01-0.41}. But we note that the line is
      weak and shifted by $\sim 5$\,MHz to lower frequency. The
      identification of this line is quite tentative.

\underline{\bf U-lines} There are five line-like features for which we can not
      find reasonable carriers in either of the line list
      databases. They are all weak lines with $3\sigma$-$4\sigma$
      detection levels in terms of integrated line intensity.
      {\it U260865} is clearly detected as a weak
      narrow line in four objects. Its average observed frequency is
      $260864.89\pm0.99$\,MHz.
      The other U-lines are only detected in a single object and thus their
      detection is even less certain. 
      {\it U260797}  has a frequency close to that of
      OS$^{17}$O\,$4_{4,1}$-$5_{3,2}$, but the non-detection of an expected
      equally strong line 
      OS$^{17}$O\,$4_{4,0}$-$5_{3,3}$ (260808.384\,MHz) in our data
      casts doubt on OS$^{17}$O as its carrier.

\section{Discussions}
\label{discuss}

In this section, we will discuss the detectability of dense gas and outflow 
tracers and demonstrate that most of the observed EGOs are associated with 
outflows and dense gas. We also compare the velocity-integrated spectral line 
luminosities and the line widths between or among our high density tracer lines 
and lower density tracer lines (isotopic CO\,1-0 lines)
and continuum fluxes from literature
to discuss the multiple subcomponent size-scale properties of the EGO clouds.
Further analysis of the complex organic species and outflows will be retained for future papers.

Supplemented literature data used in this work include
\begin{itemize}
\item{
$^{12}$CO, $^{13}$CO, and C$^{18}$O\,1-0 lines. They are from \citet{chen10} for nearly the same sample of EGOs.
Several EGOs have multiple C$^{18}$O\,1-0 line peaks in which case
only the peak closest to our H$^{13}$CO$^+$\,3-2 line is used.}
\item{
{\it Spitzer MIPS} $24\,\micron$ continuum fluxes. They are taken from \citet{cyga08} as tracers of warm dusty cloud cores. 
Note that some EGOs do have $24\,\micron$ counterparts
on their {\it Spitzer MIPS} images but their fluxes were not measured by \citet{cyga08} due to spatial confusion
issues. Thus the available $24\,\micron$ fluxes are not complete and are available only for 53 of our 89 EGO samples 
(including $24\,\micron$ upper limits). }
\item{
BGPS 1.1\,mm continuum fluxes. They are taken from the CSO Bolocam Galactic Plane Survey 
by \citet{roso10} as tracers of dense molecular clumps. We adopt the flux
measurements made with an aperture diameter of $40\arcsec$, 
because it is the closest to our beam size
($29\arcsec$). The cross identification is done by cone search in
VizieR\footnote{VizieR:
  \url{http://vizier.u-strasbg.fr/viz-bin/VizieR}} database within a radius of $40\arcsec$
around the EGO positions. 
The 1.1\,mm counterpart are found for 67 of our 89 EGOs and 
the resulting position discrepancies are $\sim 18\arcsec$ on average.  }
\end{itemize}

\subsection{Dense molecular cloud cores in EGOs}
\label{densecores}

Two dense gas tracers, H$^{13}$CO$^+$\,3-2 and
SO\,$^3\Sigma$ $6_{5}$-$5_{4}$, have been observed in our
survey. Their critical densities are  $\sim 2.11\times
10^6$\,cm$^{-3}$ and $\sim 1.34\times 10^6$\,cm$^{-3}$, respectively, 
which are computed using the molecular transition 
data from the LAMDA database
\footnote{LAMDA database: \url{http://www.strw.leidenuniv.nl/~moldata/}} \citep{scho05}.
The H$^{13}$CO$^+$\,1-0 and SO\,$^3\Sigma$\,$6_{5}$-$5_{4}$ are always the first and second strongest lines 
in our observed frequency range for the EGOs. Although the SO line is
detected only in 43 of the 70 H$^{13}$CO$^+$ detected EGOs, all the remaining 27 SO
non-detections have the weakest H$^{13}$CO$^+$\,3-2 line strengths and also the lowest 
line luminosities (see Sect.~\ref{lineluminosity}), which 
indicates that the SO
non-detections are probably due to our limited sensitivity. Thus, we conclude that all the 70
EGOs with H$^{13}$CO$^+$ detection have dense cloud cores (with density over $10^6$\,cm$^{-3}$) detected in our
$29\arcsec$ beam. 

\begin{figure}
\centering
\epsscale{0.8}
\plotone{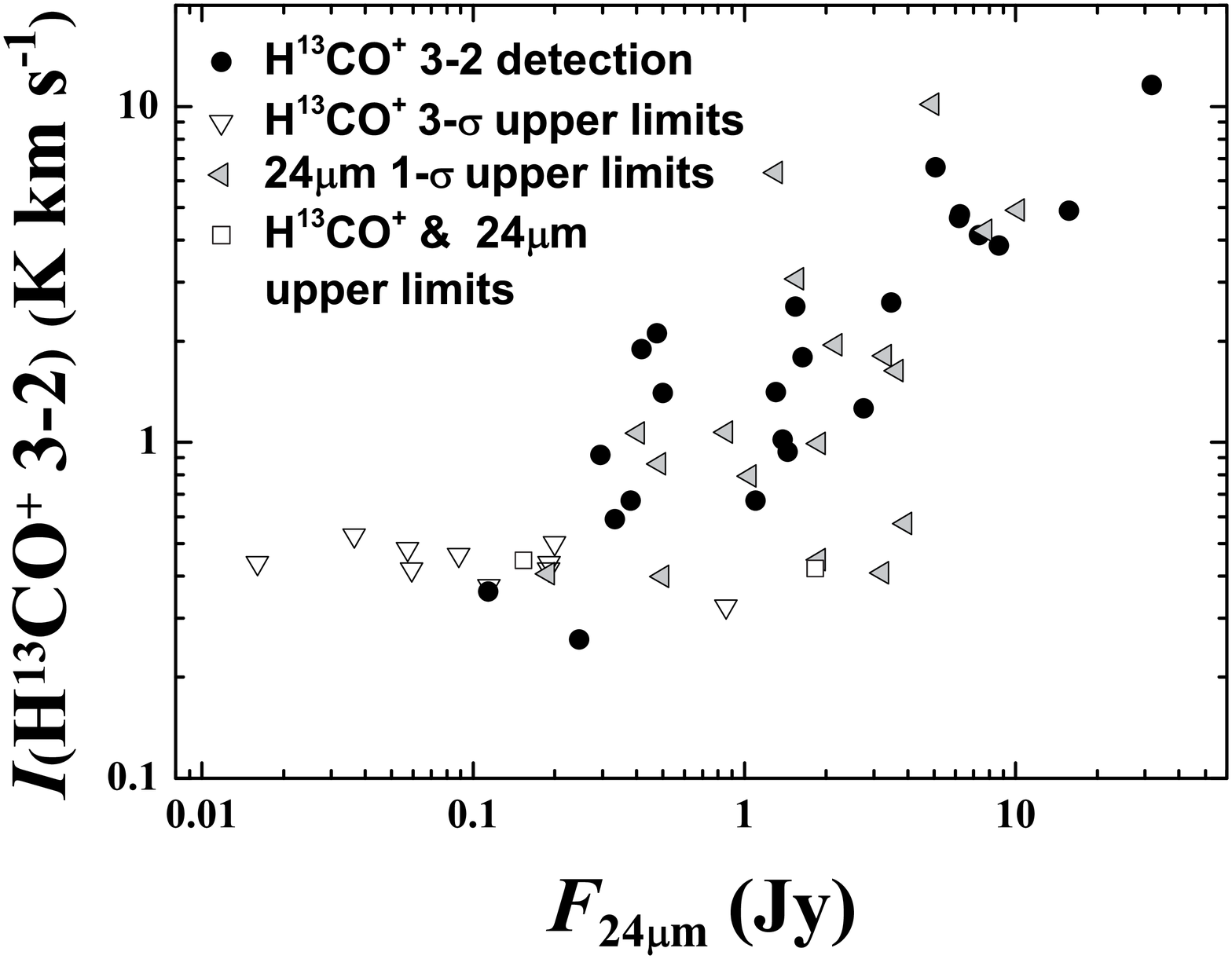}
\plotone{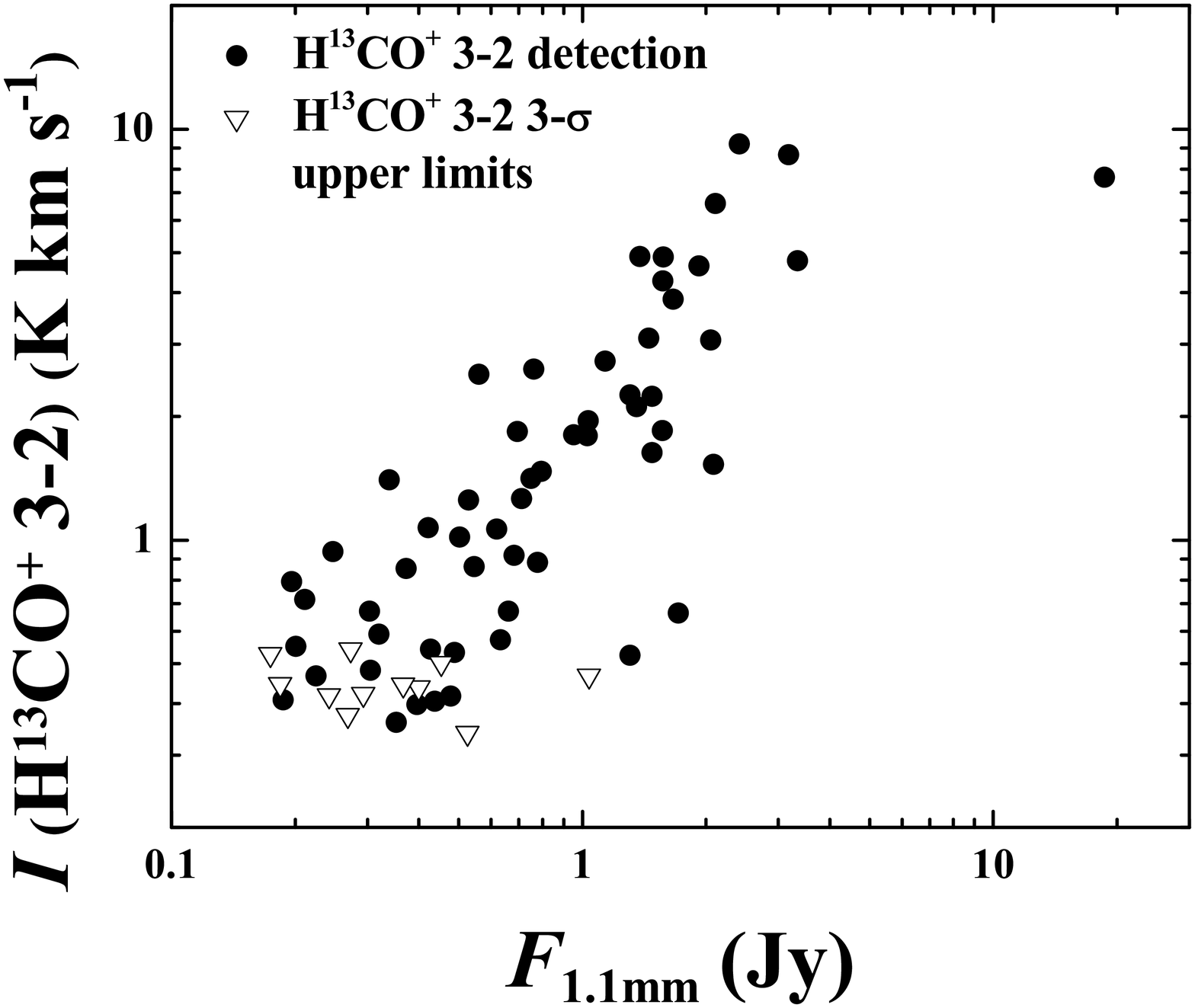}
\caption{The comparison of the H$^{13}$CO$^+$\,3-2 line strength with
  literature 1.1\,mm and 24\,$\mu$m continuum flux densities. See the analysis 
  in Sect.~\ref{densecores}.}
\label{figcont}
\end{figure}

Integrated H$^{13}$CO$^+$\,3-2 line intensities 
(or 3-$\sigma$ upper limits) are compared with available 
$24\,\micron$ and 1.1\,mm continuum flux densities in Fig.~\ref{figcont}.
Clear linear correlations are found in logarithmic scales in both panels, confirming that the H$^{13}$CO$^+$\,3-2 line 
is a good tracer of both cool and warm dense molecular gas 
(similar correlations will be discussed in detail in later sections).
Eighteen of the 19 H$^{13}$CO$^+$\,3-2 non-detection EGOs have counterparts in
the $24\,\micron$ or 1.1\,mm or both data, but their fluxes are all among the smallest, 
which indicates that these H$^{13}$CO$^+$\,3-2 non-detection EGOs are also associated with 
dense and/or warm molecular cloud cores while the non-detection of 
the line could be due to our limited sensitivity.

Thus, we conclude that most of the 89 observed EGOs are associated with dense and/or warm molecular 
cloud cores.

\subsection{Outflows in the EGOs}
\label{outflow}

\begin{figure}[ht]
\centering
%\epsscale{0.8}
\plotone{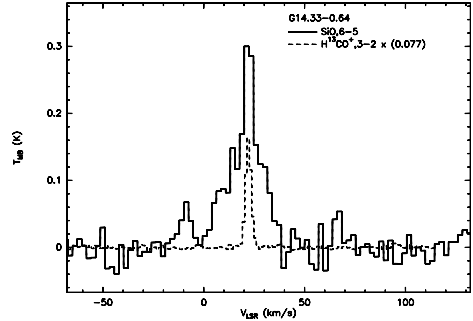}
\caption{Example of the broad line wings in the SiO\,6-5 profile (solid line) of G\,14.33-0.64. The systemic 
velocity of the object is indicated by the peak of the narrower H$^{13}$CO$^+$\,3-2 line (dashed line).}
\label{sioprofile}
\end{figure}

Our major outflow tracer SiO is a refractory 
species and its J=6-5 line has a 
high critical density of $6.22\times 10^6$\,cm$^{-3}$. It should be produced
through disruption of dust grains in active shock regions and excited 
in the post-shock region of fast outflows \citep{hart80,neuf89a,schi97,gusd08}. 
Thus, its high detection rate of 53 per cent among our H$^{13}$CO$^+$\,3-2 detected EGOs
supports the idea of \citet{cyga08} that the extended 
green ($4.5\,\micron$ excess) emission structures are produced by active outflow shocks. 
This is also in agreement to the broad line wings found in many of our SiO\,6-5 lines, as exampled 
in Fig.~\ref{sioprofile}. 

\subsection{Kinematic distance}
\label{distance}

Kinematic distances are computed from the source velocity for 80 EGOs 
\citep[70 H$^{13}$CO$^+$\,3-2 detections and another 10 C$^{18}$O\,1-0 detections from ][]{chen10} 
with the updated Galactic rotation
model of \citet{reid09}. The error in the distance only reflects 
the uncertainty of the model itself.
The velocities of 12 EGOs are new compared to Chen's
results, either because of multi-peak C$^{18}$O\,1-0 profile confusion or
non-detection of HCO$^+$\,1-0 line in their data (they did not observe C$^{18}$O
if this line was not detected). They are marked as `new' in the `NoteV' column of
Table~\ref{tab2}. Another 10 EGOs with
their H$^{13}$CO$^+$\,3-2 lines not detected by us but with 
their C$^{18}$O\,1-0 lines detected by \citet{chen10} are marked
as `chen10' in Table~\ref{tab2}. Ten of our EGOs whose H$^{13}$CO$^+$\,3-2 line is weaker than
$5\sigma$ in terms of velocity integrated line strength are marked as `weak' in the table,
because their $V_{\rm LSR}$ values are not as reliable as the others. 

Nine EGOs do not have near/far kinematic distance ambiguity: 
two located outside of the solar circle (marked as
`Outer' in the `NoteD' column in Table~\ref{tab2}) and seven 
located at the tangential points of the Galactic rotation orbit
(marked as `Tan' in the table). For the remaining objects, we
apply following criteria to differentiate the near/far
kinematic distances: 
\begin{itemize}
\item If an EGO is in front of an IRDC in the Spitzer maps \citep{cyga08}
and its near and far kinematic distances are different
by more than 4\,kpc, we adopt the near kinematic distance, because the
chance to see IRDCs far beyond the tangential point is low, due to the lack of
infrared background (marked as `IRDC' in Table~\ref{tab2}); 
\item If the perpendicular distance of an object above the Galactic plane 
is smaller than the scale hight 
(60\,pc) of the Galaxy at near kinematic distance 
but larger than or roughly equal to two times of the scale hight at the far distance, according 
to \citet{solo87}, the near kinematic distance is preferred (marked as `Z1/2' in Table~\ref{tab2});
\item If an EGO is associated with a $^{13}$CO\,1-0 molecular cloud (within $2\arcmin$ radius) 
whose kinematic distance ambiguity has been resolved with the H\,I self absorption method or 
21\,cm continuum absorption line method by \citet{roma09},
their choice of near/far distance is adopted (marked as `HISA' and `21cm', respectively, in Table~\ref{tab2});
\item If an EGO is associated with an H\,II region 
whose kinematic distance ambiguity has been resolved with the H\,I
absorption line method for H\,II regions by \citet{kolp03},
their choice of near/far distance is adopted (they are marked as `HII' in Table~\ref{tab2});
\item If the near/far kinematic distances differ by less
than 2\,kpc, we adopt the average distance and
treat the difference between the near/far distances as a part of
uncertainty (marked as `Ave' in Table~\ref{tab2}). 
\end{itemize}
Thus, in the last criterion, the distance uncertainty 
is added with $(D_{\rm far}-D_{\rm near})/2$. 
As a result, unique kinematic distance is determined for 58 EGOs in Table~\ref{tab2},
while both near and far distances are given for the rest 22 EGOs 
for which the kinematic distance ambiguity can not be lifted.

\subsection{The line luminosity correlations}
\label{lineluminosity}

EGOs should have a centrally enhanced 
gas density and temperature gradient as typical star forming clouds, 
because the extended green structures ($4.5\,\micron$ excess) 
and $24\,\micron$ point sources in the Spitzer infrared images of some or all of them
already indicate the existence of active star formation activities. 
Thus, we can expect that our four dense gas tracers (H$^{13}$CO$^+$\,3-2,
SO\,$^{3}\Sigma\,5_6$-$4_5$, SiO\,6-5 and the CH$_3$OH\,J$_3$-J$_2$\,A line series)
trace relatively smaller cloud substructure size-scales, while the 
lower density tracer lines, $^{12}$CO, $^{13}$CO and C$^{18}$O\,1-0, 
trace relatively larger substructure size-scales of the EGO clouds.
The three isotopic CO\,1-0 lines themselves also trace slightly different 
cloud densities and substructure size-scales due to their different abundances.
Thus, comparison of their line luminosities among different objects would 
shed light on how the various cloud size scales are physically related.

Line or continuum luminosity is defined as 
$$L=4 \pi D^2 F,$$
where $D$ is the distance from Sect.~\ref{distance} and $F$ is the velocity (km\,s$^{-1}$) 
integrated line flux or continuum flux density.
Nominal telescope efficiency of 46.5\,Jy/K is used for the ARO SMT (our survey) 
and 42.3\,Jy/K for the PMO 13.7-m \cite[observation of ][]{chen10} telescopes.
For CH$_3$OH\,J$_3$-J$_2$ line series, the average line strength of all the detected 
transitions is used. Note that the continuum 
luminosities are not bolometric but monochromatic luminosities.

\subsubsection{The correlations}
\label{luminosity-correlation}

\begin{figure*}
\centering
\epsscale{2.2}
\plotone{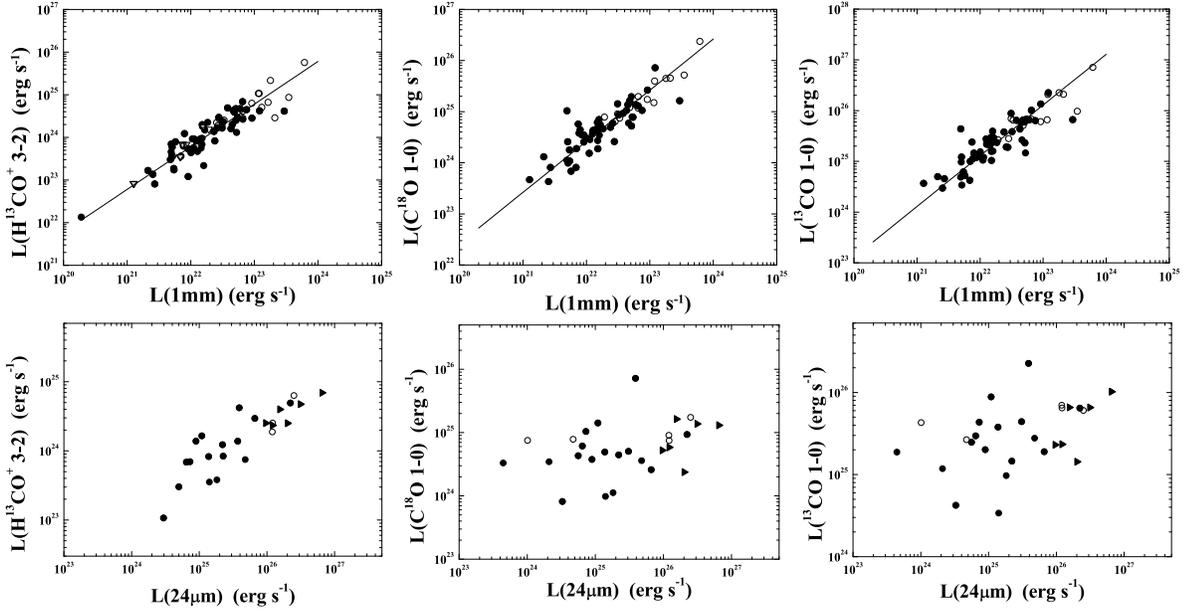}
\caption{The comparison of the luminosities of the three optically thin lines, H$^{13}$CO$^+$\,3-2, C$^{18}$O\,1-0
  and $^{13}$CO\,1-0 of EGOs with their corresponding BGPS 1.1\,mm and {\it Spitzer IRAC} $24\,\micron$ 
  continuum monochromatic luminosities.
  Upper limits of H$^{13}$CO$^+$\,3-2 and lower limits of $24\,\micron$ luminosities are shown as triangles. 
  For several EGOs whose near and far kinematic distance 
  ambiguity is not resolved, their far kinematic distance quantities are shown in open symbols, while their 
  near distance quantities are shown in filled symbols (the same as those with unique distance). 
  Only near distance quantities with certain detection of the lines are used in regression.}
\label{line-cont}
\end{figure*}

We first check the correlations of the three possibly optically thin lines, 
H$^{13}$CO$^+$ 3-2, C$^{18}$O and $^{13}$CO\,1-0, with the {\it Spitzer MIPS} 
$24\,\micron$ and BGPS 1.1\,mm continuum luminosities, as shown in Fig.~\ref{line-cont}. 
The very good correlations in the top panels demonstrates that all three lines are good tracers of dense and dusty molecular clumps.
A straight line with fixed slope of unity is fit 
to each plot in log-log scale as (the solid lines in the figure, far distance quantities shown in open circles and upper limits shown 
in triangles are not used and hence forth)
\begin{eqnarray}\nonumber
\label{eq-1mm-h13co}
\lg L_{{\rm H}^{13}{\rm CO}^+}&=&\lg L_{1.1\,{\rm mm}}+1.78(0.03),\\
                              & &R=0.79,\,\sigma=0.24\,;
\end{eqnarray}
\begin{eqnarray}\nonumber
\label{eq-1mm-c18o}
\lg L_{{\rm C}^{18}{\rm O}}&=&\lg L_{1.1\,{\rm mm}}+2.42(0.03),\\
                           & &R=0.68,\,\sigma=0.26\,;
\end{eqnarray}
\begin{eqnarray}\nonumber
\label{eq-1mm-13co}
\lg L_{^{13}{\rm CO}}&=&\lg L_{1.1\,{\rm mm}}+3.11(0.03),\\
                     & &R=0.66,\,\sigma=0.26\,,
\end{eqnarray}
where $R$ is the correlation efficient and $\sigma$ is the sample standard deviation in all fitted formulas in logarithmic space hereafter. 

Distinctively, only the H$^{13}$CO$^+$\,3-2 shows a clear correlation with the
$24\,\micron$ luminosities (but worse than with the 1.1\,mm) in the bottom panels of the figure, 
which indicates that only the H$^{13}$CO$^+$\,3-2 line is a tracer of warm molecular 
cloud cores. This is not unexpected because H$^{13}$CO$^+$ 3-2 has three orders of magnitude higher critical density than the other two lines.
However, we stress that the $24\,\micron$ data are incomplete due to spatial 
confusion issues on the {\it Spitzer MIPS} maps \citep[see in ][]{cyga08}. Furthermore, 
diverse masses and evolution status of the hidden central forming stars that dictate the $24\,\micron$ emission 
could also contribute to the badness of the correlations.

\begin{figure*}
\centering
\epsscale{2.0}
\plotone{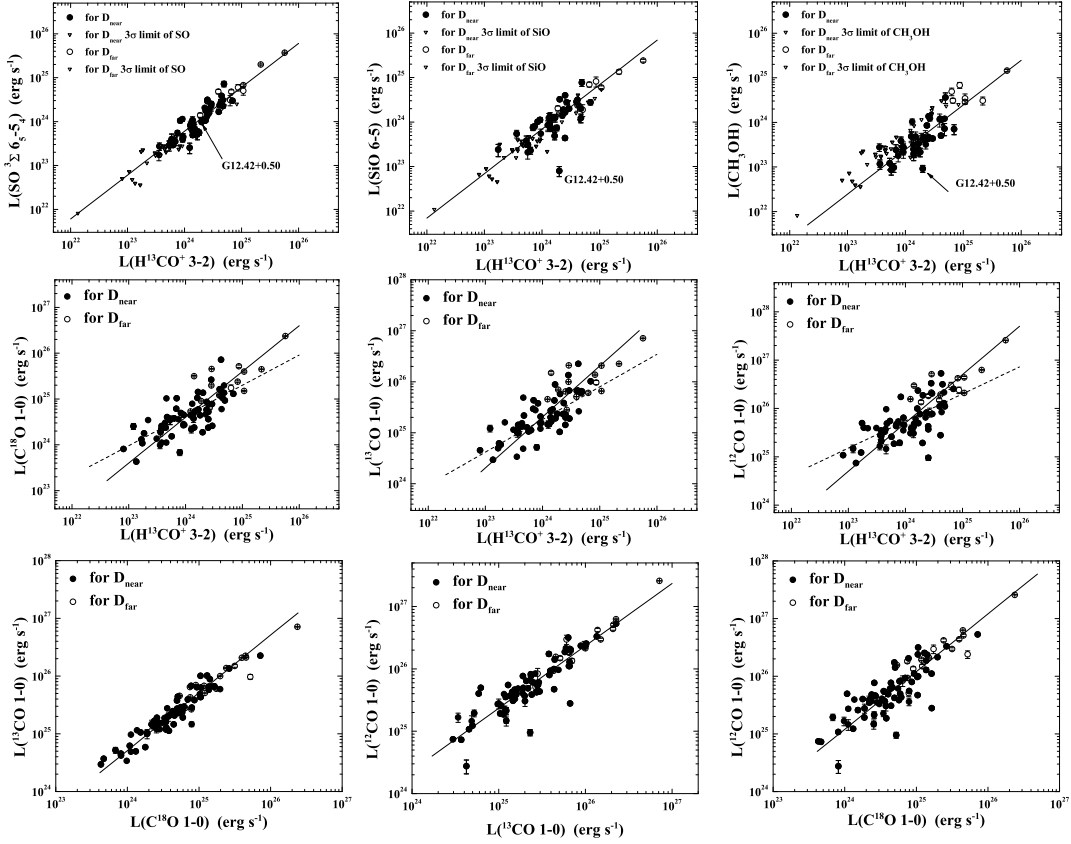}
\caption{The comparison of observed and literature line luminosities. Upper limits are shown in triangles in the top row, 
         and far distance quantities are shown in open symbols in all panels. The marked object, G12.42+0.50, has exceptionally 
         weak SiO and CH$_3$OH lines and thus is excluded from the log-linear fitting.}
\label{comp-line}
\end{figure*}

Then, the correlations among our four dense gas (small size-scale) and shock tracers and 
the three isotopic CO\,1-0 lines (large size-scale tracers) are shown in 
Fig.~\ref{comp-line}. Good log-linear correlations are found among all lines.
A straight line with fixed slope of unity 
is fit to each plot in log-log scale as (the solid lines in the figure)
\begin{eqnarray}\nonumber
\label{eq-h13co-so}
\lg L_{\rm SO}&=&\lg L_{{\rm H}^{13}{\rm CO}^+}-0.22(0.03),\\
              & &R=0.78,\,\sigma=0.18\,;
\end{eqnarray}
\begin{eqnarray}\nonumber
\label{eq-h13co-sio}
\lg L_{\rm SiO}&=&\lg L_{{\rm H}^{13}{\rm CO}^+}-0.16(0.04),\\
               & &R=0.61,\,\sigma=0.23\,;
\end{eqnarray}
\begin{eqnarray}\nonumber
\label{eq-h13co-ch3oh}
\lg L_{{\rm CH}_3{\rm OH}}&=&\lg L_{{\rm H}^{13}{\rm CO}^+}-0.61(0.04),\\
                          & &R=0.57,\,\sigma=0.23\,;
\end{eqnarray}
\begin{eqnarray}\nonumber
\label{eq-h13co-c18o}
\lg L_{{\rm C}^{18}{\rm O}}&=&\lg L_{{\rm H}^{13}{\rm CO}^+}+0.60(0.04),\\
                           & &R=0.38,\,\sigma=0.33\,;
\end{eqnarray}
\begin{eqnarray}\nonumber
\label{eq-h13co-13co}
\lg L_{^{13}{\rm CO}}&=&\lg L_{{\rm H}^{13}{\rm CO}^+}+1.31(0.04),\\
                     & &R=0.39,\,\sigma=0.31\,;
\end{eqnarray}
\begin{eqnarray}\nonumber
\label{eq-h13co-12co}
\lg L_{^{12}{\rm CO}}&=&\lg L_{{\rm H}^{13}{\rm CO}^+}+1.70(0.05),\\
                     & &R=0.16,\,\sigma=0.36\,;
\end{eqnarray}
\begin{eqnarray}\nonumber
\label{eq-c18o-13co}
\lg L_{^{13}{\rm CO}}&=&\lg L_{{\rm C}^{18}{\rm O}}+0.71(0.02),\\
                     & &R=0.87,\,\sigma=0.14\,;
\end{eqnarray}
\begin{eqnarray}\nonumber
\label{eq-13co-12co}
\lg L_{^{12}{\rm CO}}&=&\lg L_{^{13}{\rm CO}}+0.37(0.02),\\
                     & &R=0.75,\,\sigma=0.21\,;
\end{eqnarray}
\begin{eqnarray}\nonumber
\label{eq-c18o-12co}
\lg L_{^{12}{\rm CO}}&=&\lg L_{{\rm C}^{18}{\rm O}}+1.08(0.03),\\
                     & &R=0.61,\,\sigma=0.26\,.
\end{eqnarray}

The ubiquitous correlations (Eqs.~\ref{eq-h13co-so}-\ref{eq-c18o-12co}) are unexpected,
because the various lines trace different physical subcomponents of the EGO clouds due to different critical densities and/or chemistry: H$^{13}$CO$^+$\,3-2 
traces dense and warm cloud cores, SiO\,6-5 traces active outflow shocks, SO$^3\Sigma\,6_5-5_4$ \citep{wake04,char97,vand03} and 
the CH$_3$OH line series \citep{vand03,cyga11} may trace both components to some extent, while the three 
isotopic CO\,1-0 lines trace large scale cloud structures. 
As discussed by \citep{good98}, a molecular line probes a density range of usually no more than 
2 orders of magnitude above its critical density, due to opacity and chemistry 
(and perhaps beam dilution). Thus, our dense gas tracers very possibly probe completely different 
density regimes and size-scales than the three isotopic CO\,1-0 lines due to their very different critical densities. 
Therefore, the good correlations we found have a strong 
indication that not only the emission from the very different cloud subcomponent size-scales are closely 
related, but the strength of the shock-region line emission is also tightly linked to that of the cloud 
emission in most of the observed EGO clouds. 

Particularly, SiO molecules are usually highly depleted in quiescent 
cold clouds \citep{herb89,ziur89} and also deficient in photon dominated regions \citep{walm99}, but can be enriched in 
shocked regions where the silicon element is eroded or sputtered off dust grains by non-dissociative shocks 
\citep{hart80,schi97,gusd08} or fast dissociative shocks \citep{neuf89a,neuf89b}. 
Moreover, dense ambient clouds are needed to obstruct the stellar outflows to produce observable enhancement in SiO abundance. 
Otherwise, the outflows would simply escape away, leaving little SiO molecules behind, as illustrated by
the one-sided SiO outflow in the low mass star forming region \object[LDN 1157]{L1157} \citep{mika92}. 

Therefore, we argue that 
the good correlations between the shock emission and all cloud emission, as found in our data, can be explained if {\it most of the EGOs are embedded 
in molecular clouds that have very similar density and thermal structures and the stellar outflows are mainly 
obstructed by the peripheral parts of their own natal clouds} \citep[`cloud shocks' by][]{holl97}.
 
In another word, although the SiO\,6-5 and the H$^{13}$CO$^+$\,3-2 line emission comes from different regions of the EGO clouds, 
their strengths are linked through the very similar density and thermal structures of the clouds and 
similar properties of shocks, such as shock velocity and density jump across the shock front.

The worst correlations between H$^{13}$CO$^+$\,3-2 and isotopic CO\,1-0 line luminosities can be improved a little 
by loosing the slope of the log-linear fit and thus obtain (the dashed lines in the plots)
\begin{eqnarray}\nonumber
\label{eq-h13co-c18o2}
\lg L_{{\rm C}^{18}{\rm O}}&=&0.66(0.08)\lg L_{{\rm H}^{13}{\rm CO}^+}+8.8(2.0),\\
                           & &R=0.52,\,\sigma=0.29\,;
\end{eqnarray}
\begin{eqnarray}\nonumber
\label{eq-h13co-13co2}
\lg L_{^{13}{\rm CO}}&=&0.64(0.07)\lg L_{{\rm H}^{13}{\rm CO}^+}+9.9(1.8),\\
                     & &R=0.56,\,\sigma=0.26\,;
\end{eqnarray}
\begin{eqnarray}\nonumber
\label{eq-h13co-12co2}
\lg L_{^{12}{\rm CO}}&=&0.56(0.08)\lg L_{{\rm H}^{13}{\rm CO}^+}+12.3(2.0),\\
                     & &R=0.42,\,\sigma=0.30\,.
\end{eqnarray}
Here the slopes are all smaller than unity, indicating that the H$^{13}$CO$^+$\,3-2 emission 
from the warm cores tends to overshine the isotopic CO\,1-0 line emission from larger size-scales in more 
luminous (and thus more massive) EGO clouds. Note however that 
the smaller slopes can be increased if far kinematic distance
(open circles in the middle row of Fig.~\ref{comp-line}) are adopted for several objects. 

We confirm that the above luminosity correlations are not artifact due to 
variation of distances, because most of our EGOs have similar kinematic distances 
of about 2-6\,kpc.

\subsubsection{The progressive de-correlations}
\label{luminosity-decorrelation}

The quality of the luminosity correlations are diverse. Compared to error bars (statistic error) in Fig.~\ref{comp-line}, the data scatter in the 
luminosity correlation plots can not be explained by observational errors. Even if taking into account the 
nominal 20 per cent of flux calibration error when comparing H$^{13}$CO$^+$\,3-2 luminosity with that 
of the isotopic CO\,1-0 lines (obtained with different telescopes at different time), it is still far from enough to 
explain the large scatter. Thus, the data scatter should reflect some intrinsic randomness in the 
thermal and/or density structures of the EGO clouds.

The sample standard deviations ($\sigma$) of the log-linear fits are gathered and compared in Table~\ref{tab-sigma}.
The correlations are the 
best (with the smallest $\sigma$) between 
$^{13}$CO -- C$^{18}$O and 
SO -- H$^{13}$CO$^+$, progressively worse between 
$^{12}$CO -- $^{13}$CO and 
$^{12}$CO -- C$^{18}$O, and the worst between 
H$^{13}$CO$^+$ and the three isotopic CO\,1-0 lines. 
This is an evidence of a progressive de-correlation 
of line luminosities across larger and larger size-scales of the EGO clouds, 
given that the tracers are progressively tracing larger cloud subcomponent sizes along the 
H$^{13}$CO$^+$ -- SO -- C$^{18}$O -- $^{13}$CO -- $^{12}$CO sequence. It suggests 
that the randomness increases with cloud substructure size-scale.
\begin{table}[th]
\centering
%\begin{minipage}{70mm}
\footnotesize\rm
\caption{\label{tab-sigma}Compare the sample standard deviation ($\sigma$) of the log-linear fits to the luminosity correlations.}
\begin{tabular}{l@{ }l@{ }l@{ }l}
\hline
$X$             & 
$Y$             & 
$\sigma$        & 
Equation         \\
\hline
1.1\,mm         & H$^{13}$CO$^+$ & 0.24   & (\ref{eq-1mm-h13co})      \\
1.1\,mm         & C$^{18}$O      & 0.26   & (\ref{eq-1mm-c18o})       \\
1.1\,mm         & $^{13}$CO      & 0.26   & (\ref{eq-1mm-13co})       \\
H$^{13}$CO$^+$  & SO             & 0.18   & (\ref{eq-h13co-so})       \\
H$^{13}$CO$^+$  & SiO            & 0.23   & (\ref{eq-h13co-sio})      \\
H$^{13}$CO$^+$  & CH$_3$OH       & 0.23   & (\ref{eq-h13co-ch3oh})    \\
H$^{13}$CO$^+$  & C$^{18}$O      & 0.29   & (\ref{eq-h13co-c18o2})    \\
H$^{13}$CO$^+$  & $^{13}$CO      & 0.26   & (\ref{eq-h13co-13co2})    \\
H$^{13}$CO$^+$  & $^{12}$CO      & 0.30   & (\ref{eq-h13co-12co2})    \\
C$^{18}$O       & $^{13}$CO      & 0.14   & (\ref{eq-c18o-13co})      \\
$^{13}$CO       & $^{12}$CO      & 0.21   & (\ref{eq-13co-12co})      \\
C$^{18}$O       & $^{12}$CO      & 0.26   & (\ref{eq-c18o-12co})      \\
\hline
\end{tabular}
%\end{minipage}
\end{table}

In a summary for this subsection, good correlations are found between various 
line and continuum luminosities that trace various subcomponents of the EGO clouds such as low dense components, 
dense warm cores and shock regions, which requires similarity in thermal and density structures 
and perhaps also in shock properties among all EGO clouds to explain. However,  
evidence is also found for progressive de-correlation of line luminosities across larger and 
larger size-scales of the cloud subcomponents.

\subsection{Line width correlations}
\label{line-width}

\subsubsection{Dense gas and shock tracer lines}
\label{line-width-dense}

Aside from line opacity effect, millimeter line widths observed in interstellar clouds usually reflect gas kinematics in the clouds, 
including thermal motion of the observed molecules, turbulence, internal motion of subcomponents of 
the clouds, systematic motion of gas such as infall, outflow and rotation. 

\begin{figure*}[ht]
\centering
\epsscale{2.0}
\plotone{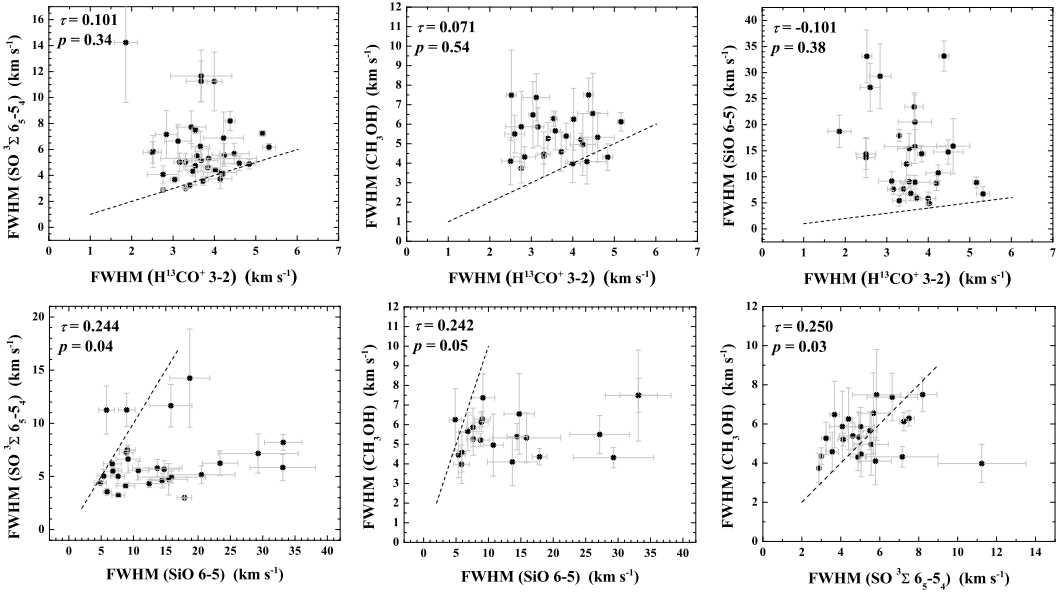}
\caption{The comparison of line widths among the four 
  dense warm gas and shock tracers observed in this survey. Only those line width data with S/N$>3$ are used and henceforth.
  The dashed lines show where the two compared line widths are just equal. The parameters $\tau$ and $p$ in the top left 
  corner of each panel are the Kendall tau coefficient and the probability to reject the null hypothesis (see in the text).}
\label{comp-v1}
\end{figure*}

The line widths (FWHM) of our dense gas and shock tracers are compared in Fig.~\ref{comp-v1}. All the lines 
are significantly broader than their typical thermal line width at representative temperatures, including 
the usually optically thin warm core tracer H$^{13}$CO$^+$\,3-2. Thus, 
non-thermal motion is ubiquitous in the high density part of the EGO clouds. 
Note that in several cases where broad line wings are 
evident and two Gaussian profiles are fit to the broad and narrow components separately, only 
the central narrow component is used. 

The H$^{13}$CO$^+$ lines are always the narrowest in Fig.~\ref{comp-v1}, while none 
of them are strongly correlated with each other. 
The results are reminiscent of what \citet{saka10} found among 20 massive clumps.
To quantitatively illustrate the lack of correlation in these plots, a Kendall rank correlation 
test is performed to each pair of quantities in these plots by ignoring the error bars. 
The null hypothesis is 
`no correlation', which is equivalent to Kendall tau coefficient $\tau=0$. If the probability to reject 
the null hypothesis is $p<0.05$, the null hypothesis is rejected and the opposite 
is accepted. The resulting $\tau$ and $p$ are shown in the top left corner of each panel of the figure.
The large probability $p$ in the top row of the figure clearly demonstrate the lack of correlations.
Although marginal correlations can be concluded from the smaller $p$ values in the bottom row of the figure,
the correlations become questionable when the large error bars in these plots are taken back into account. 

The single dish H$^{13}$CO$^+$\,3-2 line data should not be dominated by outflows due to its high critical density and low optical depth, 
as indicated by the dominant disk like morphology and kinematics in the H$^{13}$CO$^+$\,1-0 line map of an outflow source by \citet{taka06}. 
Unless outflow powered turbulence is functioning in the EGO clouds, gravity induced systematic motions such as infall, cloud 
rotation, and the motion of multiple protostars should be the most probable contributors to the observed H$^{13}$CO$^+$ line widths. 

The lack of correlation with SiO line, which is also true if the broader component 
of the two-Gaussian fit of SiO lines is used, rules out outflow as a major contributor to 
the line width of the SO and CH$_3$OH lines. 
The latter two species could be strongly affected by large errors in the data (the large error bars in the figure)
and opacity effect which has veiled their correlation with SiO line width. 
\citet{reit11} also found that the widths of molecular line tracers such as SO\,$6_7$-$5_6$ 
are significantly affected by line opacity in high mass star forming clumps.

\subsubsection{Low density gas tracers}
\label{line-width-co}

\begin{figure*}[ht]
\centering
\epsscale{2.0}
\plotone{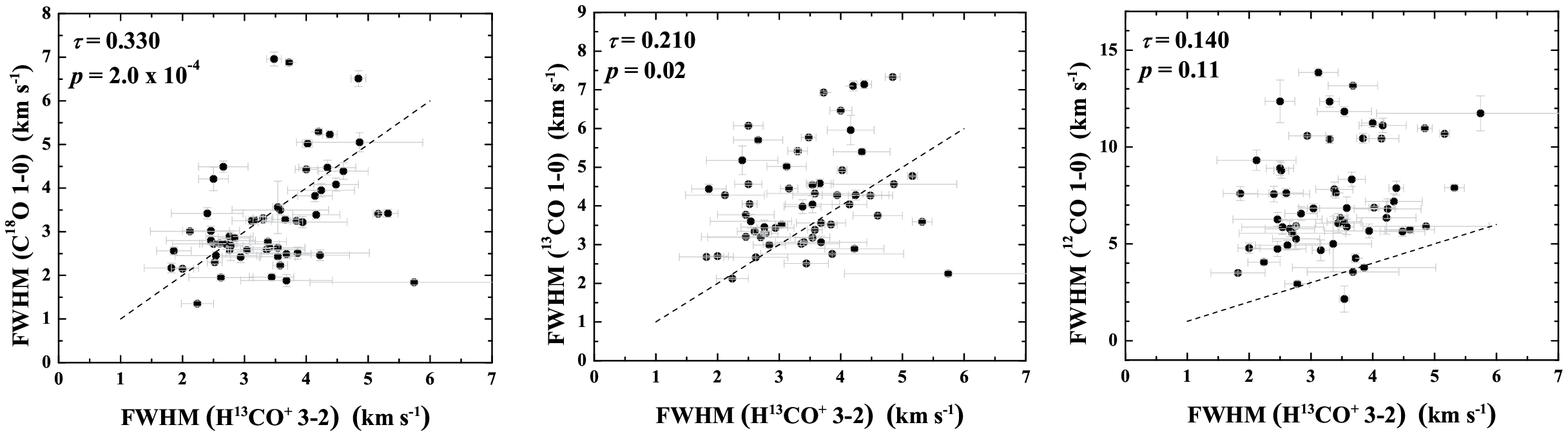}
\caption{The same as Fig.~\ref{comp-v1} but between the dense warm gas tracer line (H$^{13}$CO$^+$\,3-2)
  and the three lower density tracer lines (isotopic CO\,1-0). The parameters $\tau$ and $p$ in the top left 
  corner of each panel are the Kendall tau coefficient and the probability to reject the null hypothesis (see in the text).}
\label{comp-v2}
\end{figure*}

To check if the non-thermal motion in the small scale dense cloud cores 
is linked to that in larger size-scale low density parts of the EGO clouds,  
the H$^{13}$CO$^+$\,3-2 line width are compared with that of the three isotopic CO\,1-0 lines
in Fig.~\ref{comp-v2}.
The C$^{18}$O\,1-0 line widths are comparable to that of the H$^{13}$CO$^+$\,3-2 lines and show a 
rough positive correlation, 
while the $^{13}$CO and $^{12}$CO\,1-0 lines are progressively broader and show no correlation. 
Similarly, a Kendall rank correlation test is performed to each plot in the figure. The resulting 
$p$ values shown in the top left corner of each plot demonstrate a clear correlation in the left panel, 
a marginal correlation in the middle panel and no correlation in the right panel.
It indicates that, although the non-thermal velocity fields are still correlated to some degree across the 
different size-scales traced by the H$^{13}$CO$^+$ and C$^{18}$O lines, the correlation is already very loose 
and progressively worse from C$^{18}$O to $^{13}$CO to $^{12}$CO\,1-0 lines.

\begin{figure*}[ht]
\centering
\epsscale{2.0}
\plotone{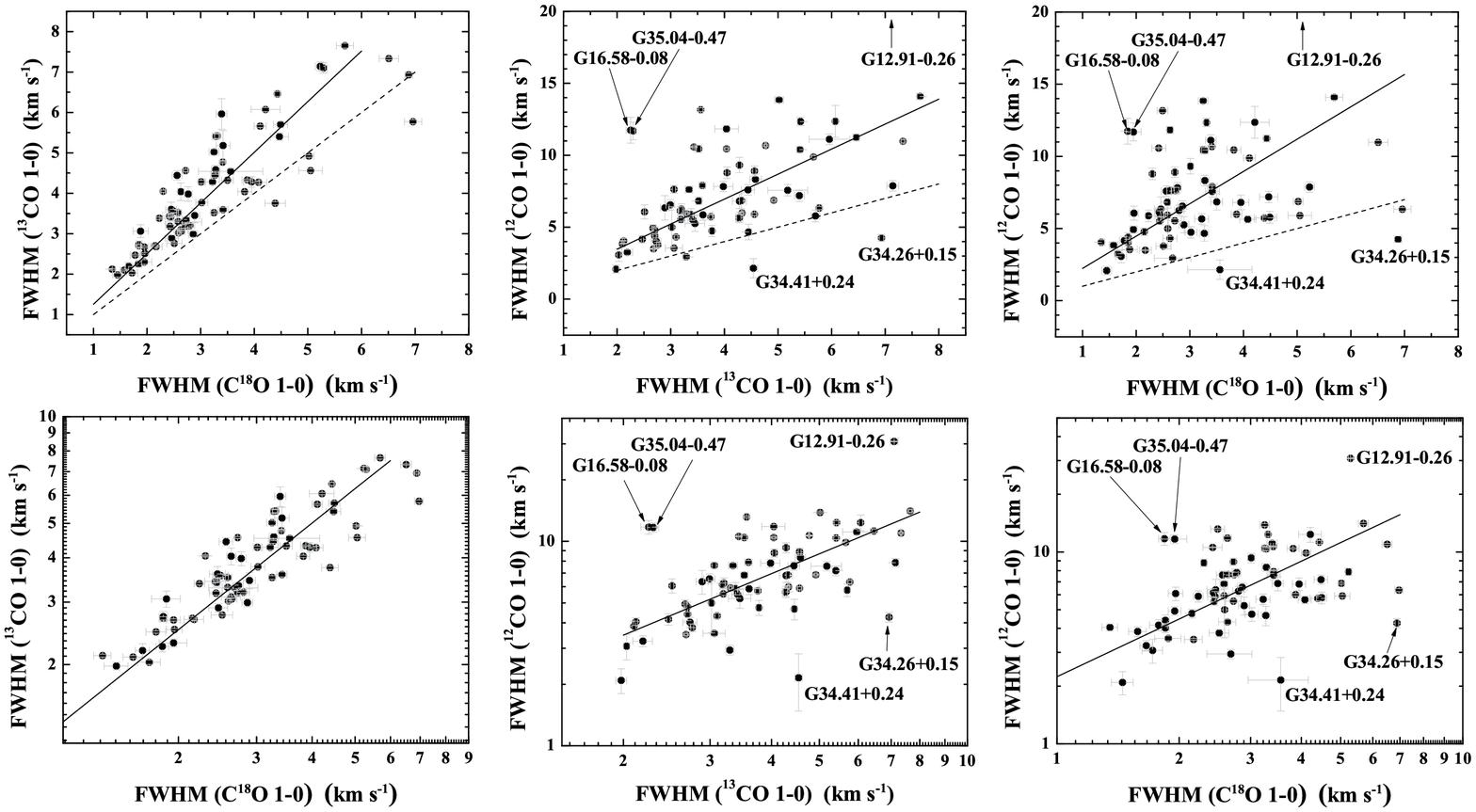}
\caption{The same as Fig.~\ref{comp-v1} but among the three lower density tracer isotopic CO\,1-0 lines in linear (top row)
  and logarithmic (bottom row) scales. The solid lines 
  are linear fits in logarithmic space to the bottom row with a fixed slope of unity. Several objects whose 
  $^{12}$CO\,1-0 line widths are abnormally too broad or too narrow and thus are excluded from the fittings 
  are marked with their object names. The object G12.91-0.26 is outside of the plot limit in the top middle and top right panels.}
\label{comp-v3}
\end{figure*}

However, the line width correlations among the three isotopic CO\,1-0 line themselves are much better, 
as shown in Fig.~\ref{comp-v3} (in linear scales in the top row and logarithmic scales in the bottom row).
The data scatter looks more homogeneous in the log-log plots and a log-linear fit yields (see the solid line)
\begin{eqnarray}\nonumber
\label{eq-fwhm-c18o-13co}
\lg \Delta V_{^{13}{\rm CO}}&=&\lg \Delta V_{{\rm C}^{18}{\rm O}}+0.098(0.008),\\
                              & &R=0.77,\,\sigma=0.07.
\end{eqnarray}
The equations for the other two worse correlations are omitted. Hereafter, we use $\Delta V$ to represent the FWHM of line profiles. 
Eq.~(\ref{eq-fwhm-c18o-13co}) is equivalent to $\Delta V$($^{13}$CO)$\approx 1.25(\pm 0.02)\times \Delta V$(C$^{18}$O). 

The broader $^{13}$CO\,1-0 lines than the C$^{18}$O\,1-0 lines are clearly not caused by higher opacity in the 
former lines, because the opacities in the $^{13}$CO\,1-0 lines are not much larger than unity, 
which can be verified by multiplying a factor of 7.3 to the C$^{18}$O\,1-0 opacities in the Table 7 of \citet{chen10}. 
Instead, it should be understood as the property of interstellar turbulent velocity fields. According to 
\citet{solo87}, turbulence dominated line width scales with cloud size ($l$ in pc) and 
mass ($M$ in M$_\odot$) as 
$\Delta V$ (km\,s$^{-1}$)$= 1.0\times l^{0.5}$ and 
$M = 2000 \times \Delta V^4$ 
in interstellar clouds in virial equilibrium 
and it is also applicable to substructures of a single molecular 
cloud \citep{lars81}. Thus, the above log-linear line width correlation also
suggests that the $^{13}$CO\,1-0 emission regions are universally $1.56(\pm 0.05)$ times larger in size
and $2.4(\pm 0.2)$ times larger in mass than the C$^{18}$O\,1-0 emission regions. This holds 
over a large cloud size range of 2-49\,pc and a large cloud mass range of 
$7.7\times 10^3-4.8\times 10^6$\,M$_\odot$, corresponding to the C$^{18}$O\,1-0 line width range of 1.4-7.0\,km\,s$^{-1}$.

The data scatter in the line width correlations can not be explained by observational errors. 
Instead, they could reflect some randomness in the velocity fields across cloud substructures of different 
size-scales, similar to the randomness in the luminosity correlations discussed in Sect.~\ref{luminosity-decorrelation}. 
The progressively worse line width 
correlations with the H$^{13}$CO$^+$ from C$^{18}$O to $^{13}$CO to $^{12}$CO\,1-0 lines and the similar trends 
among the three isotopic CO\,1-0 lines themselves further hint that the randomness is stronger across larger substructure 
size-scales.

\subsubsection{The inter-cloud randomness in the line width correlations}
\label{line-width-random}

The log-linear nature of the line width correlations among the three isotopic 
CO\,1-0 lines (increasing data scatter toward broader lines in the top row 
of Fig.~\ref{comp-v3}) is surprising, because such behavior is expected usually when the quantities 
vary across several orders of magnitude, while the line widths shown here vary by merely a factor of about five. 
The isotopic CO\,1-0 line widths are dominated by turbulence velocity fields at the whole cloud size-scale, 
therefore the data scatter in these line width correlations can be understood as some kind of randomness in the 
turbulence velocity fields. The increase of data scatter toward broader line widths thus means the increase of the randomness 
among more massive and bigger EGO clouds. Because the trends are manifested by comparing EGO clouds of different 
size-scales (line widths), we call it inter-cloud randomness, in contrast to the randomness revealed by comparing molecular lines 
of very different critical densities.

\begin{figure}
\centering
\epsscale{1.0}
\plotone{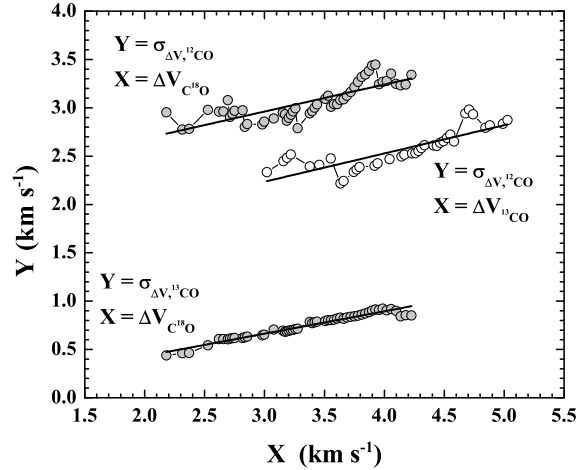}
\caption{The data scatter ($\sigma_{\Delta V}$) of the line width correlations as a function of the FWHM line width ($\Delta V$) itself.
  The thick solid lines are linear fits to the points (see Eqs.~\ref{eq-dvsig-c18o-13co},
  \ref{eq-dvsig-13co-12co} and \ref{eq-dvsig-c18o-12co}).}
\label{dvsig}
\end{figure}

To quantify the trends, 
we compute local sample standard deviation 
(with respect to the fitted log-linear correlation 
with fixed slope of unity) for smaller velocity sub-ranges and investigate 
how it varies with the mean line width. (Fig.~\ref{dvsig}). 
After some tests, the velocity sub-range 
is chosen to be 0.3 dex in the logarithmic scale, which gives the best presentation 
in Fig.~\ref{dvsig} (our conclusions are not sensitive to the choice, however).

The computed local standard deviations ($\sigma_{\Delta V}$) show up 
as linear functions of relevant line widths ($\Delta V$) in Fig.~\ref{dvsig}, or 
\begin{eqnarray}\nonumber
\label{eq-dvsig-c18o-13co}
\sigma_{\Delta V,^{13}{\rm CO}}&=&0.232(0.008)\,\Delta V_{{\rm C}^{18}{\rm O}}\\
                                 & &-0.03(0.02),\,R=0.95\,;
\end{eqnarray}
\begin{eqnarray}\nonumber
\label{eq-dvsig-13co-12co}
\sigma_{\Delta V,^{12}{\rm CO}}&=&0.29(0.03)\,\Delta V_{^{13}{\rm CO}}\\
                                 & &+1.4(0.1),\,R=0.67\,;
\end{eqnarray}
\begin{eqnarray}\nonumber
\label{eq-dvsig-c18o-12co}
\sigma_{\Delta V,^{12}{\rm CO}}&=&0.28(0.03)\,\Delta V_{{\rm C}^{18}{\rm O}}\\
                                 & &+2.1(0.1),\,R=0.65\,.
\end{eqnarray}
The interception of Eq.~(\ref{eq-dvsig-c18o-13co}) is comparable to its uncertainty 
and much smaller than the linear term and thus can be treated as zero. 
Therefore, Eq.~(\ref{eq-dvsig-c18o-13co}) 
demonstrates that the scatter of data points is exactly proportional to the C$^{18}$O\,1-0 line width, which
exactly agrees to a proportionality between the $^{13}$CO and 
C$^{18}$O\,1-0 line widths in logarithmic space, instead of in linear space. 
In another word, the correlation between the two lines
has a constant relative error of $0.232\pm 0.008$. 

The slopes in the other two equations that involve the $^{12}$CO\,1-0 line are about 0.3, not too different from
that in Eq.~(\ref{eq-dvsig-c18o-13co}). However,
the latter two equations both possess a non-zero constant term. 
This constant term actually becomes progressively larger from 
Eq.~(\ref{eq-dvsig-c18o-13co}) to (\ref{eq-dvsig-13co-12co}) to (\ref{eq-dvsig-c18o-12co}), which just
corresponds the progressive worsening of the line width correlations among the three isotopic CO\,1-0 lines 
in Fig.~\ref{comp-v3} (from left to right). Interpretation of this term may need a more detailed investigation 
of line blending and non-Gaussianity in the original CO line profiles, which is beyond the scope of this paper.

Combining the inter-cloud randomness of the turbulence velocity fields with 
earlier discussed progressive worsening trends of luminosity (Sect.~\ref{luminosity-decorrelation}) 
and line width (Sect.~\ref{line-width-co}) correlations
that suggests the increase of randomness across larger cloud subcomponent size-scales,
we envision a full view of the randomness: despite of the ubiquitous similarity of cloud thermal and density 
structure and shock properties, the density and thermal structure and velocity fields in the EGO clouds all 
possess some degree of randomness that increases over larger cloud substructure size-scales and increases toward 
larger EGO clouds. The better correlations among line luminosities than among line widths hints 
that the growth of the randomness might be faster in velocity fields 
than in the thermal and density structures. 
The nice linear relation between the line widths and their data scatter, as discussed in this subsection, further 
reveals some regularity in this randomness. Our EGO sample are biased to molecular clouds/clumps that harbor 
massive YSOs and outflows. It is interesting to check with further observations whether these properties 
(similarity and randomness) still hold for more general molecular clouds.

\subsection{Line width -- luminosity relations}
\label{line-width-L}

\begin{figure*}
\centering
\epsscale{1.6}
\plotone{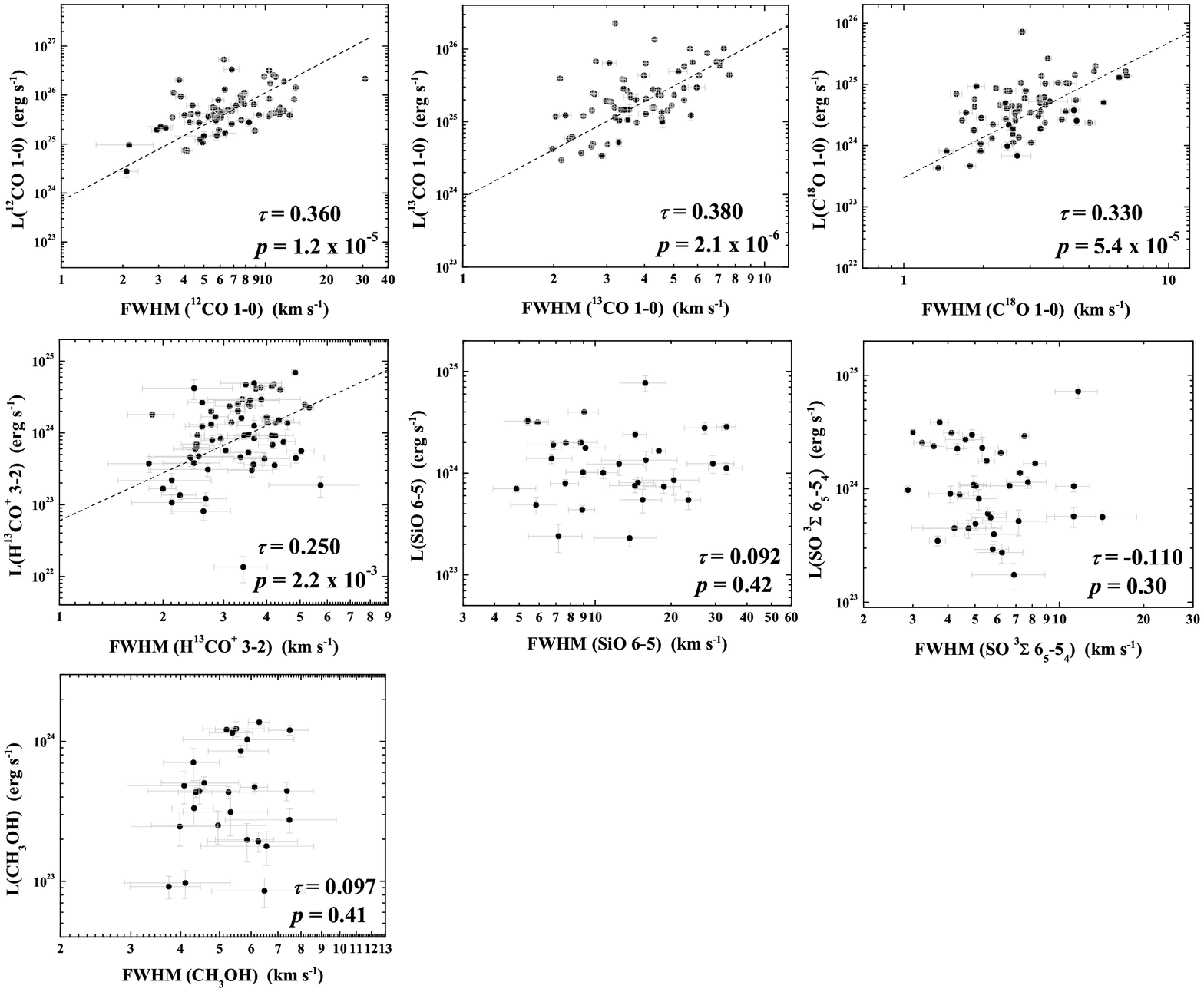}
\caption{The comparison of line luminosities to line widths for all discussed lines. 
  The dashed line in some panels shows an arbitrarily positioned straight line with a fixed slope of 2.2 in logarithmic space 
  for the ease of comparison. The parameters $\tau$ and $p$ in the bottom right 
  corner of each panel are the Kendall tau coefficient and the probability to reject the null hypothesis (see in the text).}
\label{fig-v-l}
\end{figure*}

The different behavior between the line width and luminosity correlations can be double 
checked by directly comparing the line luminosities with their corresponding line widths, as
shown in Fig.~\ref{fig-v-l}. Diverse degrees of correlation are found in the figure. Thus, 
a similar Kendall rank correlation test is performed to each plot. The resulting $p$ values in the 
bottom right corner of each panel indicate strong correlations in the first four panels and the 
lack of correlation in the rest three panels.
An arbitrarily placed dashed straight line with a fixed slope of 2.2 in logarithmic space is drawn 
in those panels with correlation to ease qualitative comparison. The figure simply confirms that, 
the line luminosities of the three isotopic CO\,1-0 lines are clearly correlated with their line widths (top panels), as expected 
in turbulence dominated clouds under virial equilibrium \citep{solo87}. A similar correlation of H$^{13}$CO$^+$\,3-2 line 
looks worse and seems to have a slightly steeper slope than the dashed line (middle left panel). 
The other three shock-related tracer lines (SO, SiO and CH$_3$OH) have their luminosities totally 
uncorrelated with their line widths, which confirms our previous argument that their line widths could be either broadened by 
outflows or dominated by observational errors and opacity effects, but not by canonical interstellar turbulence.

The diverse degrees of line width-luminosity correlations support that the different lines trace different dynamical subcomponents of the EGO clouds.

Multi-tracer multi-cloud intercomparison study has been systematically discussed by \citet{good98}. However, they 
only discussed the study of cloud size-velocity dispersion relation with single dish mapping data, and thus is 
not directly applicable to our study of line width and luminosity correlations with single dish single pointing observations.
Future extension of their discussion to the kind of analysis done in this work will be helpful, but will involve more physics 
such as energy level excitation, randomness in line widths and luminosities, and probably also the astrochemistry of 
the various species in various cloud subcomponents. Extension of the current study to single dish mapping could be 
another direction of further steps.

\section{Summary}
\label{summary}

We have surveyed two 1-GHz bands in 1.1\,mm toward a northern 
subsample of 89 {\it Spitzer GLIMPSE}
extended green objects (EGOs) using a single dish and present a comprehensive catalog of observed
molecular line data (Sect.~\ref{paraplot}). Eight molecular species are undoubtedly detected: 
H$^{13}$CO$^+$, SiO, SO, CH$_3$OH, CH$_3$OCH$_3$, CH$_3$CH$_2$CN,
HCOOCH$_3$, and HN$^{13}$C. Line of sight velocity is determined for 
70 EGOs and kinematic distance is estimated for 80 EGOs \citep[with the help
of literature C$^{18}$O observations from][]{chen10}. The main conclusions are

\begin{enumerate}
\item{
The high detection rate of 79 per cent of H$^{13}$CO$^+$\,3-2 line demonstrates the detection of 
dense cores toward most of the EGOs. All of these detected lines 
are broader than the thermal line width, suggesting the common existence of
non-thermal supersonic motions in the EGO clouds.}
\item{
The high detection rate of 53
per cent of SiO\,6-5 line among the H$^{13}$CO$^+$ detected EGOs supports the idea that
these EGOs are experiencing active outflow shocks. }
\item{
The line luminosities are found to be strongly log-linearly correlated among all considered lines, including 
our four high density or shock tracers and three lower density tracer isotopic 
CO\,1-0 lines. Distinctively, diverse degrees of correlations are found among the line widths, 
which supports that they trace quite different dynamical components of the clouds. 
Such ubiquitous luminosity correlations require {\it a universal similarity of density and thermal structures 
and perhaps of shock properties among all EGO clouds. This is possible if all the observed shocks are produced 
within the natal clouds of the forming stars in the EGOs.}}
\item{
The data scatter in the line width and luminosity correlations are interpreted by randomness in 
the density and thermal structures and velocity fields in the EGO clouds. 
Both the line width and luminosity correlations are found to be 
progressively worse when the pairs of compared lines trace more different cloud subcomponent size-scales, which 
indicates the increase of the randomness across larger cloud subcomponent size-scales.}
\item{
The data scatters in the line widths correlations among the three isotopic 
CO\,1-0 lines appear as nice linear functions of line width itself, which hints both the increase of 
velocity randomness toward larger whole-cloud sizes and some regularity behind.}
\end{enumerate}


%% Included in this acknowledgments section are examples of the
%% AASTeX hypertext markup commands. Use \url without the optional [HREF]
%% argument when you want to print the url directly in the text. Otherwise,
%% use either \url or \anchor, with the HREF as the first argument and the
%% text to be printed in the second.

\acknowledgments

J.H. thanks Dr. Tatsuhiko Hasegawa for some comments on the chemistry of the detected 
species and Dr. Robert Simon for some helpful
discussions upon the kinematic distances. J.H. also thanks
the support of the `Western Light' project of China and
the Chinese National Science Foundation (Grant No. 11173056).
S.T. is financially supported by a postdoctoral fellowship 
at the Institute of Astronomy and Astrophysics, Academia Sinica, Taiwan. 
X.C. acknowledges the finantial support from the Chinese National Science 
Foundation (Grants No. 11073041 and 11133008).


%% in parentheses, after the keyword, but they will not be verified.

{\it Facilities:} \facility{HHT ()}



\appendix

%\section{Appendix}

 % (produced by maketabapp1.f)
\begin{table*}
\caption{\label{tabapp1}Line parameters.}
\centering
\begin{minipage}{160mm}
\footnotesize\rm
\begin{tabular*}{\textwidth}{l@{ }l@{ }l@{ }l@{ }r@{ }l@{ }r@{ }r@{ }r@{ }r@{ }r@{ }c}
\hline
       &            & $\nu$ & $\sigma_\nu$\tablenotemark{b} & FWHM         & $\sigma_{\rm FWHM}$\tablenotemark{c}  & $T_{\rm MB}$    & $\sigma_{\rm Tmb}$  & $I_{\rm int}$\tablenotemark{d}    & $\sigma_I$\tablenotemark{d}      &    &   \\
Object & Transition\tablenotemark{a} & (MHz)   & (MHz)   & (km\,s$^{-1}$) & (km\,s$^{-1}$)  & (mK)  & (mK)  & (K\,km\,s$^{-1}$) & (K\,km\,s$^{-1}$) & SNR\tablenotemark{d} & Note\tablenotemark{e}    \\
  (1)  &   (2)      & (3)   & (4)          & (5)          & (6)                  & (7)             & (8)                 &    (9)          & (10)            & (11)  & (12)   \\
\hline \hline
G10.29-0.13 &                          H$^{13}$CO$^+$,3-2 & 260255.34 &  0.14 &  2.66 &  0.40 &   176 &   23 &  0.488 & 0.068 &   7.2 &   \\                                        
G10.34-0.14 &                CH$_3$OH,$7_{3,4}-7_{2,5}-+$ & 251641.74 &  0.28 &  3.80 &  0.70 &    91 &   20 &  0.317 & 0.075 &   4.2 &   \\                                        
            &                CH$_3$OH,$6_{3,3}-6_{2,4}-+$ & 251738.39 &  0.38 &  5.14 &  1.08 &    78 &   19 &  0.401 & 0.079 &   5.1 &   \\                                        
            &                CH$_3$OH,$5_{3,2}-5_{2,3}-+$ & 251812.21 &  0.31 &  2.22 &  0.78 &    56 &   18 &  1.042 & 0.109 &   9.6 &   \\                                        
            &                CH$_3$OH,$5_{3,2}-5_{2,3}-+$ & 251812.21 &       & 23.46 &  5.18 &    52 &   18 &        &       &       & s \\                                        
            &                 SO,$^3\Sigma$,$6_{5}-5_{4}$ & 251825.99 &  0.23 &  6.24 &  1.16 &   164 &   20 &  1.242 & 0.089 &  14.0 &   \\                                        
            &                CH$_3$OH,$4_{3,1}-4_{2,2}-+$ & 251866.85 &  0.26 &  4.58 &  0.86 &   104 &   20 &  0.568 & 0.079 &   7.2 &   \\                                        
            &                CH$_3$OH,$5_{3,3}-5_{2,4}+-$ & 251890.90 &       &  6.24 &  0.54 &    86 &   20 &  0.574 & 0.074 &   7.8 & p \\                                        
            &                CH$_3$OH,$6_{3,4}-6_{2,5}+-$ & 251895.73 &       &  6.24 &       &    72 &   20 &  0.480 & 0.079 &   6.1 & p \\                                        
            &                CH$_3$OH,$4_{3,2}-4_{2,3}+-$ & 251900.49 &       &  6.24 &       &   124 &   20 &  0.826 & 0.082 &  10.1 & p \\                                        
            &                CH$_3$OH,$3_{3,0}-3_{2,1}-+$ & 251905.81 &       &  6.24 &       &    63 &   20 &  0.416 & 0.068 &   6.1 & p \\                                        
            &                CH$_3$OH,$3_{3,1}-3_{2,2}+-$ & 251917.04 &       &  6.50 &  0.74 &    69 &   22 &  0.477 & 0.081 &   5.9 & p \\                                        
            &                CH$_3$OH,$7_{3,5}-7_{2,6}+-$ & 251923.63 &       &  6.50 &       &    47 &   22 &  0.323 & 0.075 &   4.3 & p \\                                        
            &                CH$_3$OH,$8_{3,6}-8_{2,7}+-$ & 251983.71 &  0.70 &  5.94 &  2.04 &    50 &   21 &  0.326 & 0.089 &   3.7 &   \\                                        
            &                CH$_3$OH,$9_{3,7}-9_{2,8}+-$ & 252089.14 &  2.66 & 25.16 &  7.92 &    25 &   21 &  0.607 & 0.163 &   3.7 &   \\                                        
            &                          H$^{13}$CO$^+$,3-2 & 260255.31 &  0.07 &  3.66 &  0.22 &   401 &   23 &  1.644 & 0.091 &  18.1 &   \\                                        
            &                                     SiO,6-5 & 260517.27 &  0.82 & 23.42 &  2.72 &    96 &   23 &  2.473 & 0.194 &  12.7 &   \\                                        
G11.11-0.11 &                 SO,$^3\Sigma$,$6_{5}-5_{4}$ & 251825.36 &  0.28 &  2.18 &  1.12 &   119 &   33 &  0.383 & 0.103 &   3.7 &   \\                                        
            &                          H$^{13}$CO$^+$,3-2 & 260255.38 &  0.09 &  2.60 &  0.28 &   424 &   38 &  1.859 & 0.188 &   9.9 &   \\                                        
            &                          H$^{13}$CO$^+$,3-2 & 260255.38 &       & 31.66 & 18.58 &    30 &   38 &        &       &       & s \\                                        
G11.92-0.61 &                CH$_3$OH,$8_{3,5}-8_{2,6}-+$ & 251517.35 &  0.70 &  7.64 &  1.70 &    70 &   21 &  0.530 & 0.094 &   5.6 &   \\                                        
            &                CH$_3$OH,$7_{3,4}-7_{2,5}-+$ & 251642.54 &  0.44 &  7.22 &  1.22 &    76 &   19 &  0.568 & 0.094 &   6.0 &   \\                                        
            &                CH$_3$OH,$6_{3,3}-6_{2,4}-+$ & 251738.27 &  0.34 &  7.10 &  0.82 &    98 &   20 &  0.675 & 0.091 &   7.4 &   \\                                        
            &                CH$_3$OH,$5_{3,2}-5_{2,3}-+$ & 251811.89 &  0.42 &  9.24 &  1.06 &    93 &   19 &  0.902 & 0.088 &  10.3 &   \\                                        
            &                 SO,$^3\Sigma$,$6_{5}-5_{4}$ & 251826.16 &  0.25 & 21.12 &  5.12 &    37 &   18 &  1.383 & 0.091 &  15.2 &   \\                                        
            &                 SO,$^3\Sigma$,$6_{5}-5_{4}$ & 251826.16 &       &  6.64 &  1.30 &   112 &   18 &        &       &       & s \\                                        
            &                CH$_3$OH,$4_{3,1}-4_{2,2}-+$ & 251866.38 &  0.30 &  6.08 &  1.08 &   101 &   19 &  0.708 & 0.088 &   8.0 &   \\                                        
            &                CH$_3$OH,$5_{3,3}-5_{2,4}+-$ & 251890.90 &       &  6.22 &  0.50 &    94 &   20 &  0.620 & 0.072 &   8.6 & p \\                                        
            &                CH$_3$OH,$6_{3,4}-6_{2,5}+-$ & 251895.73 &       &  6.22 &       &   108 &   20 &  0.715 & 0.081 &   8.8 & p \\                                        
            &                CH$_3$OH,$4_{3,2}-4_{2,3}+-$ & 251900.49 &       &  6.22 &       &    52 &   20 &  0.345 & 0.079 &   4.4 & p \\                                        
            &                CH$_3$OH,$3_{3,0}-3_{2,1}-+$ & 251905.81 &       &  6.22 &       &    80 &   20 &  0.528 & 0.073 &   7.2 & p \\                                        
            &                CH$_3$OH,$3_{3,1}-3_{2,2}+-$ & 251917.04 &       &  5.14 &  0.34 &   101 &   20 &  0.551 & 0.067 &   8.2 & p \\                                        
            &                CH$_3$OH,$7_{3,5}-7_{2,6}+-$ & 251923.63 &       &  5.14 &       &    80 &   20 &  0.438 & 0.063 &   7.0 & p \\                                        
            &                CH$_3$OH,$8_{3,6}-8_{2,7}+-$ & 251985.15 &  0.39 &  6.42 &  1.30 &    80 &   19 &  0.564 & 0.091 &   6.2 &   \\                                        
            &                CH$_3$OH,$9_{3,7}-9_{2,8}+-$ & 252088.69 &  0.87 & 11.20 &  1.92 &    54 &   19 &  0.570 & 0.110 &   5.2 &   \\                                        
            &              CH$_3$OH,$10_{3,8}-10_{2,9}+-$ & 252252.36 &  0.66 &  8.28 &  1.48 &    54 &   19 &  0.430 & 0.097 &   4.4 &   \\                                        
            &             CH$_3$OH,$11_{3,9}-11_{2,10}+-$ & 252485.31 &  0.41 &  6.56 &  0.98 &    82 &   21 &  0.474 & 0.091 &   5.2 &   \\                                        
            &                          H$^{13}$CO$^+$,3-2 & 260255.33 &  0.05 &  3.12 &  0.32 &   520 &   24 &  3.040 & 0.117 &  26.0 &   \\                                        
            &                          H$^{13}$CO$^+$,3-2 & 260255.33 &       &  8.32 &  1.98 &   146 &   24 &        &       &       & s \\                                        
            &                                     SiO,6-5 & 260518.21 &  0.41 &  9.16 &  1.82 &    91 &   23 &  2.294 & 0.158 &  14.5 &   \\                                        
            &                                     SiO,6-5 & 260518.21 &       & 26.94 &  4.66 &    53 &   23 &        &       &       & s \\                                        
G12.02-0.21 &                          H$^{13}$CO$^+$,3-2 & 260255.34 &  0.24 &  2.46 &  0.72 &   103 &   23 &  0.264 & 0.081 &   3.3 &   \\                                        
G12.20-0.03 &                CH$_3$OH,$8_{3,5}-8_{2,6}-+$ & 251515.65 &  0.47 &  5.08 &  1.30 &    87 &   28 &  0.531 & 0.138 &   3.8 &   \\                                        
            &                CH$_3$OH,$6_{3,3}-6_{2,4}-+$ & 251738.71 &  0.32 &  3.08 &  0.96 &   124 &   33 &  0.427 & 0.113 &   3.8 &   \\                                        
            &                 SO,$^3\Sigma$,$6_{5}-5_{4}$ & 251825.87 &  0.43 &  4.48 &  1.74 &   119 &   37 &  0.636 & 0.132 &   4.8 &   \\                                        
            &                CH$_3$OH,$6_{3,4}-6_{2,5}+-$ & 251893.26 &       &  4.76 &       &   114 &   36 &  0.577 & 0.107 &   5.4 & p \\                                        
\hline
\end{tabular*}
\tablenotetext{a}{The string '(E+A)' after the transition CH$_3$OCH$_3$,$10_{2,9}-9_{1,8}$ means it is a blending of four transitions: EA, AE, EE, and AA.}
\tablenotetext{b}{If the value of $\sigma_{\nu}$ is missing, it means the line is either fit with fixed catalogue frequency or fit with fix line frequency increment in a multi-Gaussian profile fitting.}
\tablenotetext{c}{If the value of $\sigma_{\rm HWHM}$ is missing, it means the line is fit with the same line width as other lines in a multi-Gaussian profile fitting.}
\tablenotetext{d}{If the values of $I_{\rm int}$, $\sigma_I$ and SNR are missing, it means the line is the second component of a double Gaussian fit to the wide line wing of a broad line feature.}
\tablenotetext{e}{{\bf p} = partially blended with neigbouring lines of which blended lines are fit together by a multi-Gaussian profile with fix catalogue frequencies and identical line width; {\bf b} = totally blended lines of which member lines are fit independently with a single Gaussian profile, so that the contribution from other blended lines are not separated yet. {\bf s} = the second Gaussian component of the double-Gaussian fit to a line with broad line wings.}
\end{minipage}
\end{table*}
\begin{table*}
\addtocounter{table}{-1}
\caption{(continued) Line parameters.}
\centering
\begin{minipage}{160mm}
\footnotesize\rm
\begin{tabular*}{\textwidth}{l@{ }l@{ }l@{ }l@{ }r@{ }l@{ }r@{ }r@{ }r@{ }r@{ }r@{ }c}
\hline
       &            & $\nu$ & $\sigma_\nu$\tablenotemark{b} & FWHM         & $\sigma_{\rm FWHM}$\tablenotemark{c}  & $T_{\rm MB}$    & $\sigma_{\rm Tmb}$  & $I_{\rm int}$\tablenotemark{d}    & $\sigma_I$\tablenotemark{d}      &    &   \\
Object & Transition\tablenotemark{a} & (MHz)   & (MHz)   & (km\,s$^{-1}$) & (km\,s$^{-1}$)  & (mK)  & (mK)  & (K\,km\,s$^{-1}$) & (K\,km\,s$^{-1}$) & SNR\tablenotemark{d} & Note\tablenotemark{e}    \\
  (1)  &   (2)      & (3)   & (4)          & (5)          & (6)                  & (7)             & (8)                 &    (9)          & (10)            & (11)  & (12)   \\
\hline \hline
            &                CH$_3$OH,$3_{3,0}-3_{2,1}-+$ & 251903.34 &       &  4.76 &       &    75 &   36 &  0.382 & 0.109 &   3.5 & p \\                                        
            &                CH$_3$OH,$7_{3,5}-7_{2,6}+-$ & 251917.04 &       &  3.58 &       &    90 &   39 &  0.342 & 0.099 &   3.5 & p \\                                        
            &                CH$_3$OH,$3_{3,1}-3_{2,2}+-$ & 251923.63 &       &  3.58 &       &    91 &   39 &  0.347 & 0.099 &   3.5 & p \\                                        
            &                          H$^{13}$CO$^+$,3-2 & 260255.17 &  0.17 &  4.34 &  0.46 &   301 &   38 &  1.350 & 0.143 &   9.4 &   \\                                        
G12.42+0.50 &                CH$_3$OH,$8_{3,5}-8_{2,6}-+$ & 251517.47 &  0.42 &  4.58 &  1.20 &    55 &   17 &  0.298 & 0.061 &   4.9 &   \\
\hline
\end{tabular*}
\end{minipage}
\end{table*}
\begin{table*}
\addtocounter{table}{-1}
\caption{(continued) Line parameters.}
\centering
\begin{minipage}{160mm}
\footnotesize\rm
\begin{tabular*}{\textwidth}{l@{ }l@{ }l@{ }l@{ }r@{ }l@{ }r@{ }r@{ }r@{ }r@{ }r@{ }c}
\hline
       &            & $\nu$ & $\sigma_\nu$\tablenotemark{b} & FWHM         & $\sigma_{\rm FWHM}$\tablenotemark{c}  & $T_{\rm MB}$    & $\sigma_{\rm Tmb}$  & $I_{\rm int}$\tablenotemark{d}    & $\sigma_I$\tablenotemark{d}      &    &   \\
Object & Transition\tablenotemark{a} & (MHz)   & (MHz)   & (km\,s$^{-1}$) & (km\,s$^{-1}$)  & (mK)  & (mK)  & (K\,km\,s$^{-1}$) & (K\,km\,s$^{-1}$) & SNR\tablenotemark{d} & Note\tablenotemark{e}    \\
  (1)  &   (2)      & (3)   & (4)          & (5)          & (6)                  & (7)             & (8)                 &    (9)          & (10)            & (11)  & (12)   \\
\hline \hline
            &                CH$_3$OH,$5_{3,3}-5_{2,4}+-$ & 251890.90 &       &  5.48 &  0.14 &   317 &   24 &  1.847 & 0.082 &  22.5 & p \\                                        
            &                CH$_3$OH,$6_{3,4}-6_{2,5}+-$ & 251895.73 &       &  5.48 &       &   273 &   24 &  1.593 & 0.080 &  19.9 & p \\                                        
            &                CH$_3$OH,$4_{3,2}-4_{2,3}+-$ & 251900.49 &       &  5.48 &       &   277 &   24 &  1.612 & 0.078 &  20.7 & p \\                                        
            &                CH$_3$OH,$3_{3,0}-3_{2,1}-+$ & 251905.81 &       &  5.48 &       &   241 &   24 &  1.405 & 0.078 &  18.0 & p \\                                        
            &                CH$_3$OH,$3_{3,1}-3_{2,2}+-$ & 251917.04 &       &  6.00 &  0.16 &   239 &   24 &  1.524 & 0.082 &  18.6 & p \\
\hline
\end{tabular*}
\end{minipage}
\end{table*}
\begin{table*}
\addtocounter{table}{-1}
\caption{(continued) Line parameters.}
\centering
\begin{minipage}{160mm}
\footnotesize\rm
\begin{tabular*}{\textwidth}{l@{ }l@{ }l@{ }l@{ }r@{ }l@{ }r@{ }r@{ }r@{ }r@{ }r@{ }c}
\hline
       &            & $\nu$ & $\sigma_\nu$\tablenotemark{b} & FWHM         & $\sigma_{\rm FWHM}$\tablenotemark{c}  & $T_{\rm MB}$    & $\sigma_{\rm Tmb}$  & $I_{\rm int}$\tablenotemark{d}    & $\sigma_I$\tablenotemark{d}      &    &   \\
Object & Transition\tablenotemark{a} & (MHz)   & (MHz)   & (km\,s$^{-1}$) & (km\,s$^{-1}$)  & (mK)  & (mK)  & (K\,km\,s$^{-1}$) & (K\,km\,s$^{-1}$) & SNR\tablenotemark{d} & Note\tablenotemark{e}    \\
  (1)  &   (2)      & (3)   & (4)          & (5)          & (6)                  & (7)             & (8)                 &    (9)          & (10)            & (11)  & (12)   \\
\hline \hline
G16.59-0.05 &                CH$_3$OH,$8_{3,5}-8_{2,6}-+$ & 251517.23 &  0.20 &  5.36 &  0.56 &   153 &   19 &  0.876 & 0.086 &  10.2 &   \\                                        
            &                   c-HCCCH,$6_{2,5}-5_{1,4}$ & 251526.80 &  0.50 &  4.78 &  1.80 &    54 &   19 &  0.318 & 0.082 &   3.9 &   \\                                        
            &       CH$_3$OCH$_3$,$10_{2,9}-9_{1,8}$(E+A) & 251587.28 &  0.45 &  4.16 &  1.84 &    64 &   19 &  0.308 & 0.078 &   3.9 &   \\                                        
            &                CH$_3$OH,$7_{3,4}-7_{2,5}-+$ & 251641.63 &  0.18 &  4.54 &  0.50 &   148 &   20 &  0.711 & 0.082 &   8.7 &   \\                                        
            &                CH$_3$OH,$6_{3,3}-6_{2,4}-+$ & 251738.14 &  0.19 &  5.48 &  0.52 &   163 &   20 &  0.961 & 0.086 &  11.2 &   \\
\hline
\end{tabular*}
\end{minipage}
\end{table*}
\begin{table*}
\addtocounter{table}{-1}
\caption{(continued) Line parameters.}
\centering
\begin{minipage}{160mm}
\footnotesize\rm
\begin{tabular*}{\textwidth}{l@{ }l@{ }l@{ }l@{ }r@{ }l@{ }r@{ }r@{ }r@{ }r@{ }r@{ }c}
\hline
       &            & $\nu$ & $\sigma_\nu$\tablenotemark{b} & FWHM         & $\sigma_{\rm FWHM}$\tablenotemark{c}  & $T_{\rm MB}$    & $\sigma_{\rm Tmb}$  & $I_{\rm int}$\tablenotemark{d}    & $\sigma_I$\tablenotemark{d}      &    &   \\
Object & Transition\tablenotemark{a} & (MHz)   & (MHz)   & (km\,s$^{-1}$) & (km\,s$^{-1}$)  & (mK)  & (mK)  & (K\,km\,s$^{-1}$) & (K\,km\,s$^{-1}$) & SNR\tablenotemark{d} & Note\tablenotemark{e}    \\
  (1)  &   (2)      & (3)   & (4)          & (5)          & (6)                  & (7)             & (8)                 &    (9)          & (10)            & (11)  & (12)   \\
\hline \hline
            &                CH$_3$OH,$3_{3,0}-3_{2,1}-+$ & 251905.81 &       &  4.76 &       &    52 &   21 &  0.264 & 0.063 &   4.2 & p \\                                        
            &                CH$_3$OH,$3_{3,1}-3_{2,2}+-$ & 251917.04 &       &  3.42 &  0.36 &    66 &   21 &  0.240 & 0.058 &   4.1 & p \\                                        
            &                CH$_3$OH,$7_{3,5}-7_{2,6}+-$ & 251923.63 &       &  3.42 &       &    85 &   21 &  0.311 & 0.058 &   5.4 & p \\                                        
            &                CH$_3$OH,$8_{3,6}-8_{2,7}+-$ & 251984.77 &  0.49 &  3.78 &  2.28 &    49 &   20 &  0.226 & 0.072 &   3.1 &   \\                                        
            &                          H$^{13}$CO$^+$,3-2 & 260255.32 &       &  7.20 &  1.86 &    90 &   21 &        &       &       & s \\
\hline
\end{tabular*}
\end{minipage}
\end{table*}
\begin{table*}
\addtocounter{table}{-1}
\caption{(continued) Line parameters.}
\centering
\begin{minipage}{160mm}
\footnotesize\rm
\begin{tabular*}{\textwidth}{l@{ }l@{ }l@{ }l@{ }r@{ }l@{ }r@{ }r@{ }r@{ }r@{ }r@{ }c}
\hline
       &            & $\nu$ & $\sigma_\nu$\tablenotemark{b} & FWHM         & $\sigma_{\rm FWHM}$\tablenotemark{c}  & $T_{\rm MB}$    & $\sigma_{\rm Tmb}$  & $I_{\rm int}$\tablenotemark{d}    & $\sigma_I$\tablenotemark{d}      &    &   \\
Object & Transition\tablenotemark{a} & (MHz)   & (MHz)   & (km\,s$^{-1}$) & (km\,s$^{-1}$)  & (mK)  & (mK)  & (K\,km\,s$^{-1}$) & (K\,km\,s$^{-1}$) & SNR\tablenotemark{d} & Note\tablenotemark{e}    \\
  (1)  &   (2)      & (3)   & (4)          & (5)          & (6)                  & (7)             & (8)                 &    (9)          & (10)            & (11)  & (12)   \\
\hline \hline
            &                CH$_3$OH,$4_{3,2}-4_{2,3}+-$ & 251900.49 &       &  7.44 &       &   107 &   20 &  0.847 & 0.095 &   8.9 & p \\                                        
            &                CH$_3$OH,$3_{3,0}-3_{2,1}-+$ & 251905.81 &       &  7.44 &       &   137 &   20 &  1.088 & 0.090 &  12.1 & p \\                                        
            &                CH$_3$OH,$3_{3,1}-3_{2,2}+-$ & 251917.04 &       &  8.34 &       &   123 &   20 &  1.095 & 0.082 &  13.4 & p \\                                        
            &                CH$_3$OH,$7_{3,5}-7_{2,6}+-$ & 251923.63 &       &  8.34 &       &   123 &   20 &  1.094 & 0.082 &  13.3 & p \\                                        
            &                CH$_3$OH,$8_{3,6}-8_{2,7}+-$ & 251984.58 &  0.37 &  8.70 &  1.22 &    98 &   19 &  0.896 & 0.081 &  11.1 &   \\
\hline
\end{tabular*}
\end{minipage}
\end{table*}
\begin{table*}
\addtocounter{table}{-1}
\caption{(continued) Line parameters.}
\centering
\begin{minipage}{160mm}
\footnotesize\rm
\begin{tabular*}{\textwidth}{l@{ }l@{ }l@{ }l@{ }r@{ }l@{ }r@{ }r@{ }r@{ }r@{ }r@{ }c}
\hline
       &            & $\nu$ & $\sigma_\nu$\tablenotemark{b} & FWHM         & $\sigma_{\rm FWHM}$\tablenotemark{c}  & $T_{\rm MB}$    & $\sigma_{\rm Tmb}$  & $I_{\rm int}$\tablenotemark{d}    & $\sigma_I$\tablenotemark{d}      &    &   \\
Object & Transition\tablenotemark{a} & (MHz)   & (MHz)   & (km\,s$^{-1}$) & (km\,s$^{-1}$)  & (mK)  & (mK)  & (K\,km\,s$^{-1}$) & (K\,km\,s$^{-1}$) & SNR\tablenotemark{d} & Note\tablenotemark{e}    \\
  (1)  &   (2)      & (3)   & (4)          & (5)          & (6)                  & (7)             & (8)                 &    (9)          & (10)            & (11)  & (12)   \\
\hline \hline
            &           HCOOCH$_3$,$21_{8,14}-20_{8,13}$E & 260403.85 &  0.81 &  6.00 &  2.38 &    50 &   20 &  0.394 & 0.094 &   4.2 & b \\                                        
            &                                     SiO,6-5 & 260518.07 &  1.37 & 27.16 &  4.58 &    51 &   20 &  1.427 & 0.181 &   7.9 &   \\                                        
            &           HCOOCH$_3$,$21_{5,17}-20_{5,16}$E & 261144.25 &  0.43 &  3.00 &  1.12 &    61 &   23 &  0.233 & 0.075 &   3.1 & b \\                                        
            &      CH$_3$OCH$_3$, $17_{5,13}-17_{4,14}$EE & 261144.25 &  0.43 &  3.00 &  1.12 &    61 &   23 &  0.233 & 0.075 &   3.1 & b \\                                        
            &                              HN$^{13}$C,3-2 & 261263.39 &  0.10 &  2.72 &  0.28 &   253 &   23 &  0.821 & 0.071 &  11.6 &   \\
\hline
\end{tabular*}
\end{minipage}
\end{table*}
\begin{table*}
\addtocounter{table}{-1}
\caption{(continued) Line parameters.}
\centering
\begin{minipage}{160mm}
\footnotesize\rm
\begin{tabular*}{\textwidth}{l@{ }l@{ }l@{ }l@{ }r@{ }l@{ }r@{ }r@{ }r@{ }r@{ }r@{ }c}
\hline
       &            & $\nu$ & $\sigma_\nu$\tablenotemark{b} & FWHM         & $\sigma_{\rm FWHM}$\tablenotemark{c}  & $T_{\rm MB}$    & $\sigma_{\rm Tmb}$  & $I_{\rm int}$\tablenotemark{d}    & $\sigma_I$\tablenotemark{d}      &    &   \\
Object & Transition\tablenotemark{a} & (MHz)   & (MHz)   & (km\,s$^{-1}$) & (km\,s$^{-1}$)  & (mK)  & (mK)  & (K\,km\,s$^{-1}$) & (K\,km\,s$^{-1}$) & SNR\tablenotemark{d} & Note\tablenotemark{e}    \\
  (1)  &   (2)      & (3)   & (4)          & (5)          & (6)                  & (7)             & (8)                 &    (9)          & (10)            & (11)  & (12)   \\
\hline \hline
            &                 SO,$^3\Sigma$,$6_{5}-5_{4}$ & 251825.64 &  0.09 &  7.50 &  0.28 &   489 &   24 &  4.064 & 0.101 &  40.2 &   \\                                        
            &                CH$_3$OH,$4_{3,1}-4_{2,2}-+$ & 251866.49 &  0.09 &  5.04 &  0.26 &   405 &   25 &  2.163 & 0.088 &  24.6 &   \\                                        
            &                CH$_3$OH,$5_{3,3}-5_{2,4}+-$ & 251890.90 &       &  5.08 &  0.12 &   355 &   25 &  1.921 & 0.080 &  24.0 & p \\                                        
            &                CH$_3$OH,$6_{3,4}-6_{2,5}+-$ & 251895.73 &       &  5.08 &       &   338 &   25 &  1.829 & 0.079 &  23.2 & p \\                                        
            &                CH$_3$OH,$4_{3,2}-4_{2,3}+-$ & 251900.49 &       &  5.08 &       &   371 &   25 &  2.006 & 0.078 &  25.7 & p \\
\hline
\end{tabular*}
\end{minipage}
\end{table*}
\begin{table*}
\addtocounter{table}{-1}
\caption{(continued) Line parameters.}
\centering
\begin{minipage}{160mm}
\footnotesize\rm
\begin{tabular*}{\textwidth}{l@{ }l@{ }l@{ }l@{ }r@{ }l@{ }r@{ }r@{ }r@{ }r@{ }r@{ }c}
\hline
       &            & $\nu$ & $\sigma_\nu$\tablenotemark{b} & FWHM         & $\sigma_{\rm FWHM}$\tablenotemark{c}  & $T_{\rm MB}$    & $\sigma_{\rm Tmb}$  & $I_{\rm int}$\tablenotemark{d}    & $\sigma_I$\tablenotemark{d}      &    &   \\
Object & Transition\tablenotemark{a} & (MHz)   & (MHz)   & (km\,s$^{-1}$) & (km\,s$^{-1}$)  & (mK)  & (mK)  & (K\,km\,s$^{-1}$) & (K\,km\,s$^{-1}$) & SNR\tablenotemark{d} & Note\tablenotemark{e}    \\
  (1)  &   (2)      & (3)   & (4)          & (5)          & (6)                  & (7)             & (8)                 &    (9)          & (10)            & (11)  & (12)   \\
\hline \hline
            &              CH$_3$OH,$10_{3,8}-10_{2,9}+-$ & 252251.47 &  1.71 & 14.28 &  3.82 &    38 &   23 &  0.544 & 0.153 &   3.6 &   \\                                        
            &                          H$^{13}$CO$^+$,3-2 & 260255.45 &  0.03 &  4.02 &  0.08 &  1094 &   27 &  4.662 & 0.107 &  43.6 & b \\                                        
            &                                     SiO,6-5 & 260518.09 &  0.32 &  4.88 &  0.94 &   106 &   28 &  2.312 & 0.164 &  14.1 &   \\                                        
            &                                     SiO,6-5 & 260518.09 &       & 20.30 &  2.86 &    90 &   28 &        &       &       & s \\                                        
G35.15+0.80 &                          H$^{13}$CO$^+$,3-2 & 260255.34 &  0.16 &  1.82 &  0.44 &   146 &   25 &  0.328 & 0.084 &   3.9 &   \\
\hline
\end{tabular*}
\end{minipage}
\end{table*}
\begin{table*}
\addtocounter{table}{-1}
\caption{(continued) Line parameters.}
\centering
\begin{minipage}{160mm}
\footnotesize\rm
\begin{tabular*}{\textwidth}{l@{ }l@{ }l@{ }l@{ }r@{ }l@{ }r@{ }r@{ }r@{ }r@{ }r@{ }c}
\hline
       &            & $\nu$ & $\sigma_\nu$\tablenotemark{b} & FWHM         & $\sigma_{\rm FWHM}$\tablenotemark{c}  & $T_{\rm MB}$    & $\sigma_{\rm Tmb}$  & $I_{\rm int}$\tablenotemark{d}    & $\sigma_I$\tablenotemark{d}      &    &   \\
Object & Transition\tablenotemark{a} & (MHz)   & (MHz)   & (km\,s$^{-1}$) & (km\,s$^{-1}$)  & (mK)  & (mK)  & (K\,km\,s$^{-1}$) & (K\,km\,s$^{-1}$) & SNR\tablenotemark{d} & Note\tablenotemark{e}    \\
  (1)  &   (2)      & (3)   & (4)          & (5)          & (6)                  & (7)             & (8)                 &    (9)          & (10)            & (11)  & (12)   \\
\hline \hline
            &                CH$_3$OH,$5_{3,2}-5_{2,3}-+$ & 251811.45 &  0.14 &  4.96 &  0.38 &   197 &   19 &  1.022 & 0.086 &  11.9 &   \\                                        
            &                 SO,$^3\Sigma$,$6_{5}-5_{4}$ & 251825.76 &       &  4.62 &  0.70 &   169 &   19 &        &       &       & s \\                                        
            &                 SO,$^3\Sigma$,$6_{5}-5_{4}$ & 251825.76 &  0.14 & 18.08 &  3.78 &    68 &   19 &  1.971 & 0.119 &  16.6 &   \\                                        
            &                CH$_3$OH,$4_{3,1}-4_{2,2}-+$ & 251865.97 &  0.17 &  5.36 &  0.54 &   173 &   20 &  1.054 & 0.086 &  12.3 &   \\                                        
            &                CH$_3$OH,$5_{3,3}-5_{2,4}+-$ & 251890.90 &       &  4.82 &  0.12 &   180 &   19 &  0.924 & 0.058 &  15.9 & p \\
\hline
\end{tabular*}
\end{minipage}
\end{table*}
\begin{table*}
\addtocounter{table}{-1}
\caption{(continued) Line parameters.}
\centering
\begin{minipage}{160mm}
\footnotesize\rm
\begin{tabular*}{\textwidth}{l@{ }l@{ }l@{ }l@{ }r@{ }l@{ }r@{ }r@{ }r@{ }r@{ }r@{ }c}
\hline
       &            & $\nu$ & $\sigma_\nu$\tablenotemark{b} & FWHM         & $\sigma_{\rm FWHM}$\tablenotemark{c}  & $T_{\rm MB}$    & $\sigma_{\rm Tmb}$  & $I_{\rm int}$\tablenotemark{d}    & $\sigma_I$\tablenotemark{d}      &    &   \\
Object & Transition\tablenotemark{a} & (MHz)   & (MHz)   & (km\,s$^{-1}$) & (km\,s$^{-1}$)  & (mK)  & (mK)  & (K\,km\,s$^{-1}$) & (K\,km\,s$^{-1}$) & SNR\tablenotemark{d} & Note\tablenotemark{e}    \\
  (1)  &   (2)      & (3)   & (4)          & (5)          & (6)                  & (7)             & (8)                 &    (9)          & (10)            & (11)  & (12)   \\
\hline \hline
            &                 SO,$^3\Sigma$,$6_{5}-5_{4}$ & 251826.53 &  0.28 &  5.16 &  0.84 &   133 &   25 &  0.776 & 0.090 &   8.6 &   \\                                        
            &                CH$_3$OH,$4_{3,1}-4_{2,2}-+$ & 251866.17 &  0.81 &  6.72 &  2.28 &    51 &   24 &  0.370 & 0.098 &   3.8 &   \\                                        
            &                          H$^{13}$CO$^+$,3-2 & 260255.17 &  0.12 &  3.68 &  0.40 &   265 &   25 &  1.198 & 0.095 &  12.6 &   \\                                        
            &                                     SiO,6-5 & 260518.47 &  1.80 & 20.52 &  5.20 &    39 &   23 &  0.810 & 0.189 &   4.3 &   \\                                        
G59.79+0.63 &                 SO,$^3\Sigma$,$6_{5}-5_{4}$ & 251825.37 &  0.17 &  4.94 &  0.56 &   192 &   23 &  1.081 & 0.091 &  11.9 &   \\
\hline
\end{tabular*}
\end{minipage}
\end{table*}

\newpage
\begin{figure*}
\centering
\includegraphics[scale=.30,angle=0]{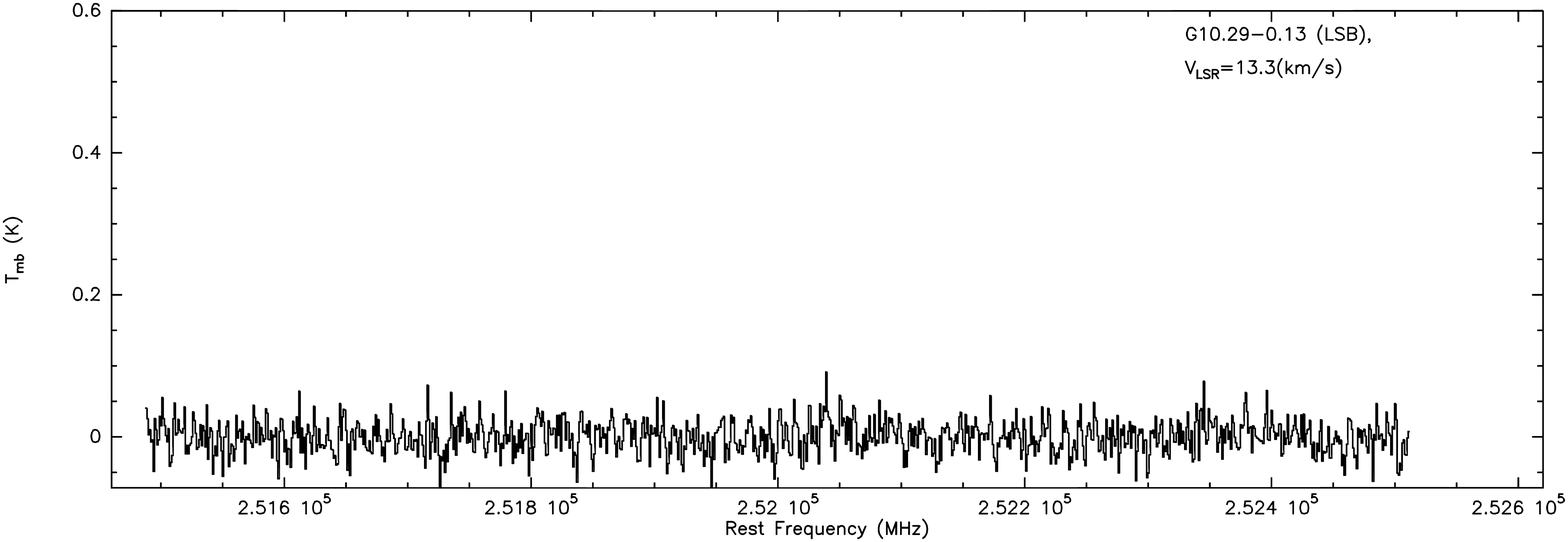}
\includegraphics[scale=.30,angle=0]{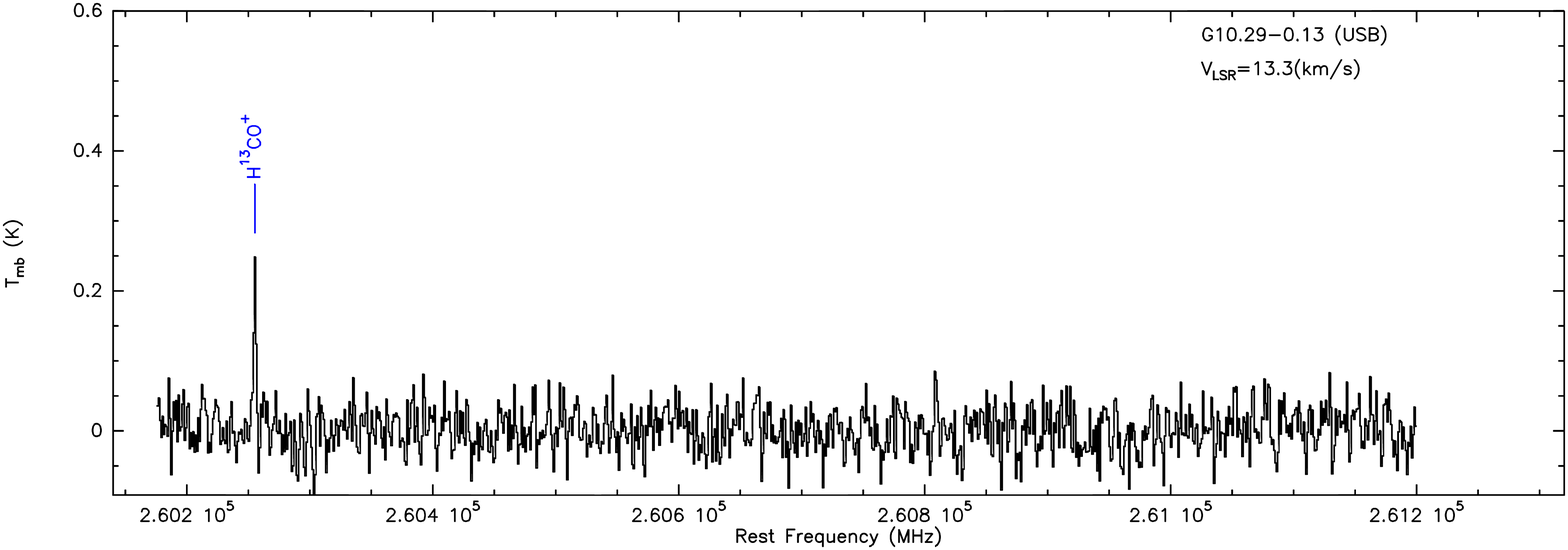}
\caption{The sideband spectral plots of all observed EGOs, with identified species labelled at the catalog frequency of each identified line. The first two plots are for G10.29-0.13.}
\label{figapp1}
\end{figure*}
 \addtocounter{figure}{-1}
\begin{figure*}
\centering
\includegraphics[scale=.30,angle=0]{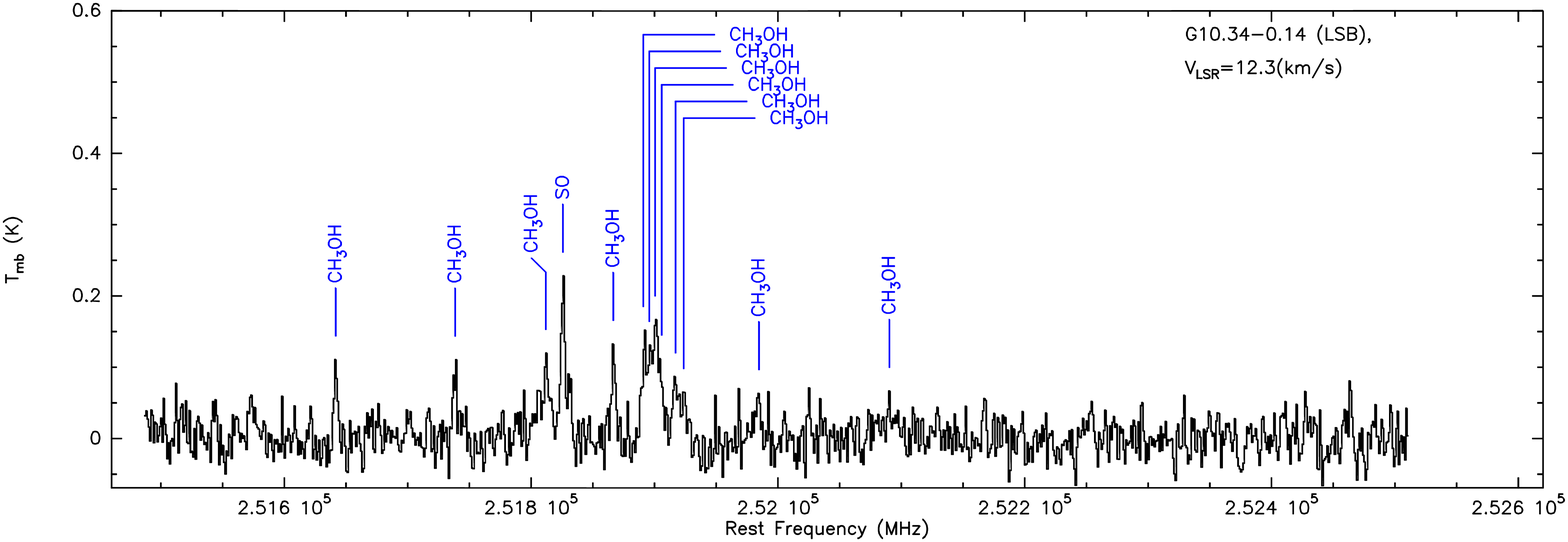}
\includegraphics[scale=.30,angle=0]{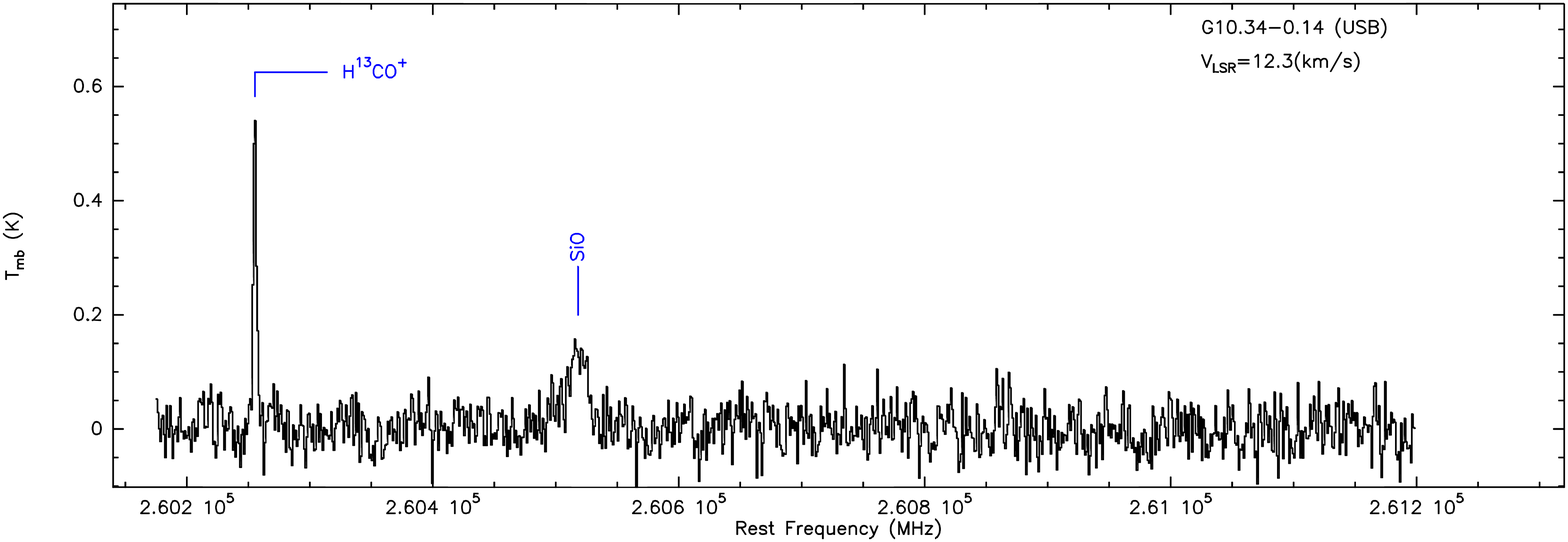}
\caption{(continued) For G10.34-0.14.}
\end{figure*}
\clearpage
 \addtocounter{figure}{-1}
\begin{figure*}
\centering
\includegraphics[scale=.30,angle=0]{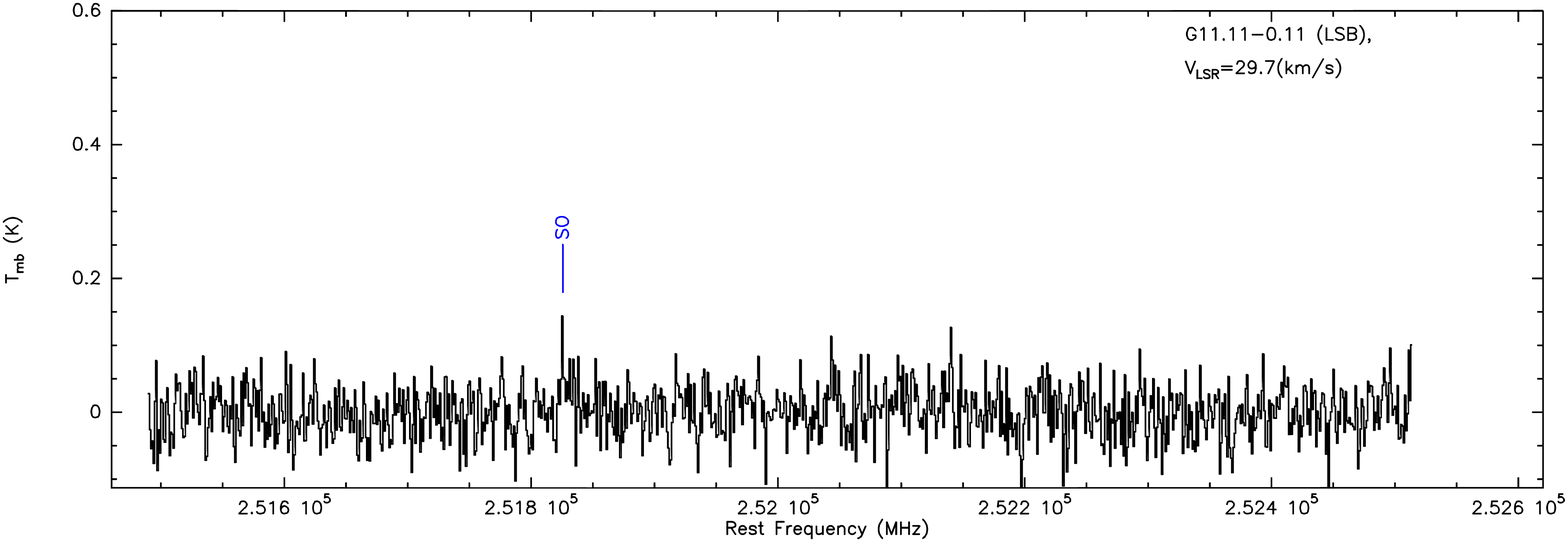}
\includegraphics[scale=.30,angle=0]{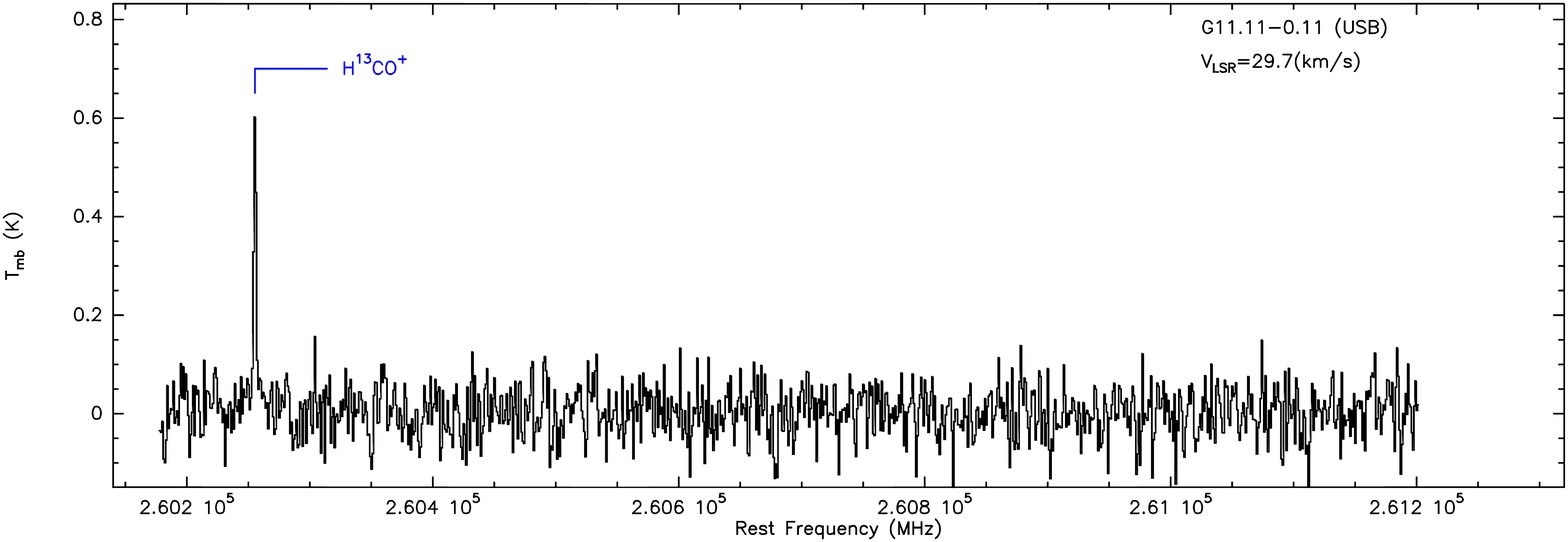}
\caption{(continued) For G11.11-0.11.}
\end{figure*}
 \addtocounter{figure}{-1}
\begin{figure*}
\centering
\includegraphics[scale=.30,angle=0]{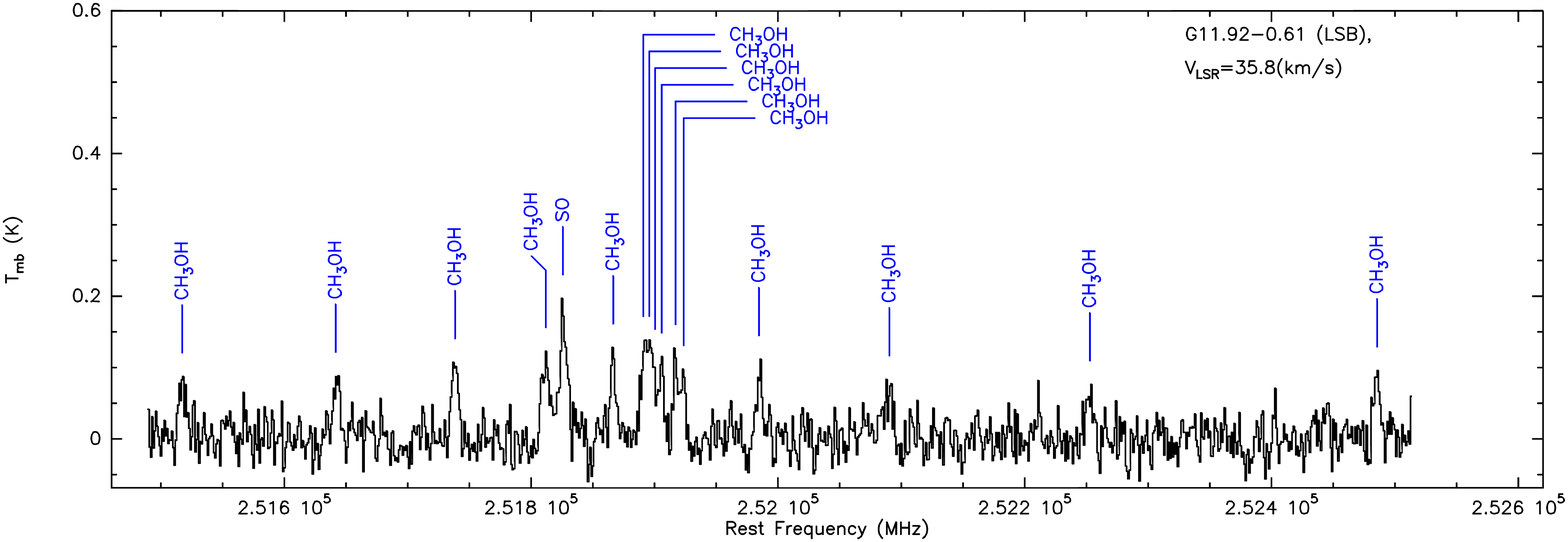}
\includegraphics[scale=.30,angle=0]{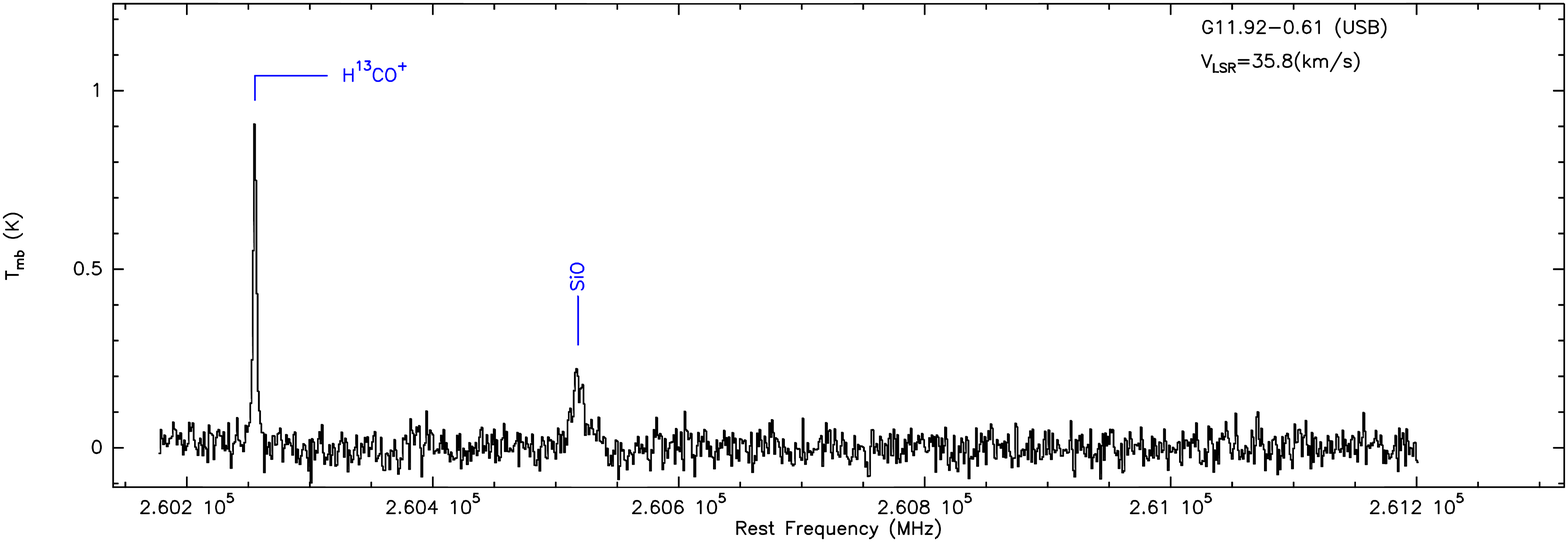}
\caption{(continued) For G11.92-0.61.}
\end{figure*}
\clearpage
 \addtocounter{figure}{-1}
\begin{figure*}
\centering
\includegraphics[scale=.30,angle=0]{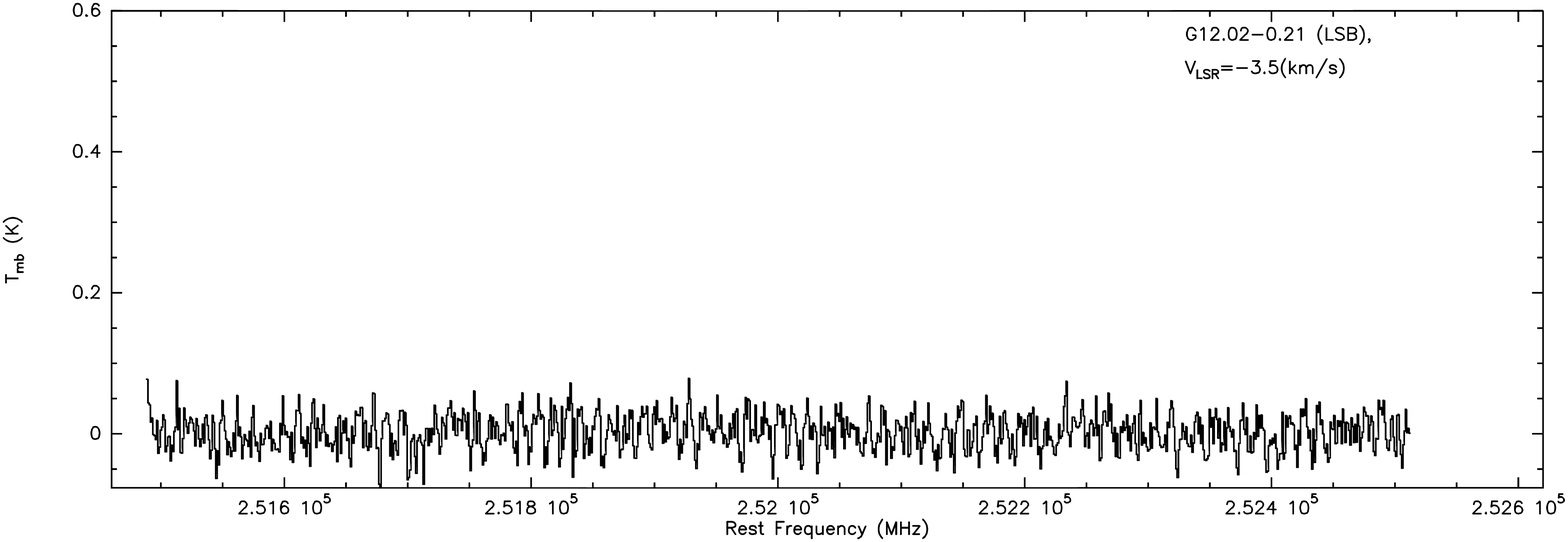}
\includegraphics[scale=.30,angle=0]{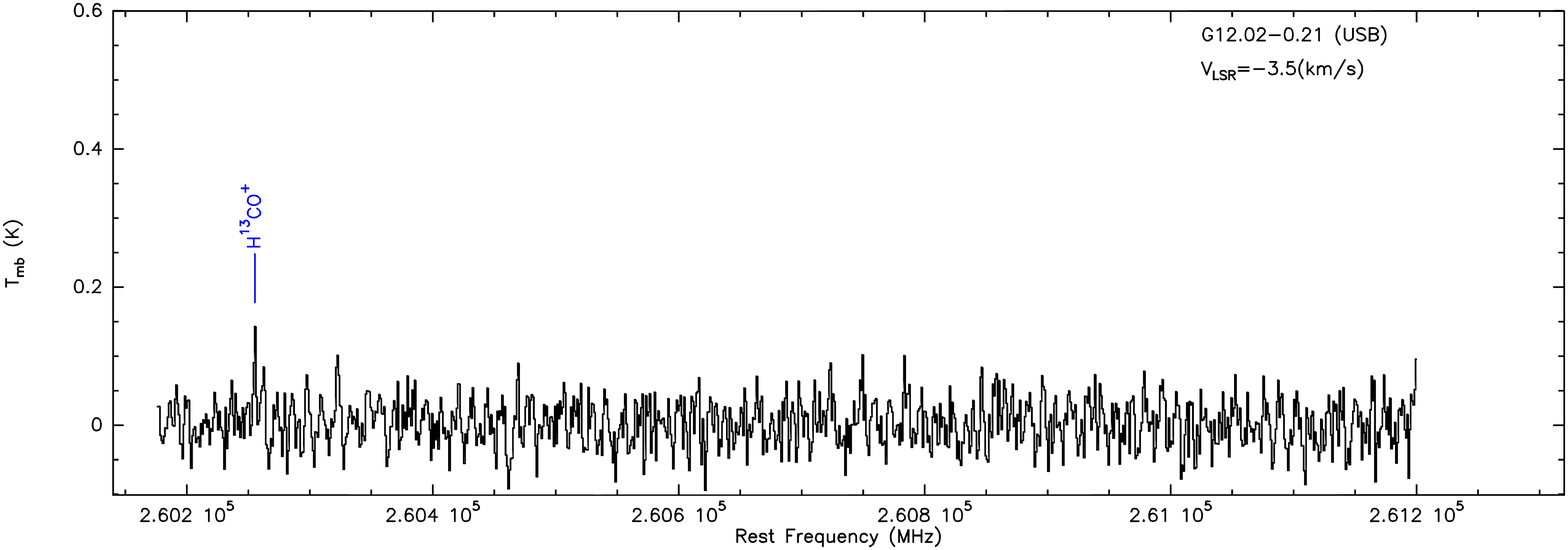}
\caption{(continued) For G12.02-0.21.}
\end{figure*}
 \addtocounter{figure}{-1}
\begin{figure*}
\centering
\includegraphics[scale=.30,angle=0]{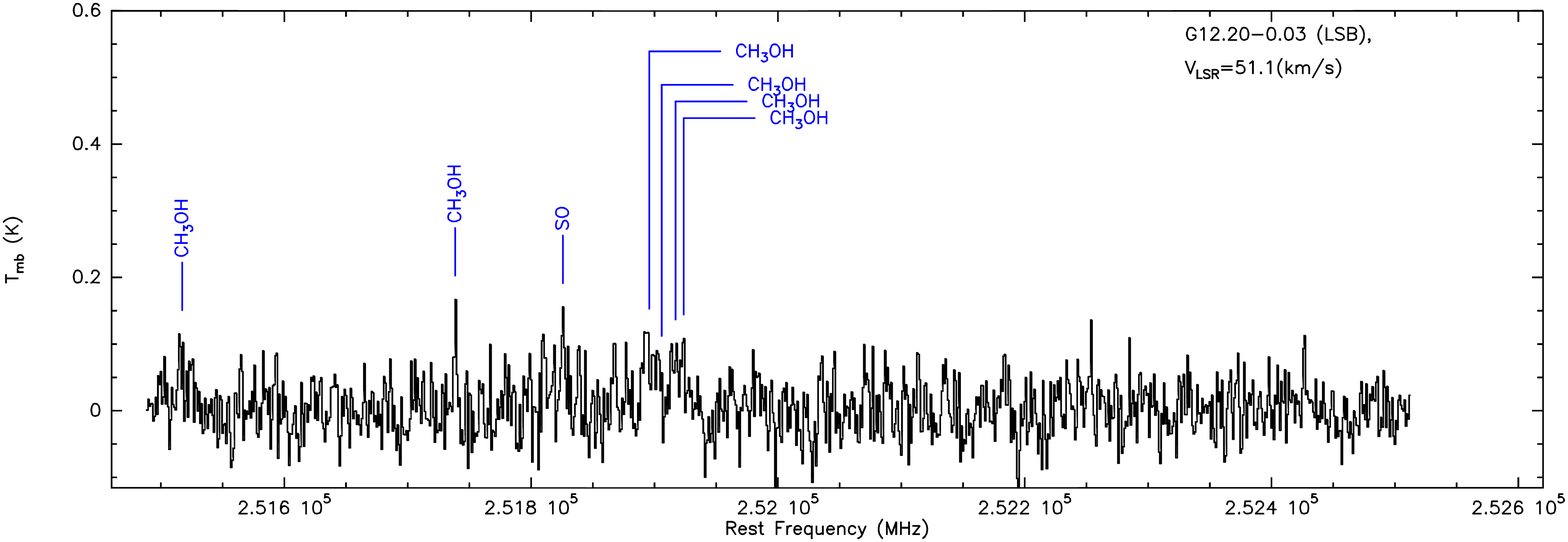}
\includegraphics[scale=.30,angle=0]{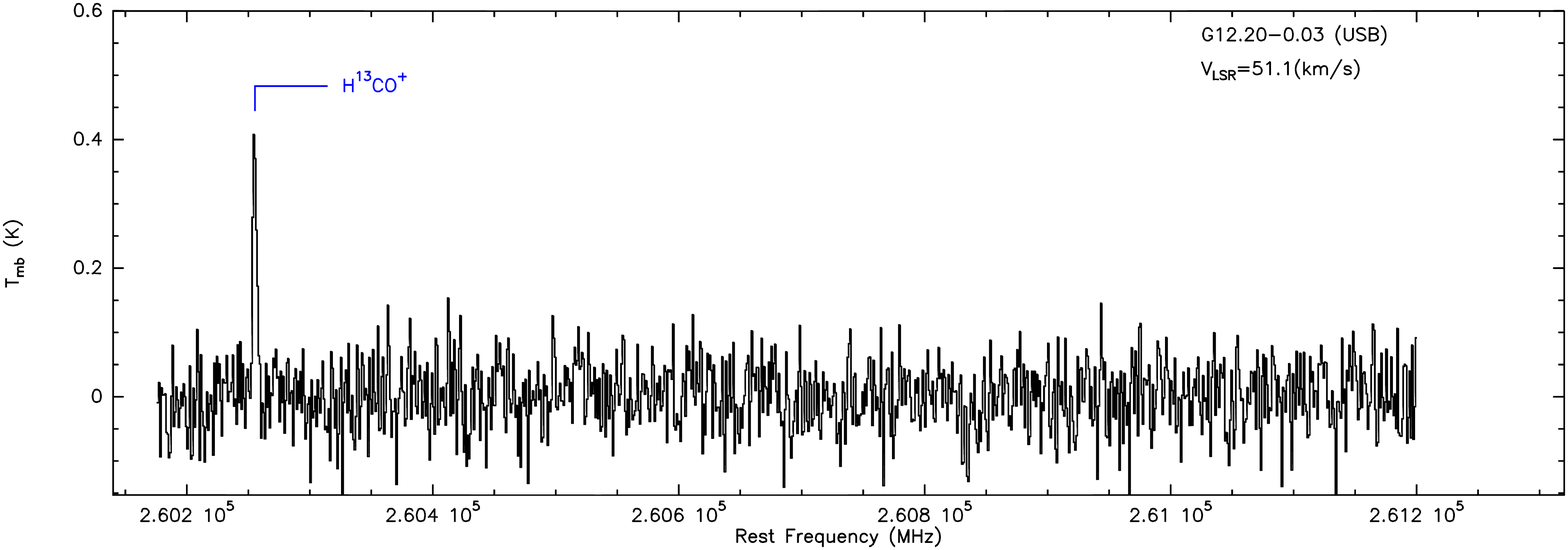}
\caption{(continued) For G12.20-0.03.}
\end{figure*}
\clearpage
 \addtocounter{figure}{-1}
\begin{figure*}
\centering
\includegraphics[scale=.30,angle=0]{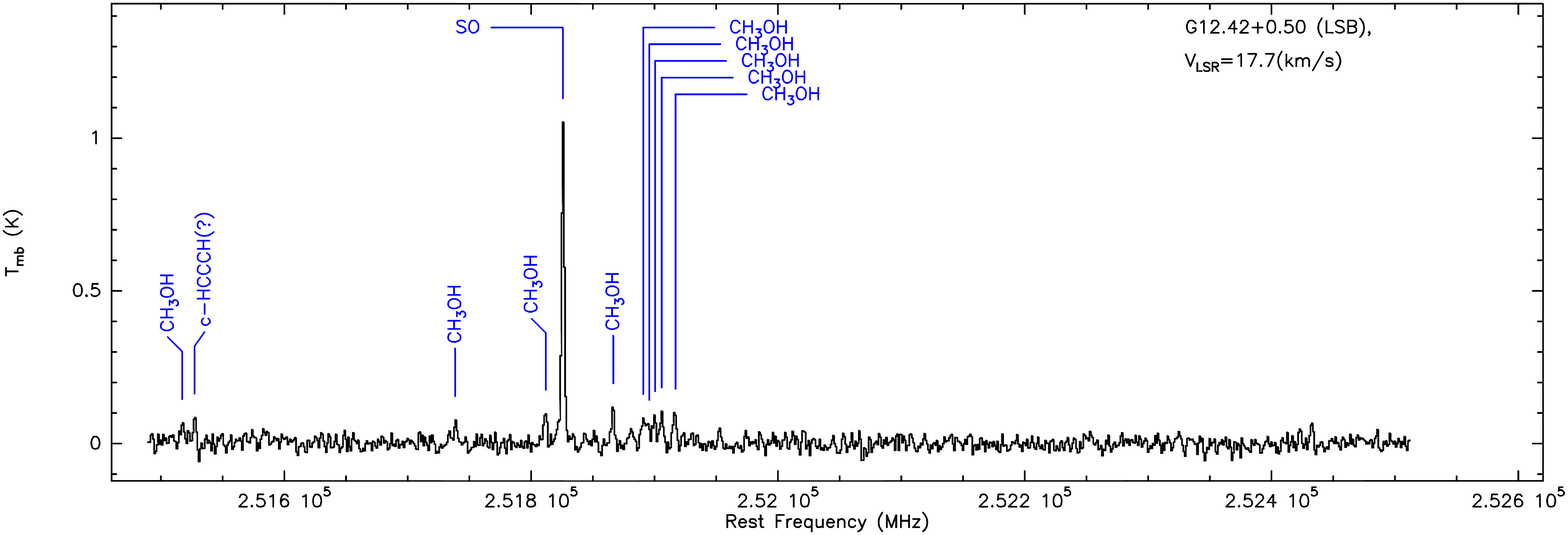}
\includegraphics[scale=.30,angle=0]{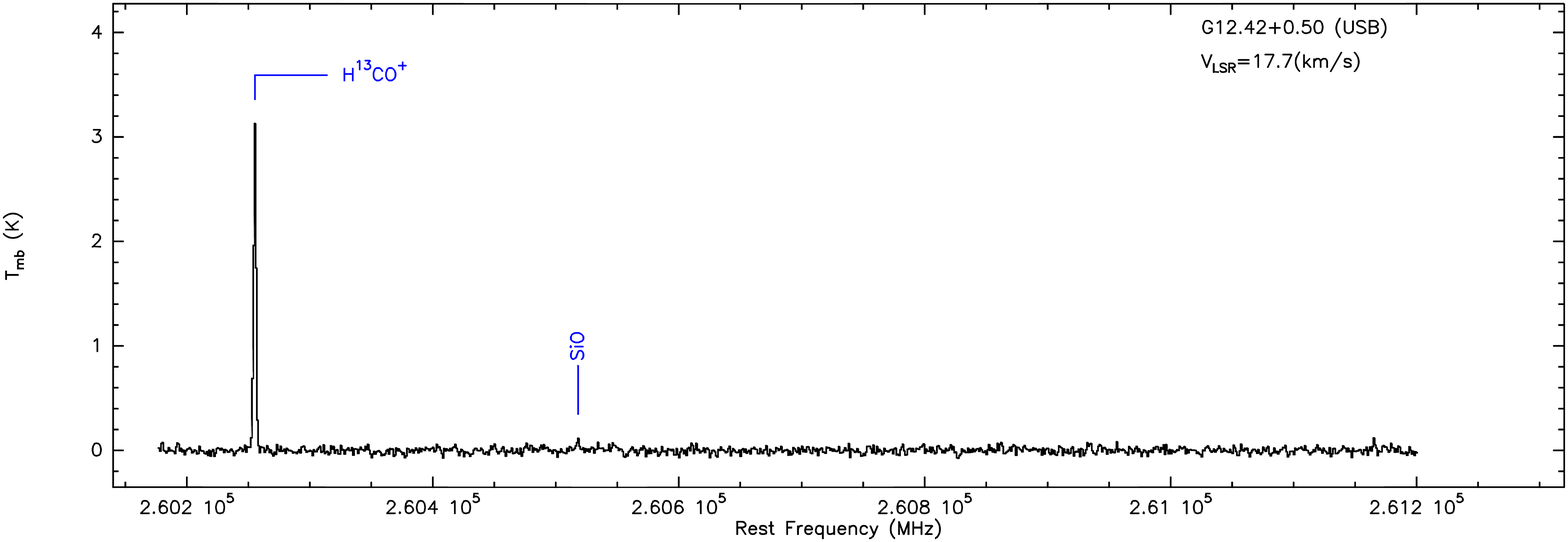}
\caption{(continued) For G12.42+0.50.}
\end{figure*}
 \addtocounter{figure}{-1}
\begin{figure*}
\centering
\includegraphics[scale=.30,angle=0]{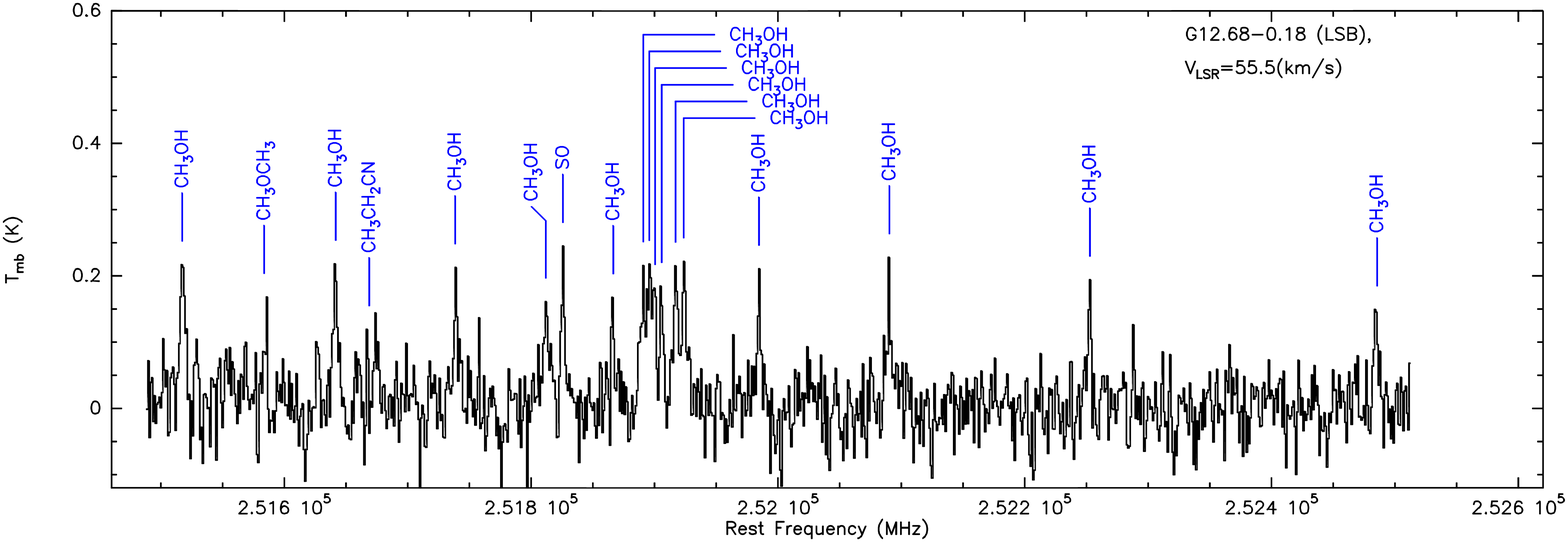}
\includegraphics[scale=.30,angle=0]{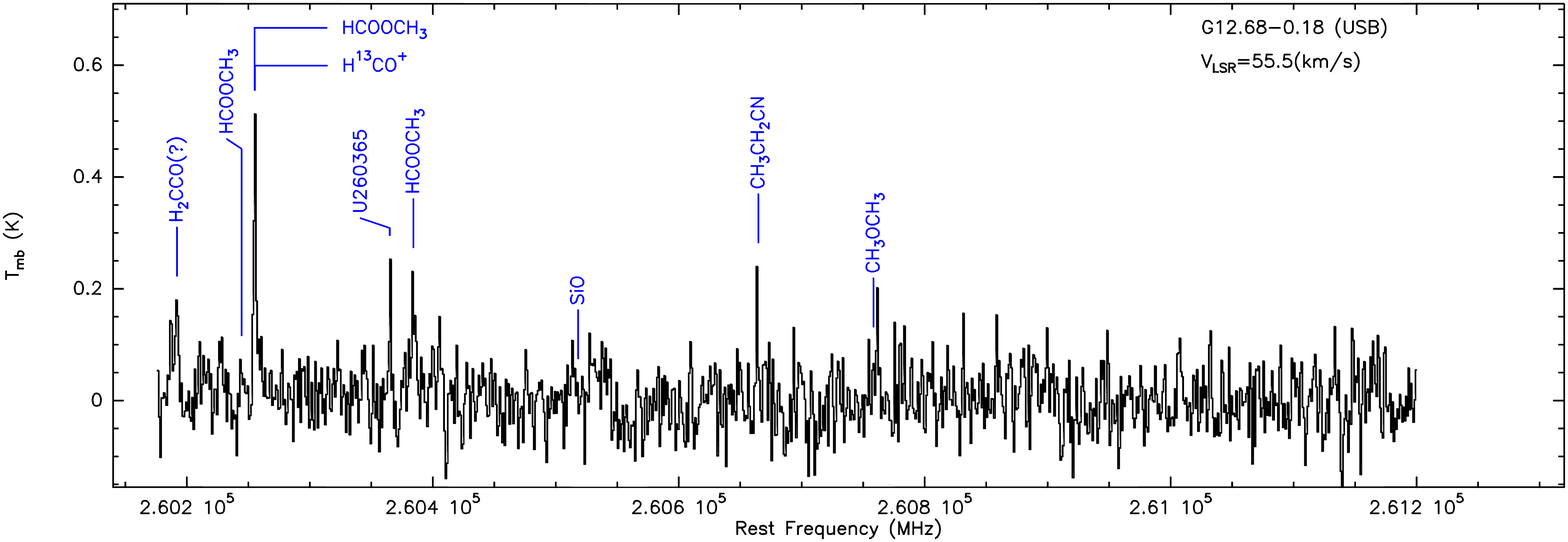}
\caption{(continued) For G12.68-0.18.}
\end{figure*}
\clearpage
 \addtocounter{figure}{-1}
\begin{figure*}
\centering
\includegraphics[scale=.30,angle=0]{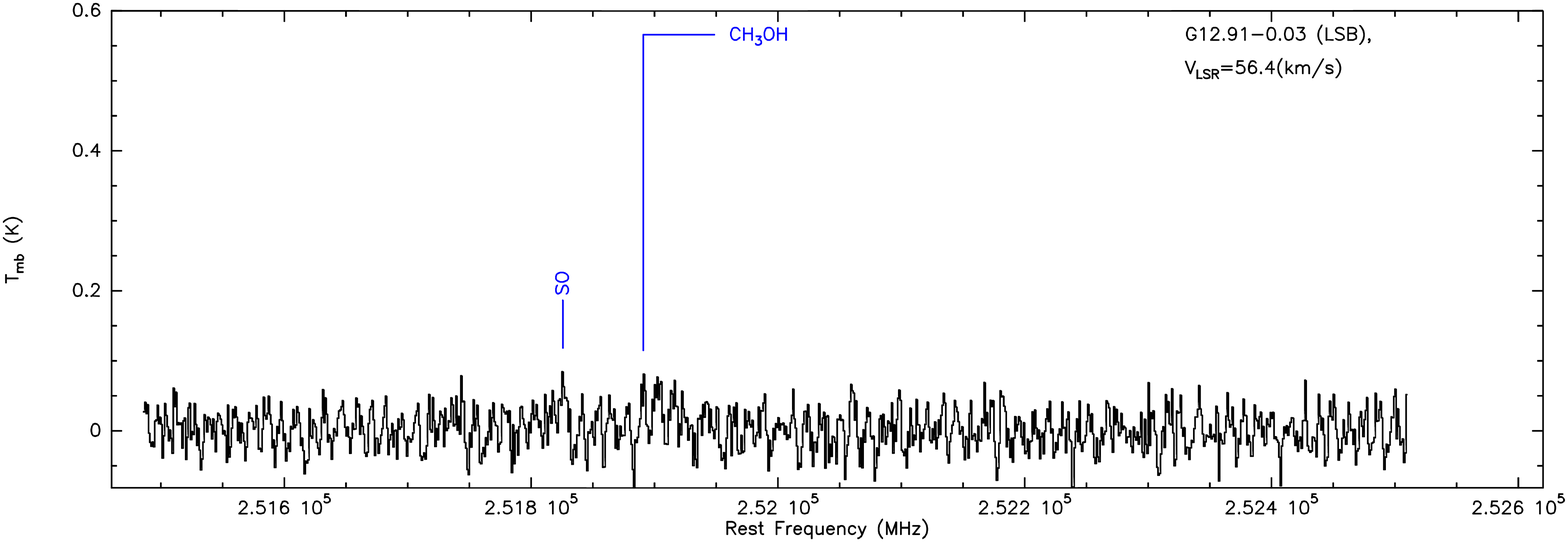}
\includegraphics[scale=.30,angle=0]{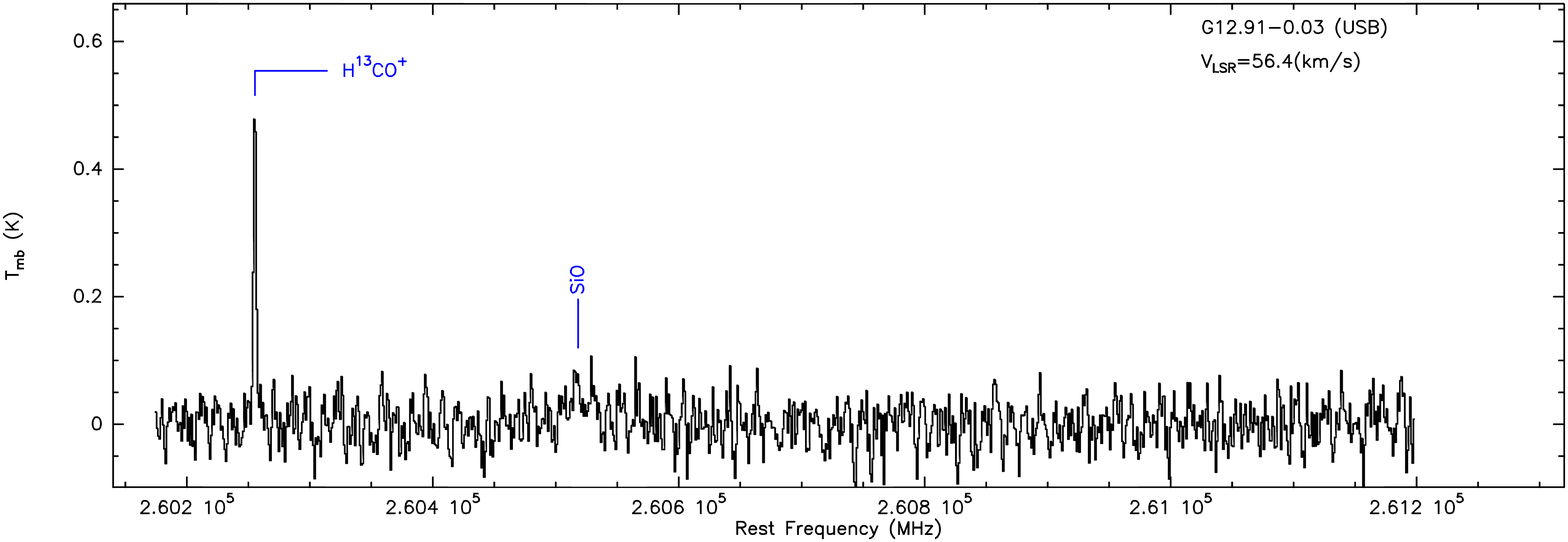}
\caption{(continued) For G12.91-0.03.}
\end{figure*}
 \addtocounter{figure}{-1}
\begin{figure*}
\centering
\includegraphics[scale=.30,angle=0]{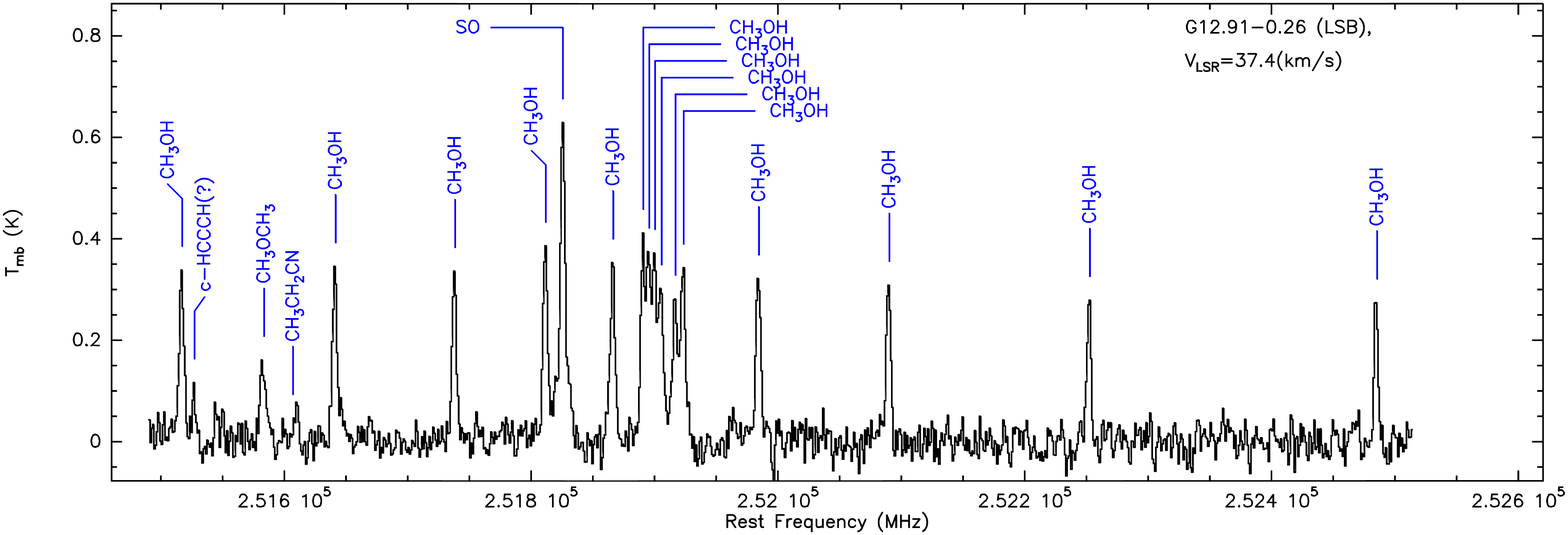}
\includegraphics[scale=.30,angle=0]{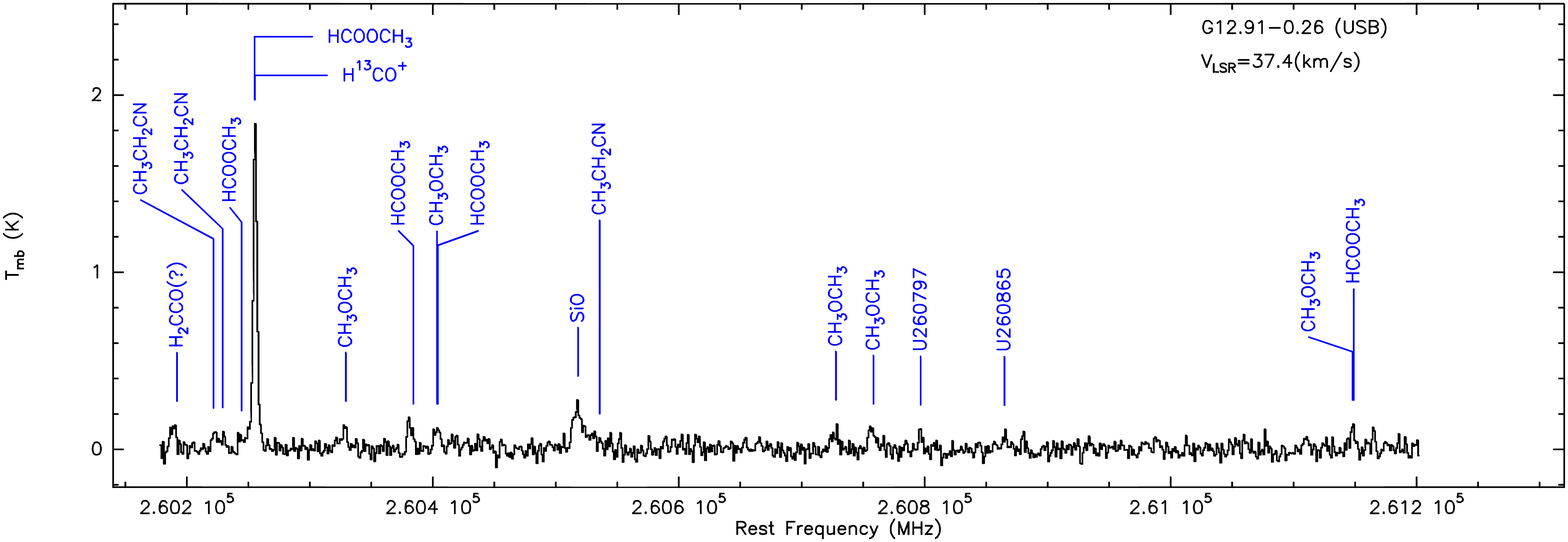}
\caption{(continued) For G12.91-0.26.}
\end{figure*}
\clearpage
 \addtocounter{figure}{-1}
\begin{figure*}
\centering
\includegraphics[scale=.30,angle=0]{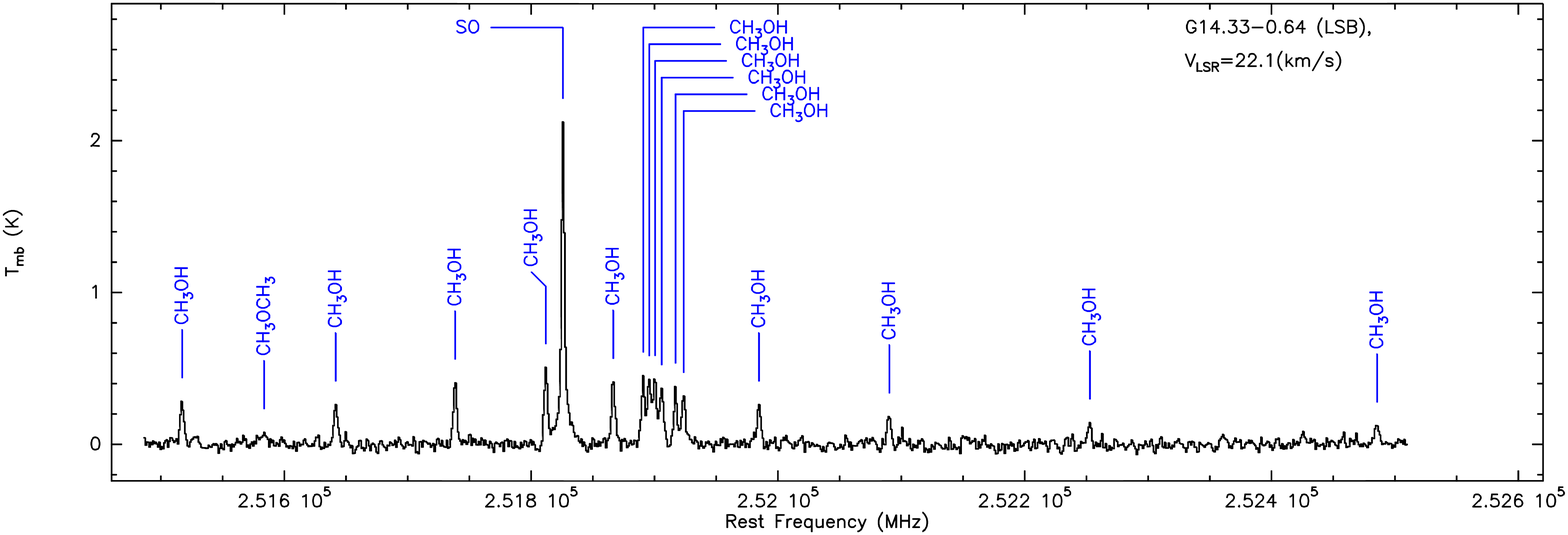}
\includegraphics[scale=.30,angle=0]{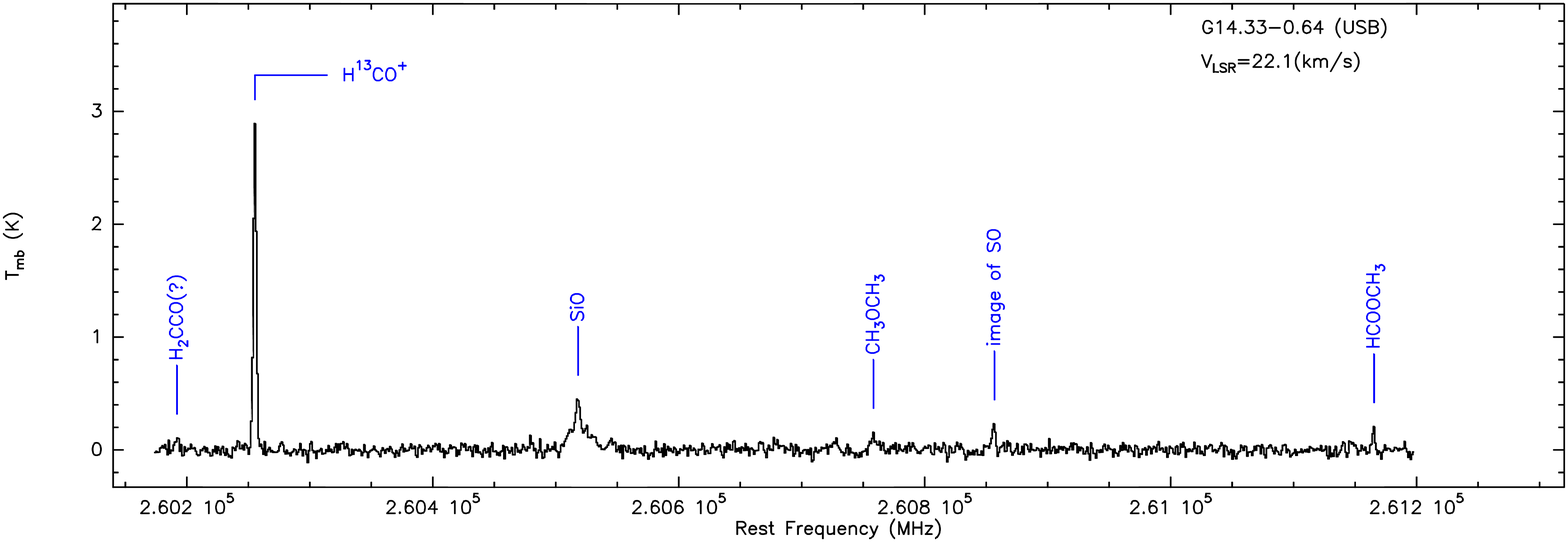}
\caption{(continued) For G14.33-0.64.}
\end{figure*}
 \addtocounter{figure}{-1}
\begin{figure*}
\centering
\includegraphics[scale=.30,angle=0]{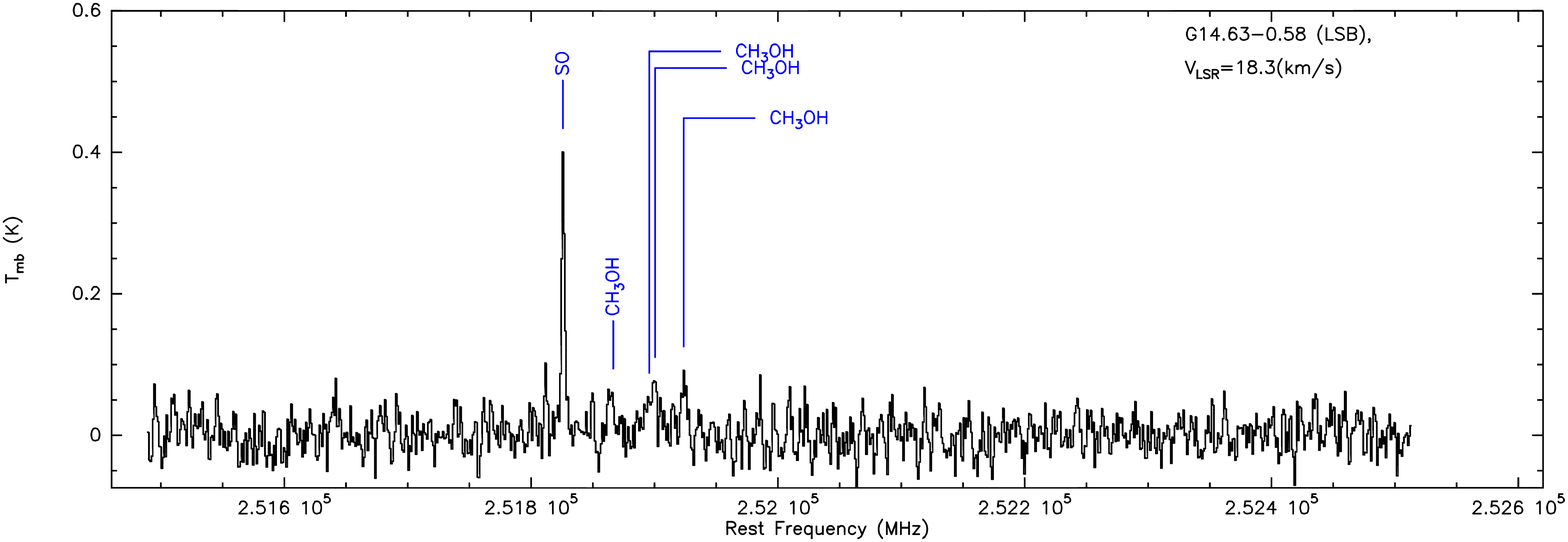}
\includegraphics[scale=.30,angle=0]{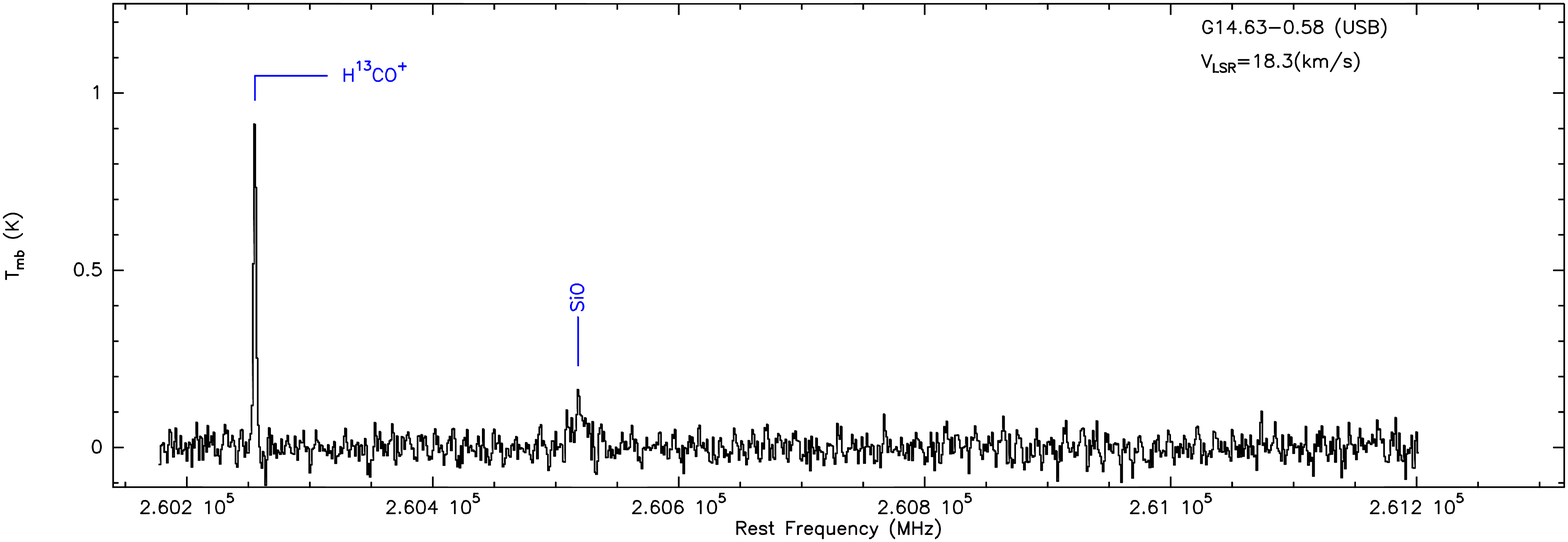}
\caption{(continued) For G14.63-0.58.}
\end{figure*}
\clearpage
 \addtocounter{figure}{-1}
\begin{figure*}
\centering
\includegraphics[scale=.30,angle=0]{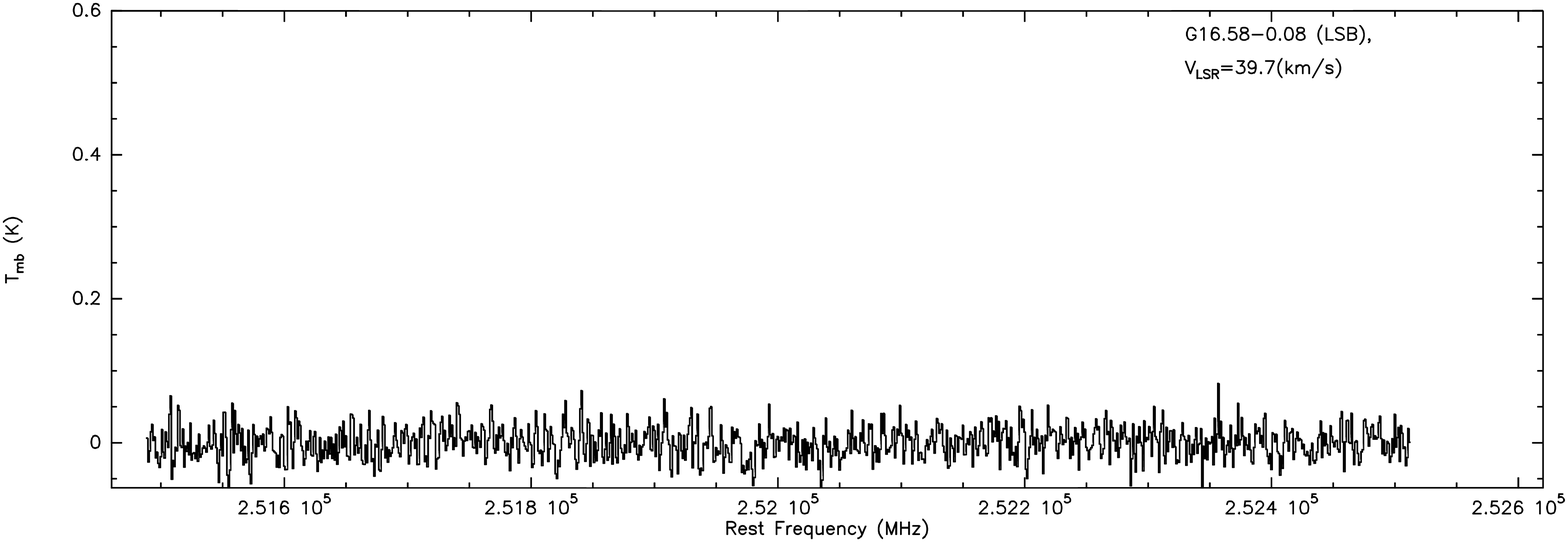}
\includegraphics[scale=.30,angle=0]{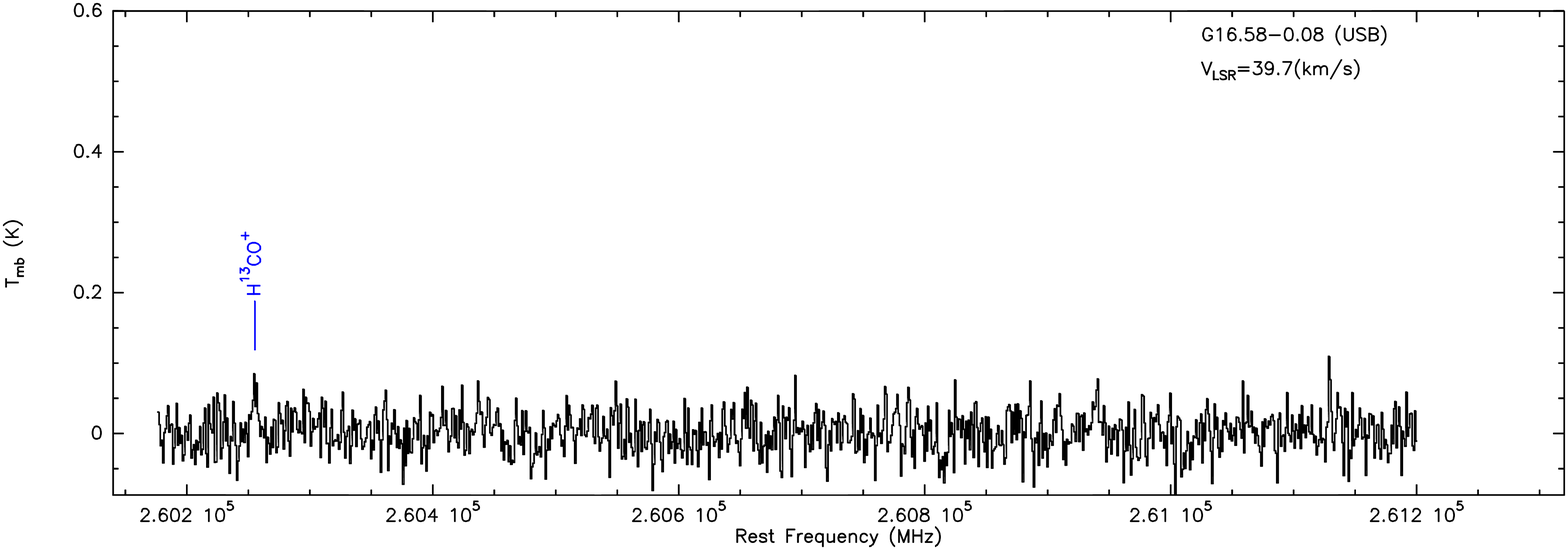}
\caption{(continued) For G16.58-0.08.}
\end{figure*}
 \addtocounter{figure}{-1}
\begin{figure*}
\centering
\includegraphics[scale=.30,angle=0]{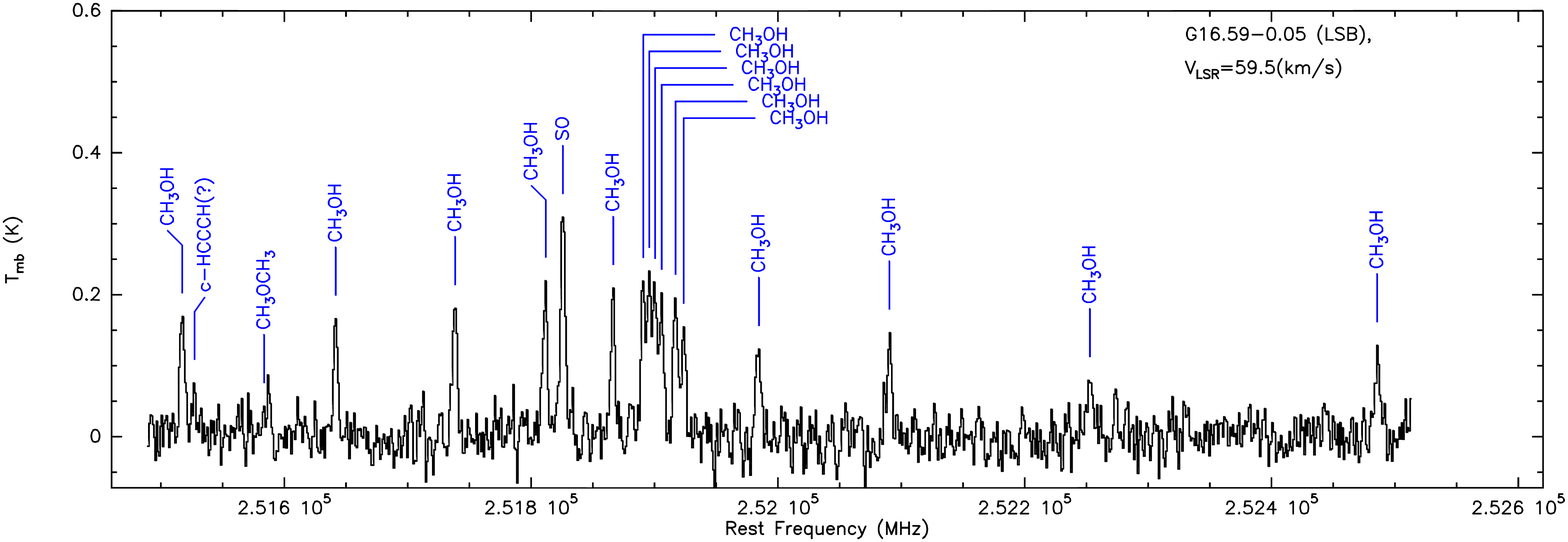}
\includegraphics[scale=.30,angle=0]{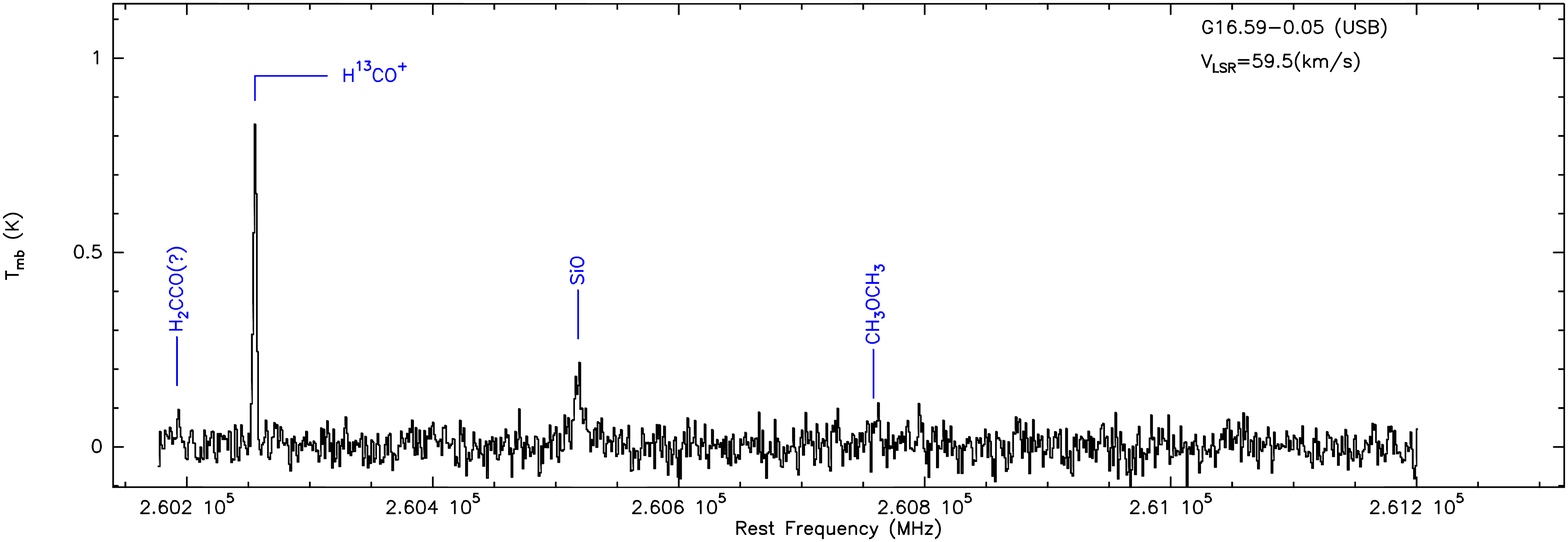}
\caption{(continued) For G16.59-0.05.}
\end{figure*}
\clearpage
 \addtocounter{figure}{-1}
\begin{figure*}
\centering
\includegraphics[scale=.30,angle=0]{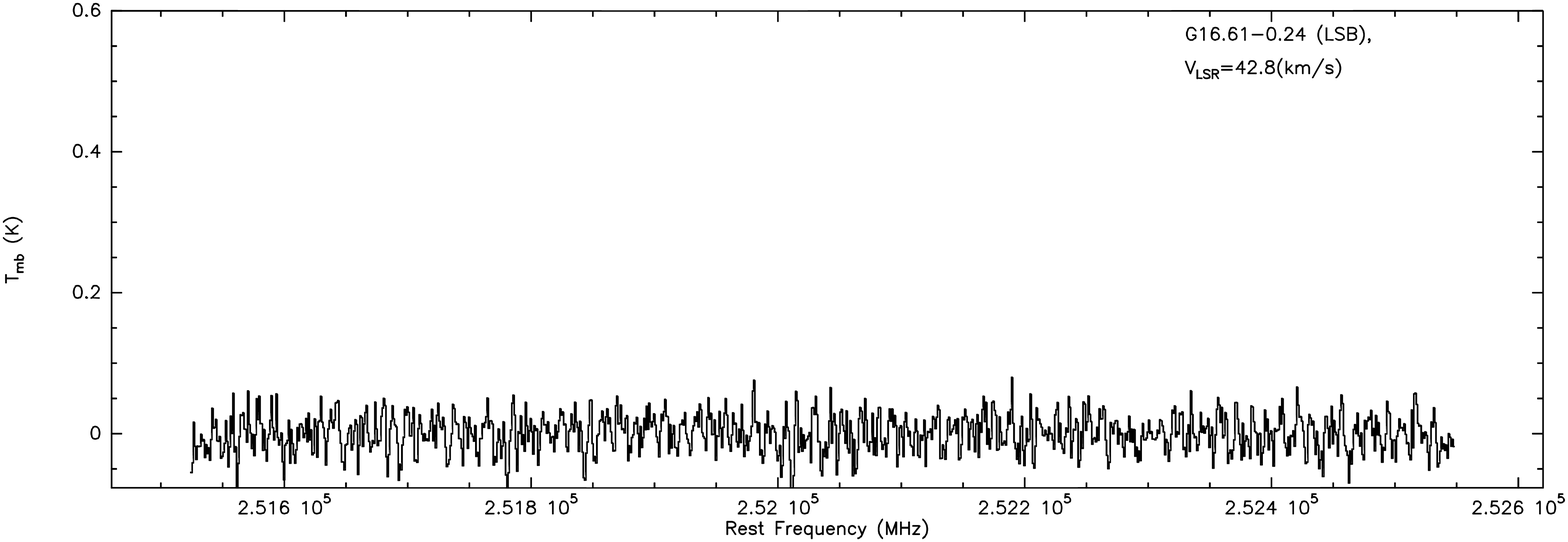}
\includegraphics[scale=.30,angle=0]{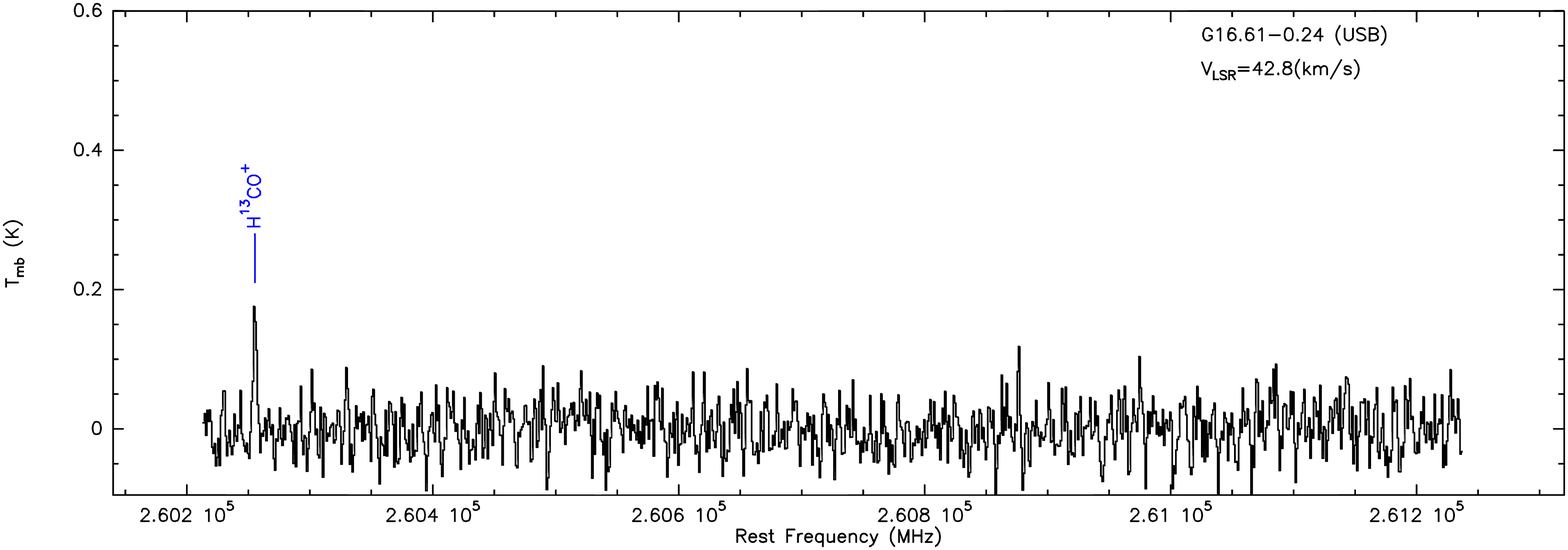}
\caption{(continued) For G16.61-0.24.}
\end{figure*}
 \addtocounter{figure}{-1}
\begin{figure*}
\centering
\includegraphics[scale=.30,angle=0]{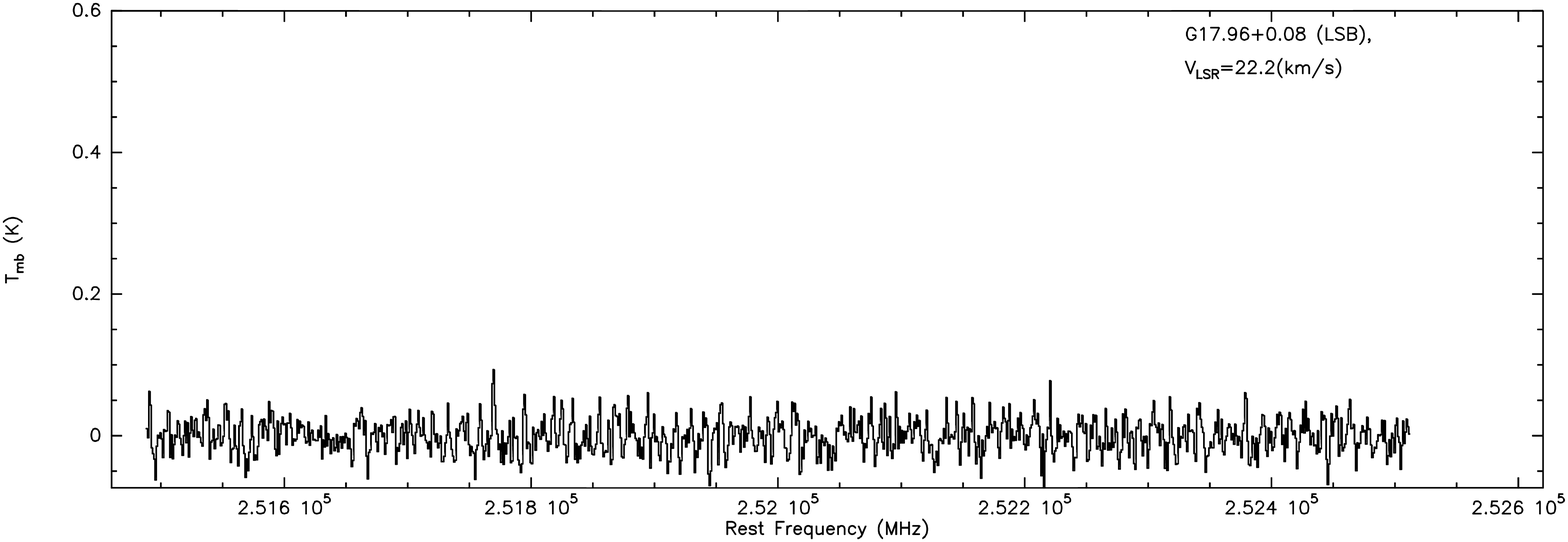}
\includegraphics[scale=.30,angle=0]{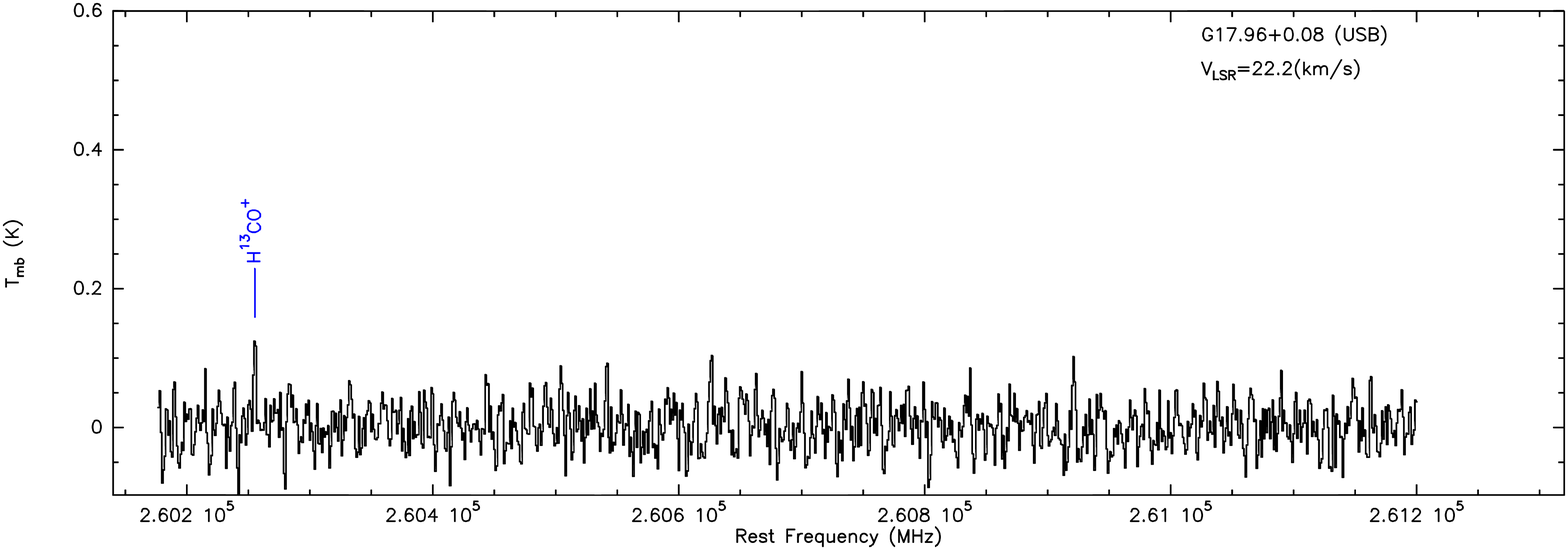}
\caption{(continued) For G17.96+0.08.}
\end{figure*}
\clearpage
 \addtocounter{figure}{-1}
\begin{figure*}
\centering
\includegraphics[scale=.30,angle=0]{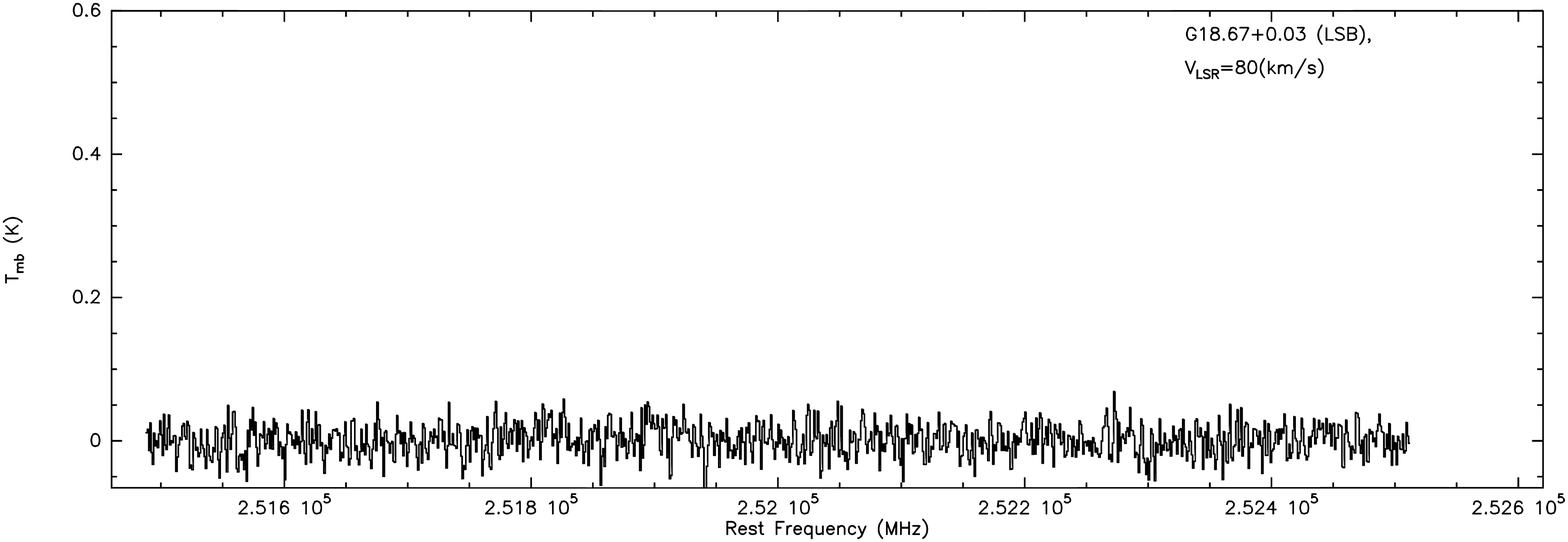}
\includegraphics[scale=.30,angle=0]{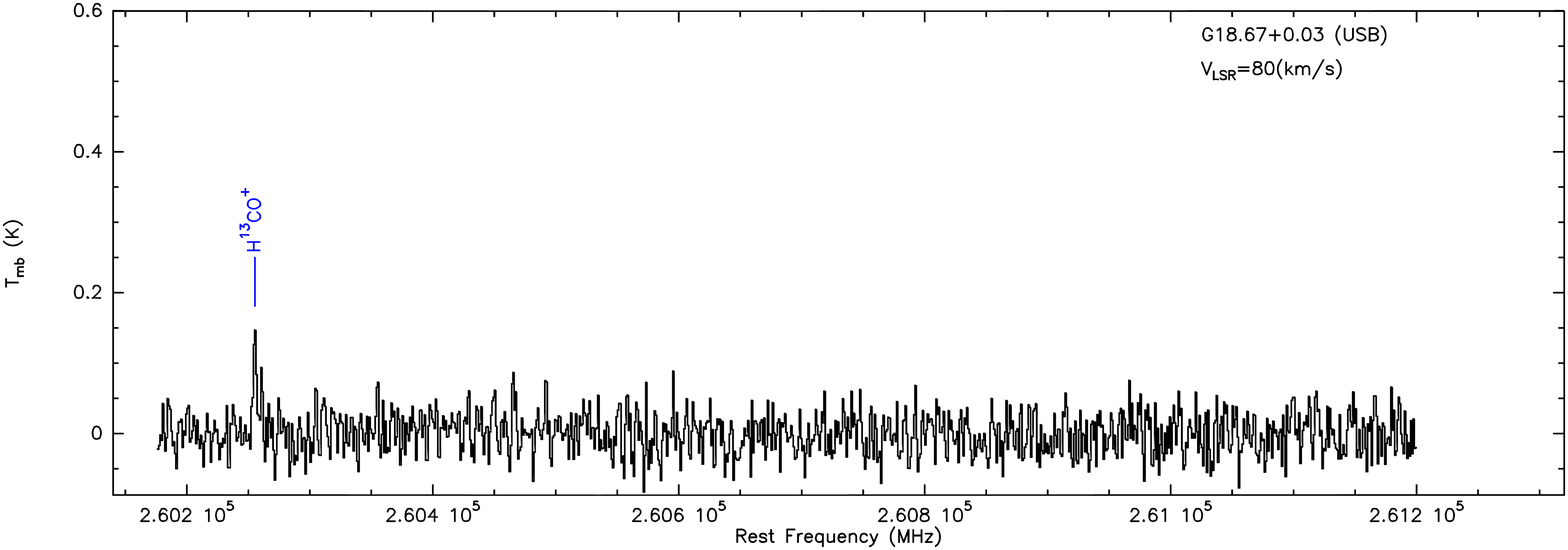}
\caption{(continued) For G18.67+0.03.}
\end{figure*}
 \addtocounter{figure}{-1}
\begin{figure*}
\centering
\includegraphics[scale=.30,angle=0]{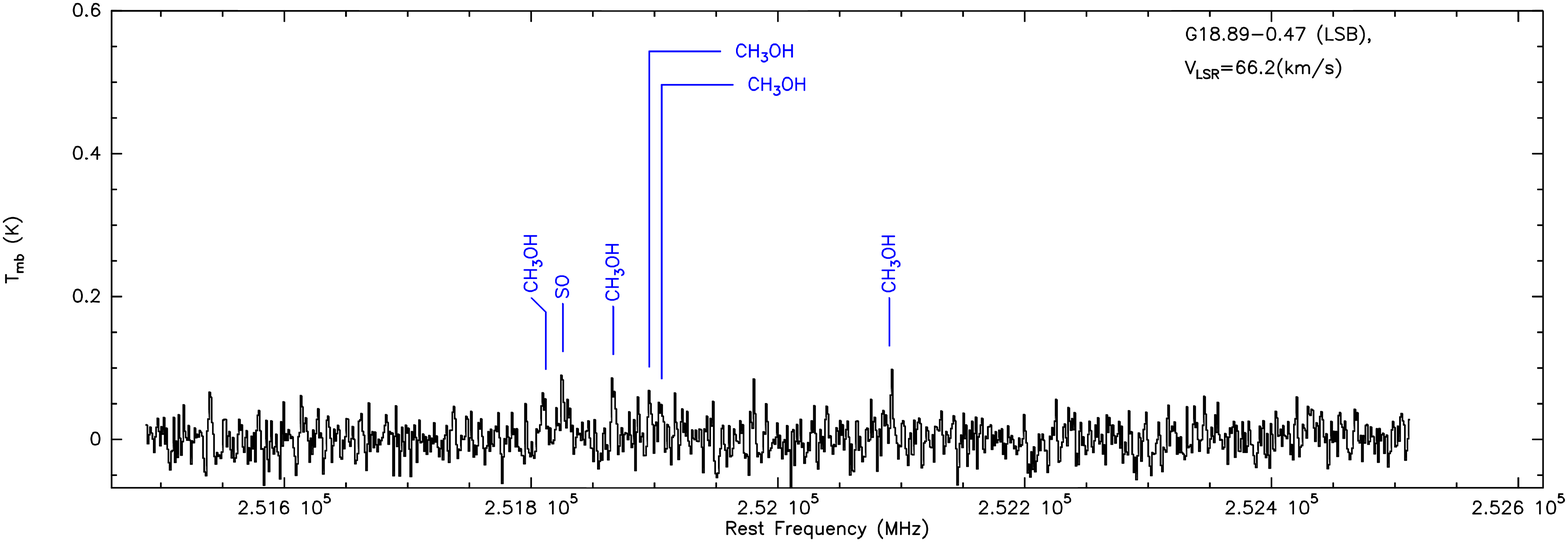}
\includegraphics[scale=.30,angle=0]{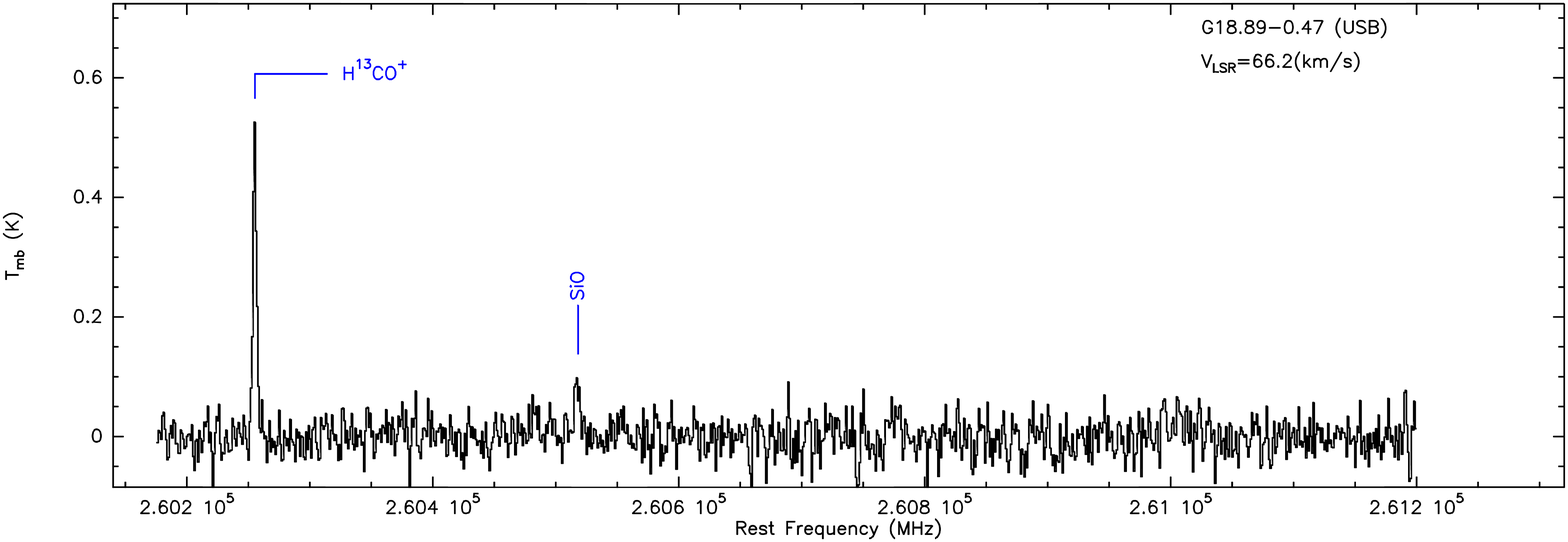}
\caption{(continued) For G18.89-0.47.}
\end{figure*}
\clearpage
 \addtocounter{figure}{-1}
\begin{figure*}
\centering
\includegraphics[scale=.30,angle=0]{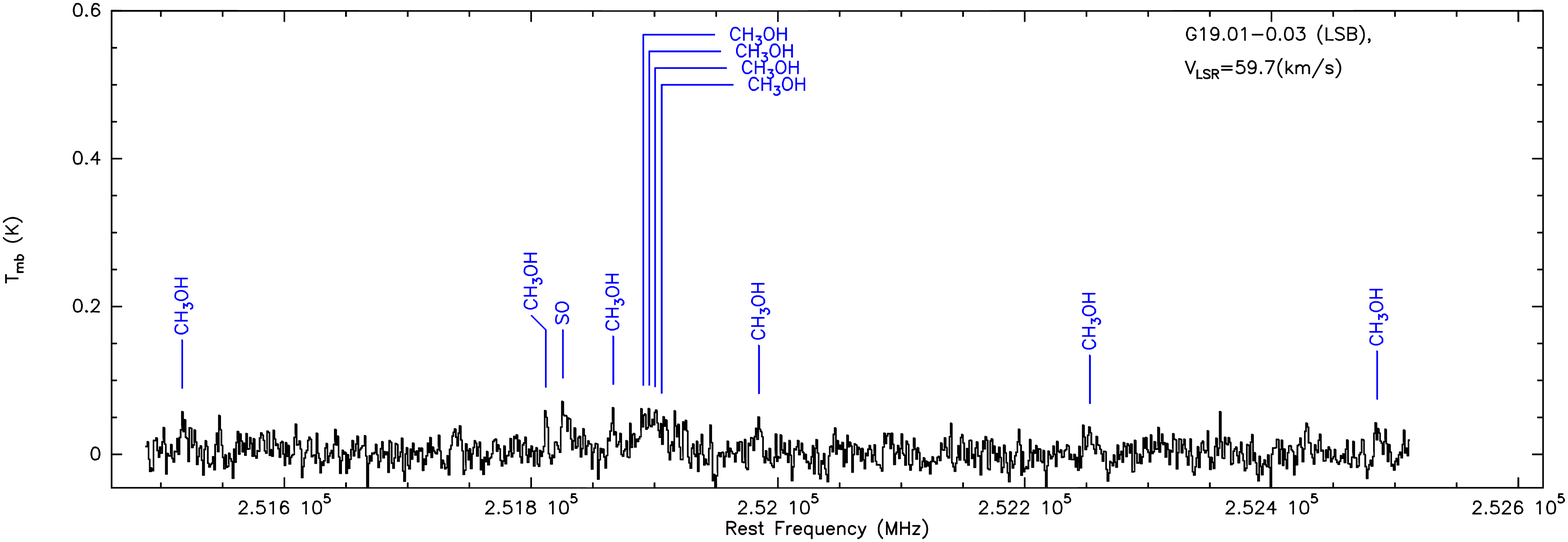}
\includegraphics[scale=.30,angle=0]{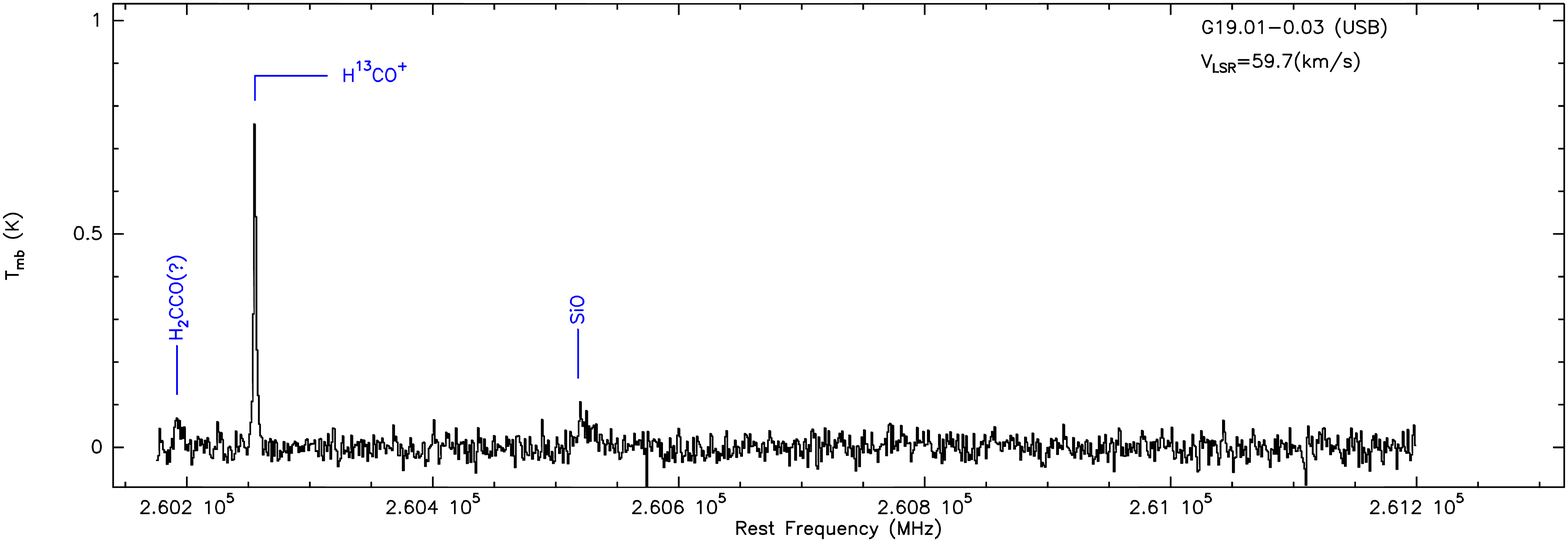}
\caption{(continued) For G19.01-0.03.}
\end{figure*}
 \addtocounter{figure}{-1}
\begin{figure*}
\centering
\includegraphics[scale=.30,angle=0]{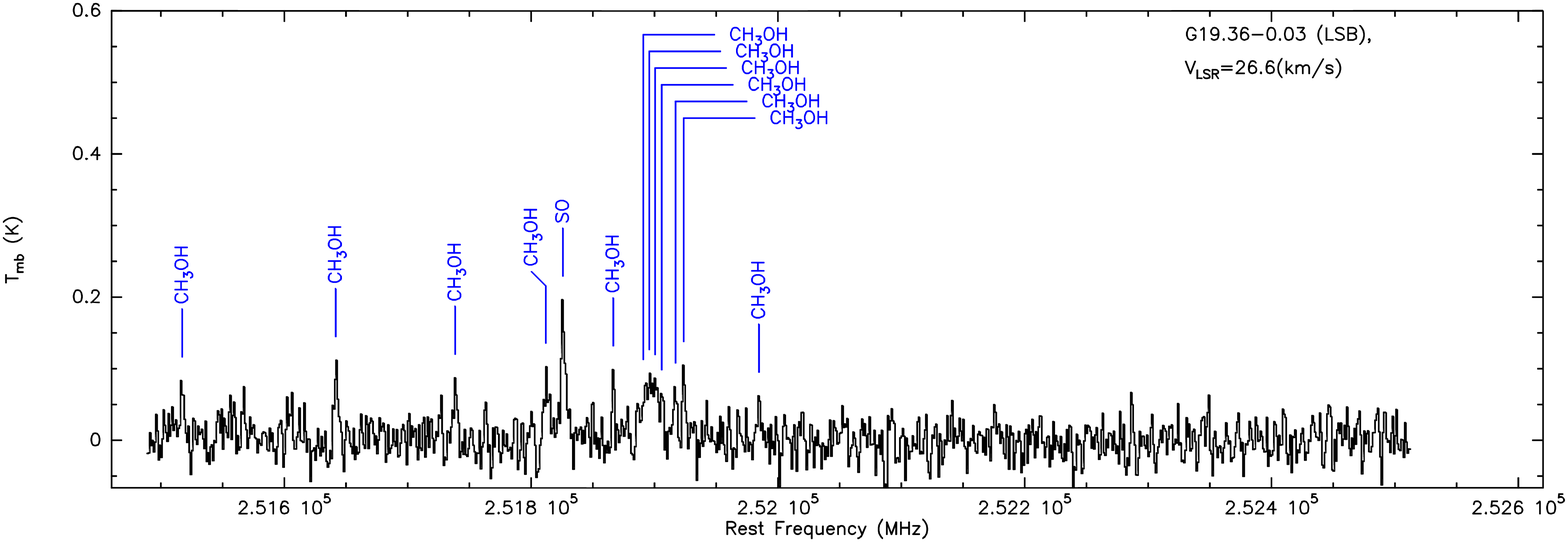}
\includegraphics[scale=.30,angle=0]{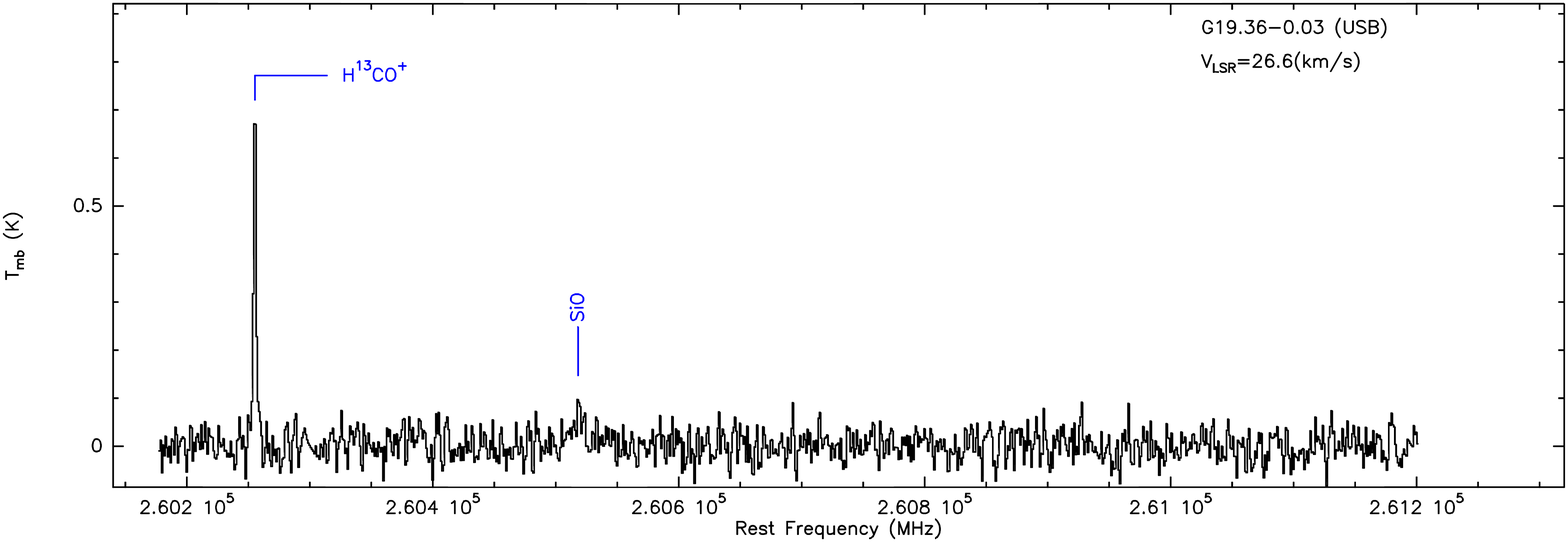}
\caption{(continued) For G19.36-0.03.}
\end{figure*}
\clearpage
 \addtocounter{figure}{-1}
\begin{figure*}
\centering
\includegraphics[scale=.30,angle=0]{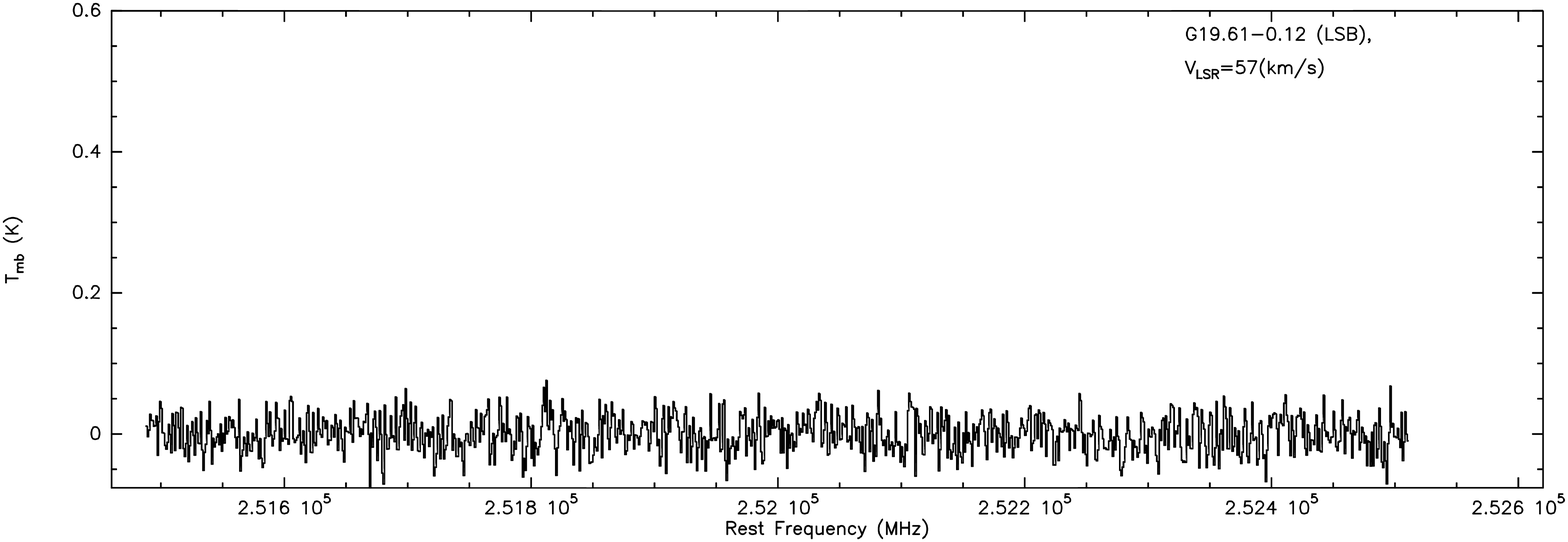}
\includegraphics[scale=.30,angle=0]{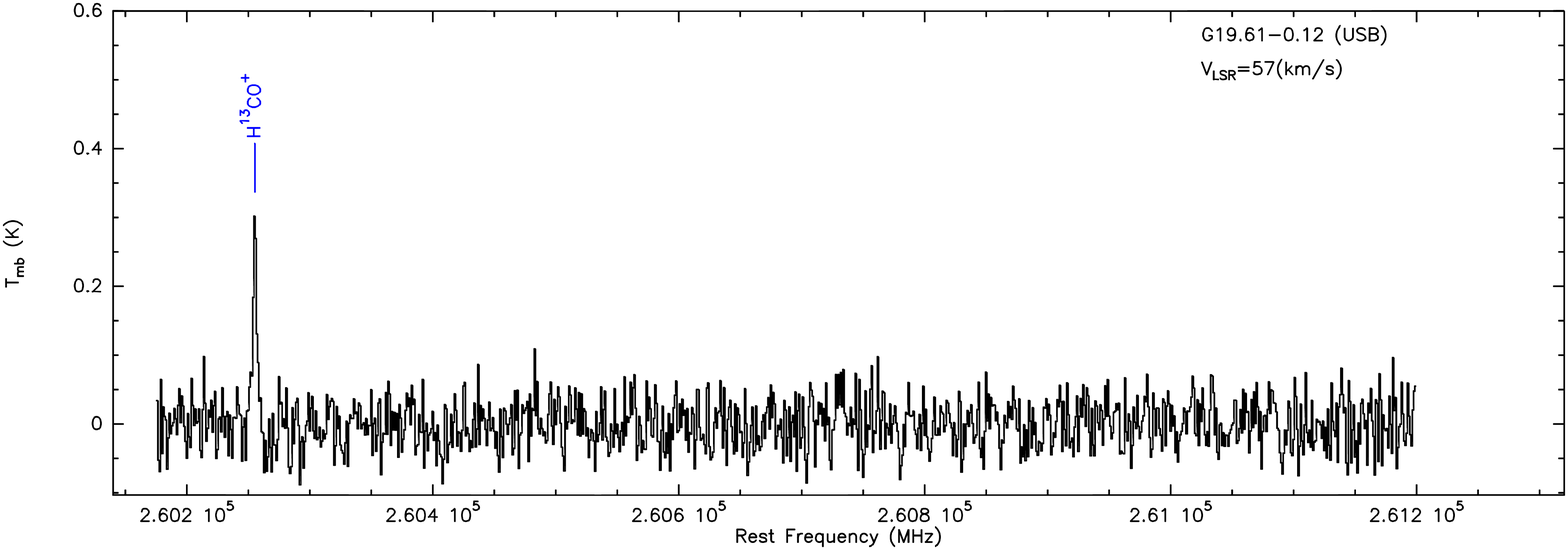}
\caption{(continued) For G19.61-0.12.}
\end{figure*}
 \addtocounter{figure}{-1}
\begin{figure*}
\centering
\includegraphics[scale=.30,angle=0]{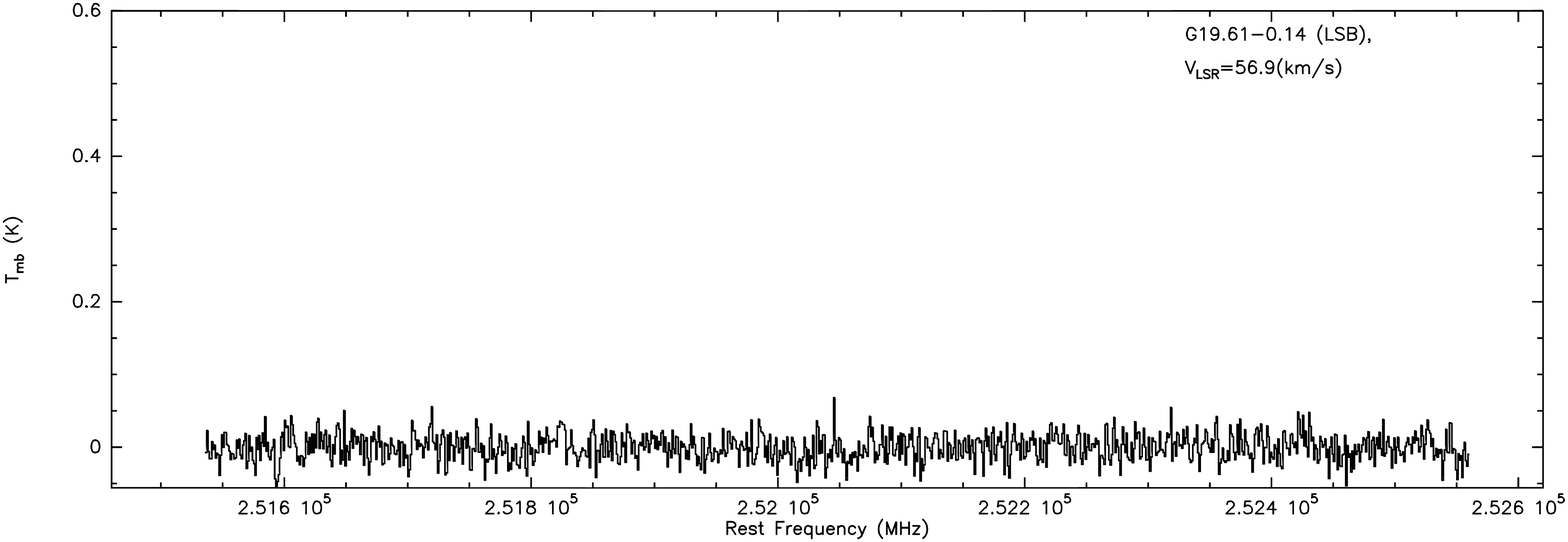}
\includegraphics[scale=.30,angle=0]{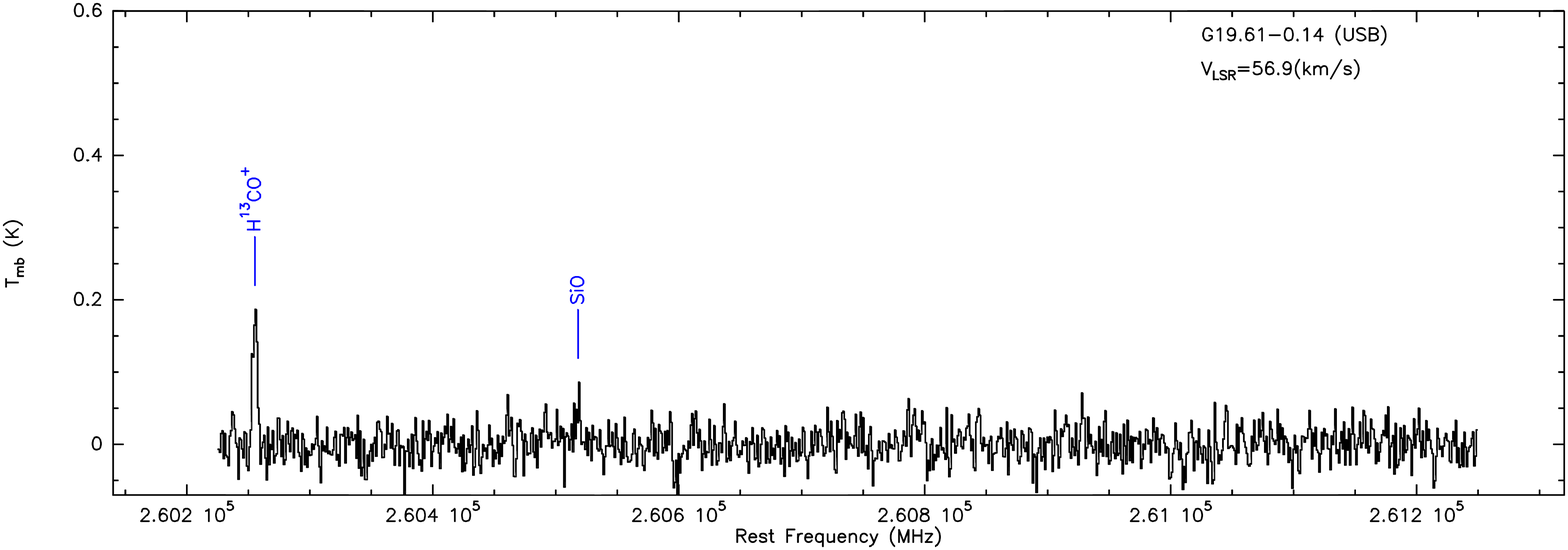}
\caption{(continued) For G19.61-0.14.}
\end{figure*}
\clearpage
 \addtocounter{figure}{-1}
\begin{figure*}
\centering
\includegraphics[scale=.30,angle=0]{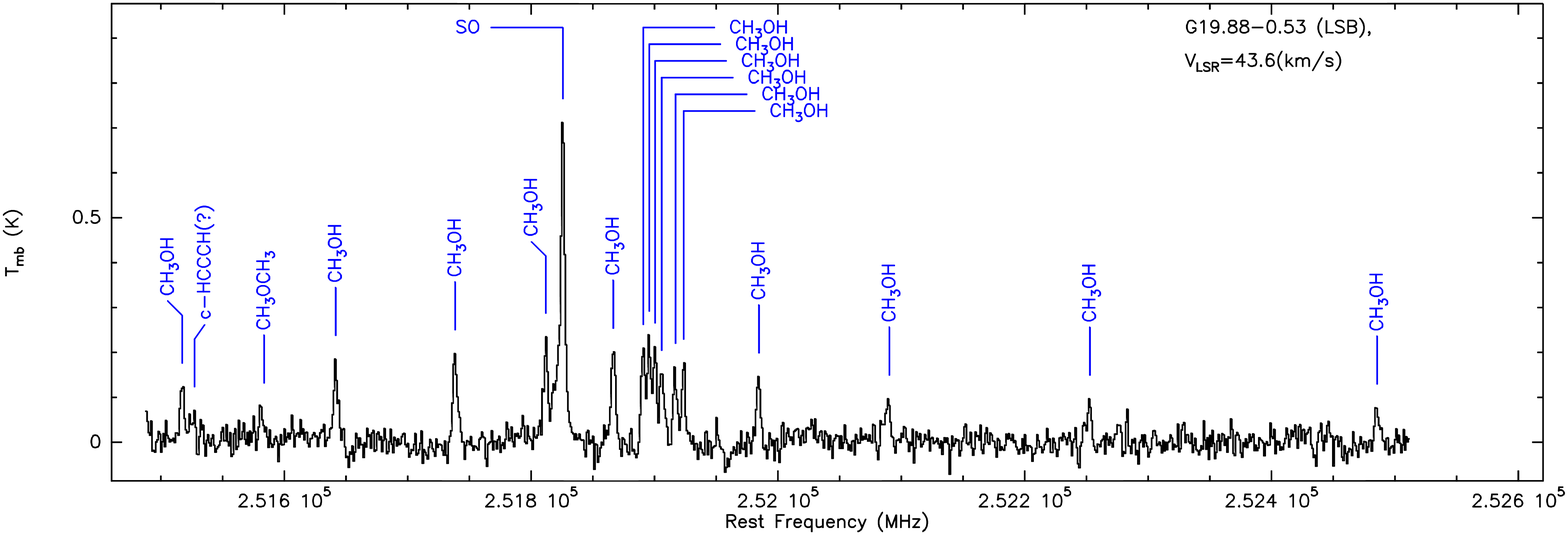}
\includegraphics[scale=.30,angle=0]{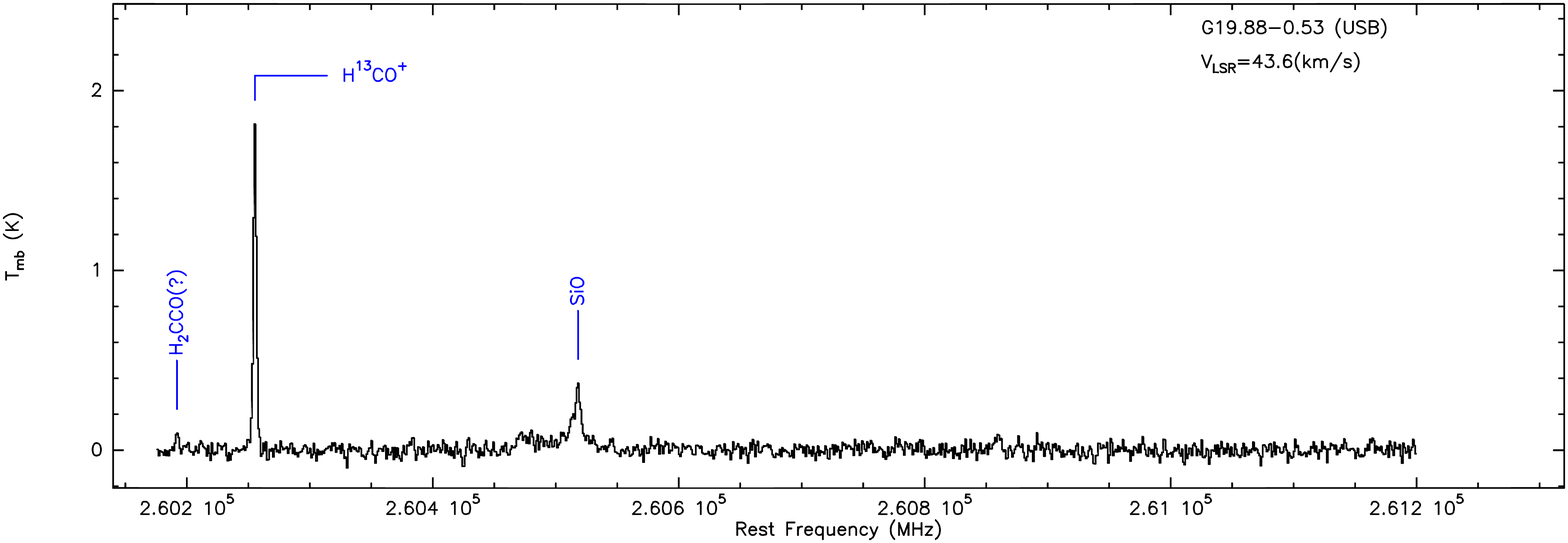}
\caption{(continued) For G19.88-0.53.}
\end{figure*}
 \addtocounter{figure}{-1}
\begin{figure*}
\centering
\includegraphics[scale=.30,angle=0]{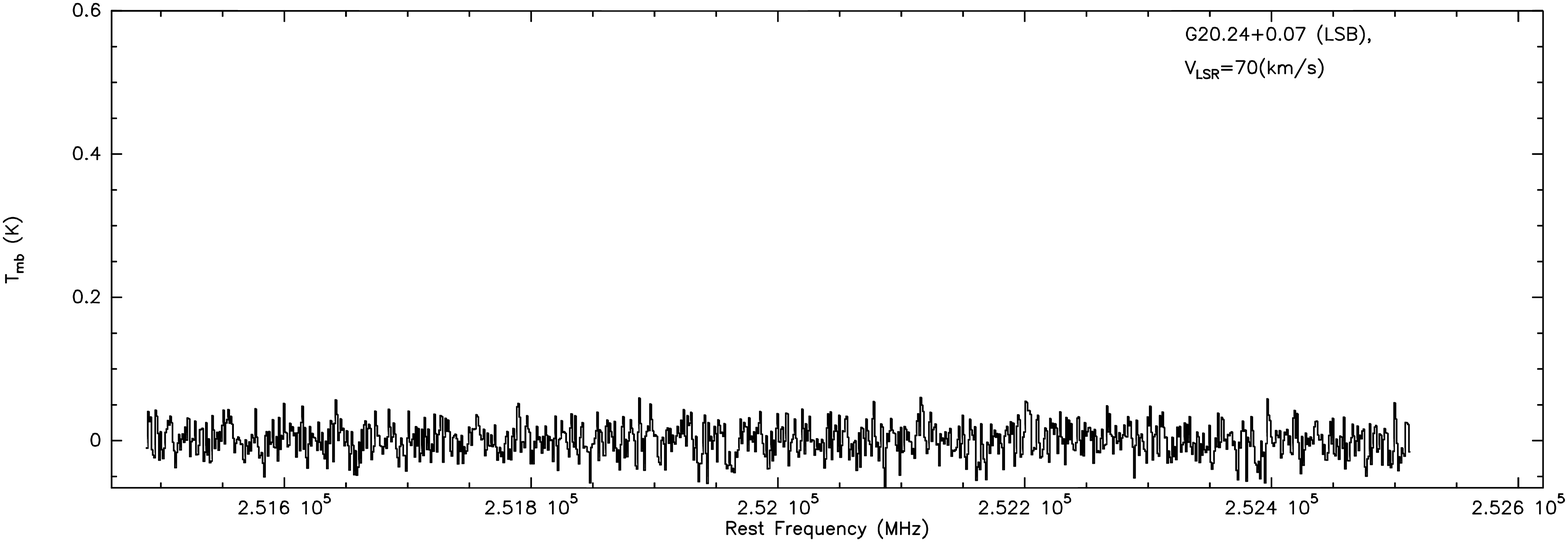}
\includegraphics[scale=.30,angle=0]{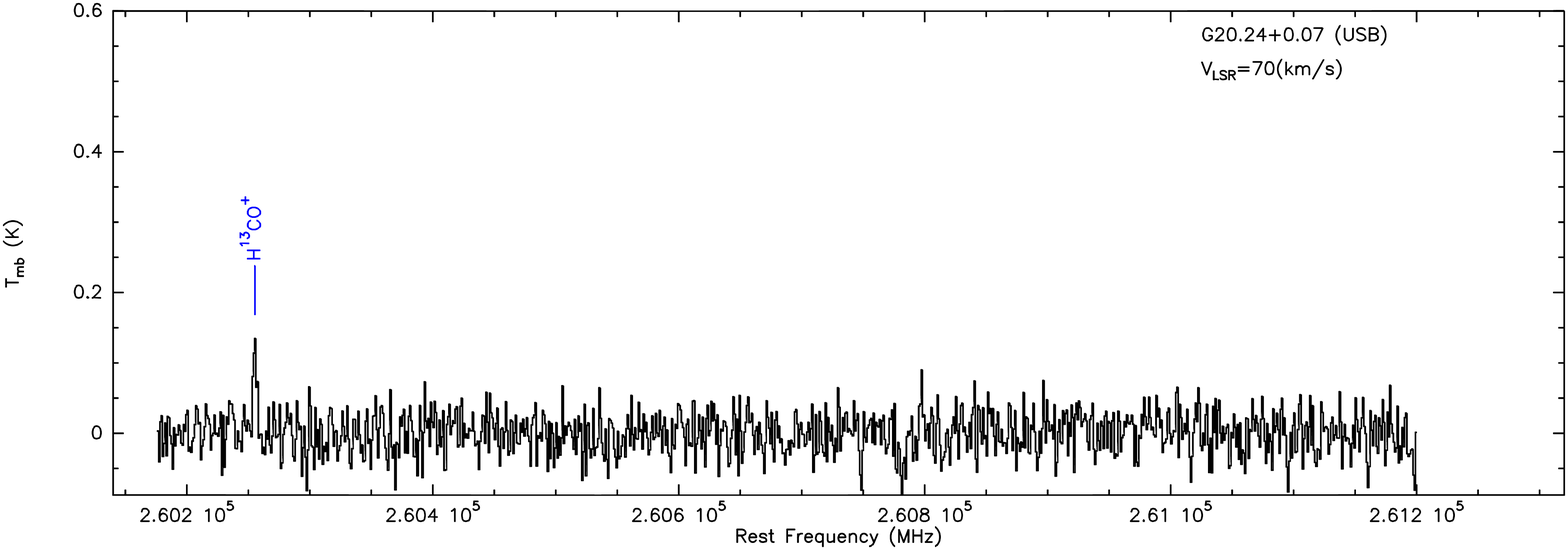}
\caption{(continued) For G20.24+0.07.}
\end{figure*}
\clearpage
 \addtocounter{figure}{-1}
\begin{figure*}
\centering
\includegraphics[scale=.30,angle=0]{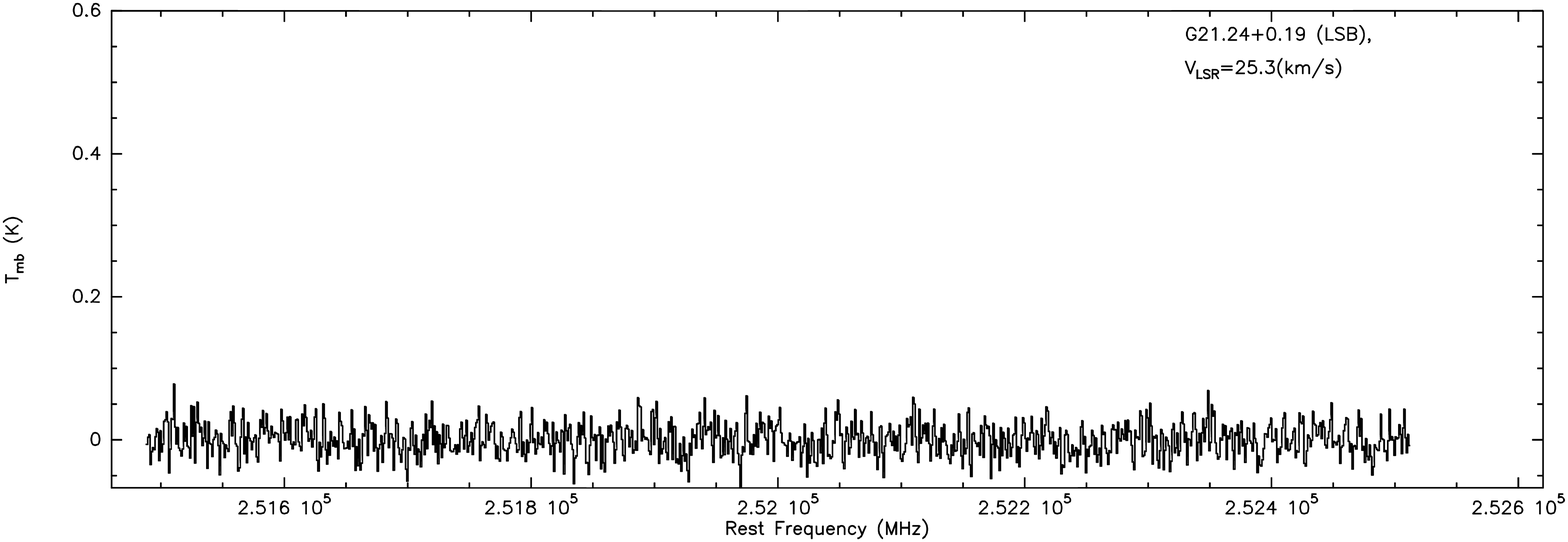}
\includegraphics[scale=.30,angle=0]{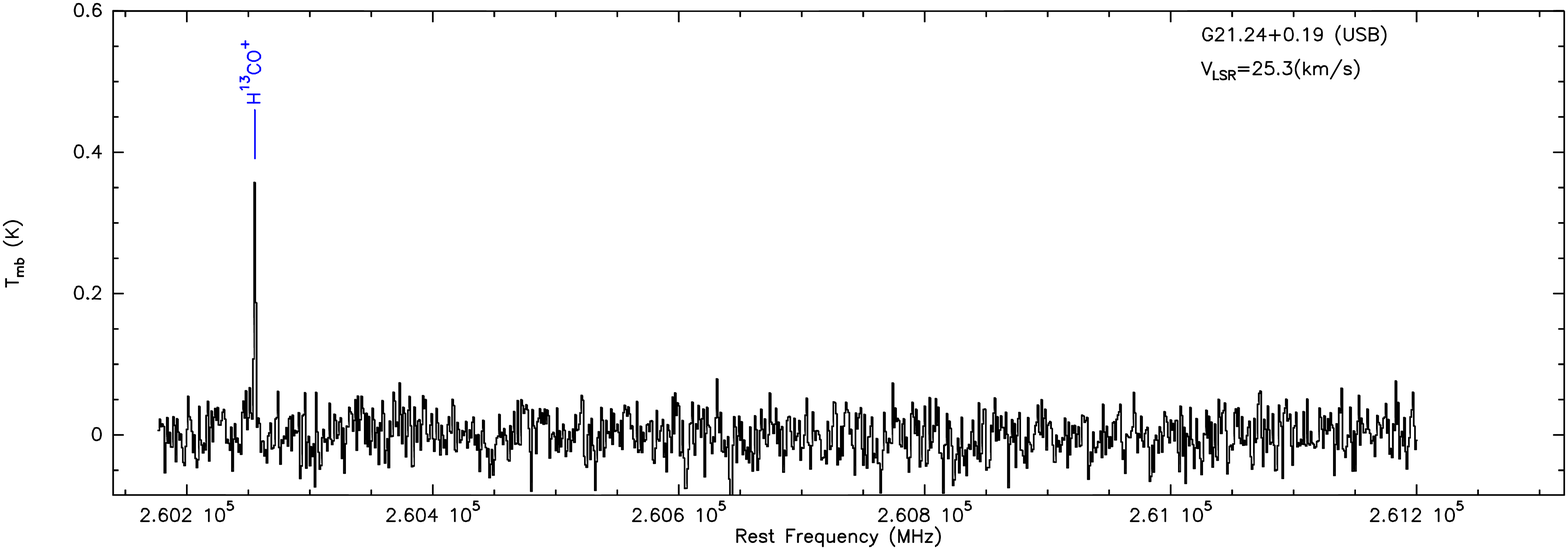}
\caption{(continued) For G21.24+0.19.}
\end{figure*}
 \addtocounter{figure}{-1}
\begin{figure*}
\centering
\includegraphics[scale=.30,angle=0]{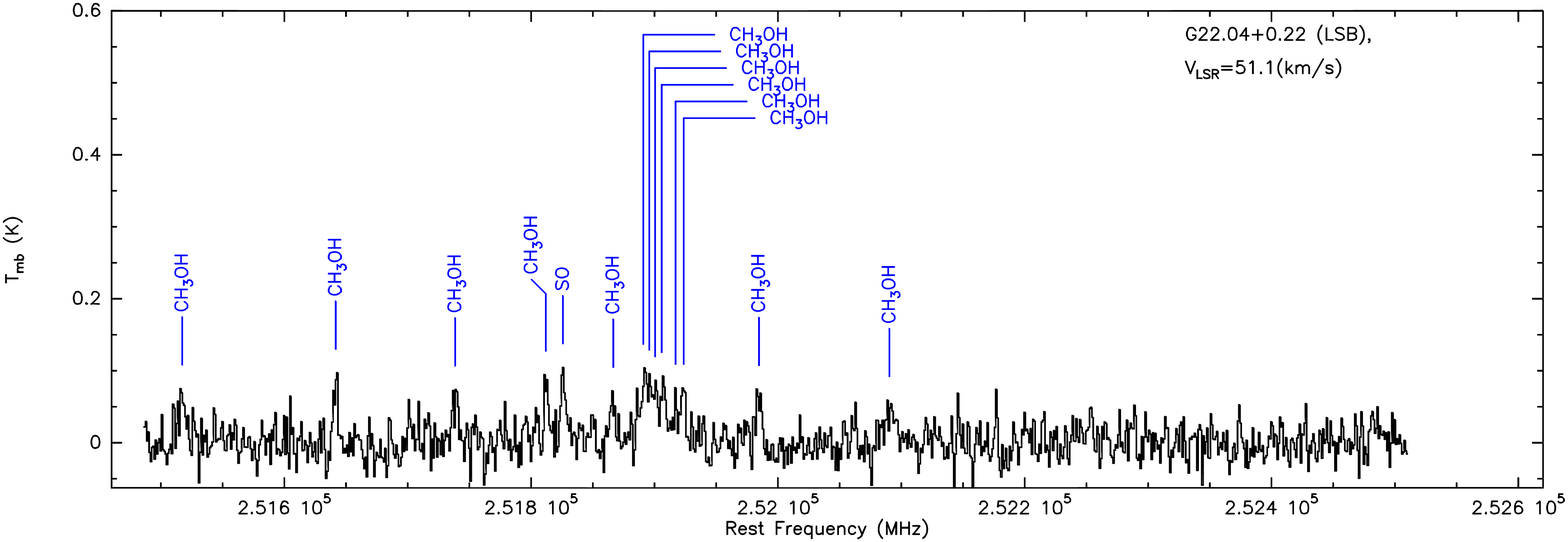}
\includegraphics[scale=.30,angle=0]{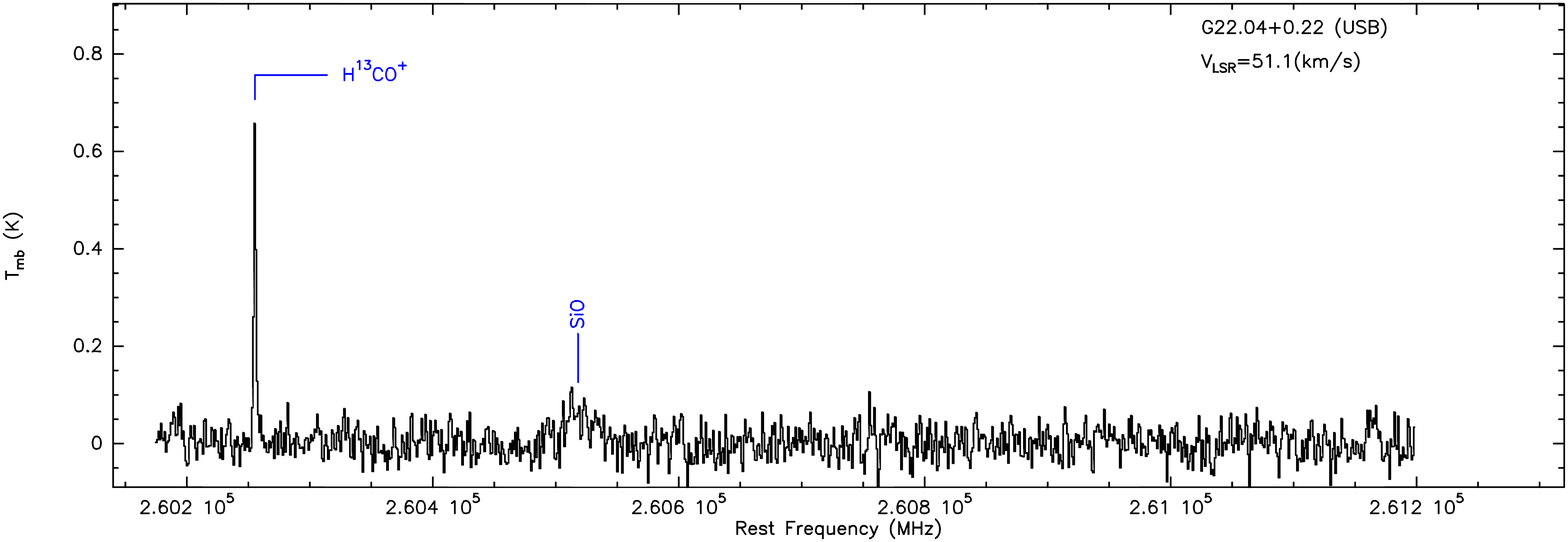}
\caption{(continued) For G22.04+0.22.}
\end{figure*}
\clearpage
 \addtocounter{figure}{-1}
\begin{figure*}
\centering
\includegraphics[scale=.30,angle=0]{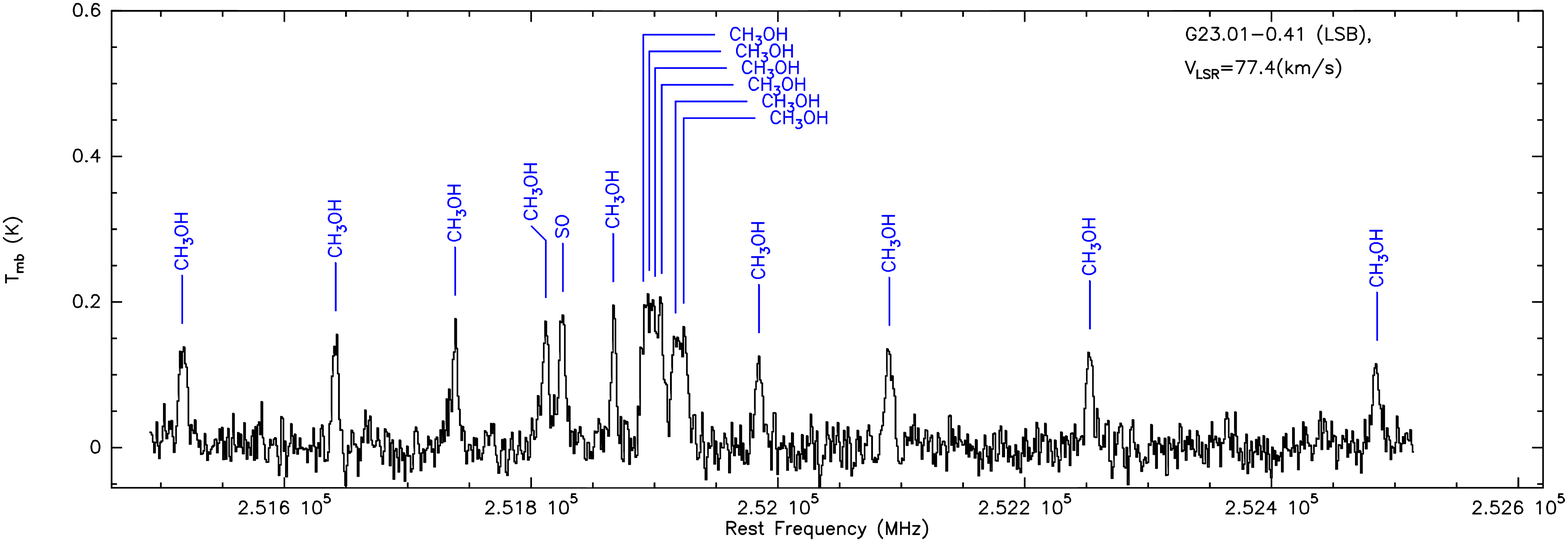}
\includegraphics[scale=.30,angle=0]{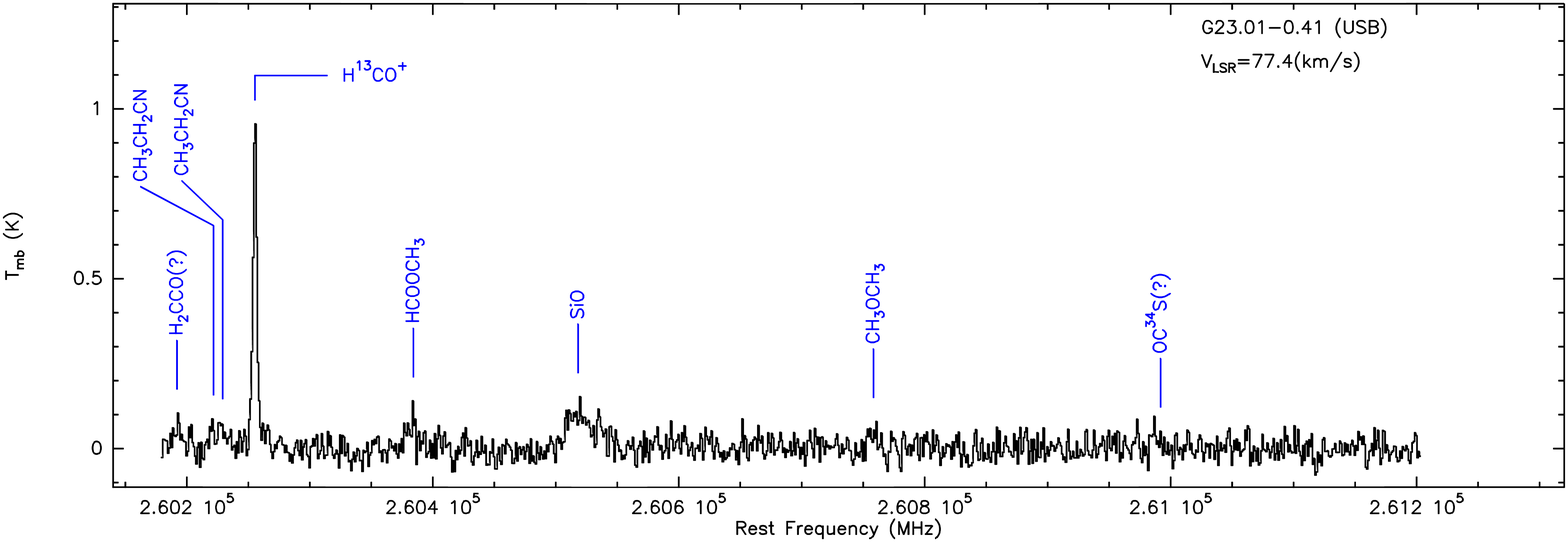}
\caption{(continued) For G23.01-0.41.}
\end{figure*}
 \addtocounter{figure}{-1}
\begin{figure*}
\centering
\includegraphics[scale=.30,angle=0]{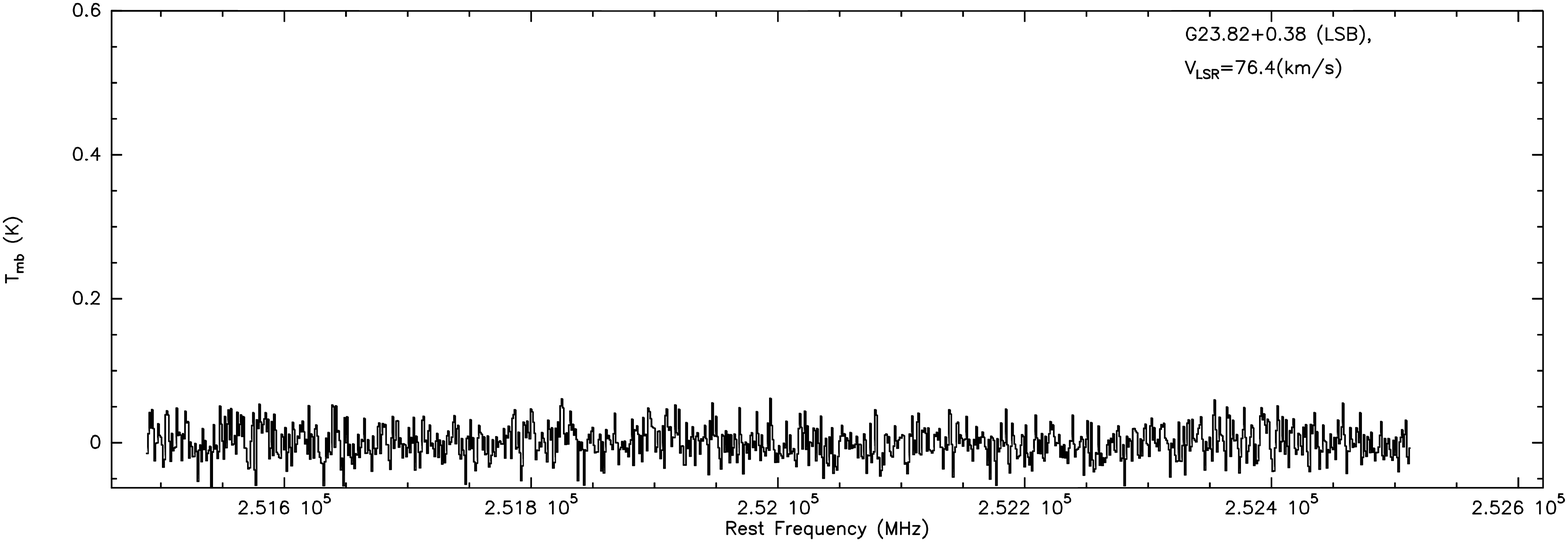}
\includegraphics[scale=.30,angle=0]{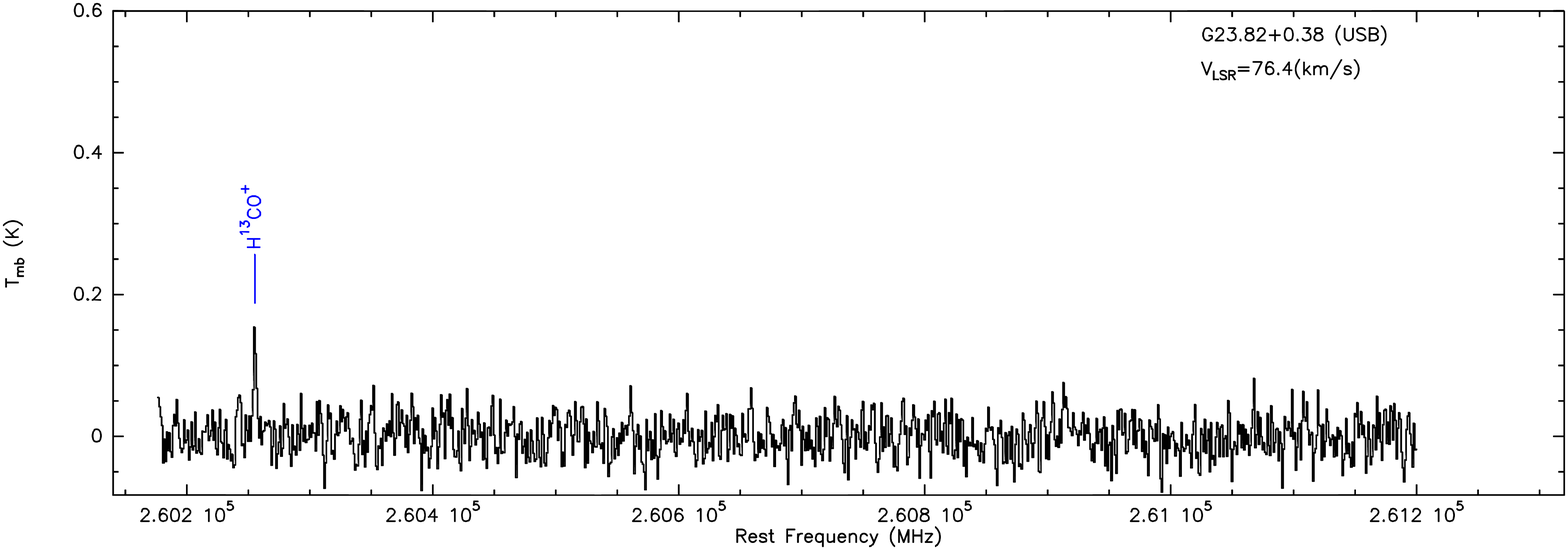}
\caption{(continued) For G23.82+0.38.}
\end{figure*}
\clearpage
 \addtocounter{figure}{-1}
\begin{figure*}
\centering
\includegraphics[scale=.30,angle=0]{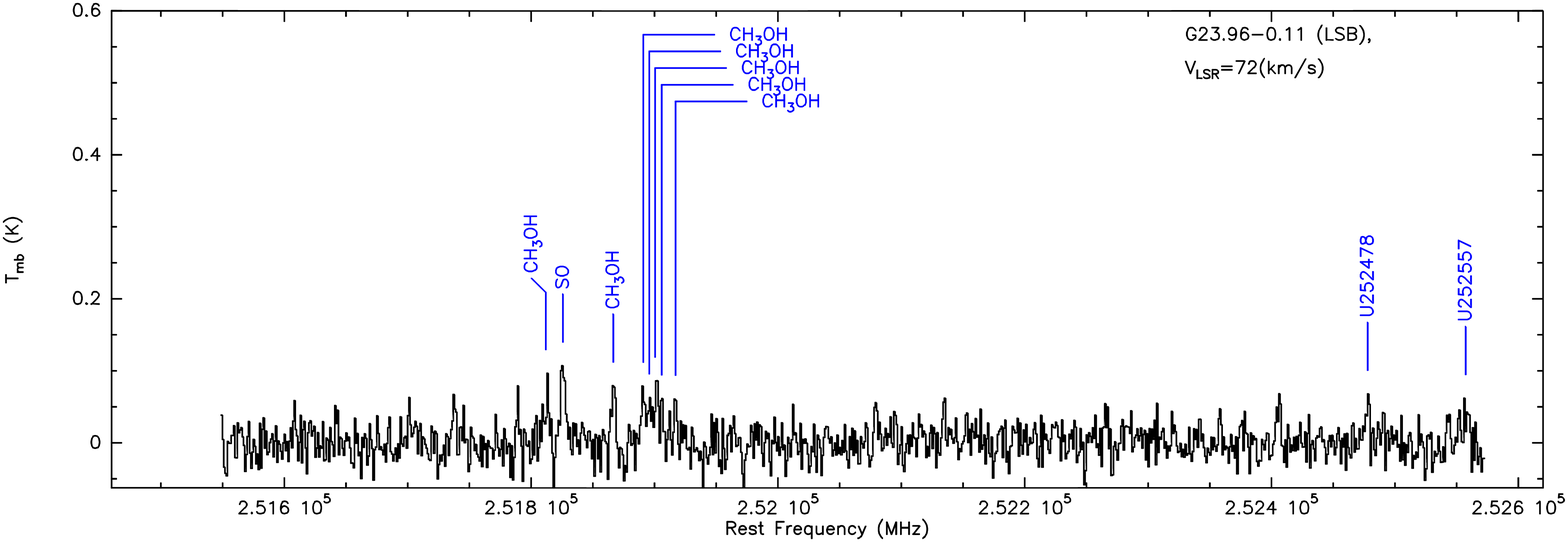}
\includegraphics[scale=.30,angle=0]{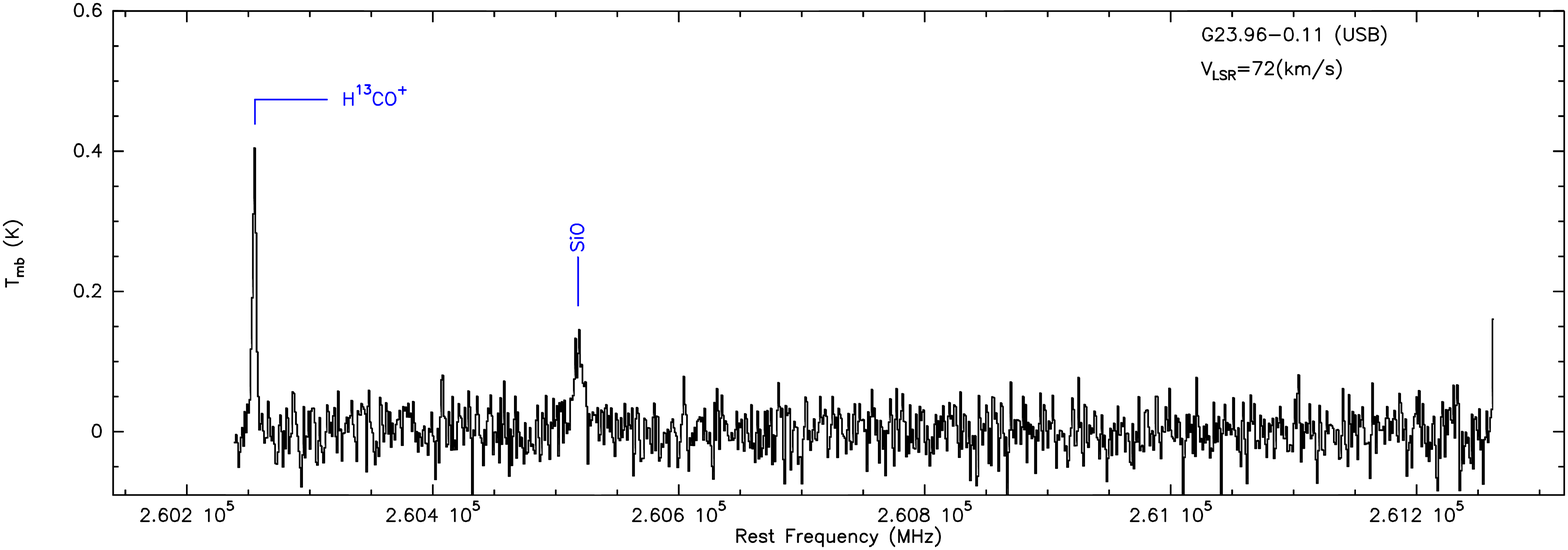}
\caption{(continued) For G23.96-0.11.}
\end{figure*}
 \addtocounter{figure}{-1}
\begin{figure*}
\centering
\includegraphics[scale=.30,angle=0]{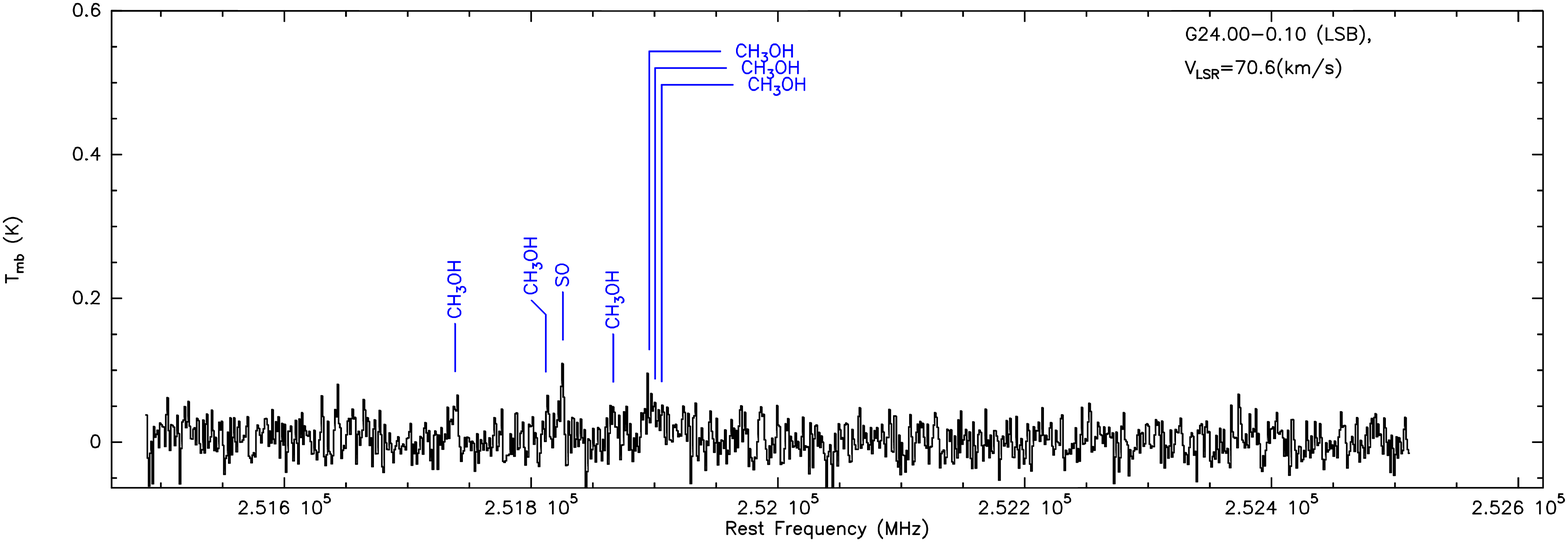}
\includegraphics[scale=.30,angle=0]{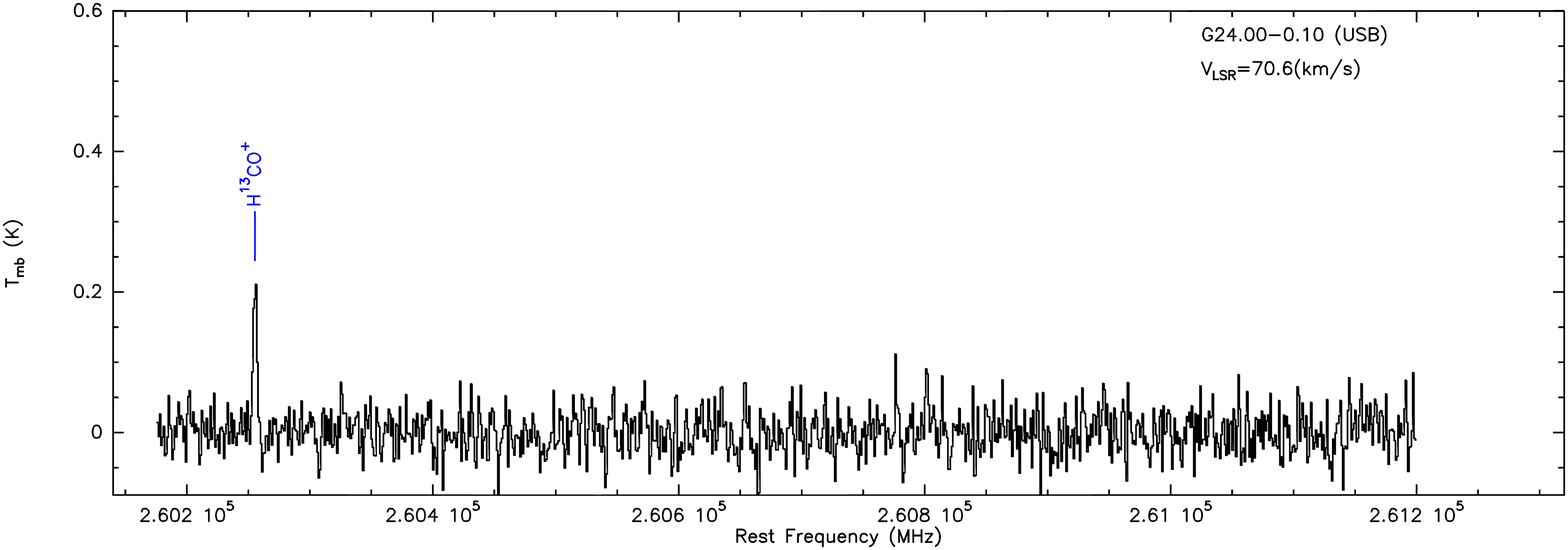}
\caption{(continued) For G24.00-0.10.}
\end{figure*}
\clearpage
 \addtocounter{figure}{-1}
\begin{figure*}
\centering
\includegraphics[scale=.30,angle=0]{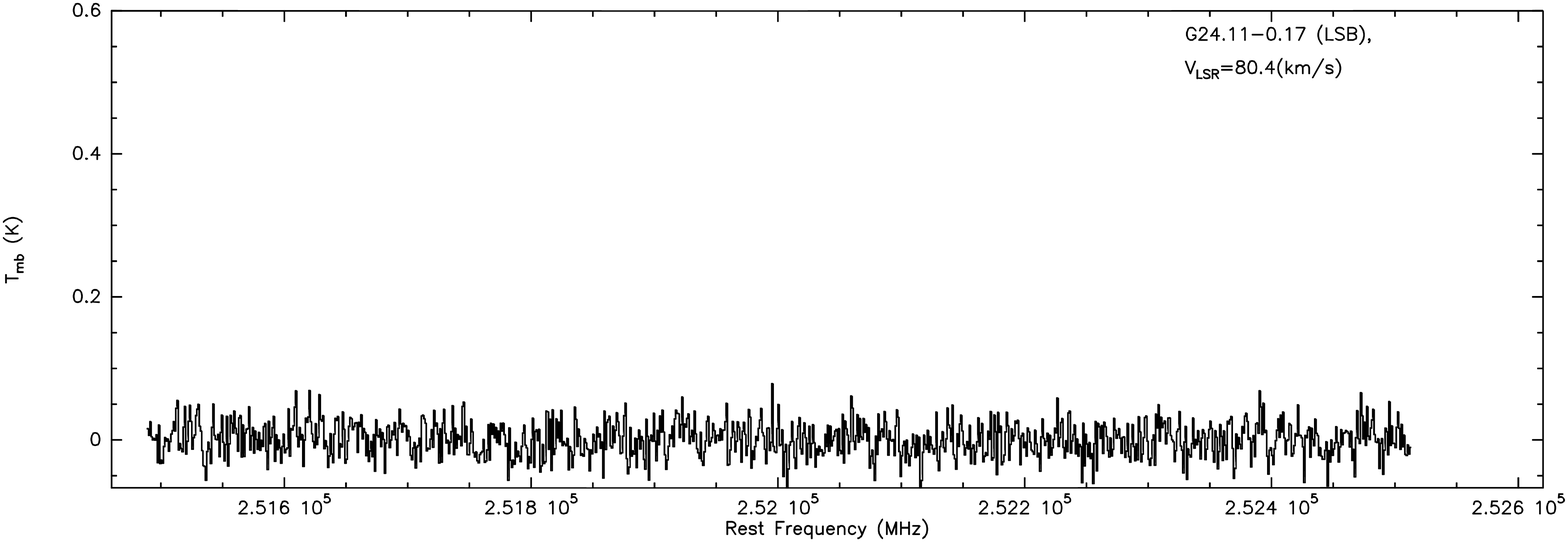}
\includegraphics[scale=.30,angle=0]{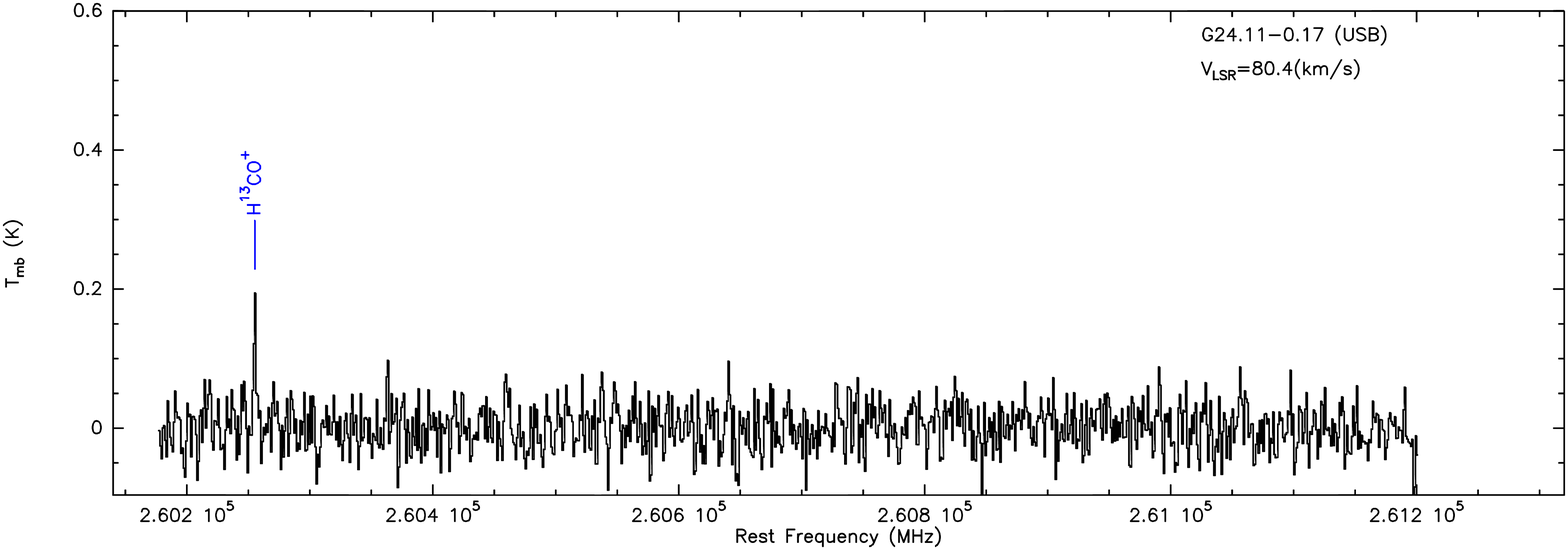}
\caption{(continued) For G24.11-0.17.}
\end{figure*}
 \addtocounter{figure}{-1}
\begin{figure*}
\centering
\includegraphics[scale=.30,angle=0]{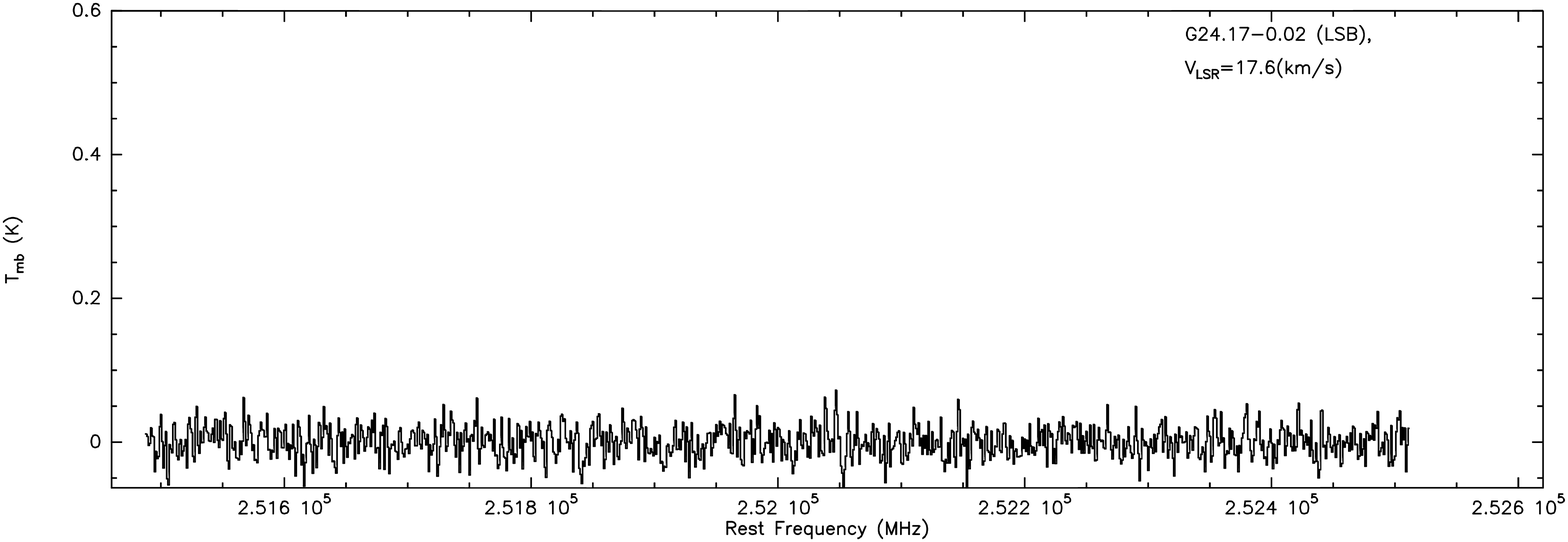}
\includegraphics[scale=.30,angle=0]{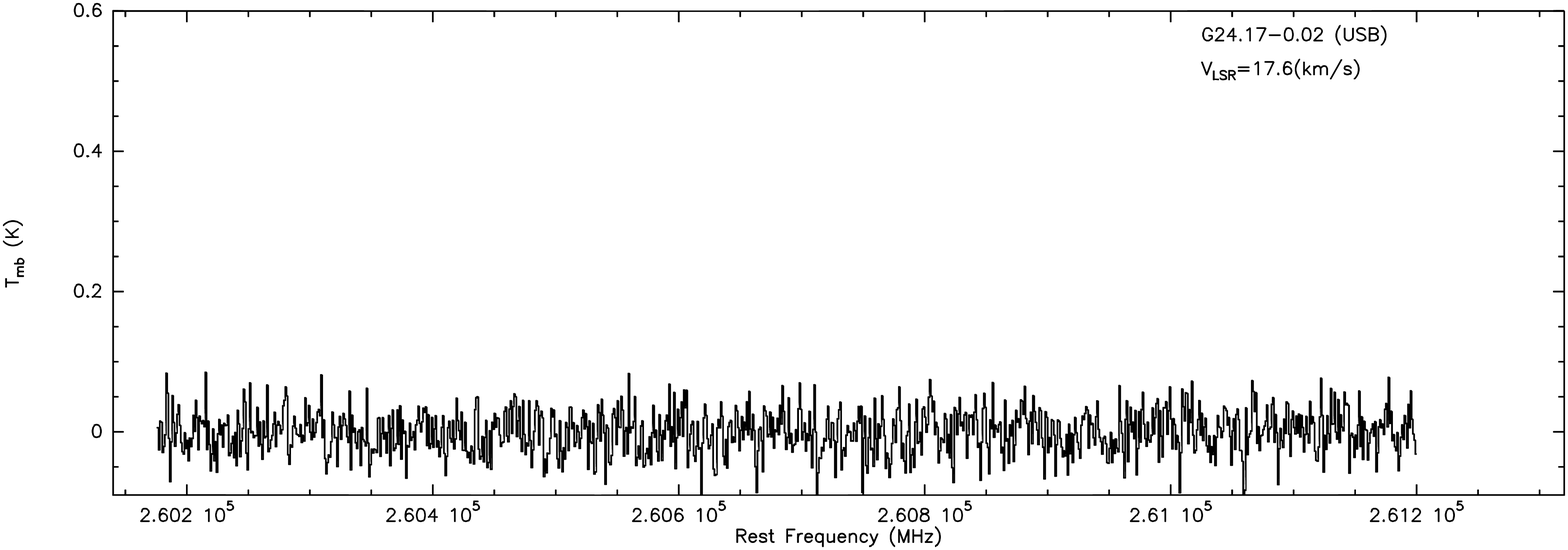}
\caption{(continued) For G24.17-0.02.}
\end{figure*}
\clearpage
 \addtocounter{figure}{-1}
\begin{figure*}
\centering
\includegraphics[scale=.30,angle=0]{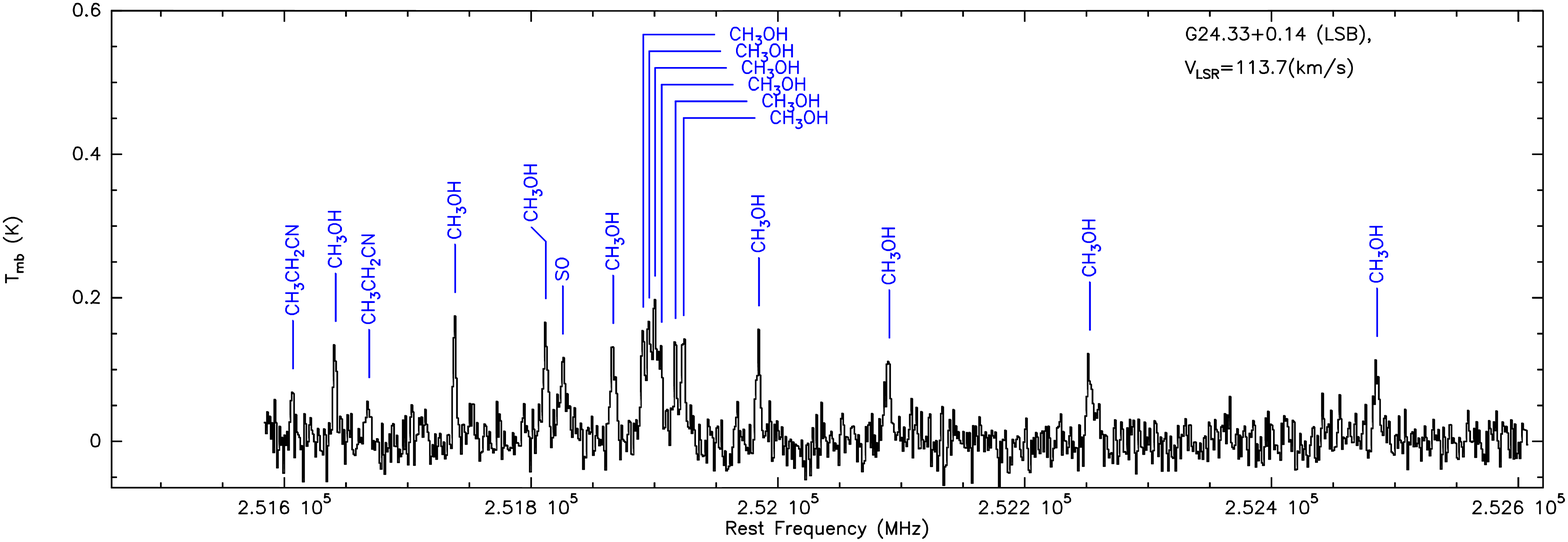}
\includegraphics[scale=.30,angle=0]{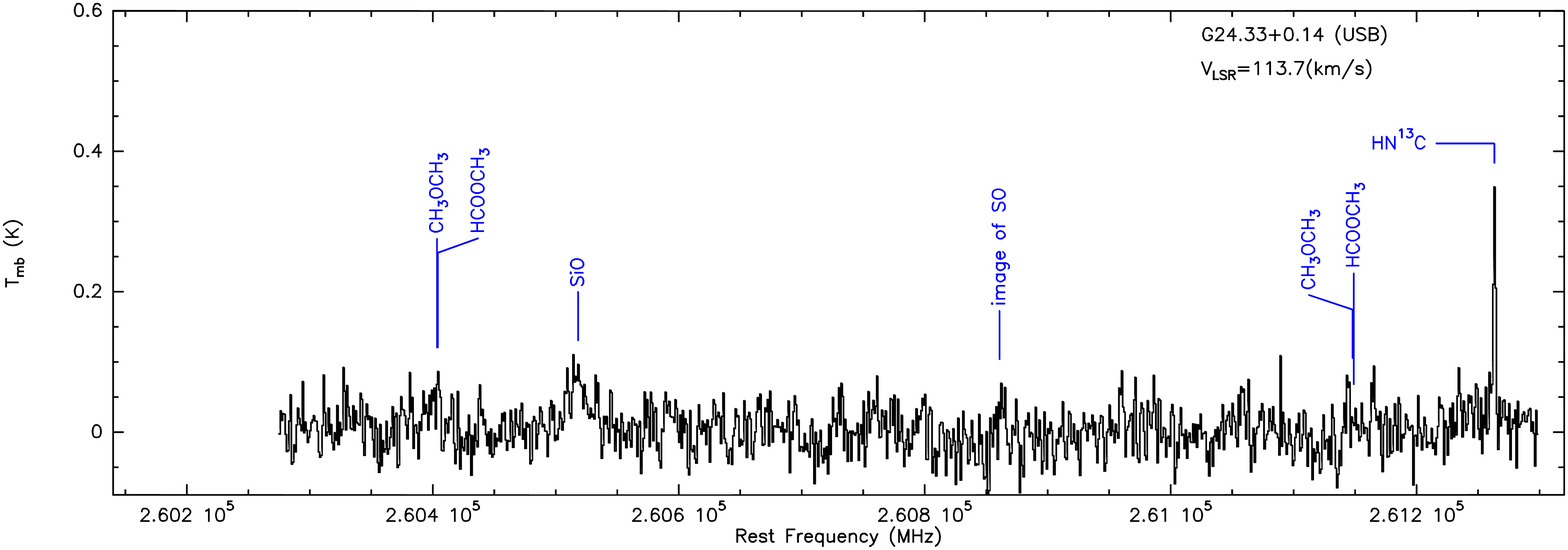}
\caption{(continued) For G24.33+0.14.}
\end{figure*}
 \addtocounter{figure}{-1}
\begin{figure*}
\centering
\includegraphics[scale=.30,angle=0]{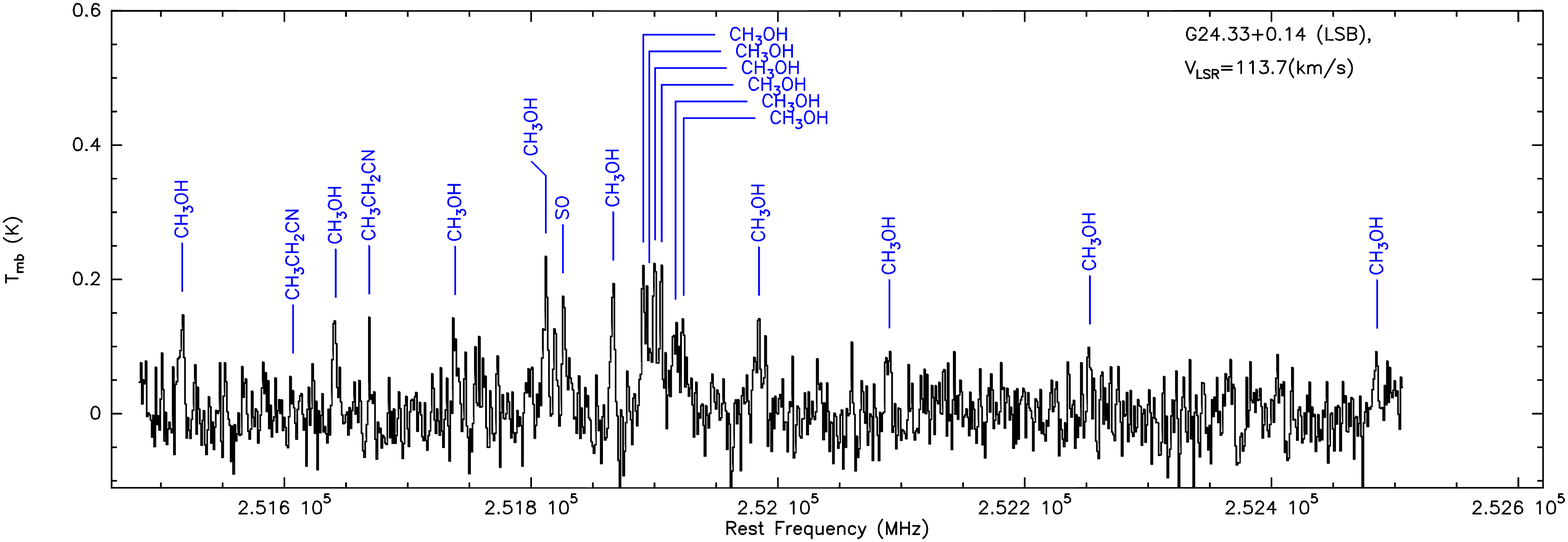}
\includegraphics[scale=.30,angle=0]{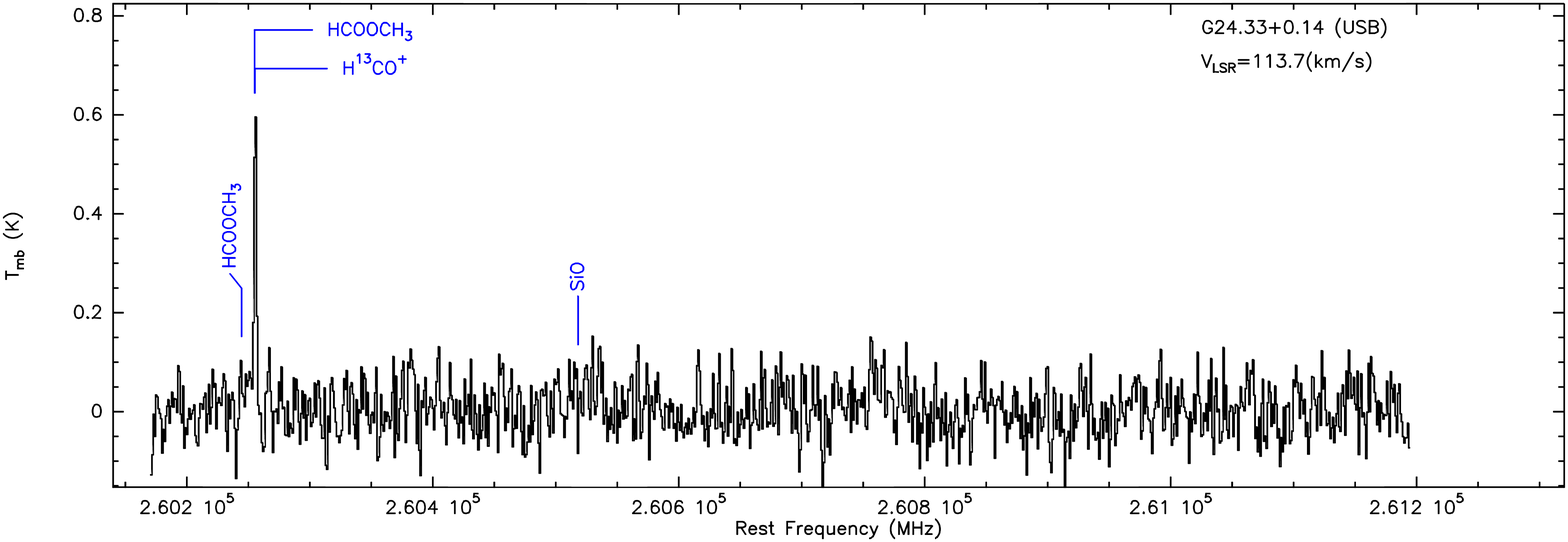}
\caption{(continued) For G24.33+0.14 (different frequency coverage).}
\end{figure*}
\clearpage
 \addtocounter{figure}{-1}
\begin{figure*}
\centering
\includegraphics[scale=.30,angle=0]{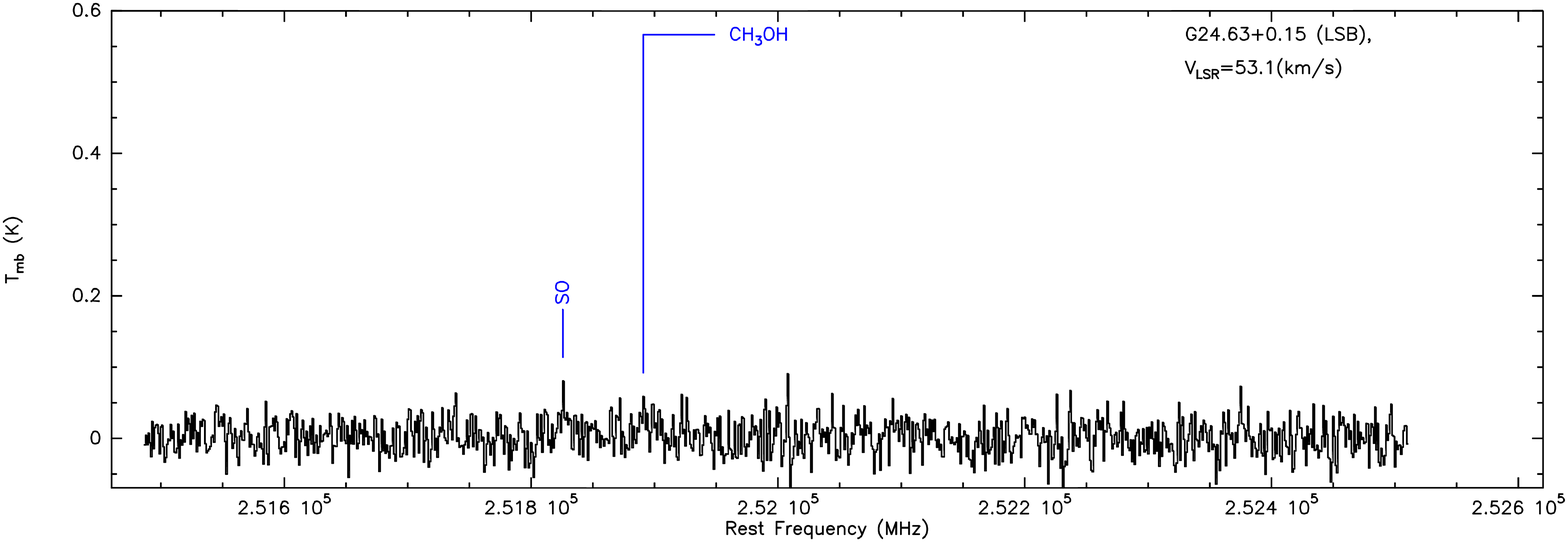}
\includegraphics[scale=.30,angle=0]{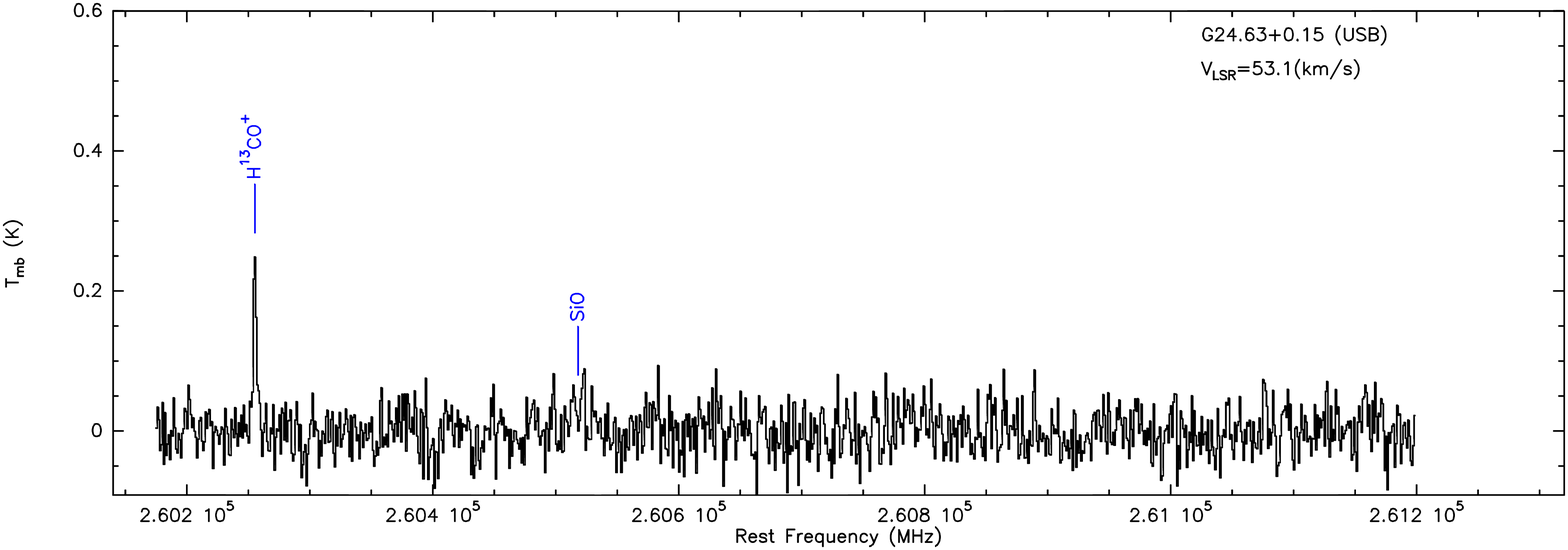}
\caption{(continued) For G24.63+0.15.}
\end{figure*}
 \addtocounter{figure}{-1}
\begin{figure*}
\centering
\includegraphics[scale=.30,angle=0]{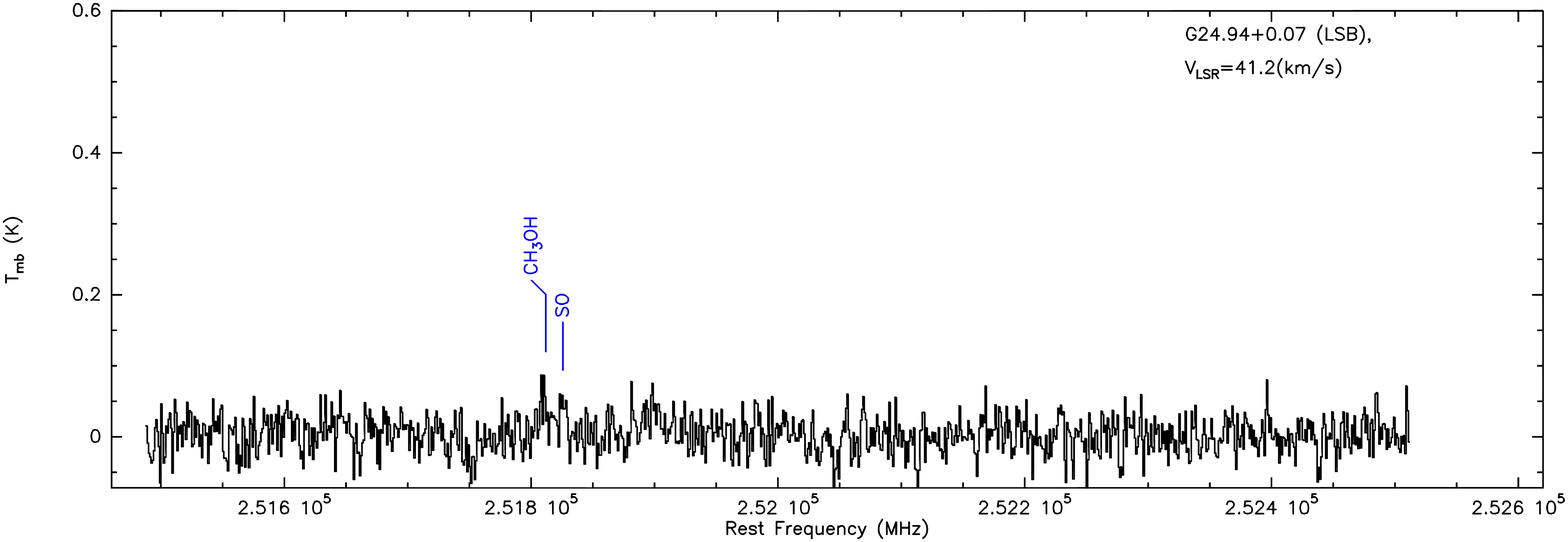}
\includegraphics[scale=.30,angle=0]{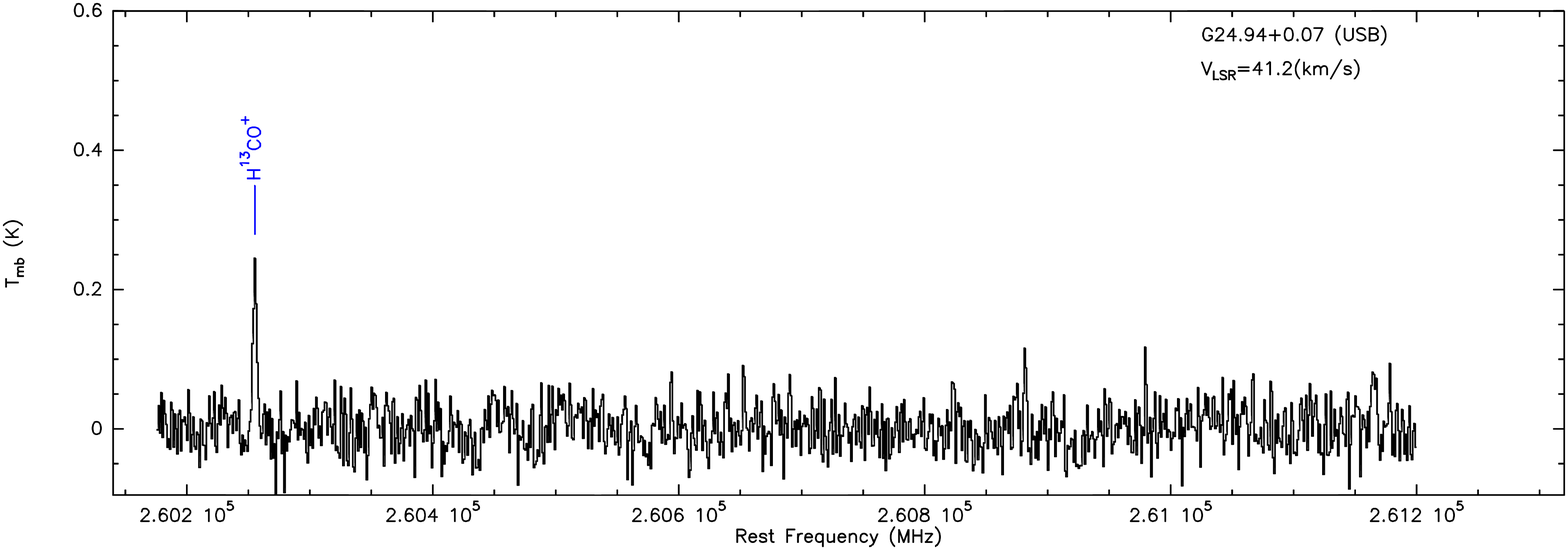}
\caption{(continued) For G24.94+0.07.}
\end{figure*}
\clearpage
 \addtocounter{figure}{-1}
\begin{figure*}
\centering
\includegraphics[scale=.30,angle=0]{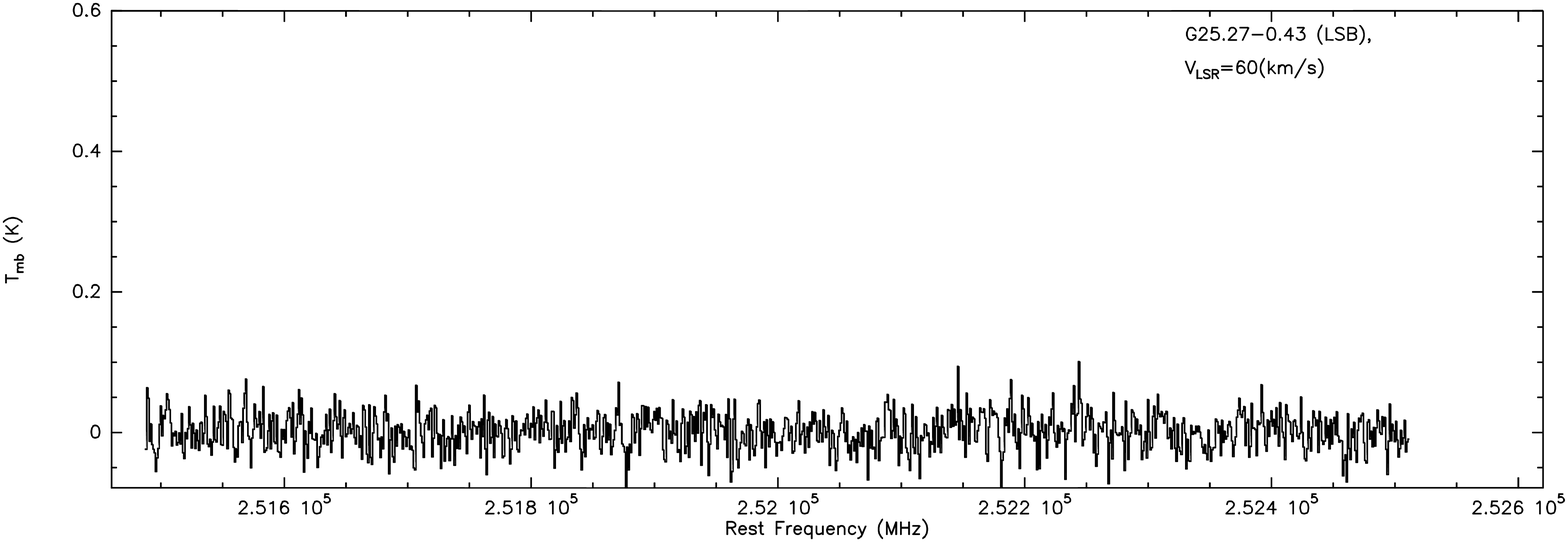}
\includegraphics[scale=.30,angle=0]{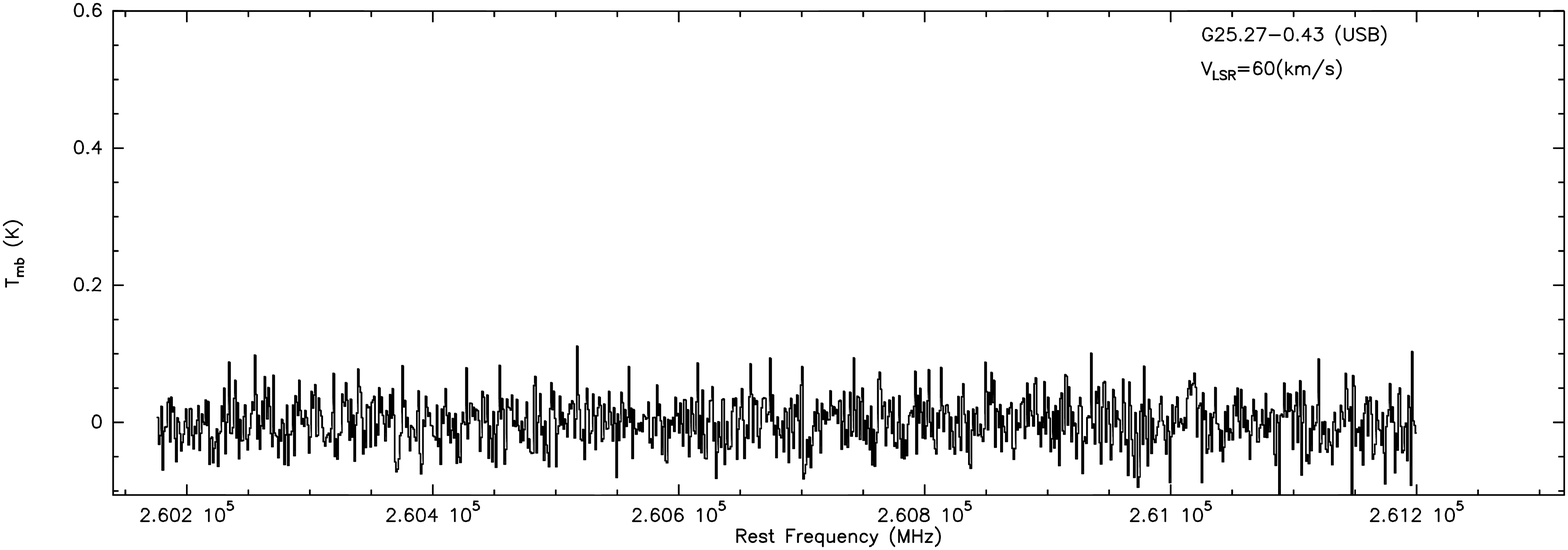}
\caption{(continued) For G25.27-0.43.}
\end{figure*}
 \addtocounter{figure}{-1}
\begin{figure*}
\centering
\includegraphics[scale=.30,angle=0]{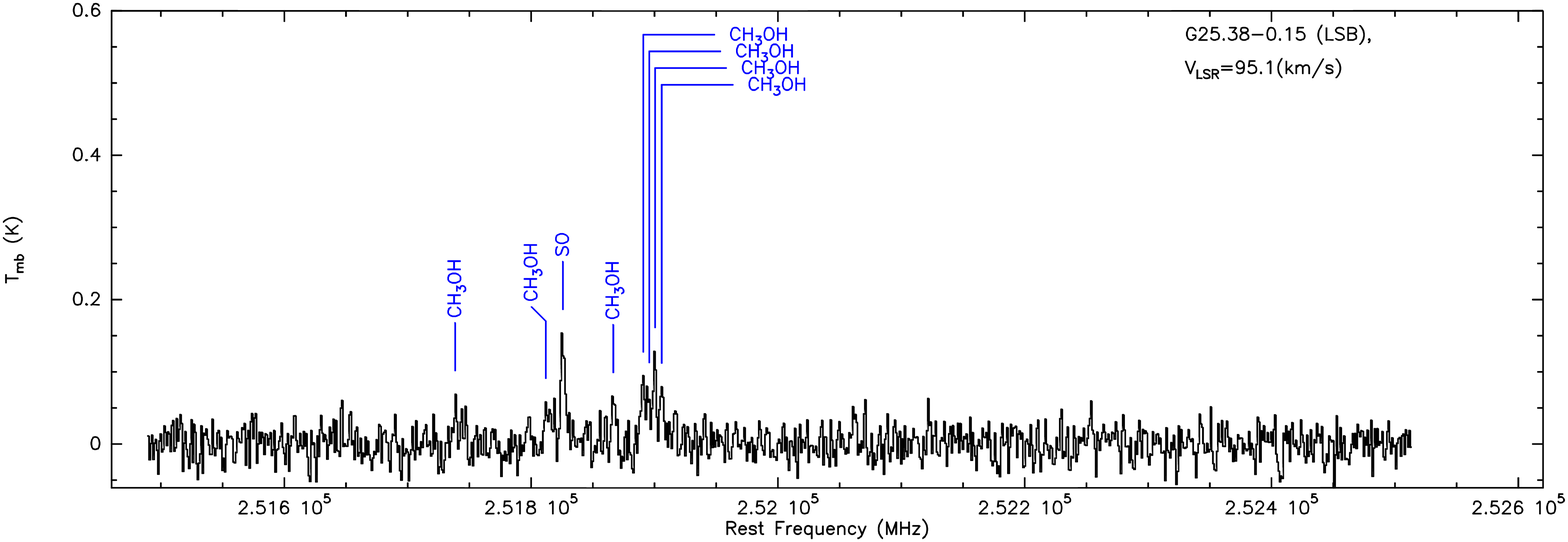}
\includegraphics[scale=.30,angle=0]{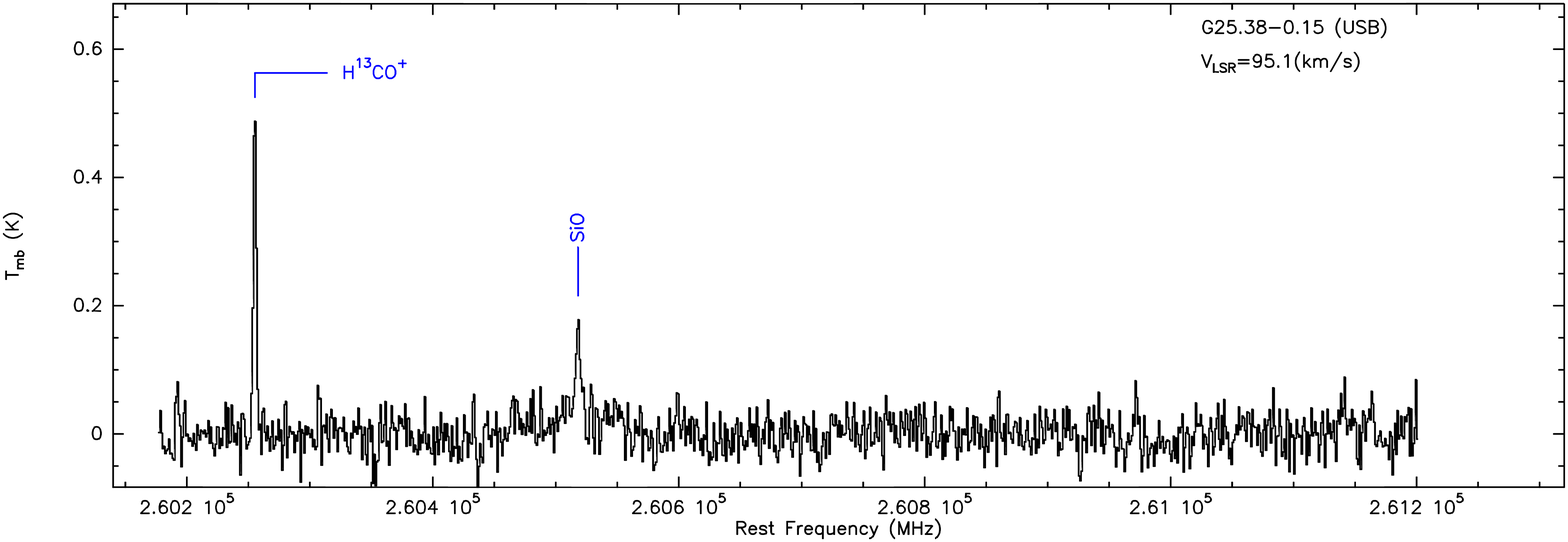}
\caption{(continued) For G25.38-0.15.}
\end{figure*}
\clearpage
 \addtocounter{figure}{-1}
\begin{figure*}
\centering
\includegraphics[scale=.30,angle=0]{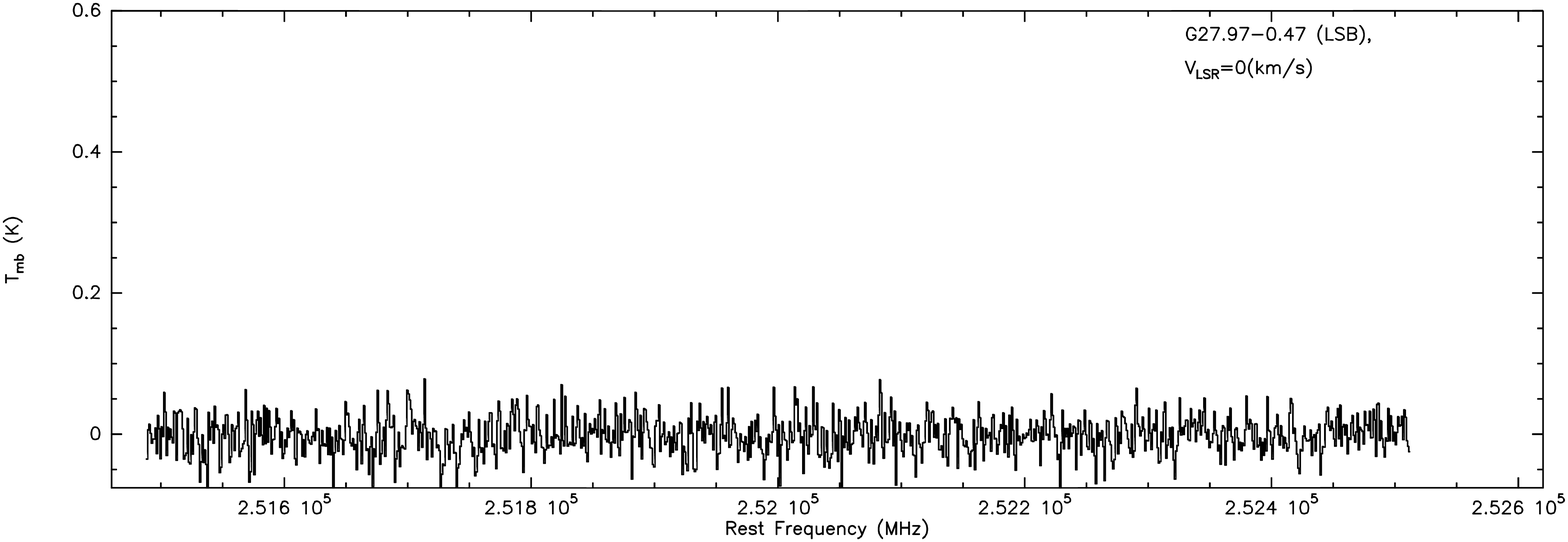}
\includegraphics[scale=.30,angle=0]{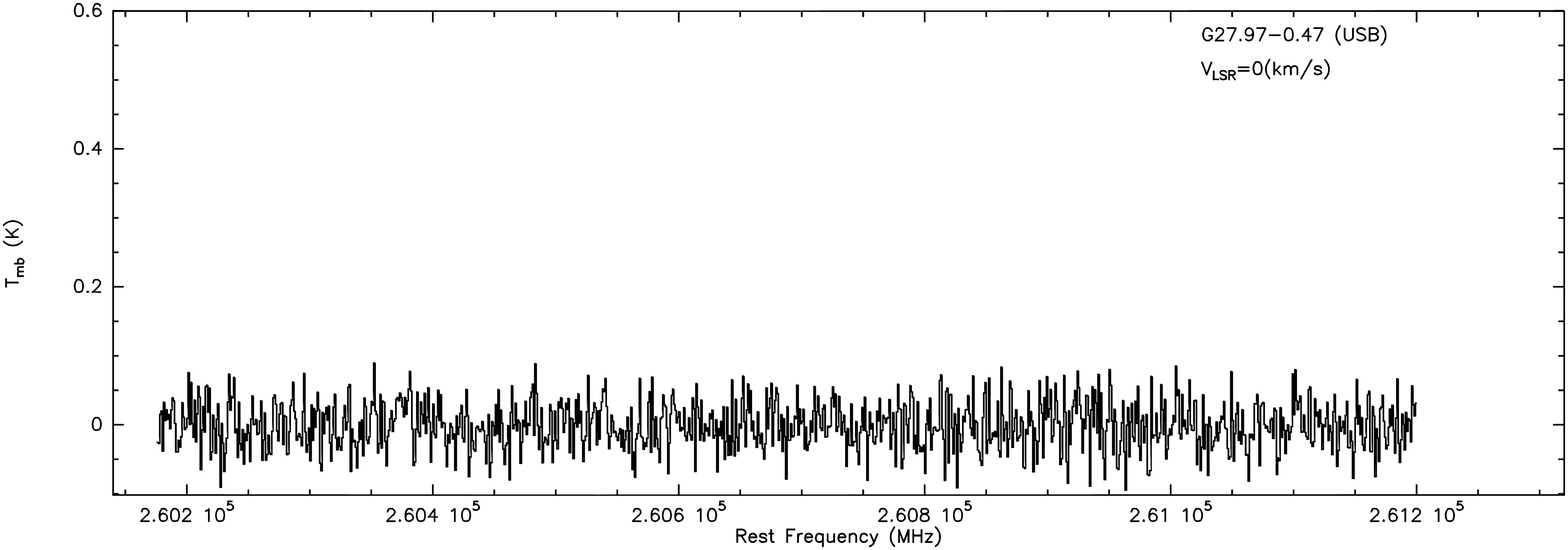}
\caption{(continued) For G27.97-0.47.}
\end{figure*}
 \addtocounter{figure}{-1}
\begin{figure*}
\centering
\includegraphics[scale=.30,angle=0]{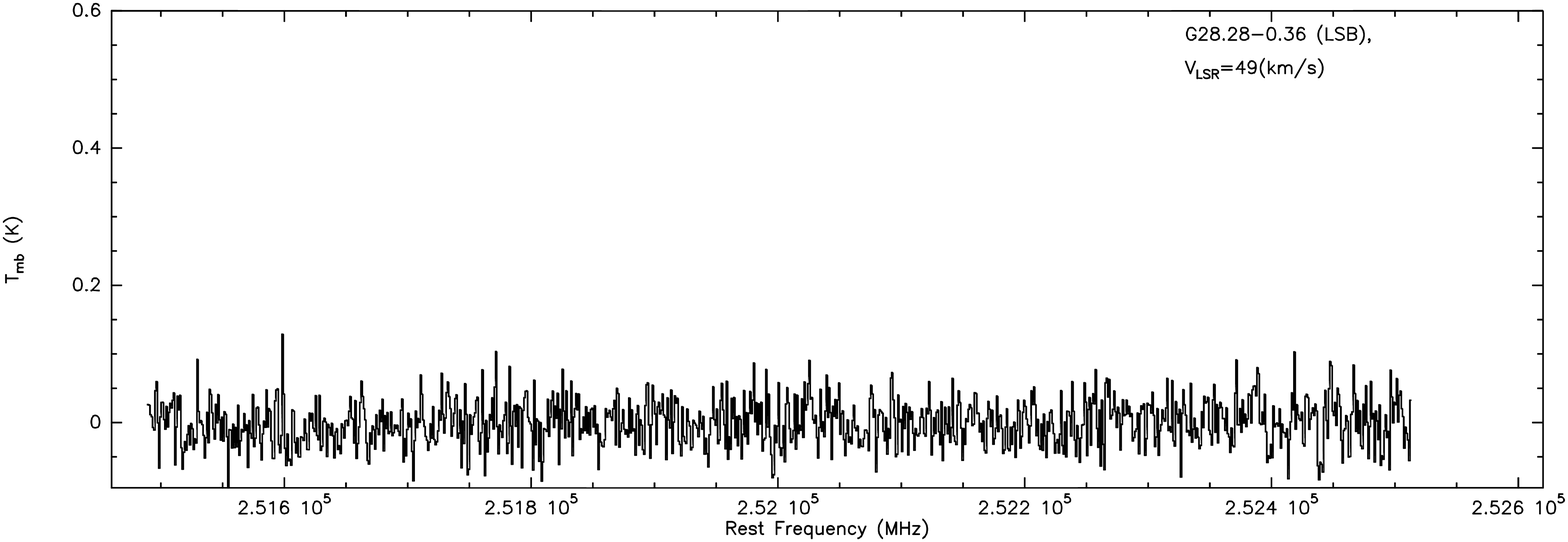}
\includegraphics[scale=.30,angle=0]{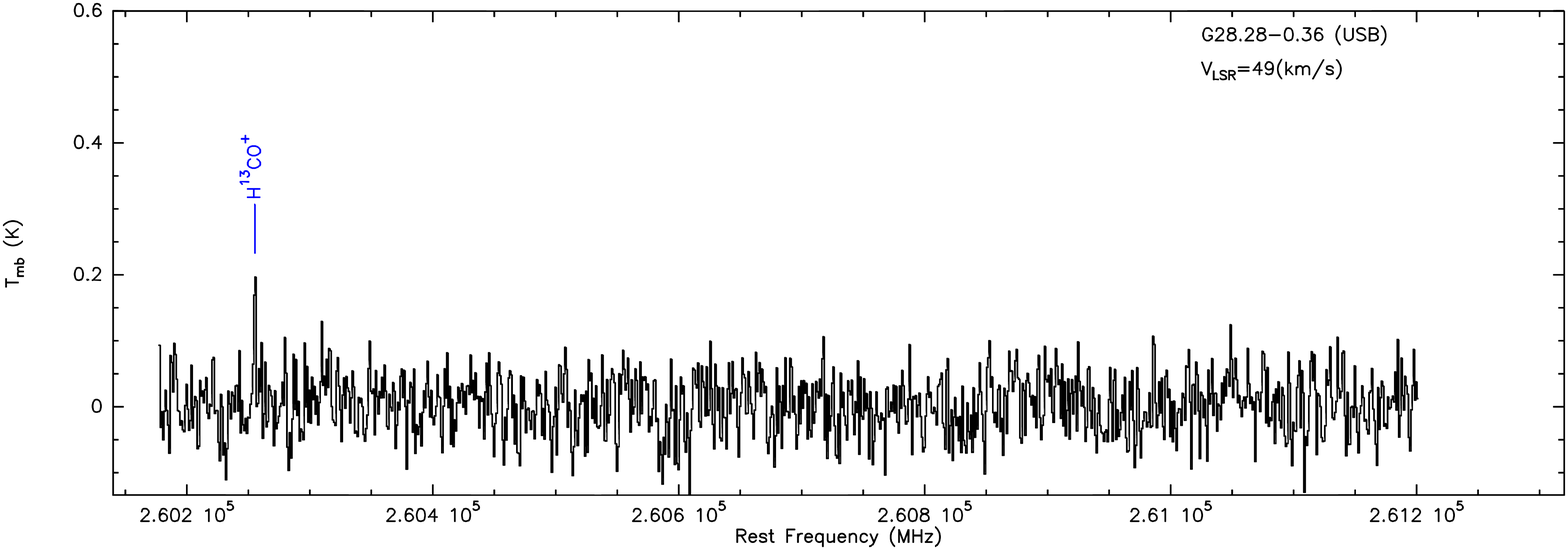}
\caption{(continued) For G28.28-0.36.}
\end{figure*}
\clearpage
 \addtocounter{figure}{-1}
\begin{figure*}
\centering
\includegraphics[scale=.30,angle=0]{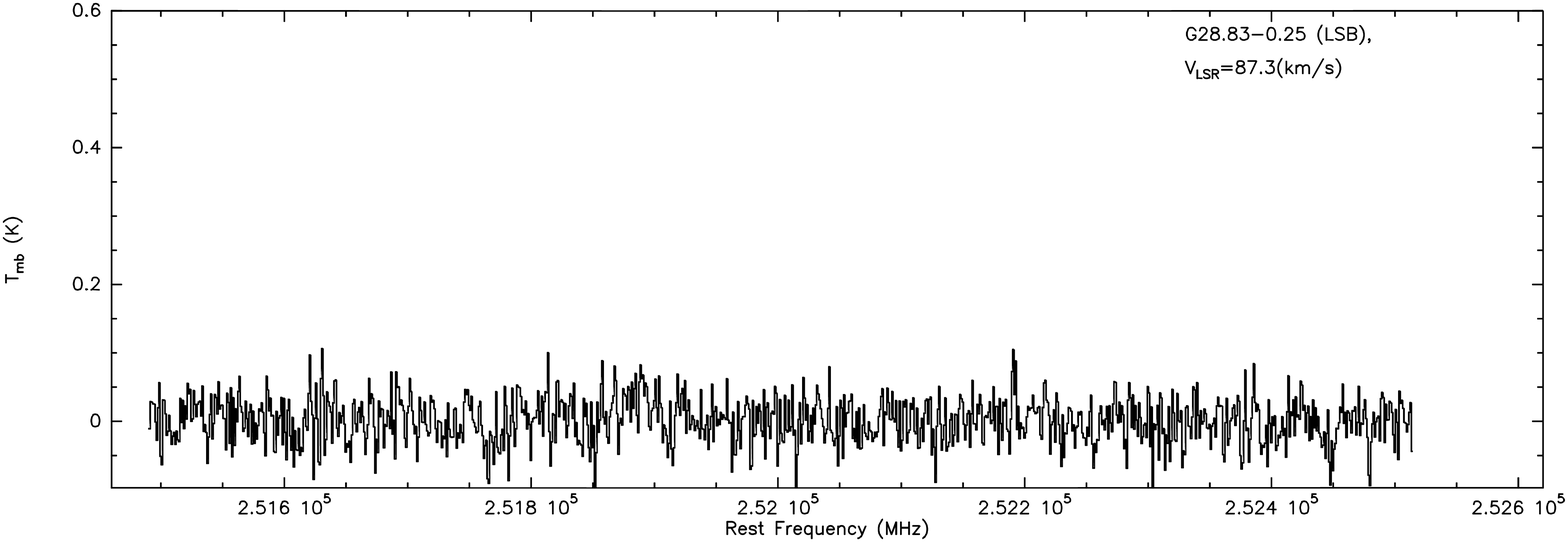}
\includegraphics[scale=.30,angle=0]{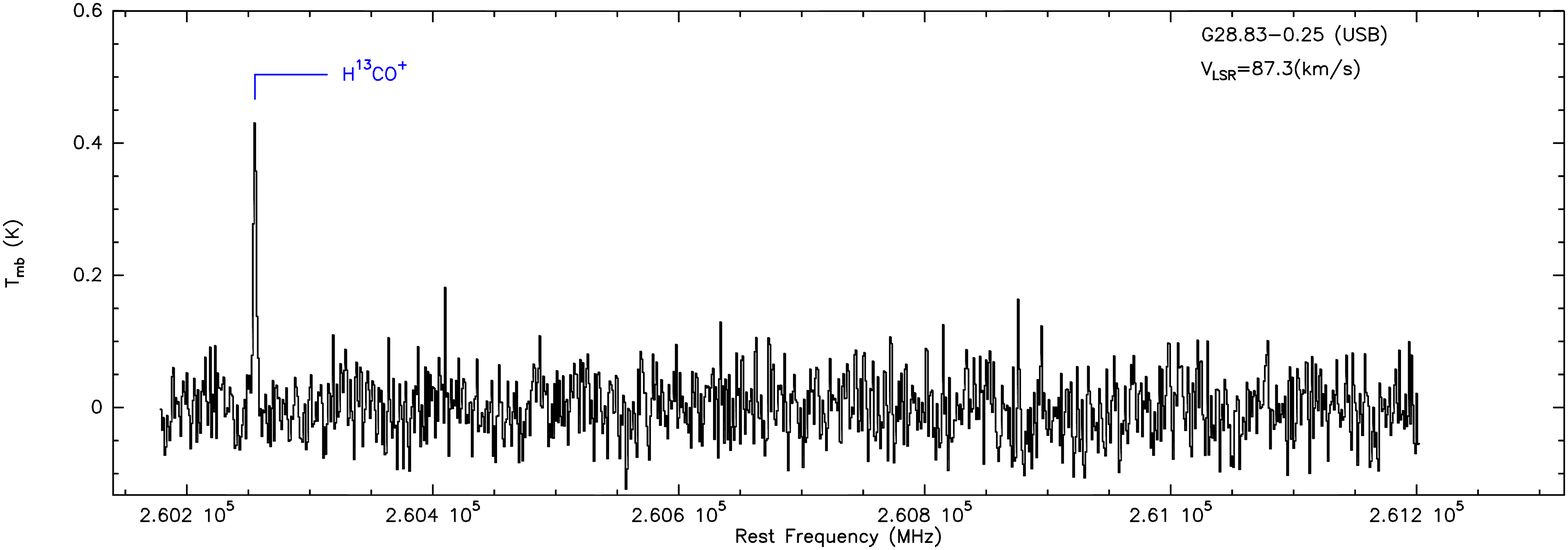}
\caption{(continued) For G28.83-0.25.}
\end{figure*}
 \addtocounter{figure}{-1}
\begin{figure*}
\centering
\includegraphics[scale=.30,angle=0]{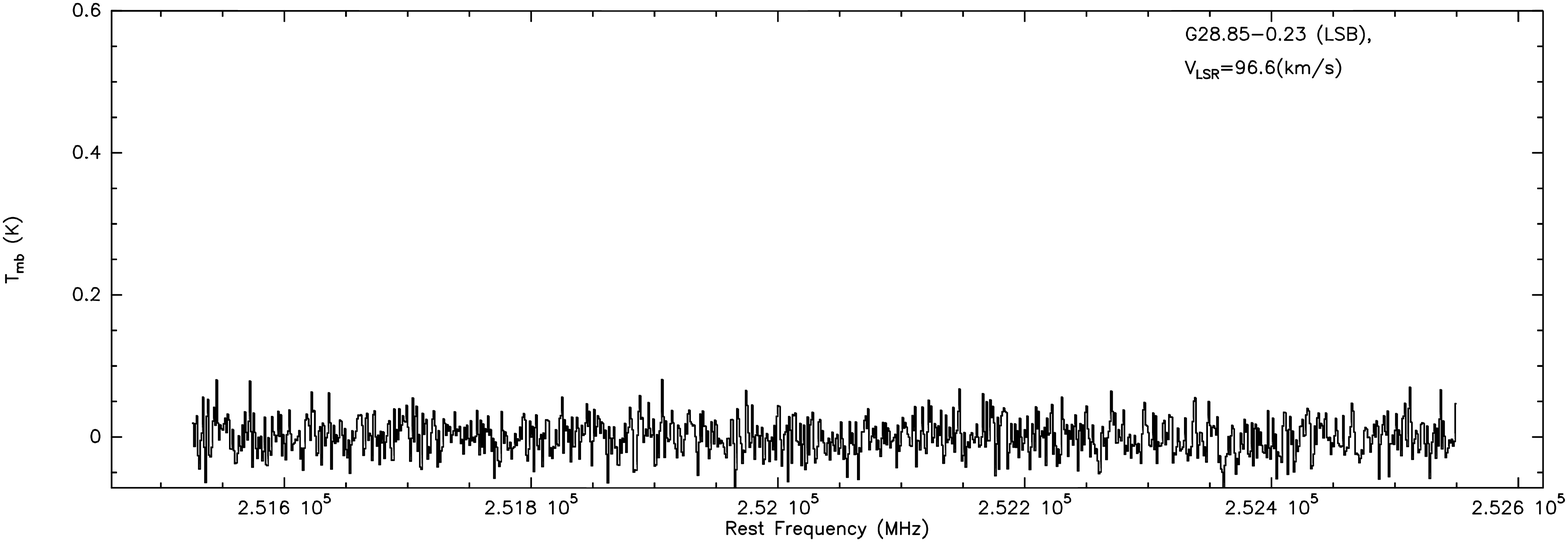}
\includegraphics[scale=.30,angle=0]{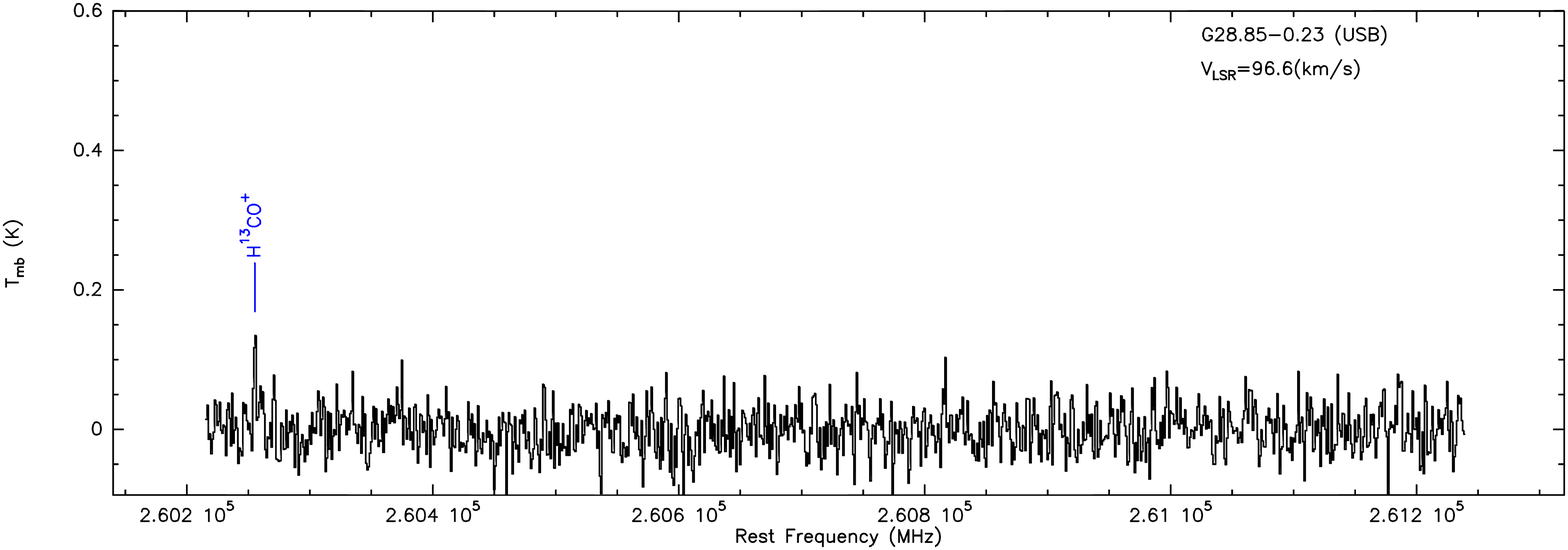}
\caption{(continued) For G28.85-0.23.}
\end{figure*}
\clearpage
 \addtocounter{figure}{-1}
\begin{figure*}
\centering
\includegraphics[scale=.30,angle=0]{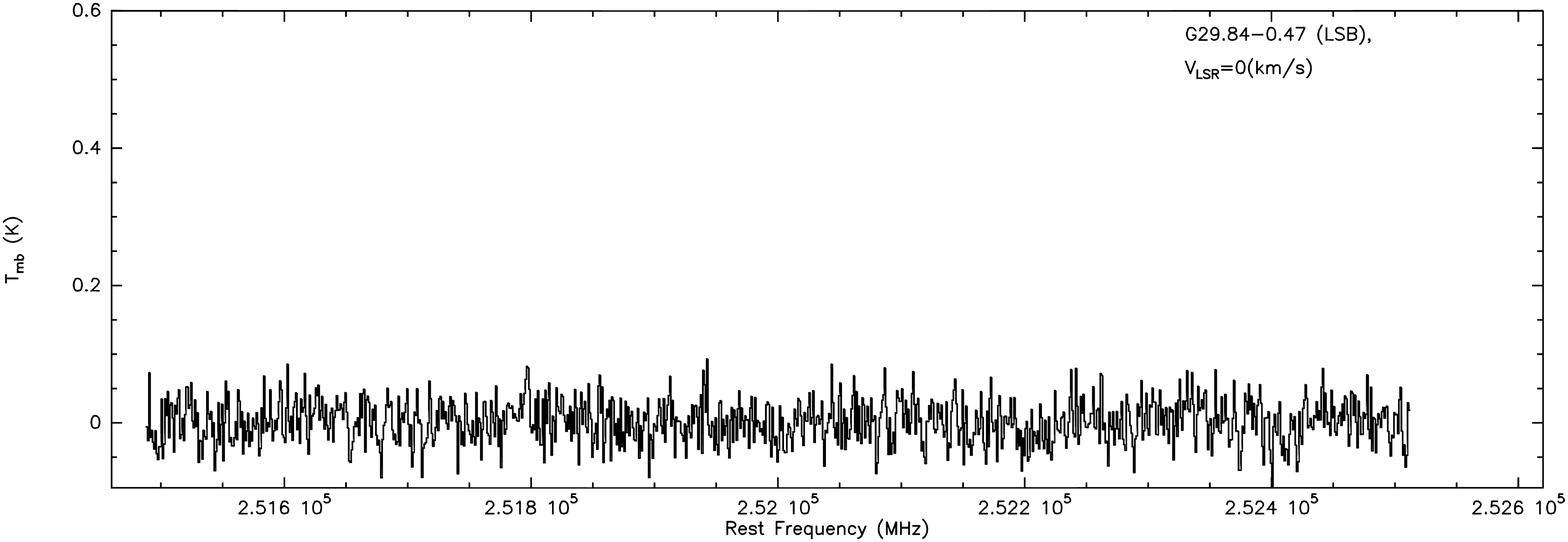}
\includegraphics[scale=.30,angle=0]{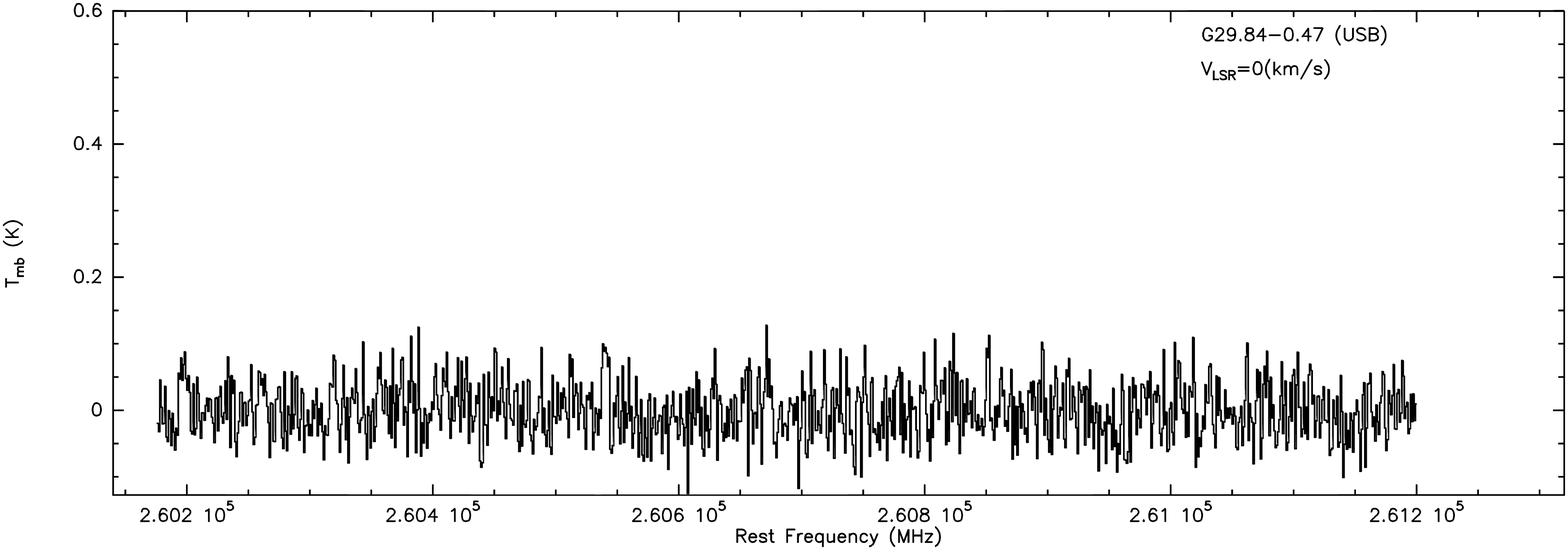}
\caption{(continued) For G29.84-0.47.}
\end{figure*}
 \addtocounter{figure}{-1}
\begin{figure*}
\centering
\includegraphics[scale=.30,angle=0]{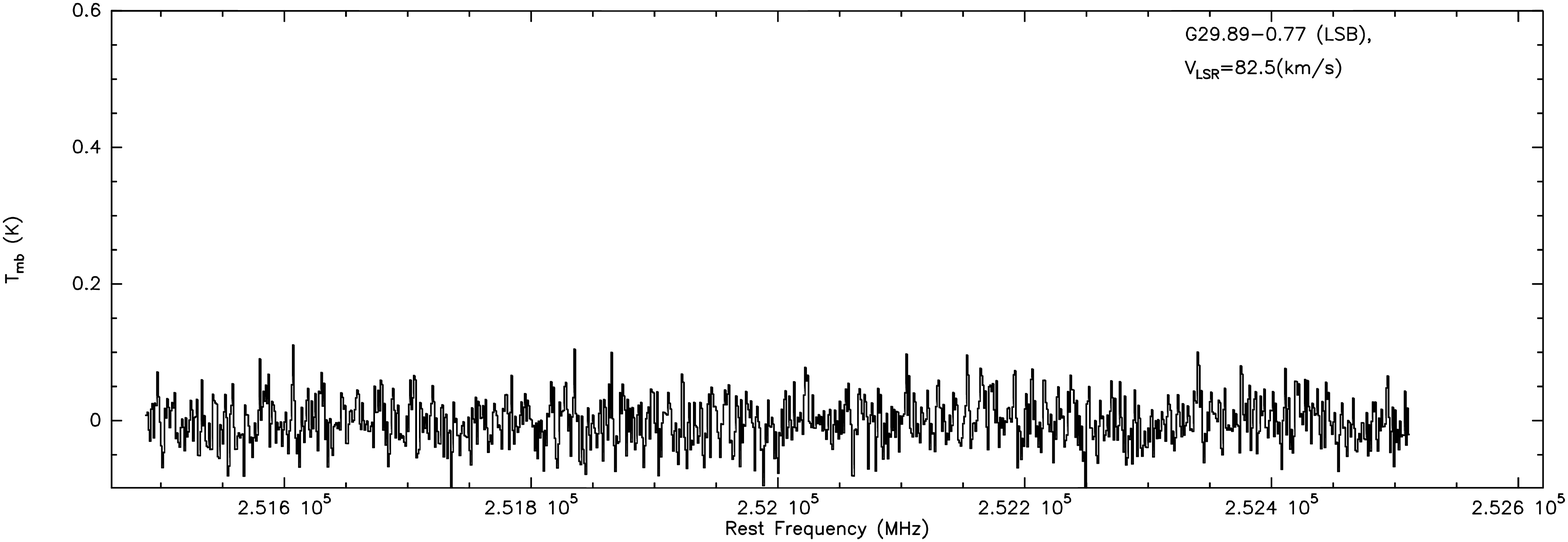}
\includegraphics[scale=.30,angle=0]{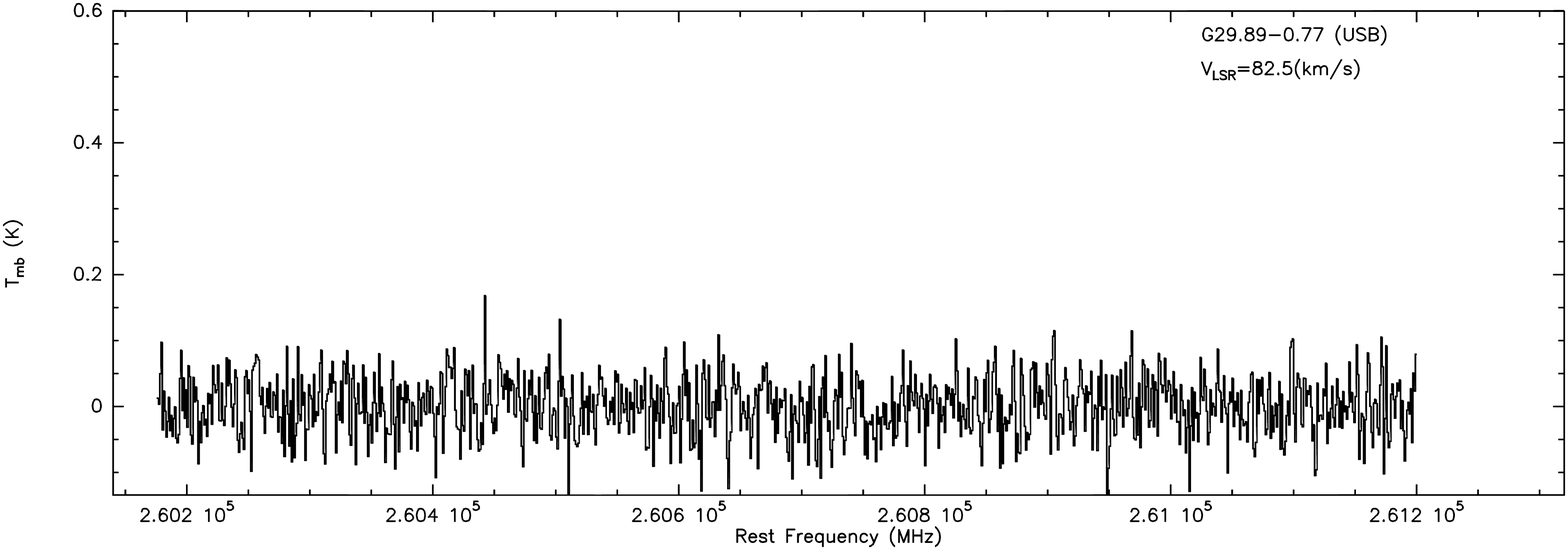}
\caption{(continued) For G29.89-0.77.}
\end{figure*}
\clearpage
 \addtocounter{figure}{-1}
\begin{figure*}
\centering
\includegraphics[scale=.30,angle=0]{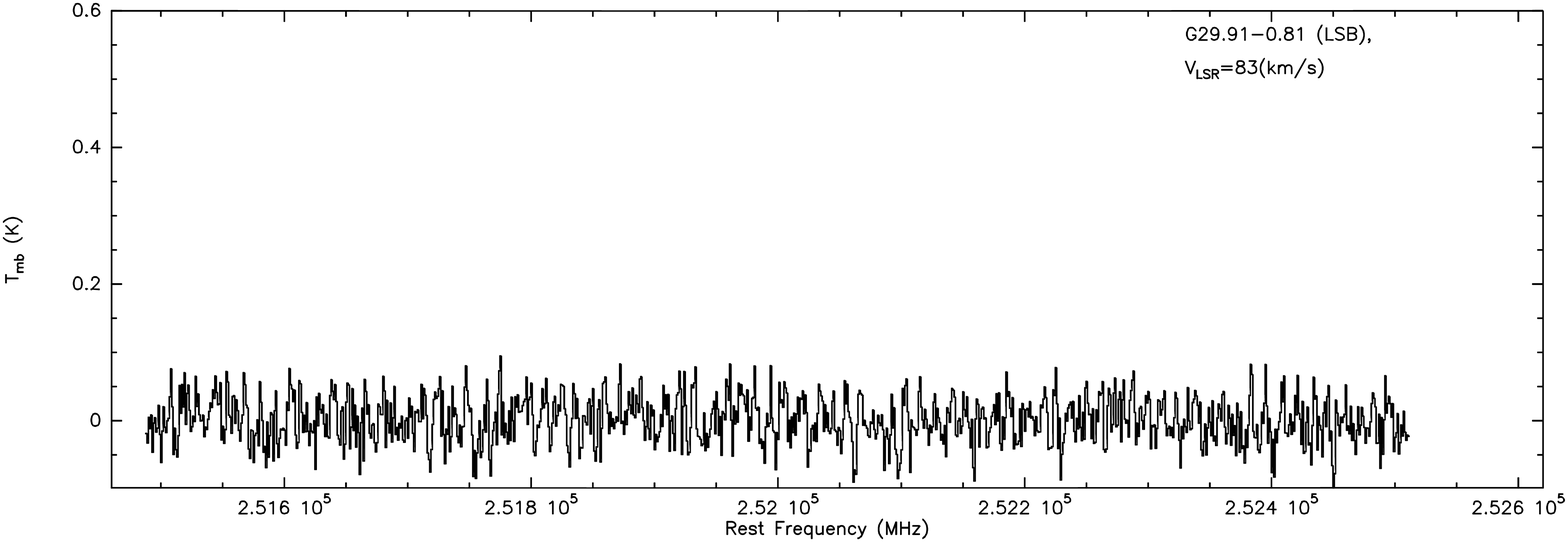}
\includegraphics[scale=.30,angle=0]{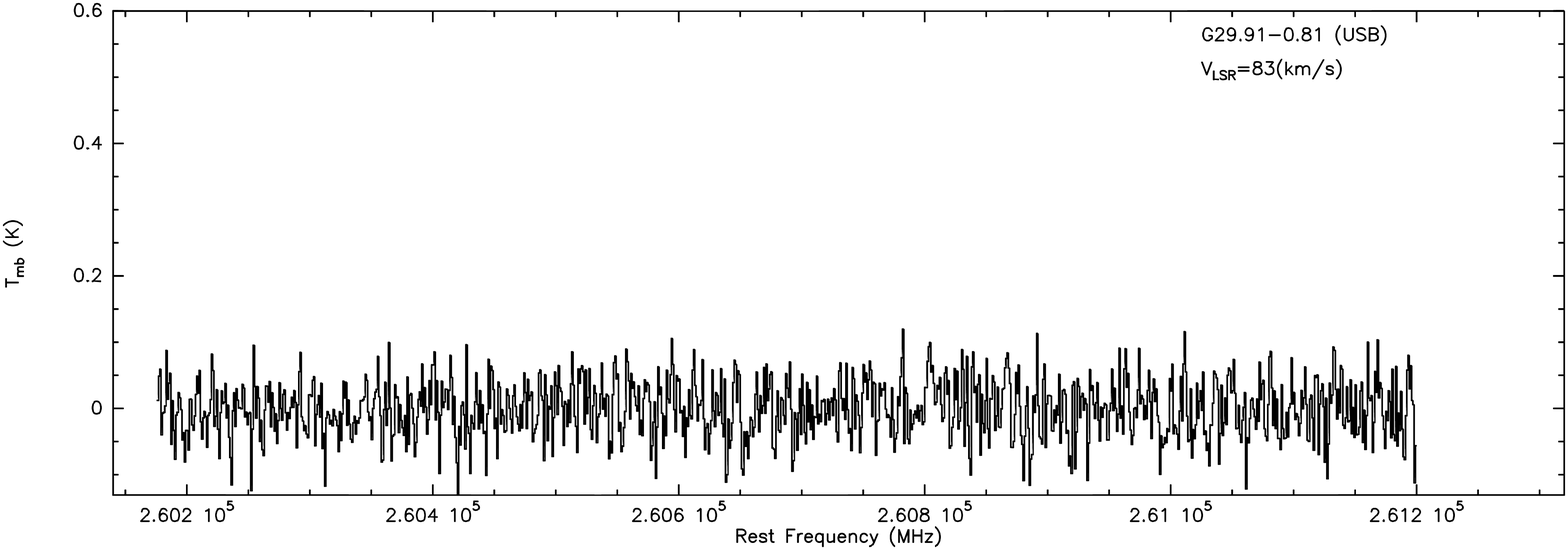}
\caption{(continued) For G29.91-0.81.}
\end{figure*}
 \addtocounter{figure}{-1}
\begin{figure*}
\centering
\includegraphics[scale=.30,angle=0]{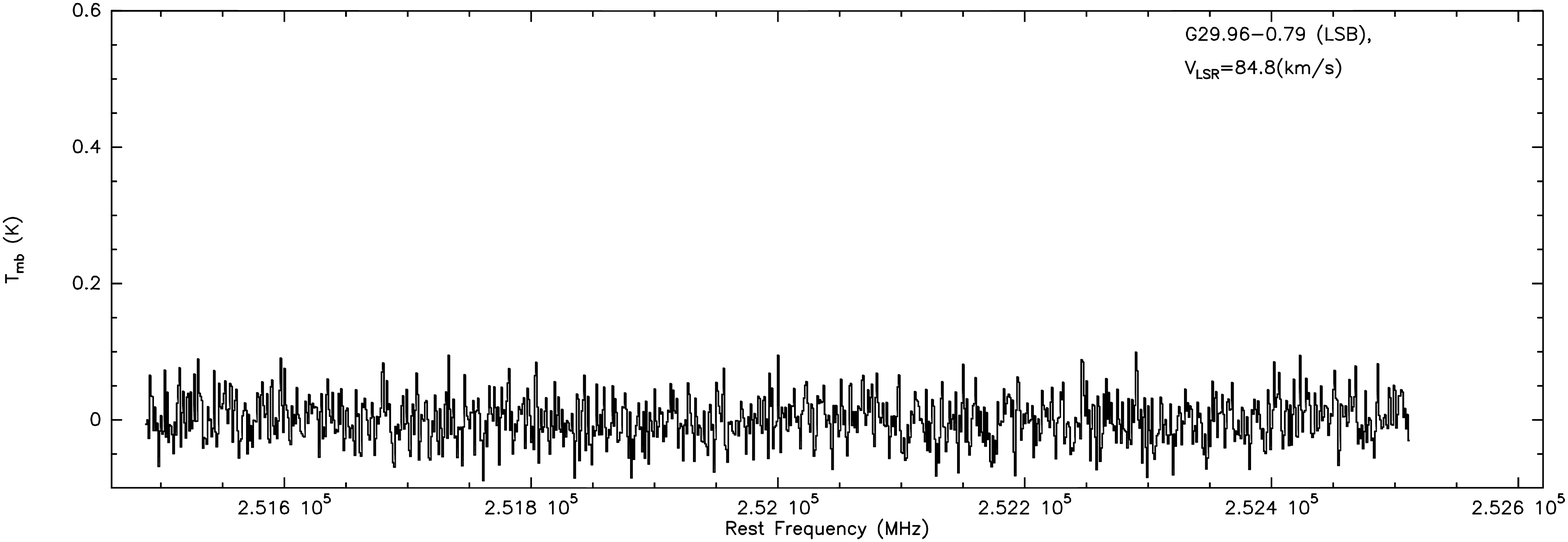}
\includegraphics[scale=.30,angle=0]{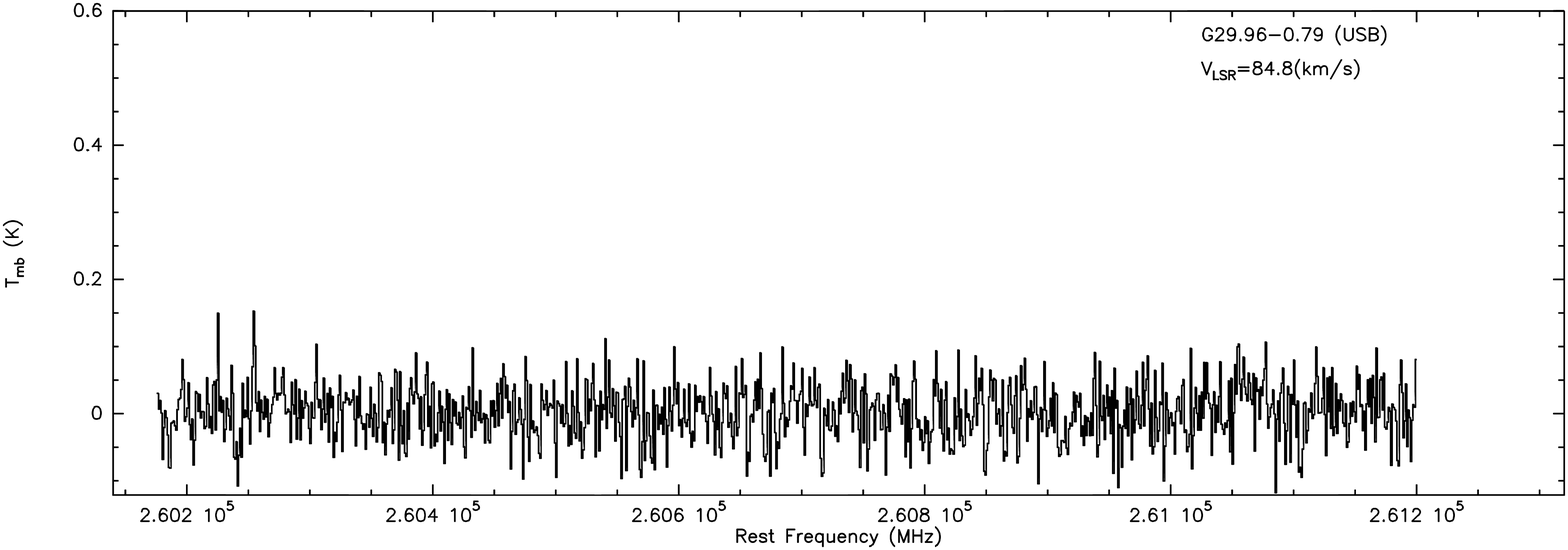}
\caption{(continued) For G29.96-0.79.}
\end{figure*}
\clearpage
 \addtocounter{figure}{-1}
\begin{figure*}
\centering
\includegraphics[scale=.30,angle=0]{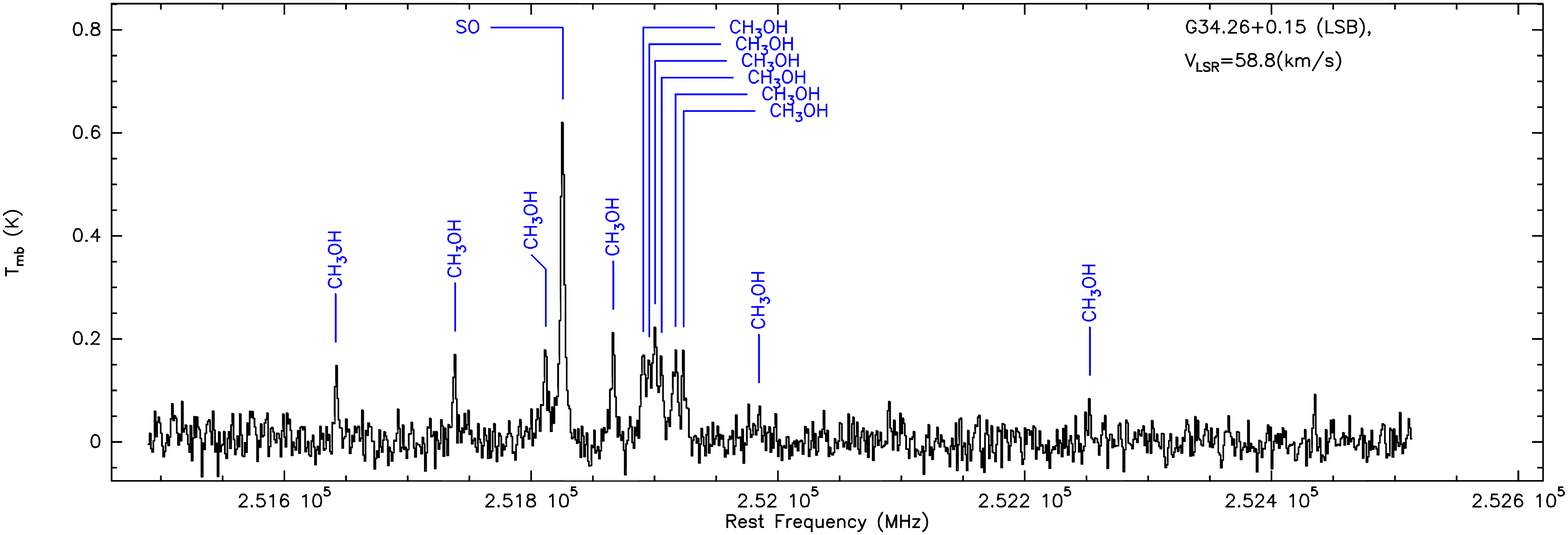}
\includegraphics[scale=.30,angle=0]{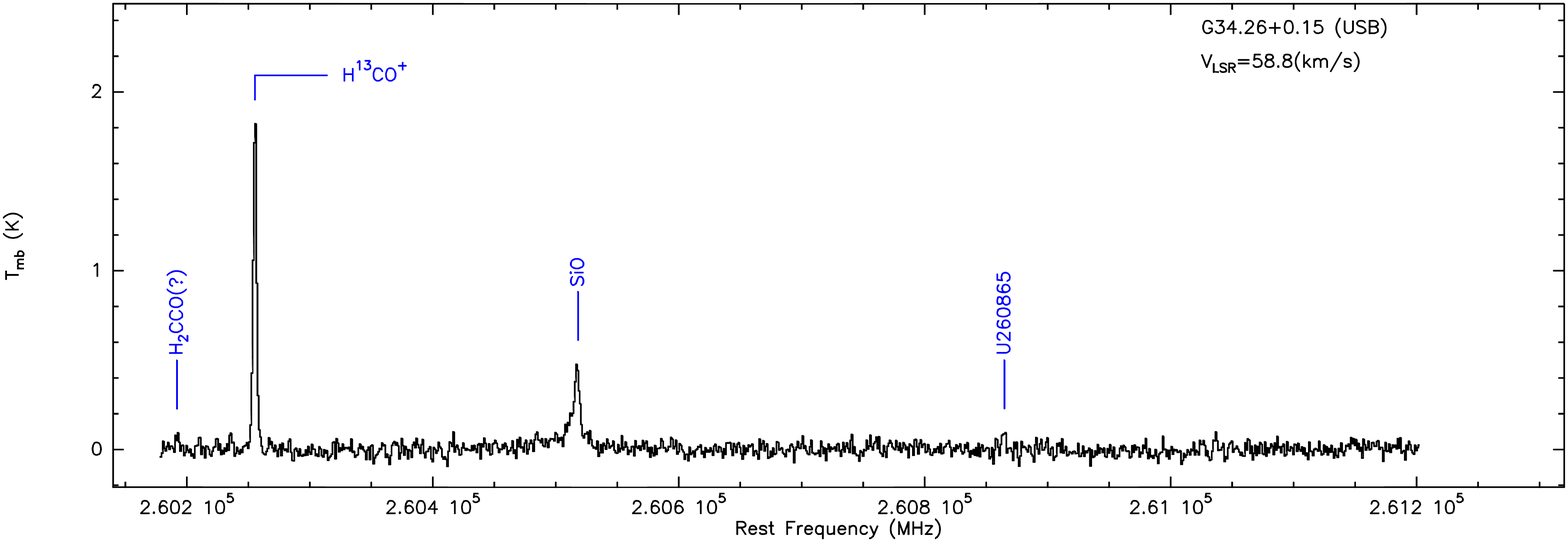}
\caption{(continued) For G34.26+0.15.}
\end{figure*}
 \addtocounter{figure}{-1}
\begin{figure*}
\centering
\includegraphics[scale=.30,angle=0]{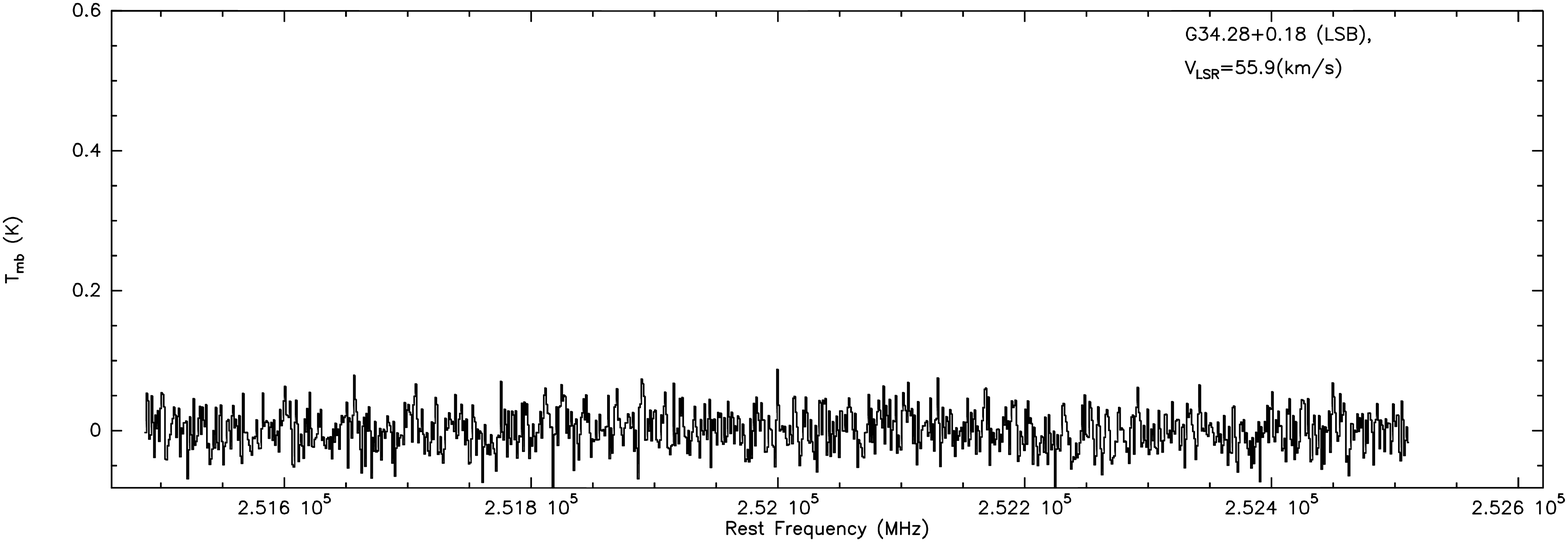}
\includegraphics[scale=.30,angle=0]{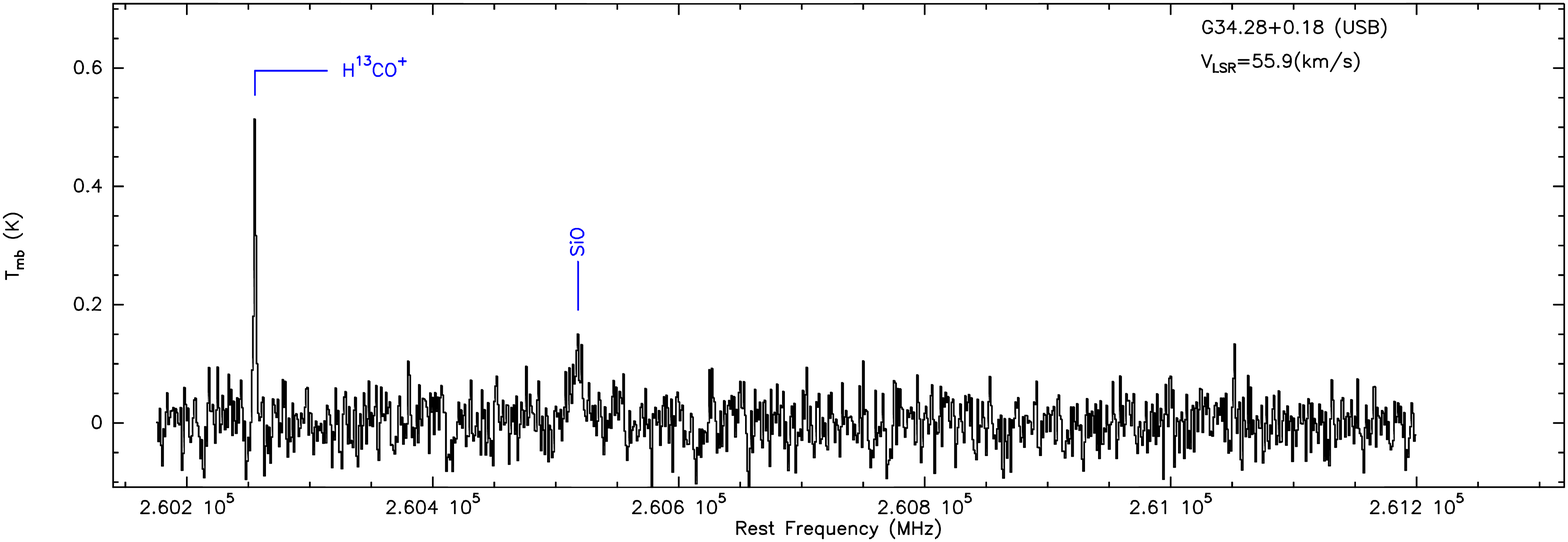}
\caption{(continued) For G34.28+0.18.}
\end{figure*}
\clearpage
 \addtocounter{figure}{-1}
\begin{figure*}
\centering
\includegraphics[scale=.30,angle=0]{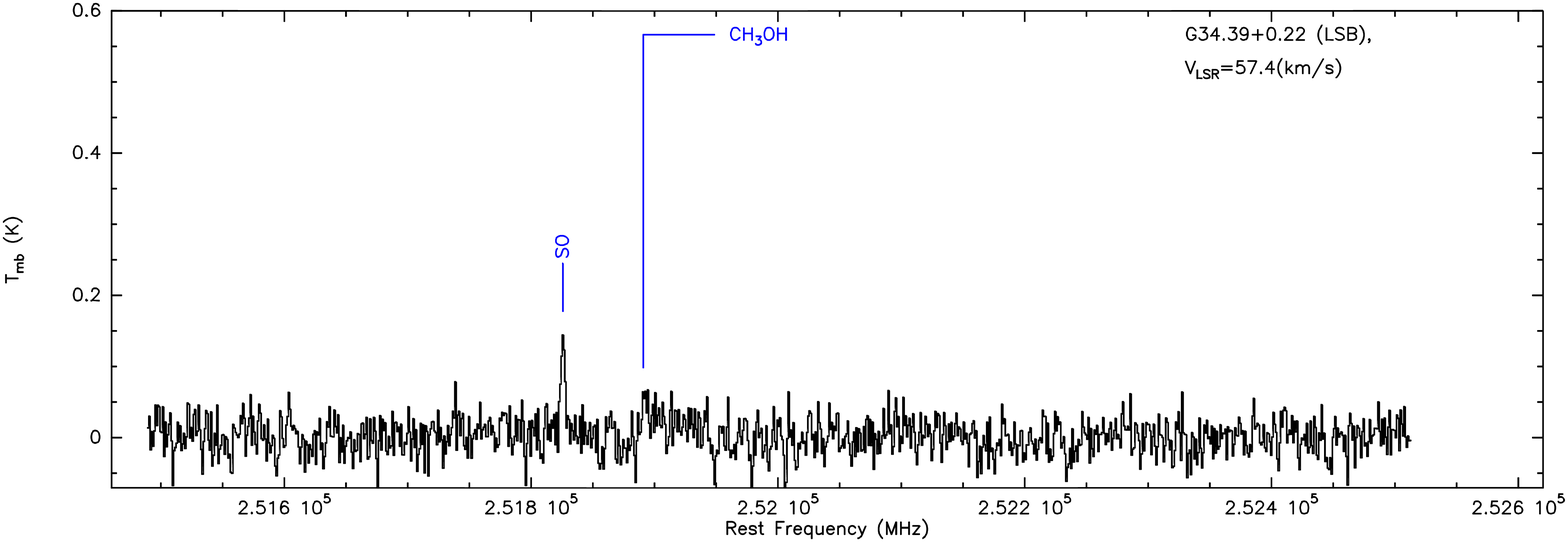}
\includegraphics[scale=.30,angle=0]{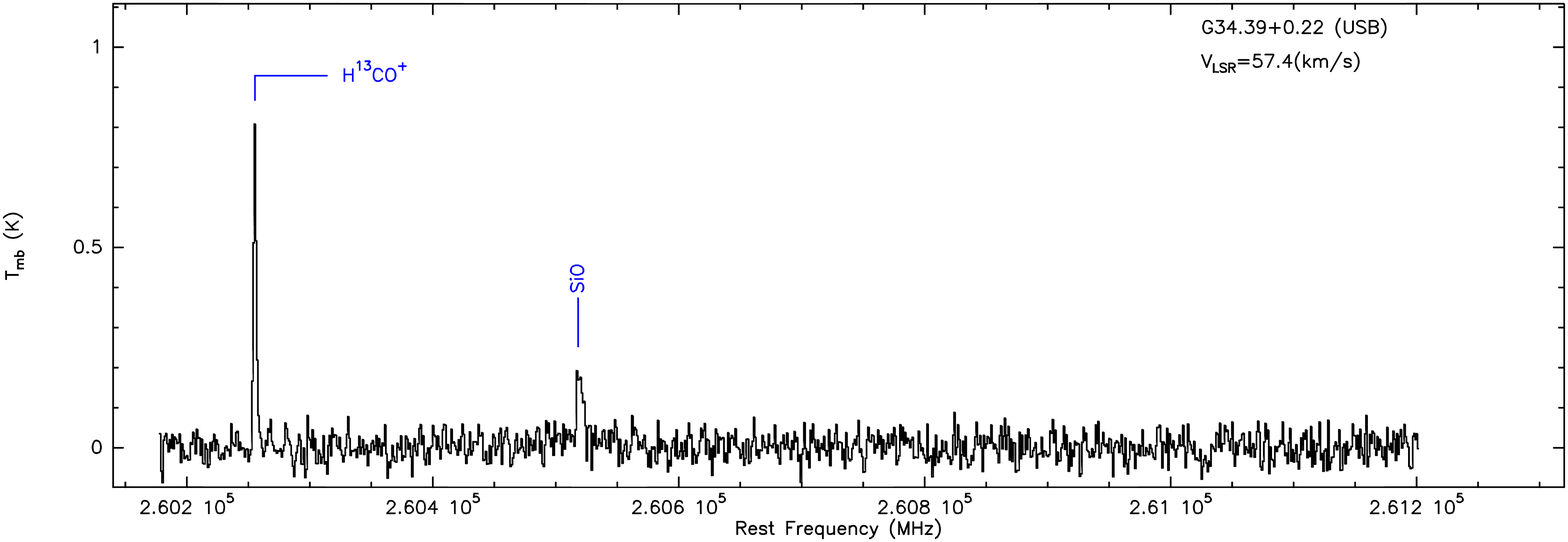}
\caption{(continued) For G34.39+0.22.}
\end{figure*}
 \addtocounter{figure}{-1}
\begin{figure*}
\centering
\includegraphics[scale=.30,angle=0]{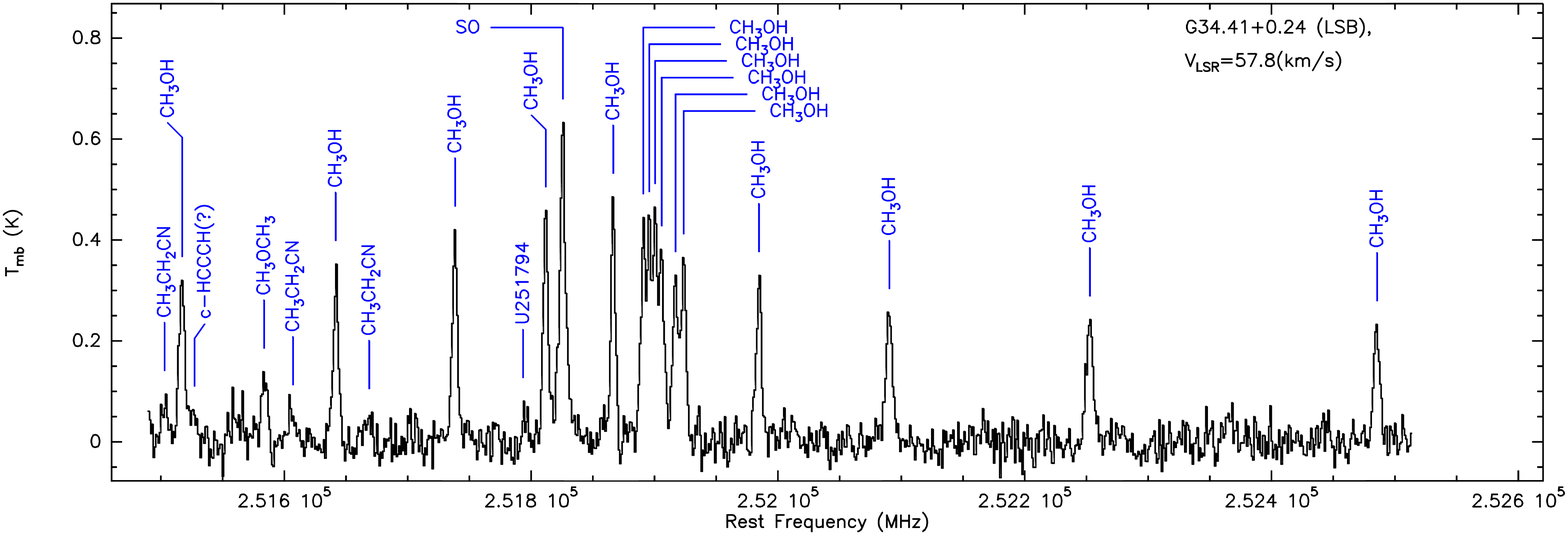}
\includegraphics[scale=.30,angle=0]{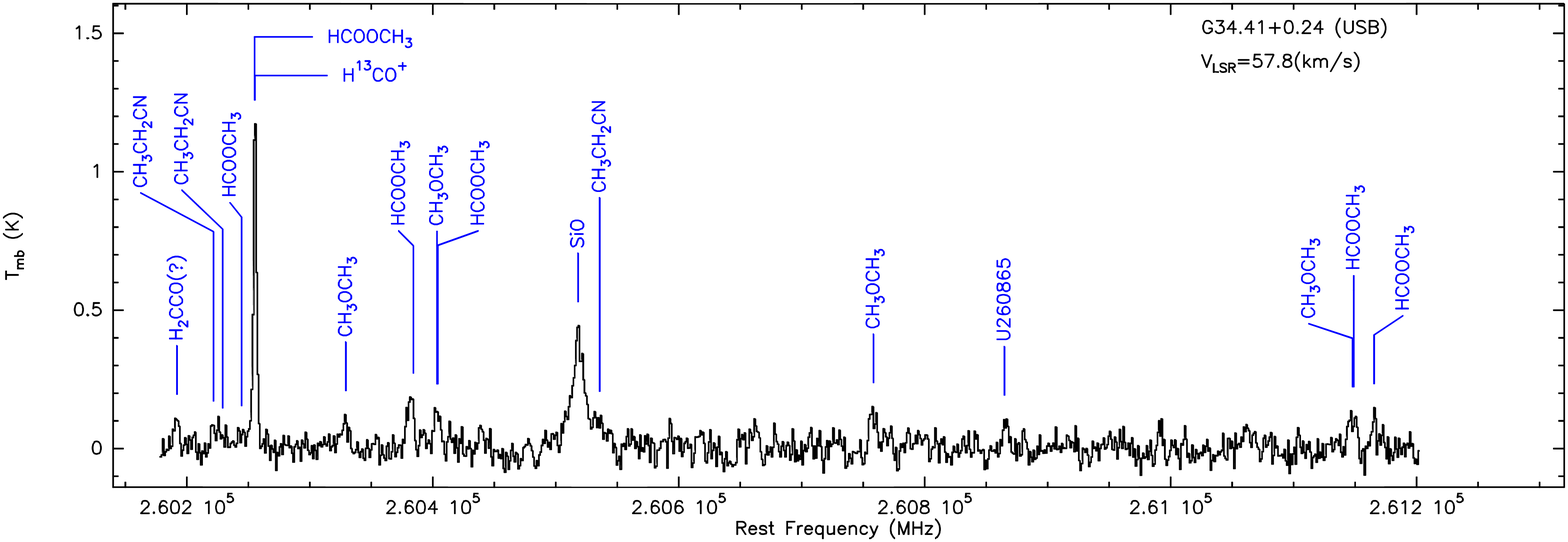}
\caption{(continued) For G34.41+0.24.}
\end{figure*}
\clearpage
 \addtocounter{figure}{-1}
\begin{figure*}
\centering
\includegraphics[scale=.30,angle=0]{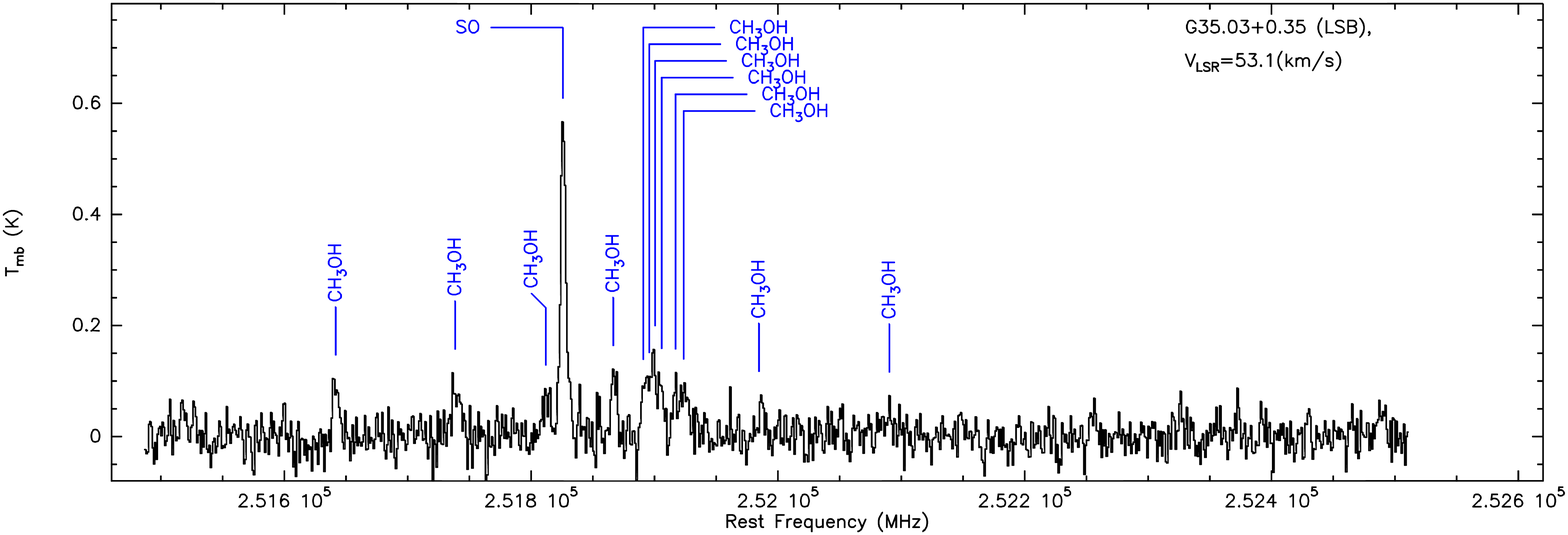}
\includegraphics[scale=.30,angle=0]{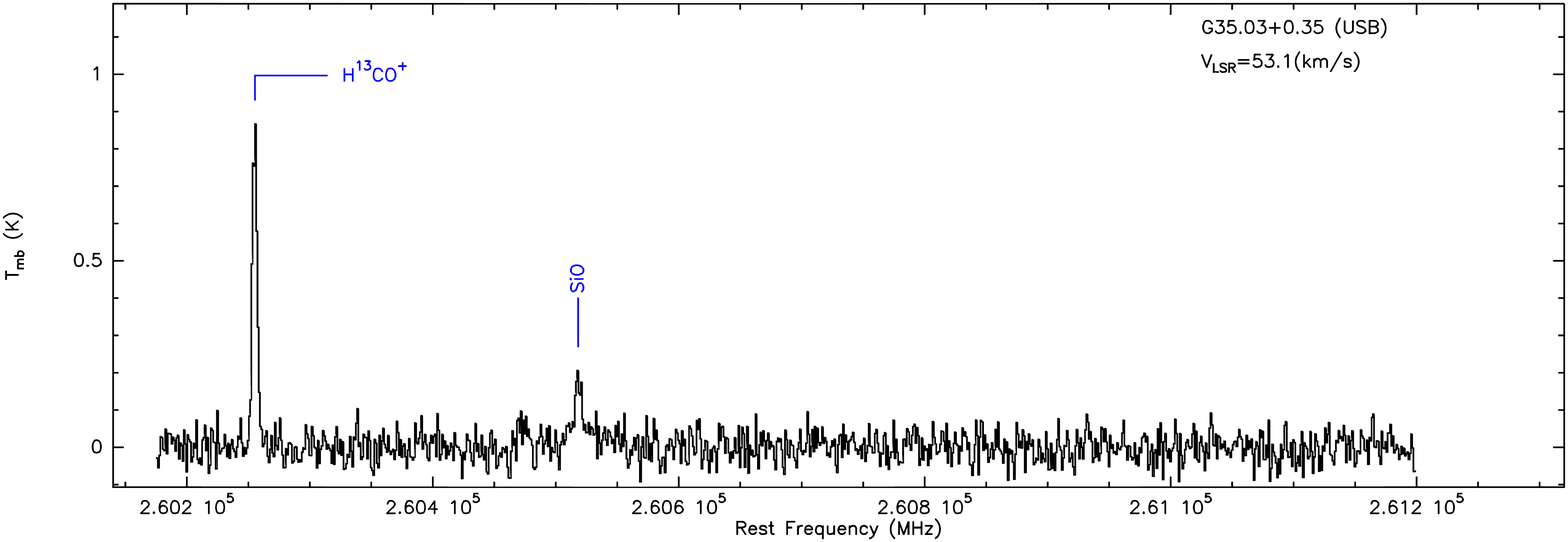}
\caption{(continued) For G35.03+0.35.}
\end{figure*}
 \addtocounter{figure}{-1}
\begin{figure*}
\centering
\includegraphics[scale=.30,angle=0]{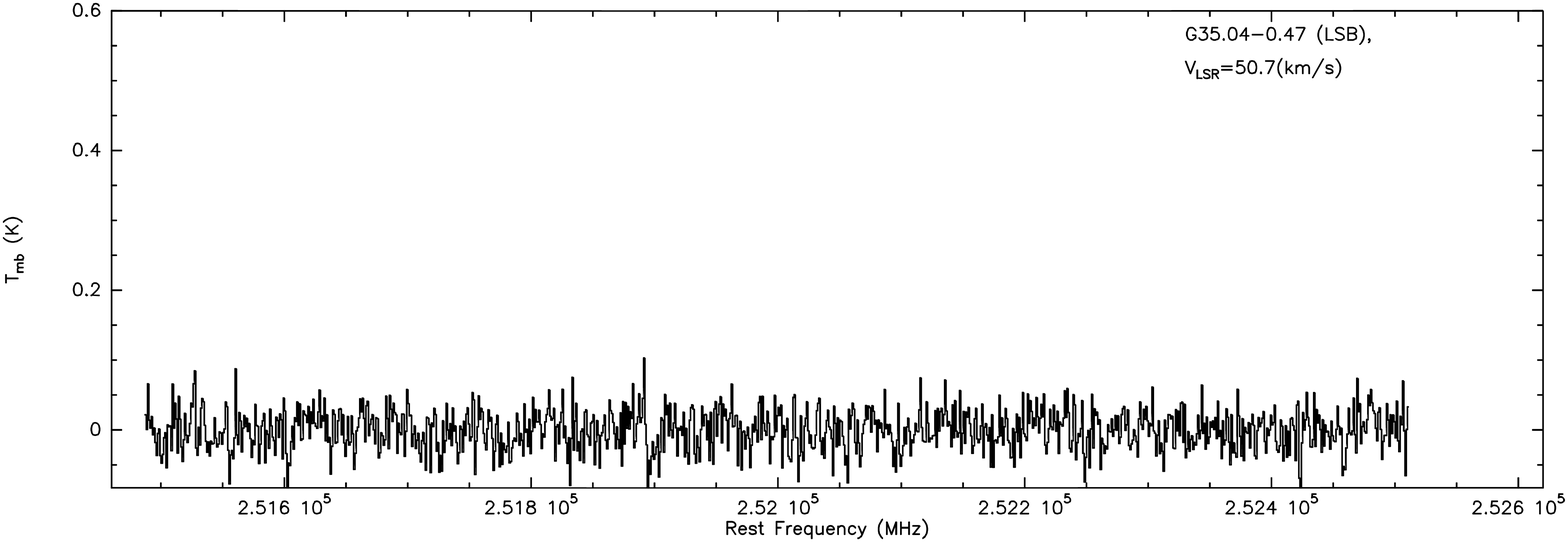}
\includegraphics[scale=.30,angle=0]{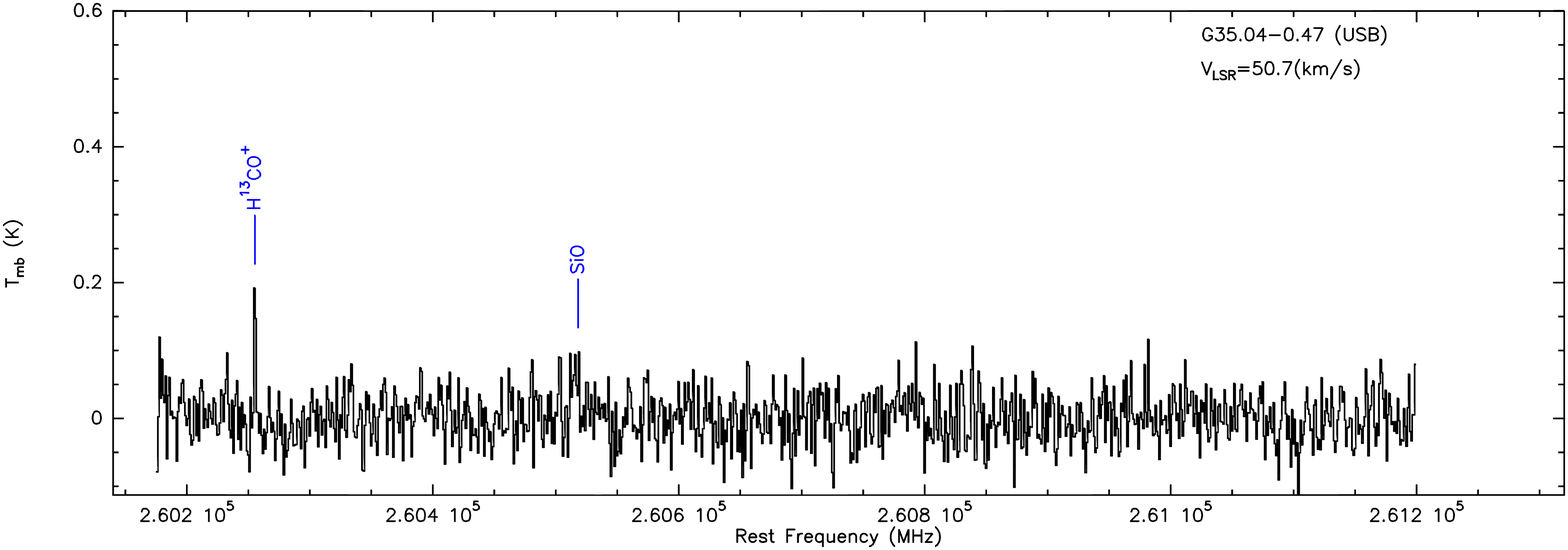}
\caption{(continued) For G35.04-0.47.}
\end{figure*}
\clearpage
 \addtocounter{figure}{-1}
\begin{figure*}
\centering
\includegraphics[scale=.30,angle=0]{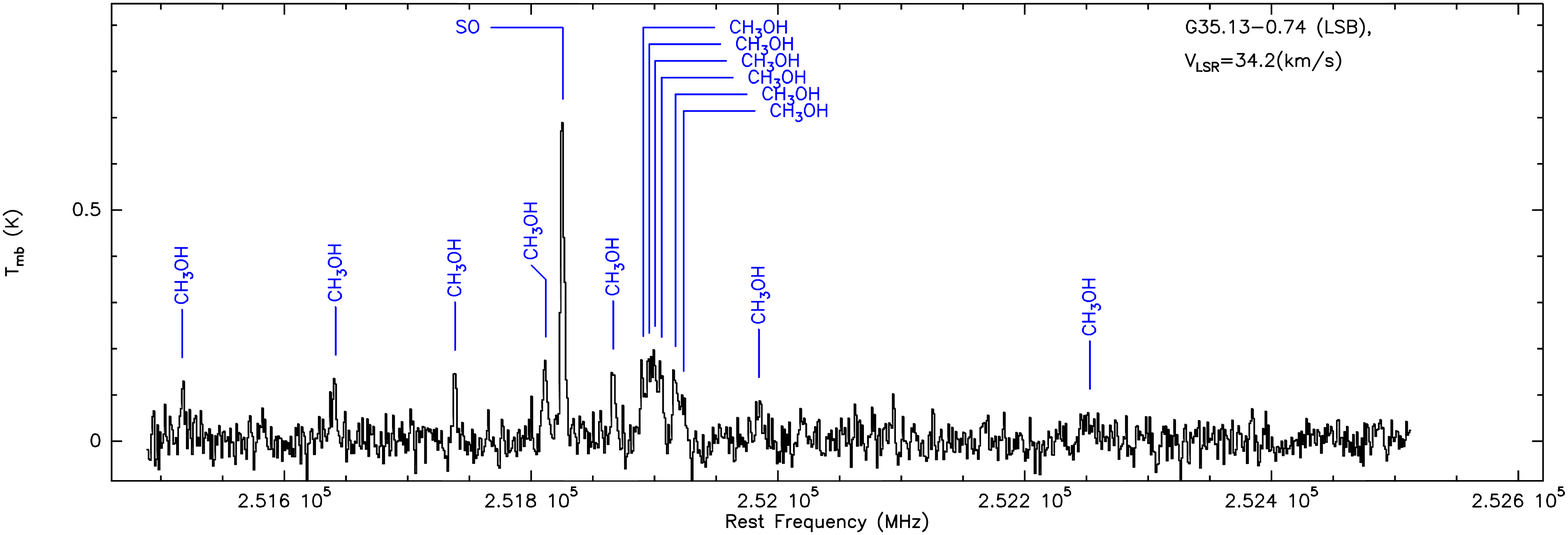}
\includegraphics[scale=.30,angle=0]{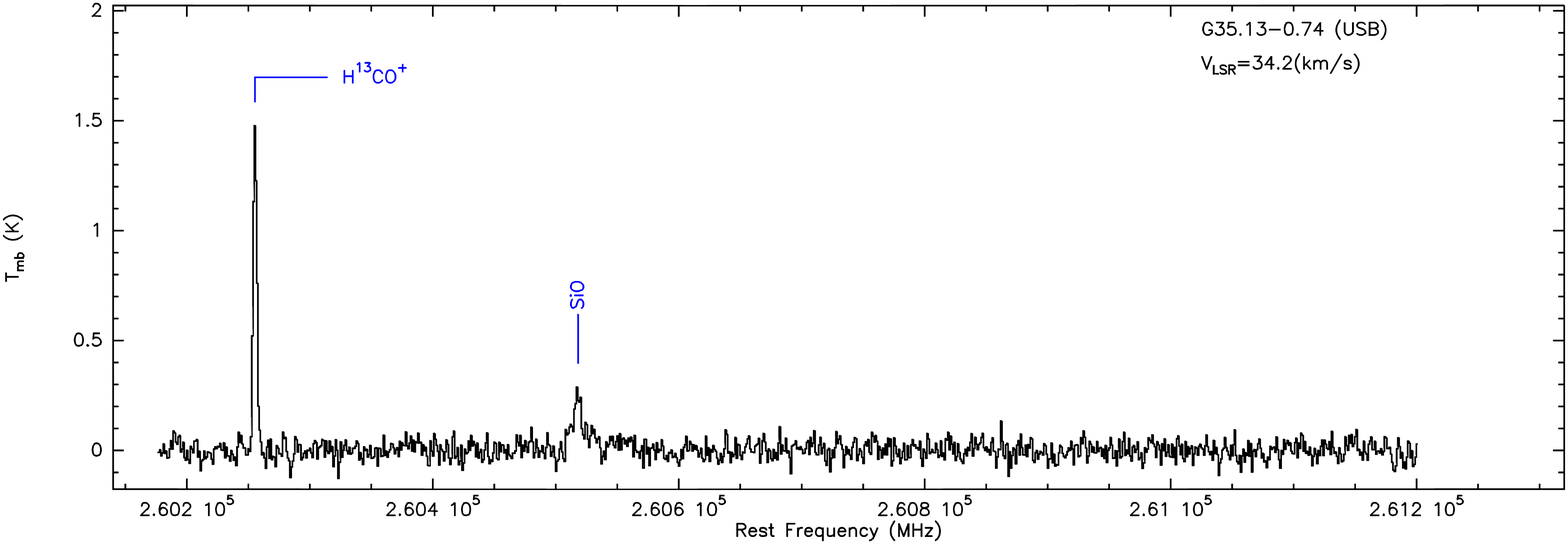}
\caption{(continued) For G35.13-0.74.}
\end{figure*}
 \addtocounter{figure}{-1}
\begin{figure*}
\centering
\includegraphics[scale=.30,angle=0]{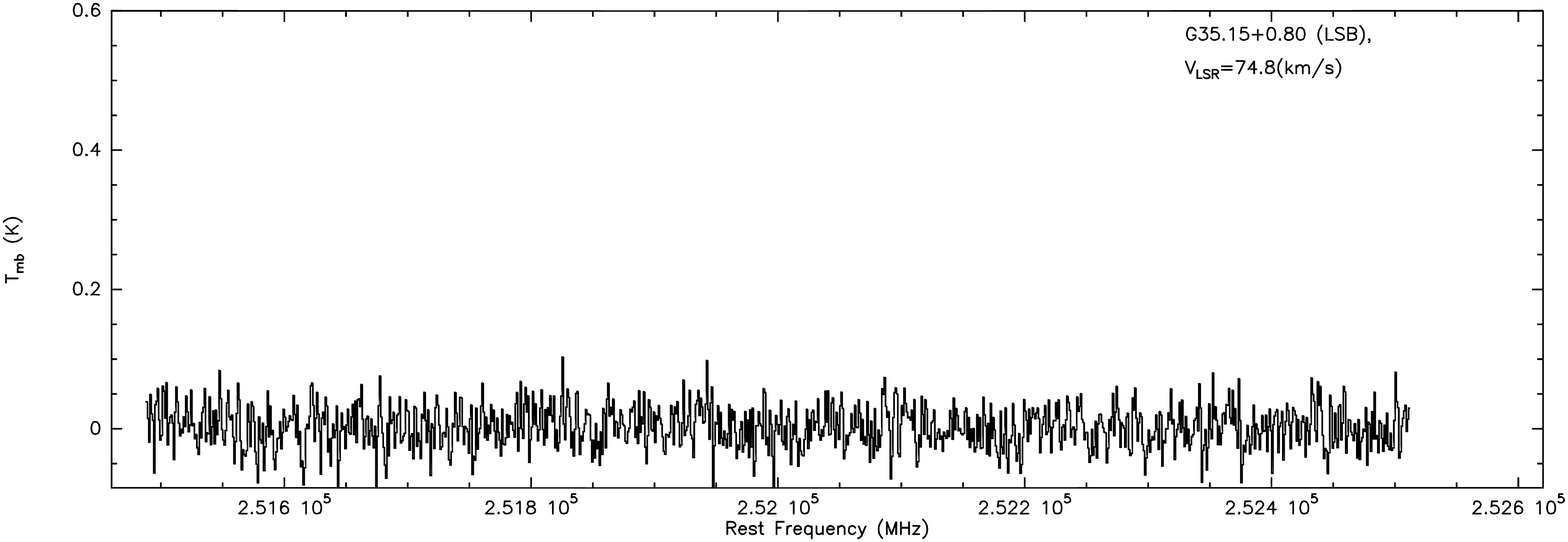}
\includegraphics[scale=.30,angle=0]{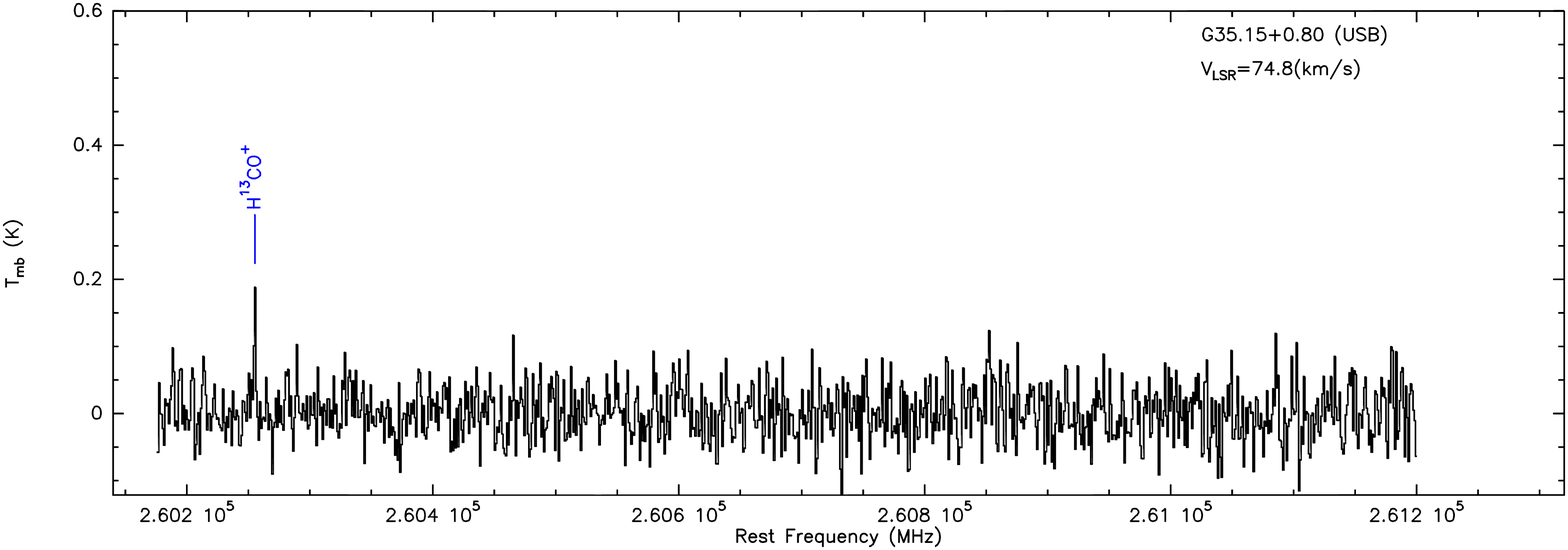}
\caption{(continued) For G35.15+0.80.}
\end{figure*}
\clearpage
 \addtocounter{figure}{-1}
\begin{figure*}
\centering
\includegraphics[scale=.30,angle=0]{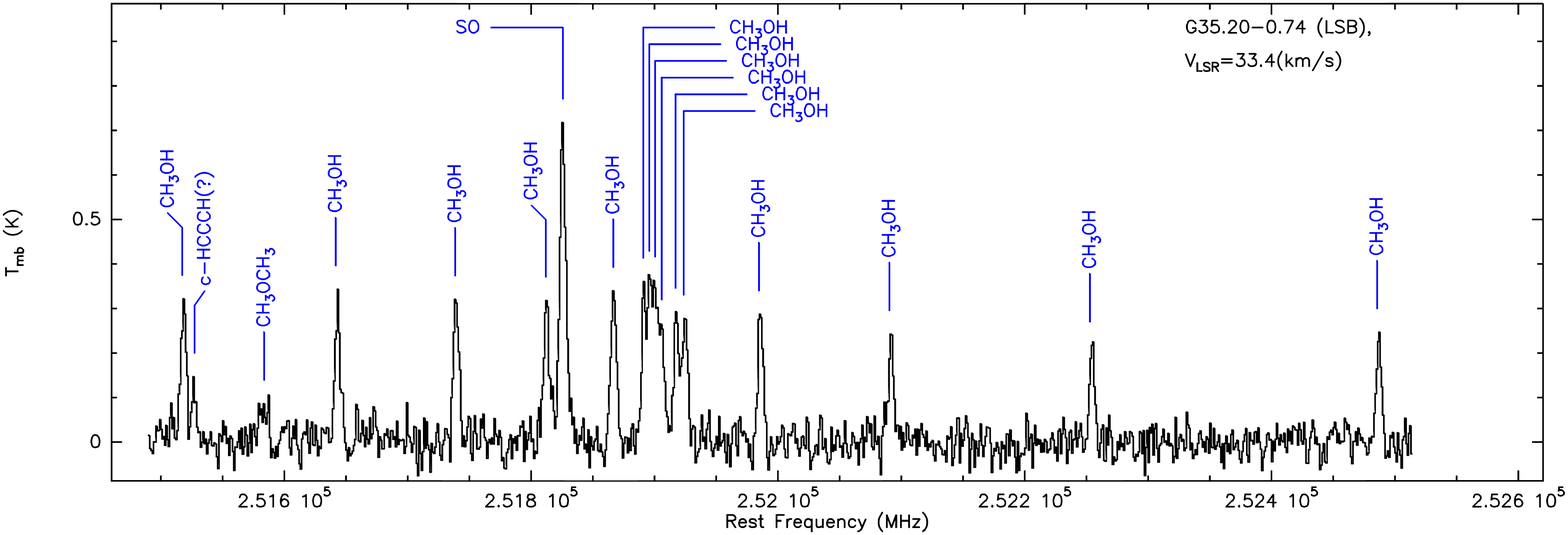}
\includegraphics[scale=.30,angle=0]{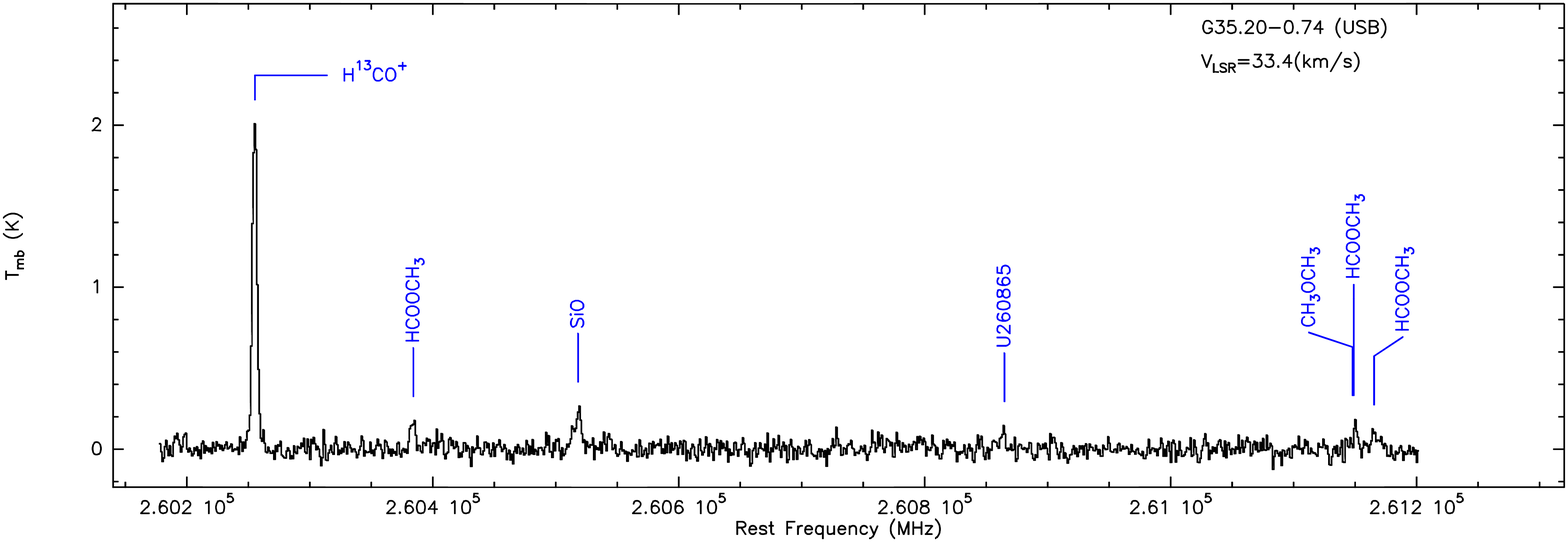}
\caption{(continued) For G35.20-0.74.}
\end{figure*}
 \addtocounter{figure}{-1}
\begin{figure*}
\centering
\includegraphics[scale=.30,angle=0]{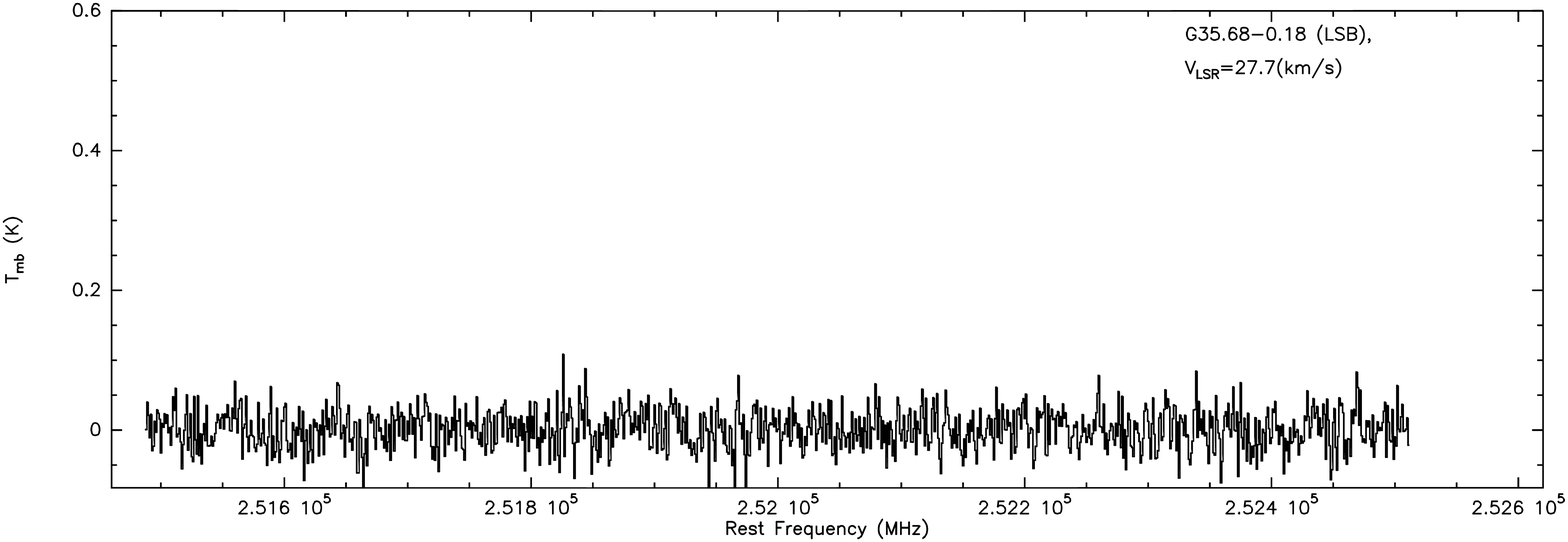}
\includegraphics[scale=.30,angle=0]{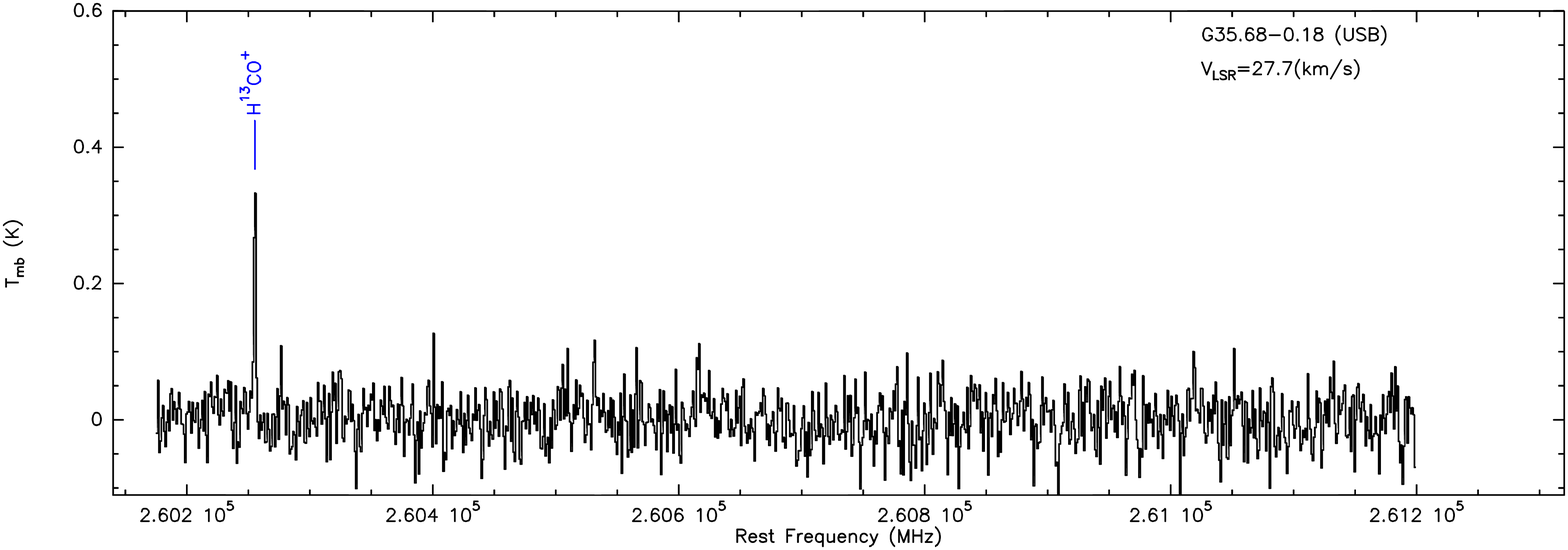}
\caption{(continued) For G35.68-0.18.}
\end{figure*}
\clearpage
 \addtocounter{figure}{-1}
\begin{figure*}
\centering
\includegraphics[scale=.30,angle=0]{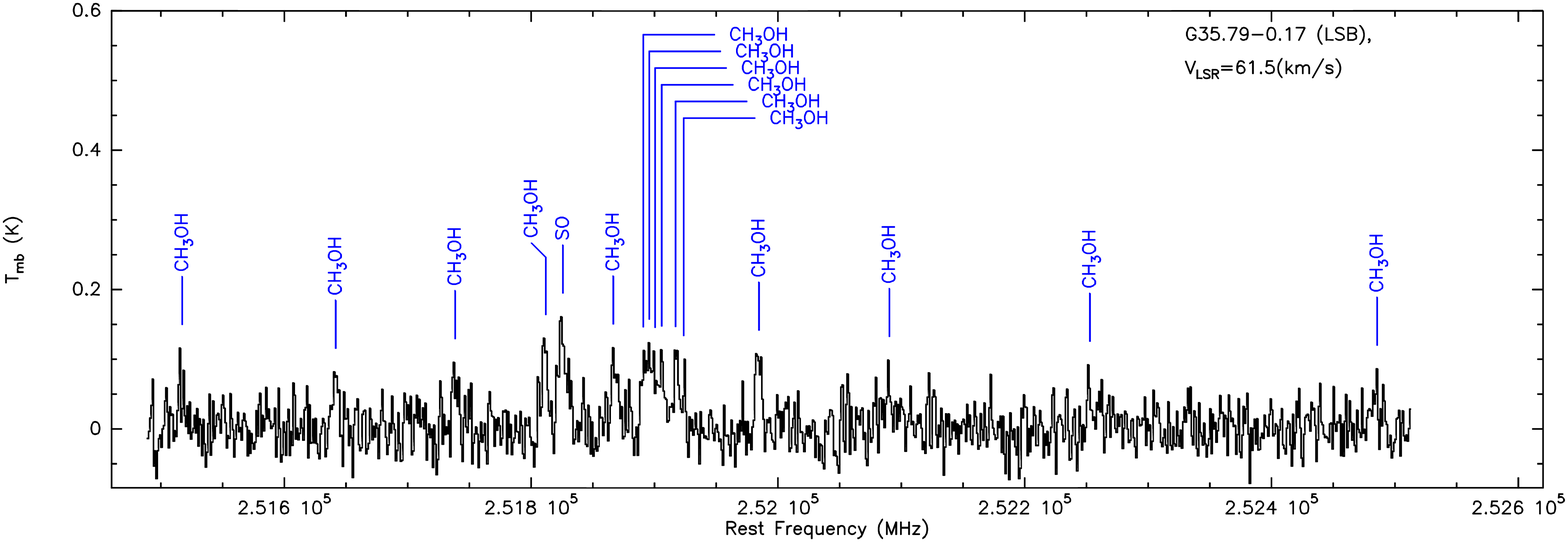}
\includegraphics[scale=.30,angle=0]{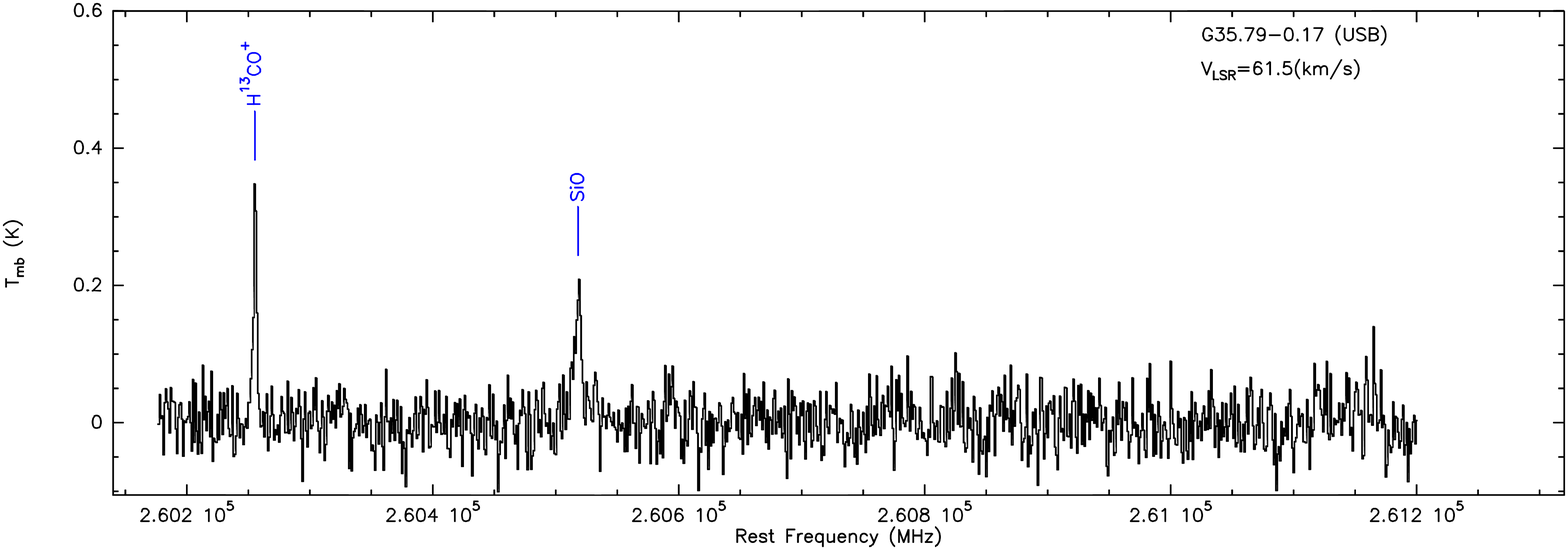}
\caption{(continued) For G35.79-0.17.}
\end{figure*}
 \addtocounter{figure}{-1}
\begin{figure*}
\centering
\includegraphics[scale=.30,angle=0]{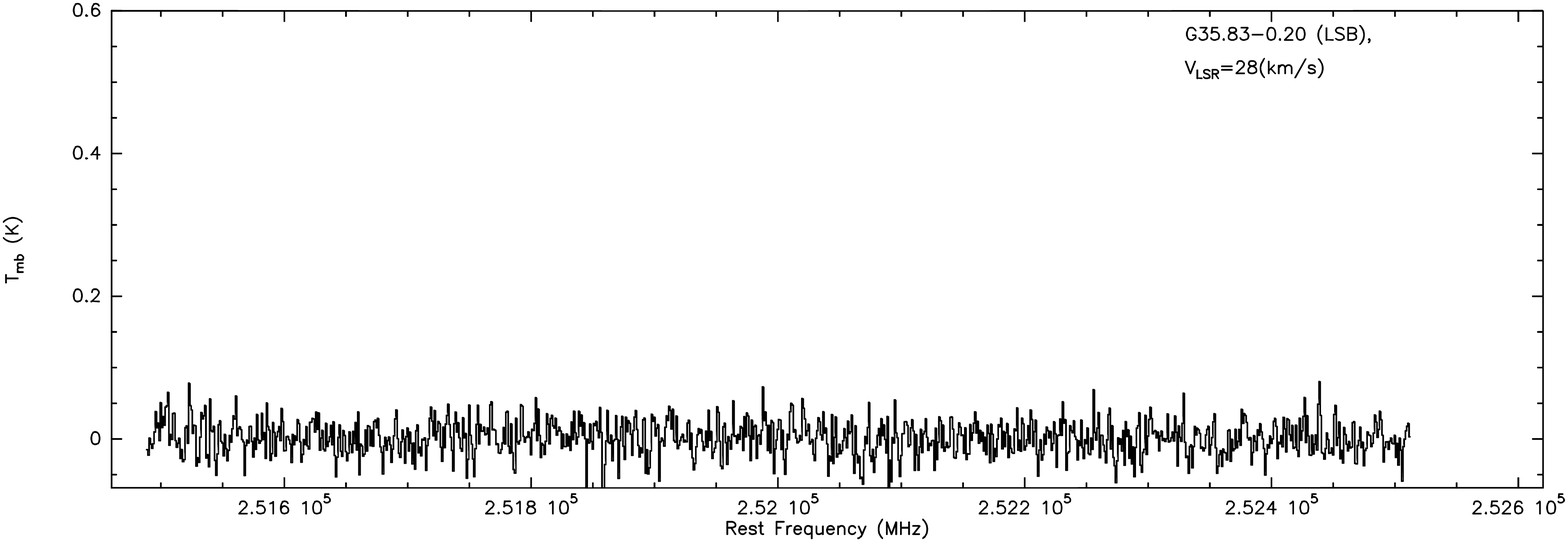}
\includegraphics[scale=.30,angle=0]{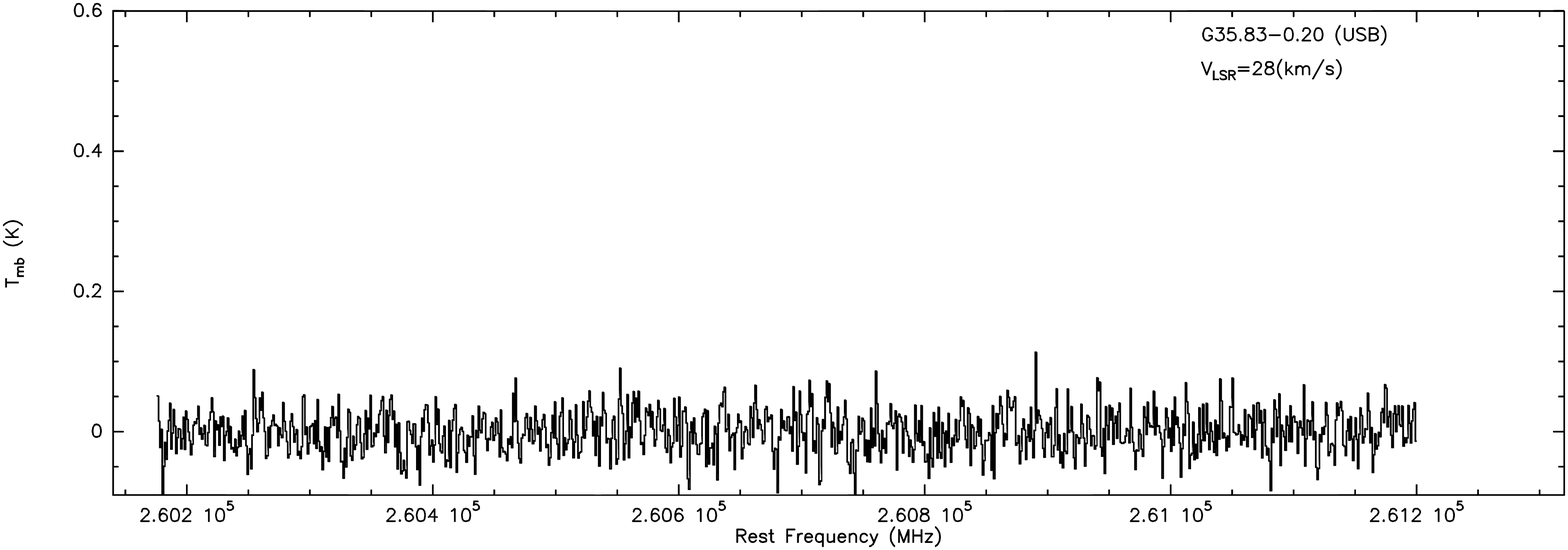}
\caption{(continued) For G35.83-0.20.}
\end{figure*}
\clearpage
 \addtocounter{figure}{-1}
\begin{figure*}
\centering
\includegraphics[scale=.30,angle=0]{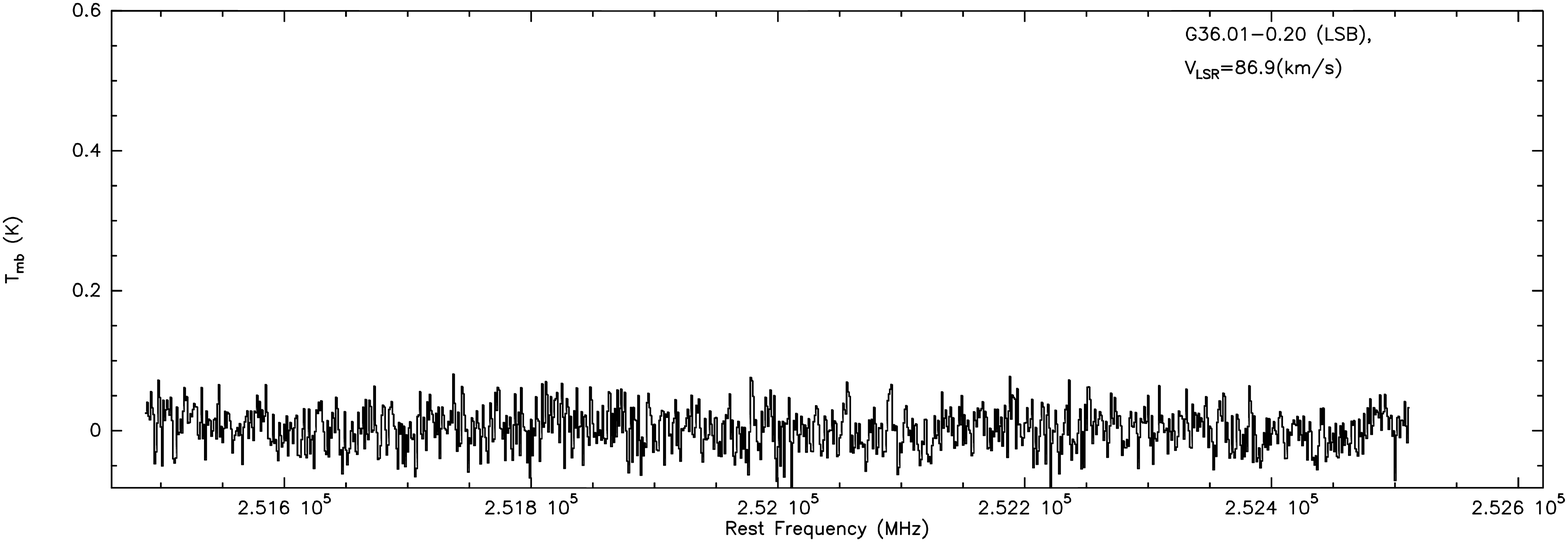}
\includegraphics[scale=.30,angle=0]{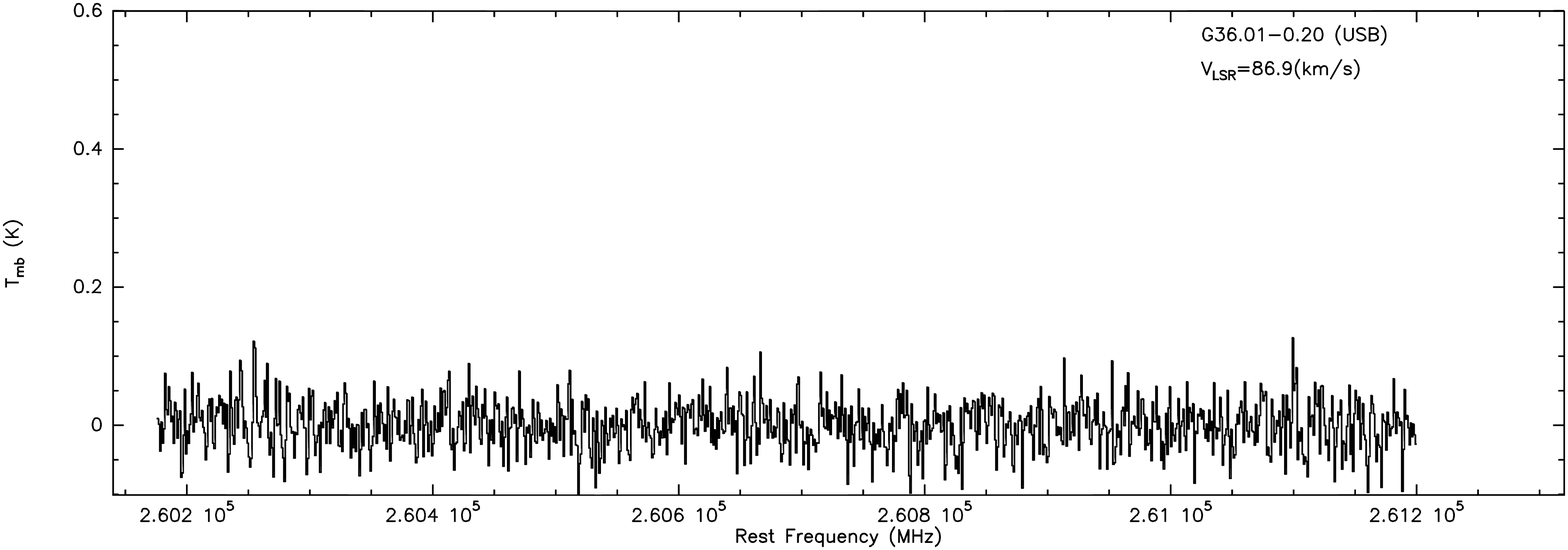}
\caption{(continued) For G36.01-0.20.}
\end{figure*}
 \addtocounter{figure}{-1}
\begin{figure*}
\centering
\includegraphics[scale=.30,angle=0]{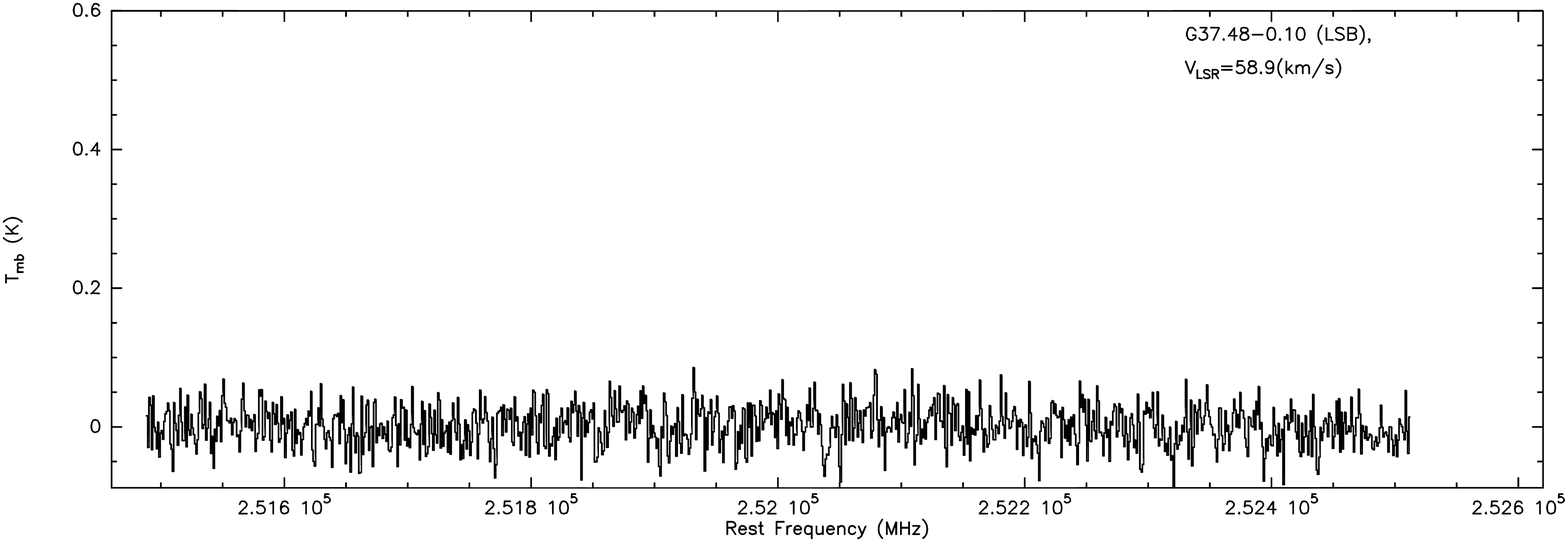}
\includegraphics[scale=.30,angle=0]{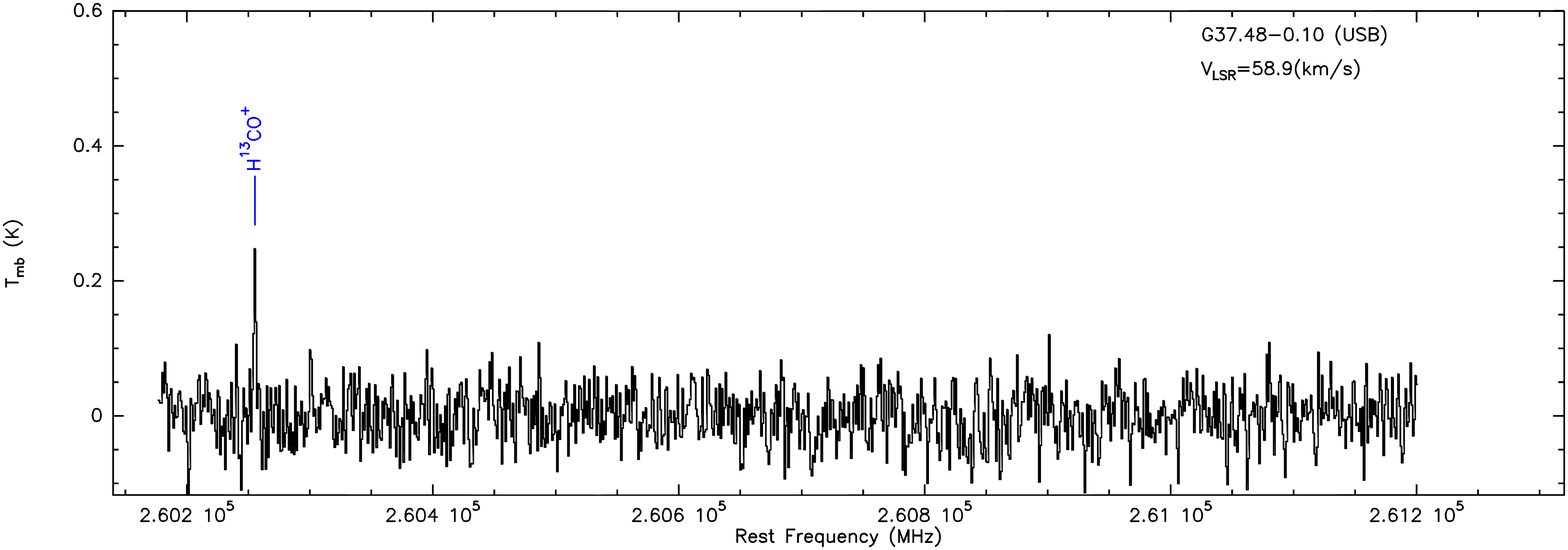}
\caption{(continued) For G37.48-0.10.}
\end{figure*}
\clearpage
 \addtocounter{figure}{-1}
\begin{figure*}
\centering
\includegraphics[scale=.30,angle=0]{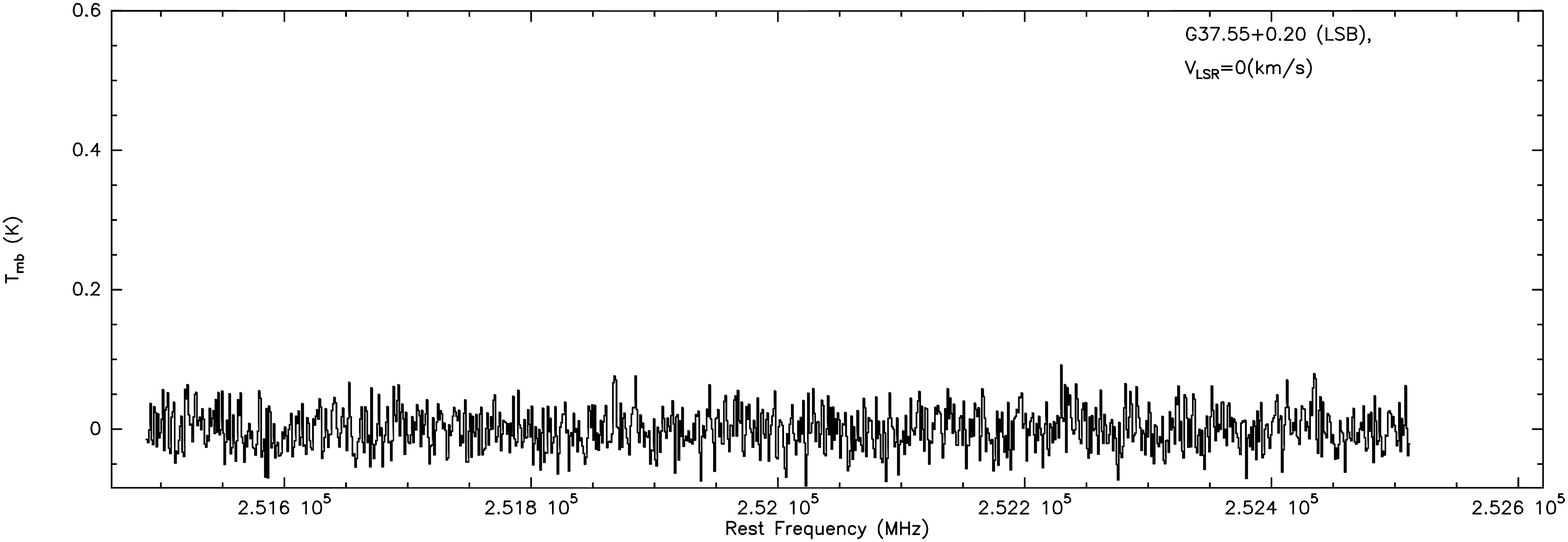}
\includegraphics[scale=.30,angle=0]{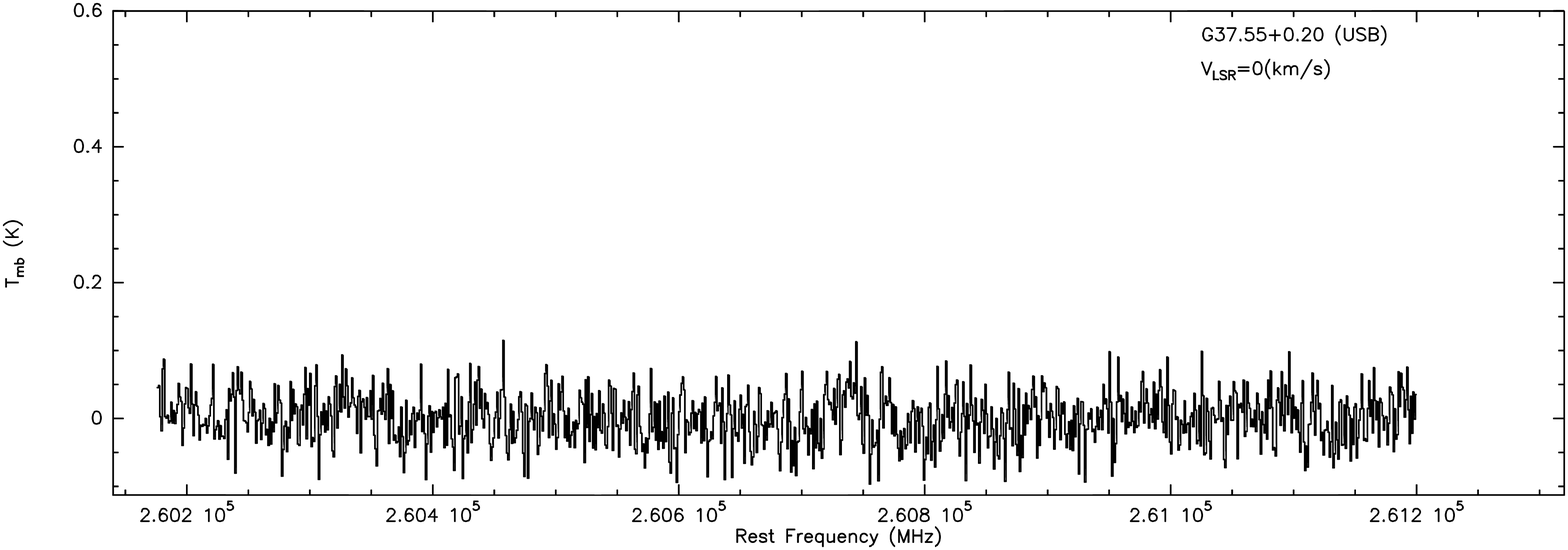}
\caption{(continued) For G37.55+0.20.}
\end{figure*}
 \addtocounter{figure}{-1}
\begin{figure*}
\centering
\includegraphics[scale=.30,angle=0]{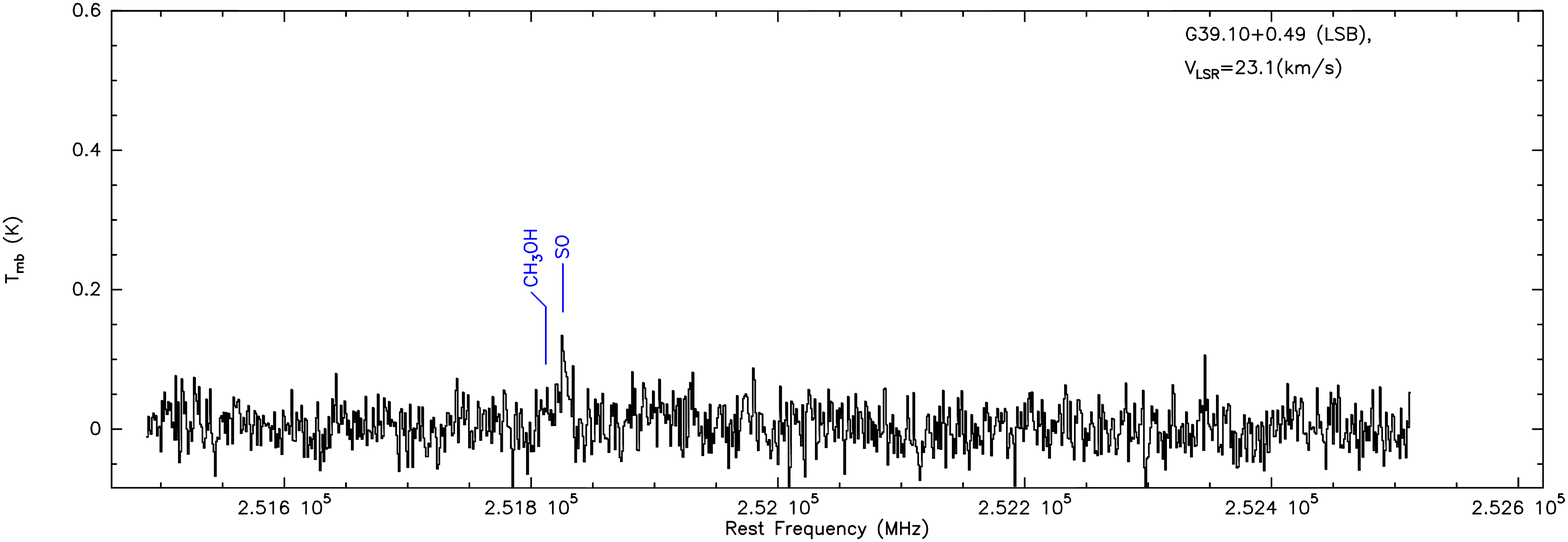}
\includegraphics[scale=.30,angle=0]{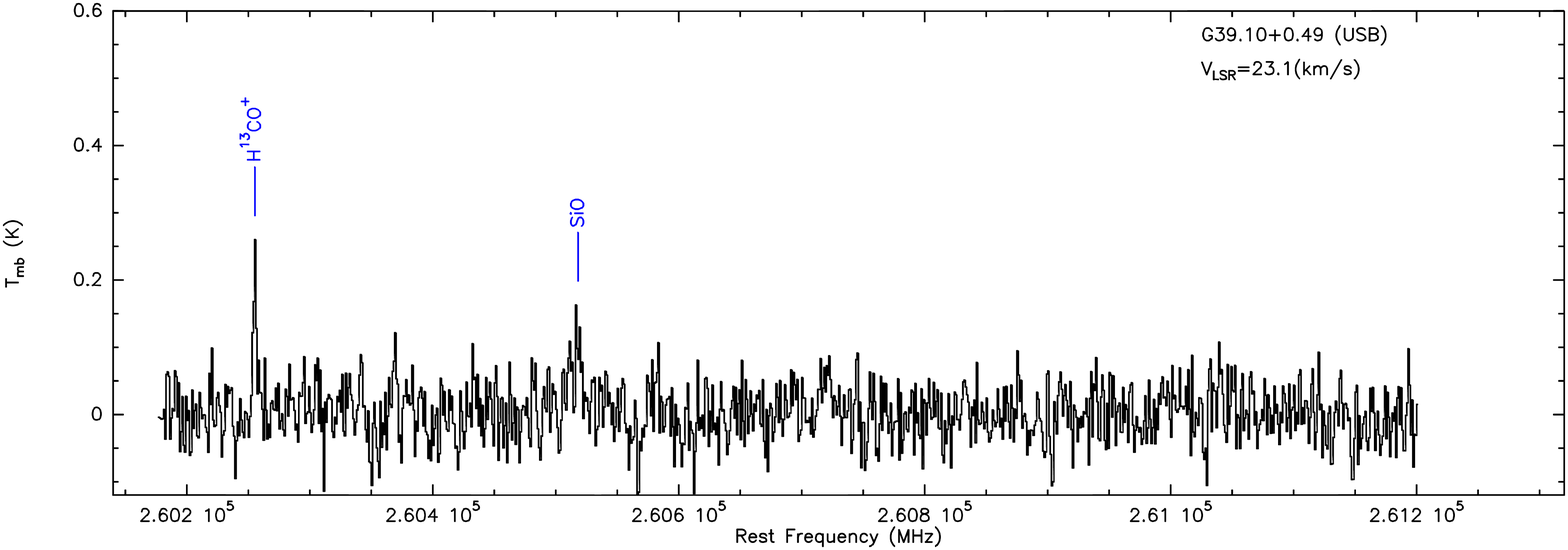}
\caption{(continued) For G39.10+0.49.}
\end{figure*}
\clearpage
 \addtocounter{figure}{-1}
\begin{figure*}
\centering
\includegraphics[scale=.30,angle=0]{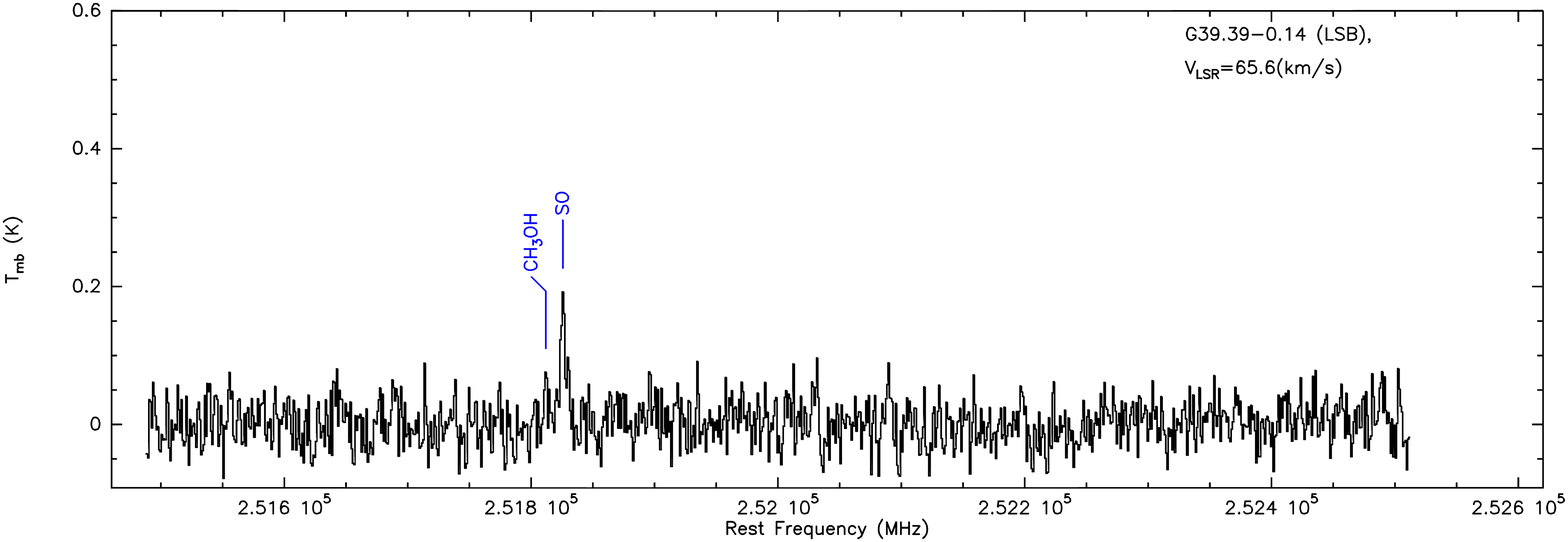}
\includegraphics[scale=.30,angle=0]{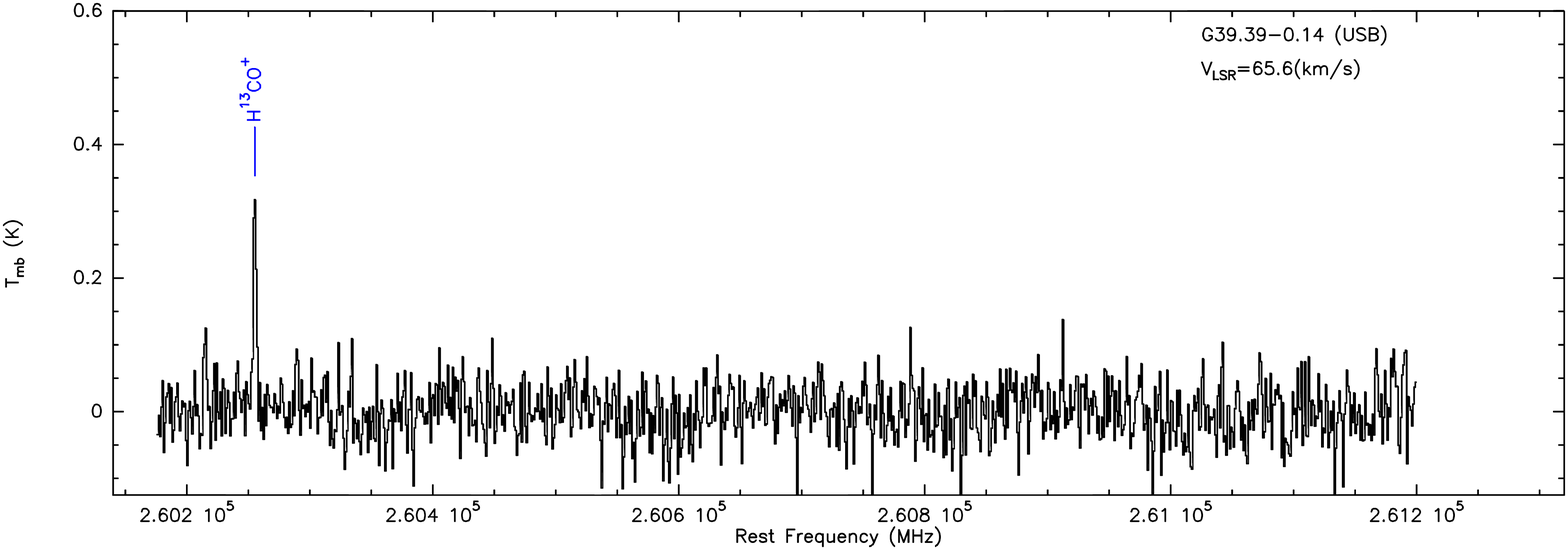}
\caption{(continued) For G39.39-0.14.}
\end{figure*}
 \addtocounter{figure}{-1}
\begin{figure*}
\centering
\includegraphics[scale=.30,angle=0]{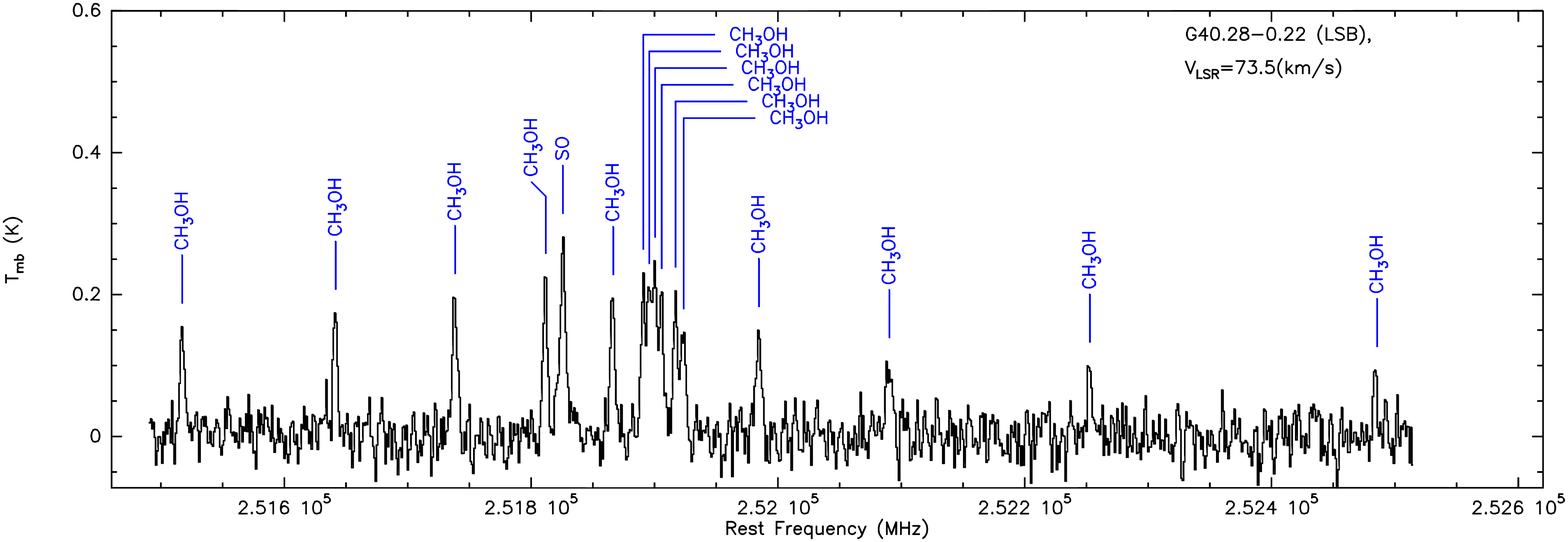}
\includegraphics[scale=.30,angle=0]{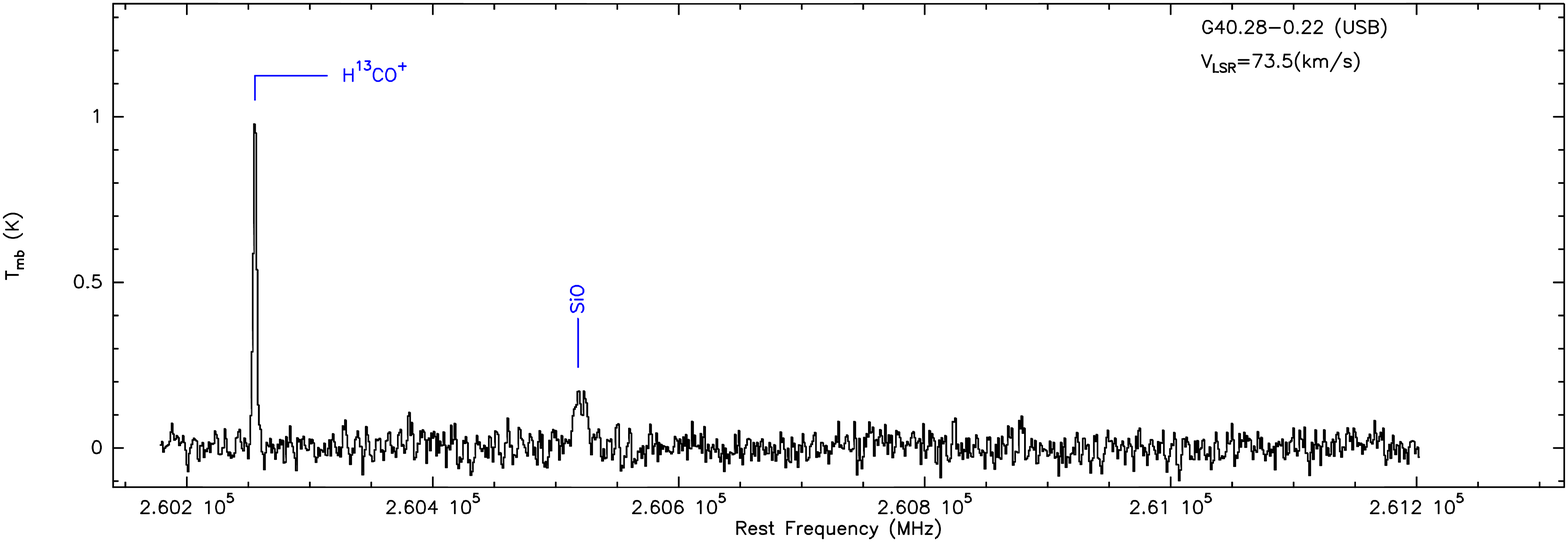}
\caption{(continued) For G40.28-0.22.}
\end{figure*}
\clearpage
 \addtocounter{figure}{-1}
\begin{figure*}
\centering
\includegraphics[scale=.30,angle=0]{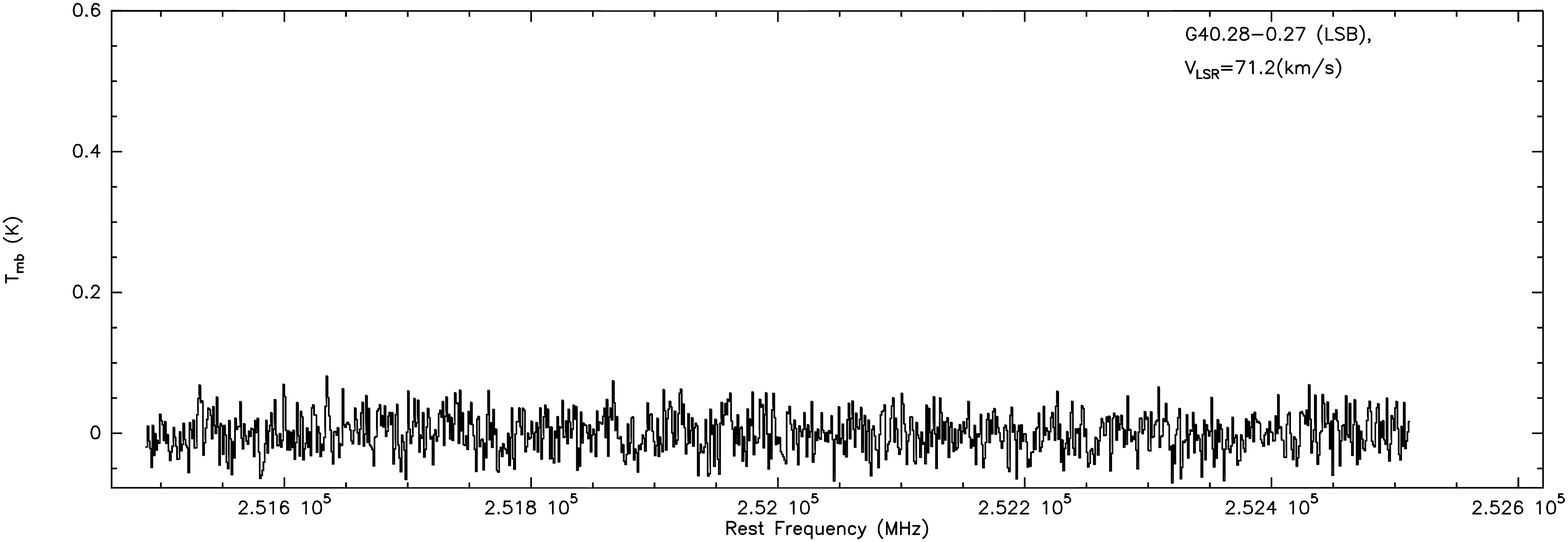}
\includegraphics[scale=.30,angle=0]{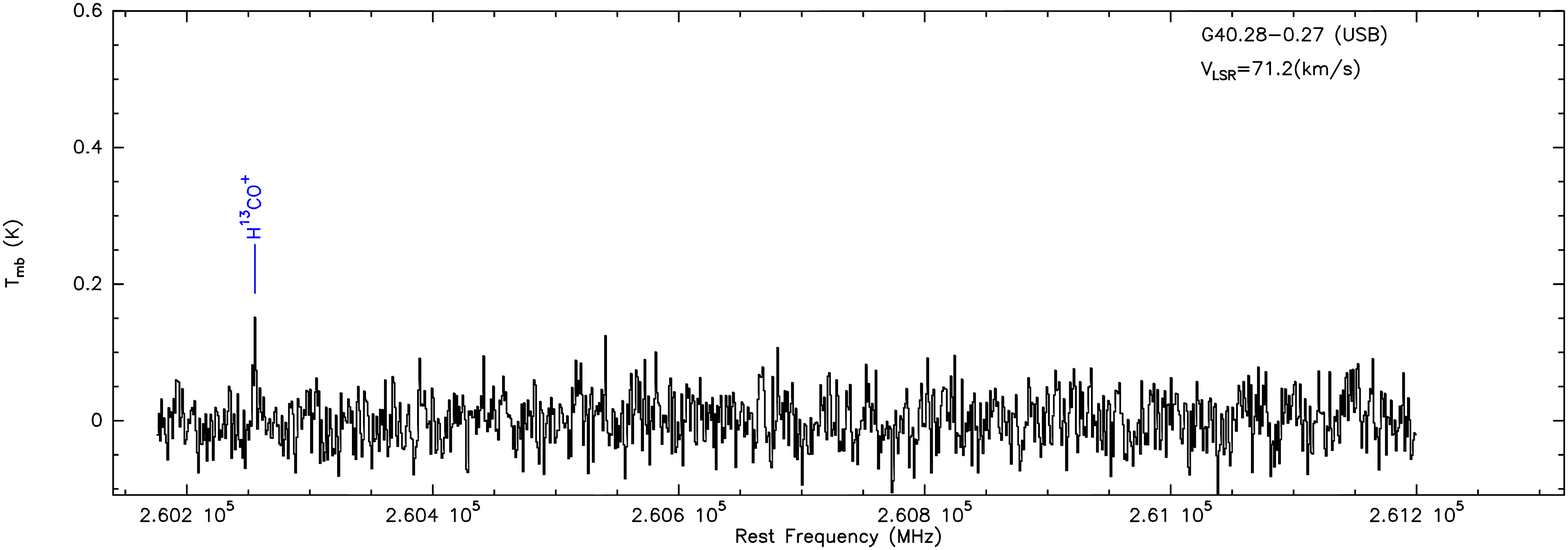}
\caption{(continued) For G40.28-0.27.}
\end{figure*}
 \addtocounter{figure}{-1}
\begin{figure*}
\centering
\includegraphics[scale=.30,angle=0]{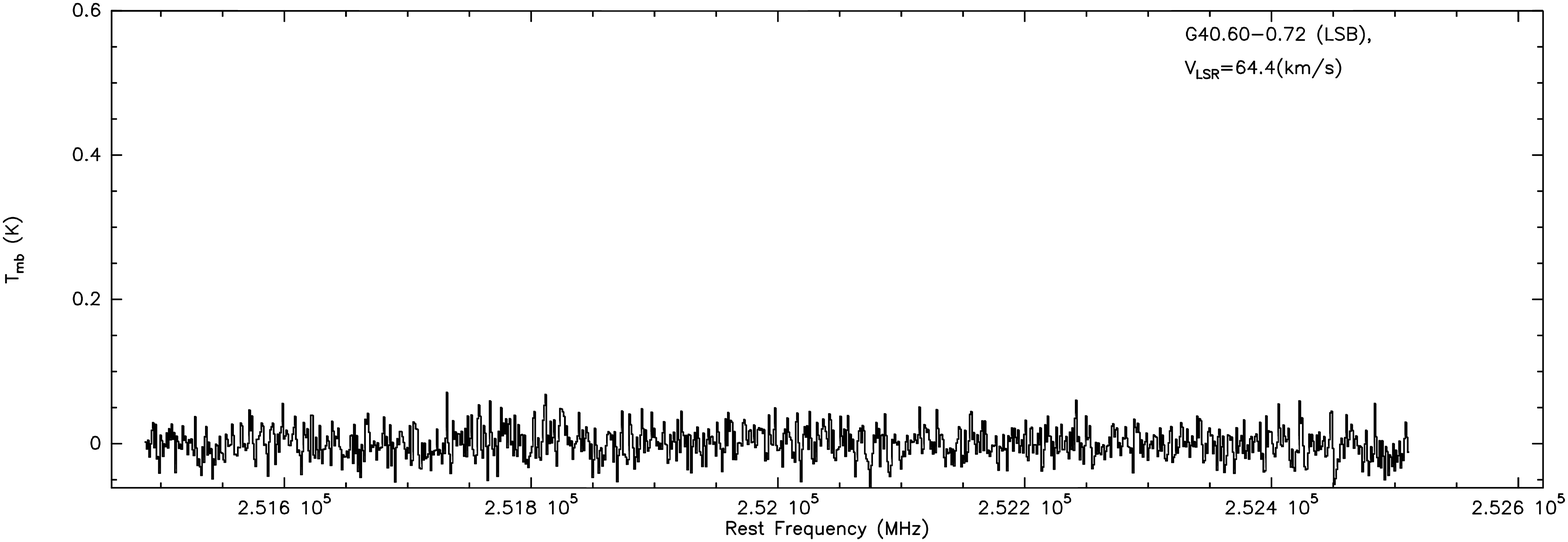}
\includegraphics[scale=.30,angle=0]{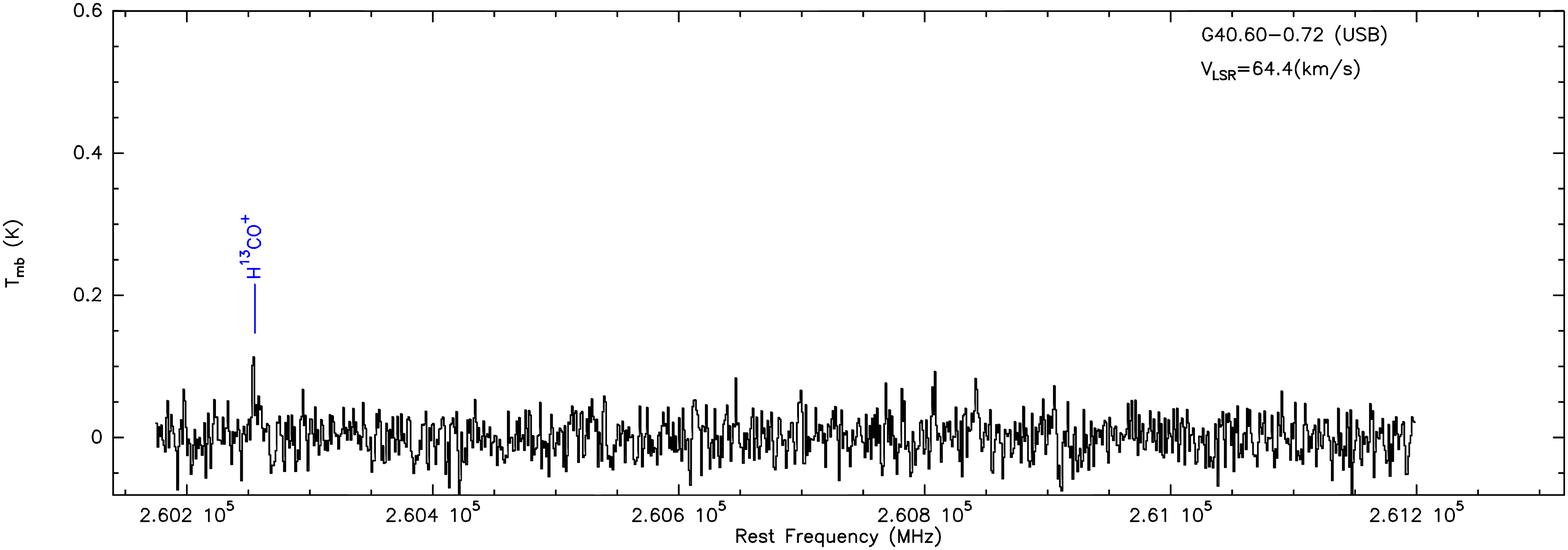}
\caption{(continued) For G40.60-0.72.}
\end{figure*}
\clearpage
 \addtocounter{figure}{-1}
\begin{figure*}
\centering
\includegraphics[scale=.30,angle=0]{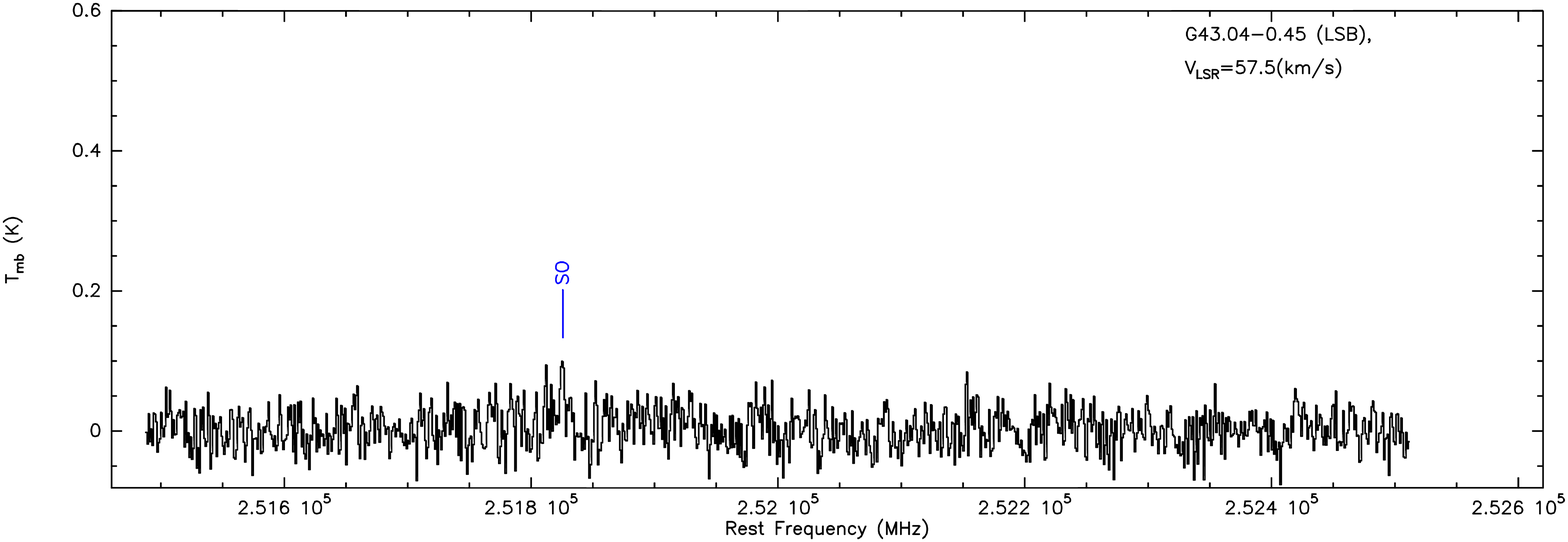}
\includegraphics[scale=.30,angle=0]{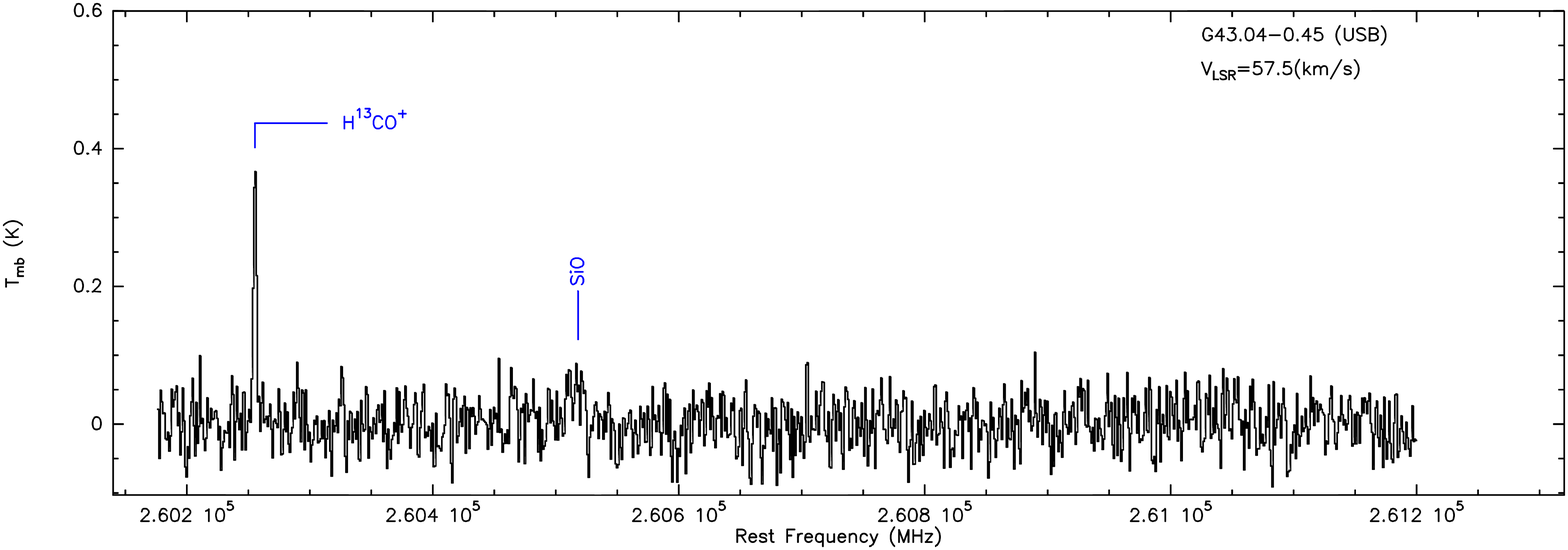}
\caption{(continued) For G43.04-0.45.}
\end{figure*}
 \addtocounter{figure}{-1}
\begin{figure*}
\centering
\includegraphics[scale=.30,angle=0]{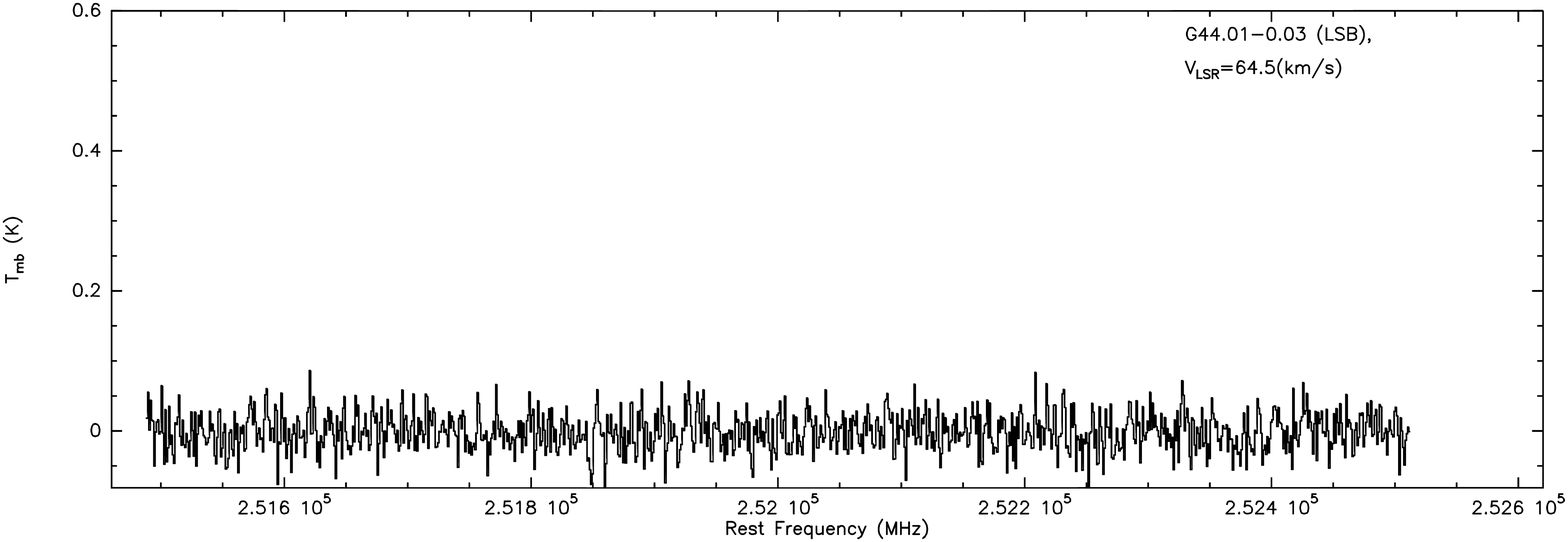}
\includegraphics[scale=.30,angle=0]{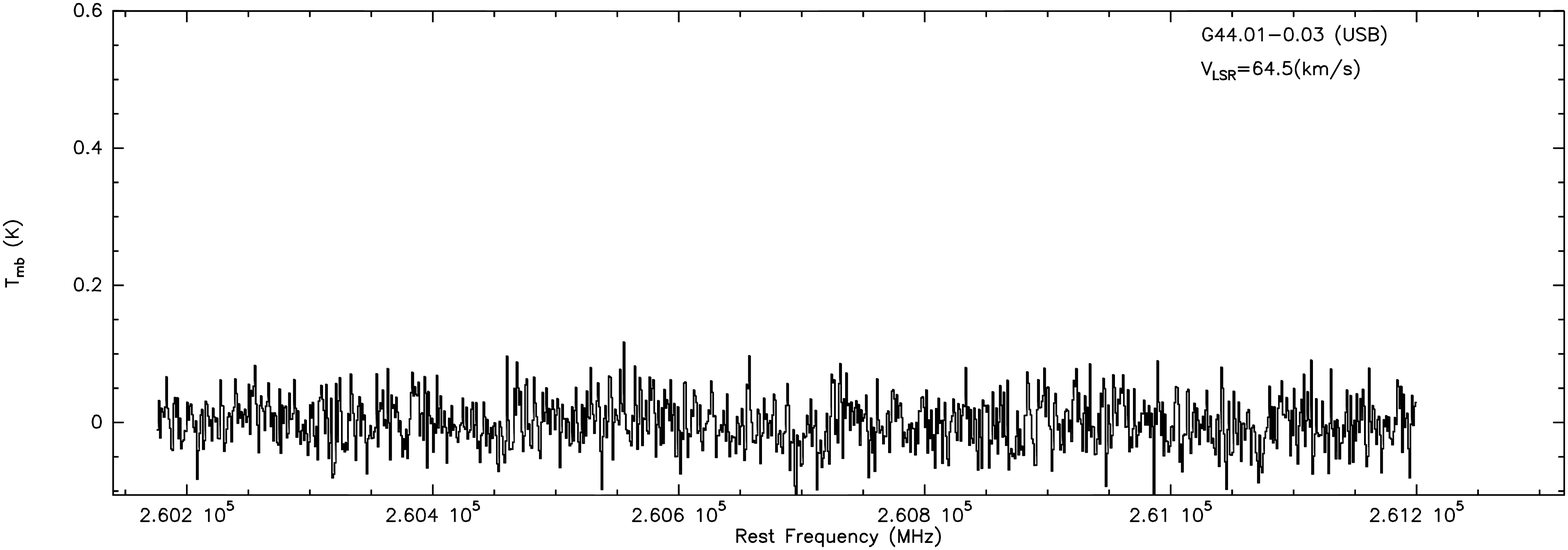}
\caption{(continued) For G44.01-0.03.}
\end{figure*}
\clearpage
 \addtocounter{figure}{-1}
\begin{figure*}
\centering
\includegraphics[scale=.30,angle=0]{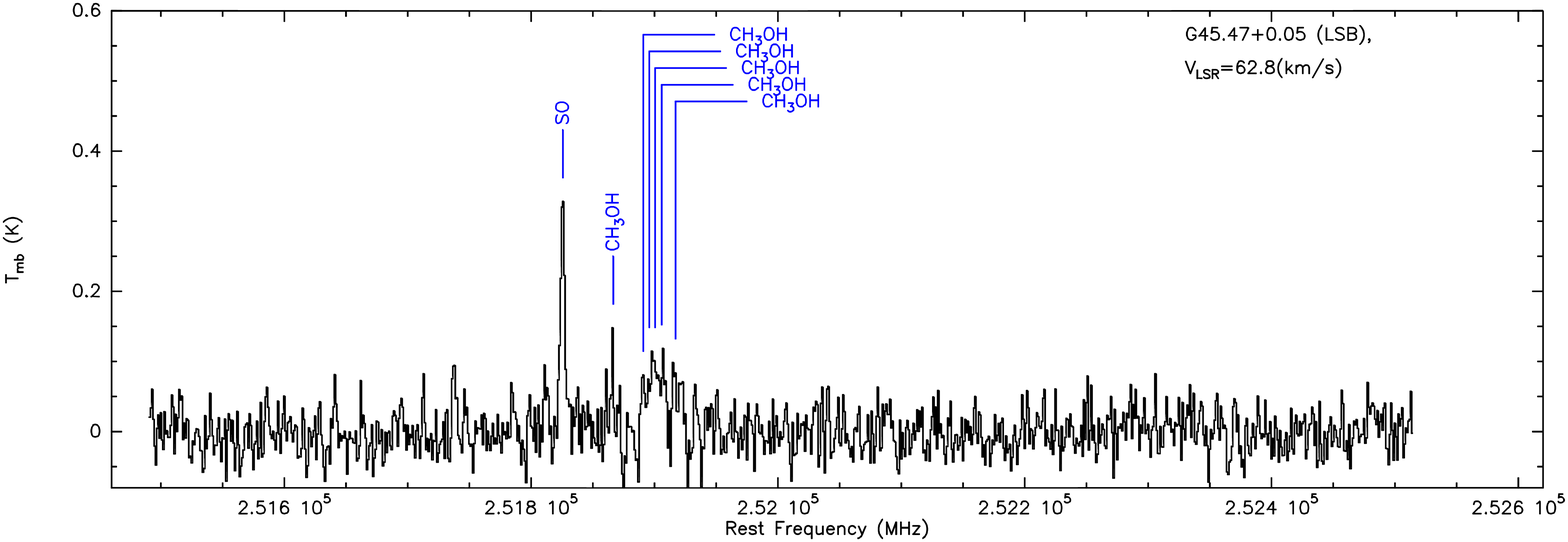}
\includegraphics[scale=.30,angle=0]{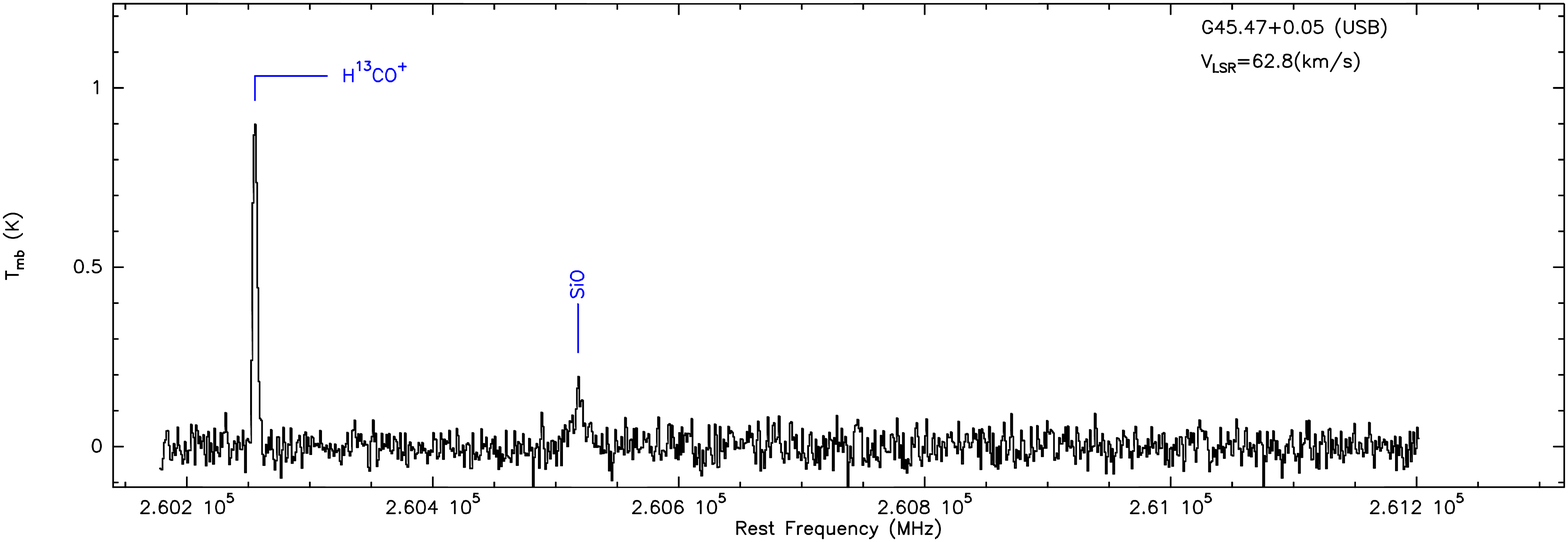}
\caption{(continued) For G45.47+0.05.}
\end{figure*}
 \addtocounter{figure}{-1}
\begin{figure*}
\centering
\includegraphics[scale=.30,angle=0]{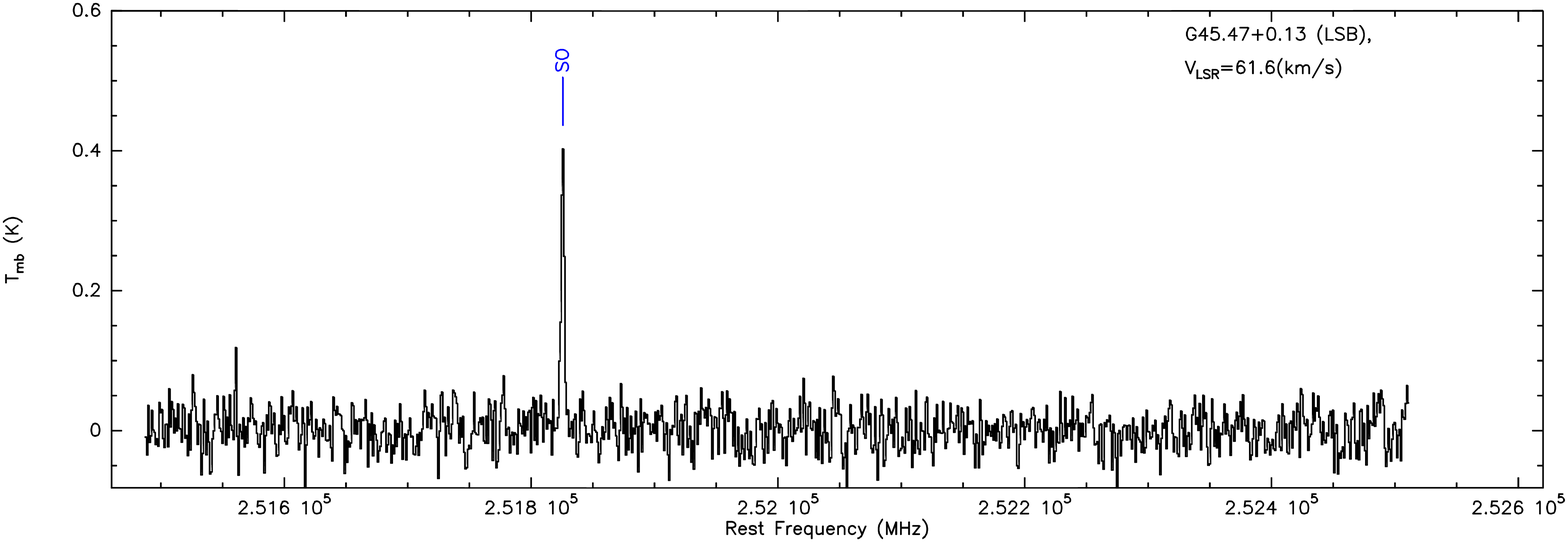}
\includegraphics[scale=.30,angle=0]{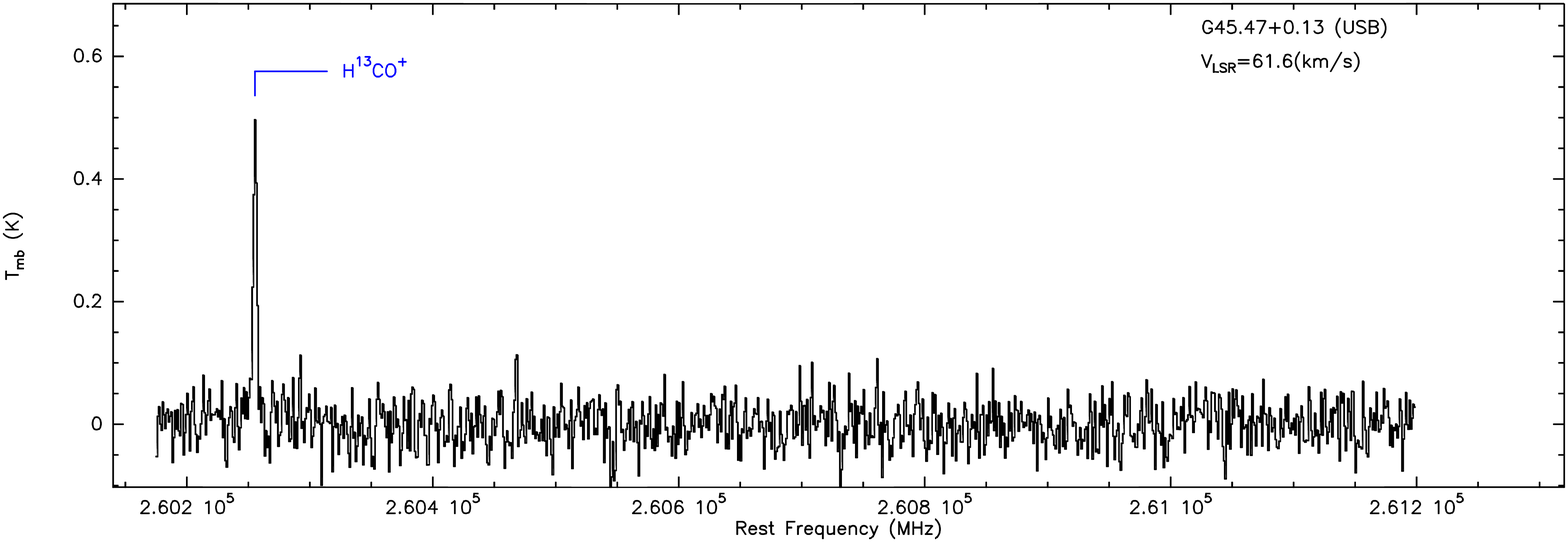}
\caption{(continued) For G45.47+0.13.}
\end{figure*}
\clearpage
 \addtocounter{figure}{-1}
\begin{figure*}
\centering
\includegraphics[scale=.30,angle=0]{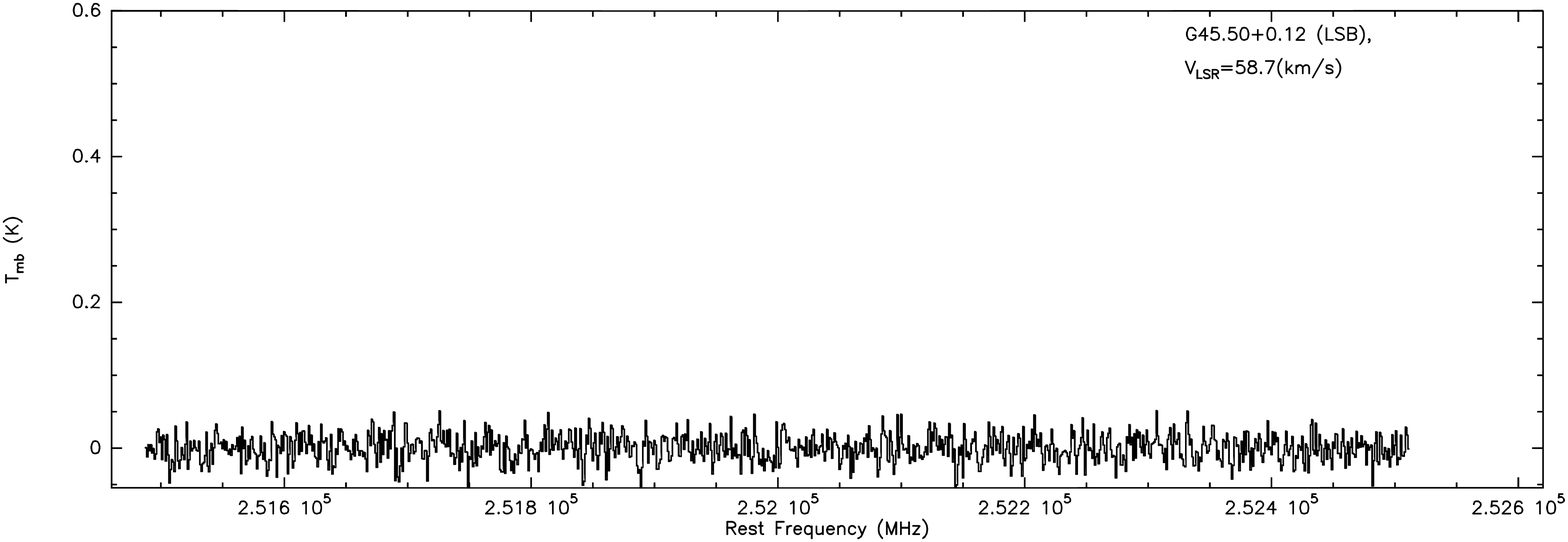}
\includegraphics[scale=.30,angle=0]{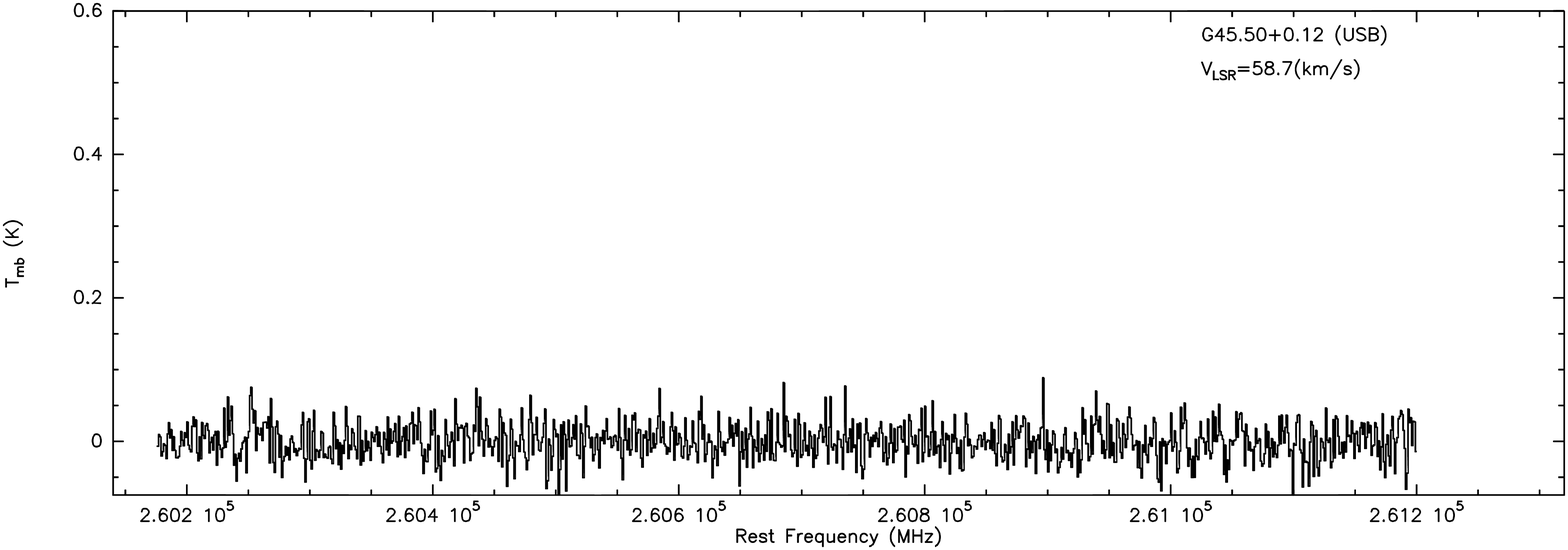}
\caption{(continued) For G45.50+0.12.}
\end{figure*}
 \addtocounter{figure}{-1}
\begin{figure*}
\centering
\includegraphics[scale=.30,angle=0]{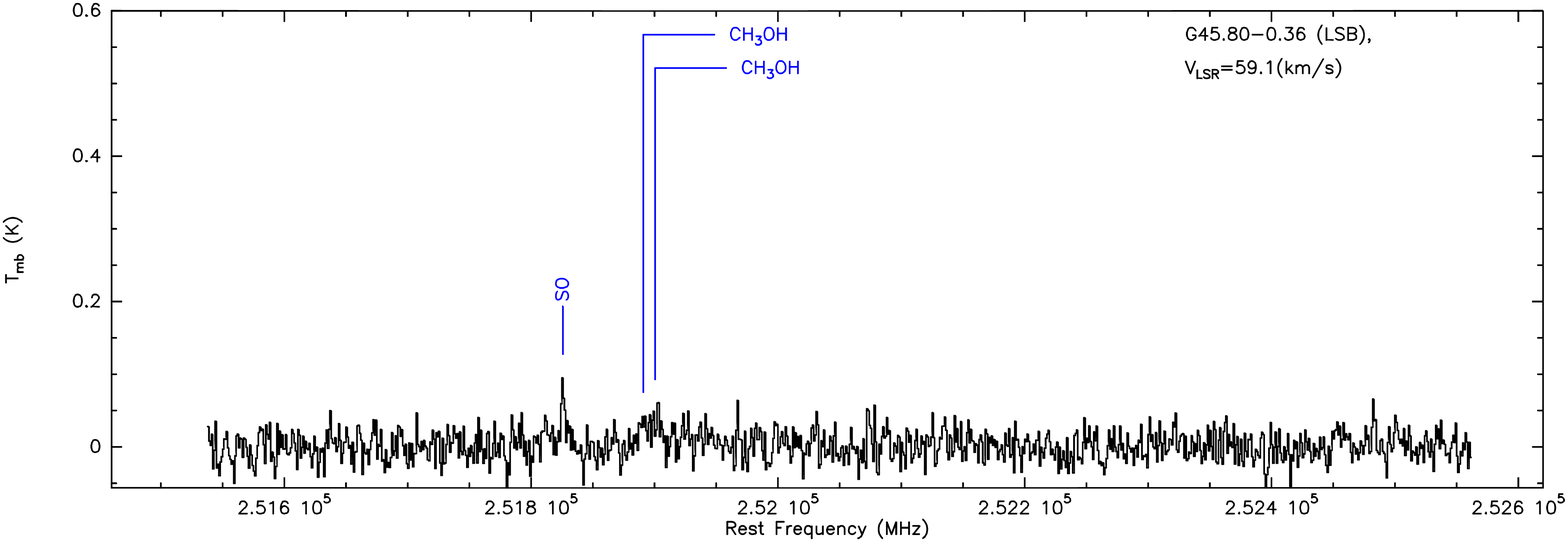}
\includegraphics[scale=.30,angle=0]{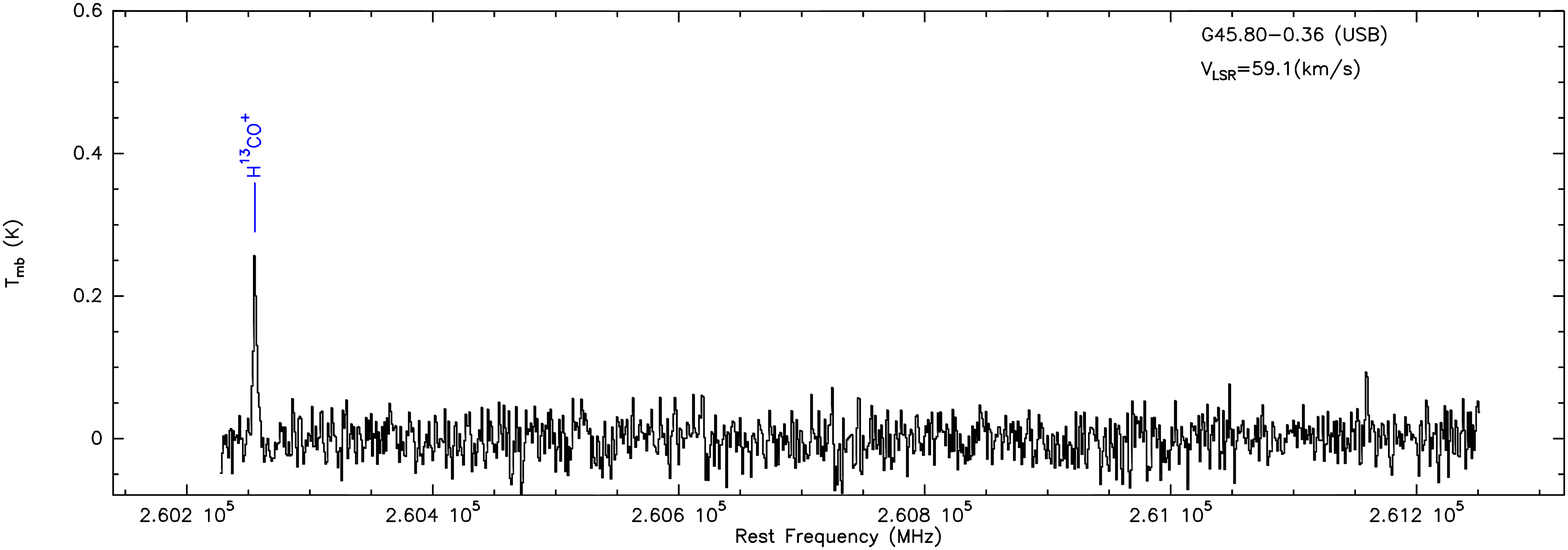}
\caption{(continued) For G45.80-0.36.}
\end{figure*}
\clearpage
 \addtocounter{figure}{-1}
\begin{figure*}
\centering
\includegraphics[scale=.30,angle=0]{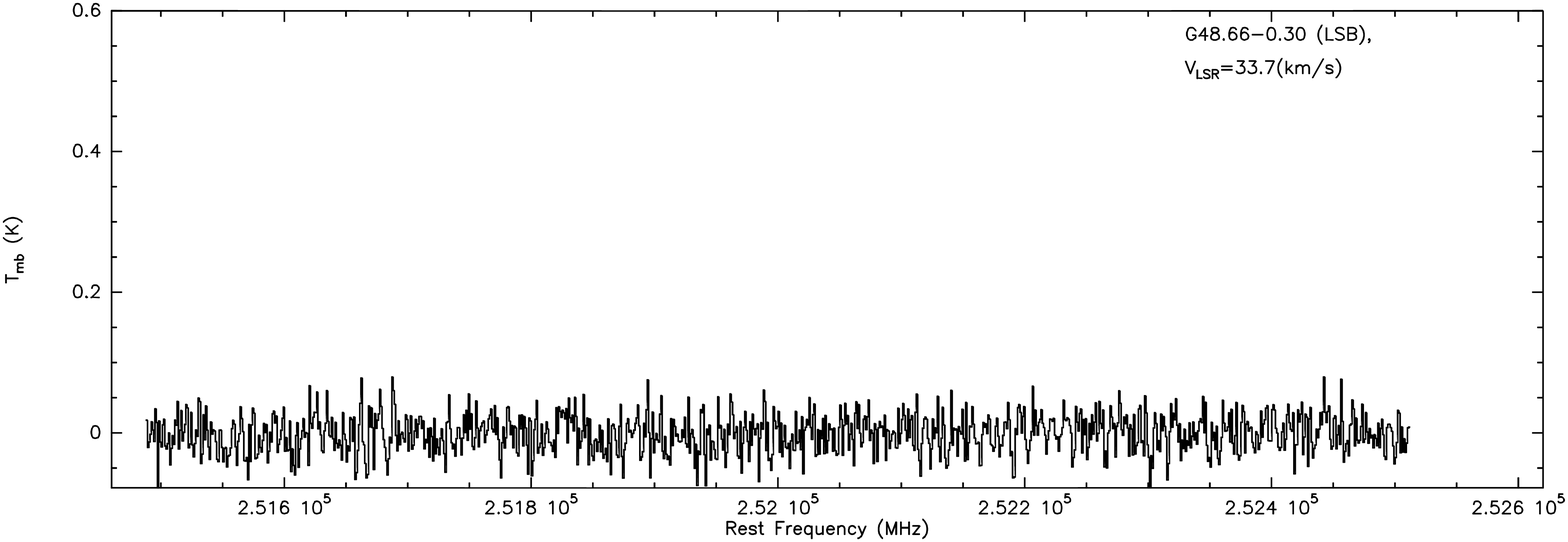}
\includegraphics[scale=.30,angle=0]{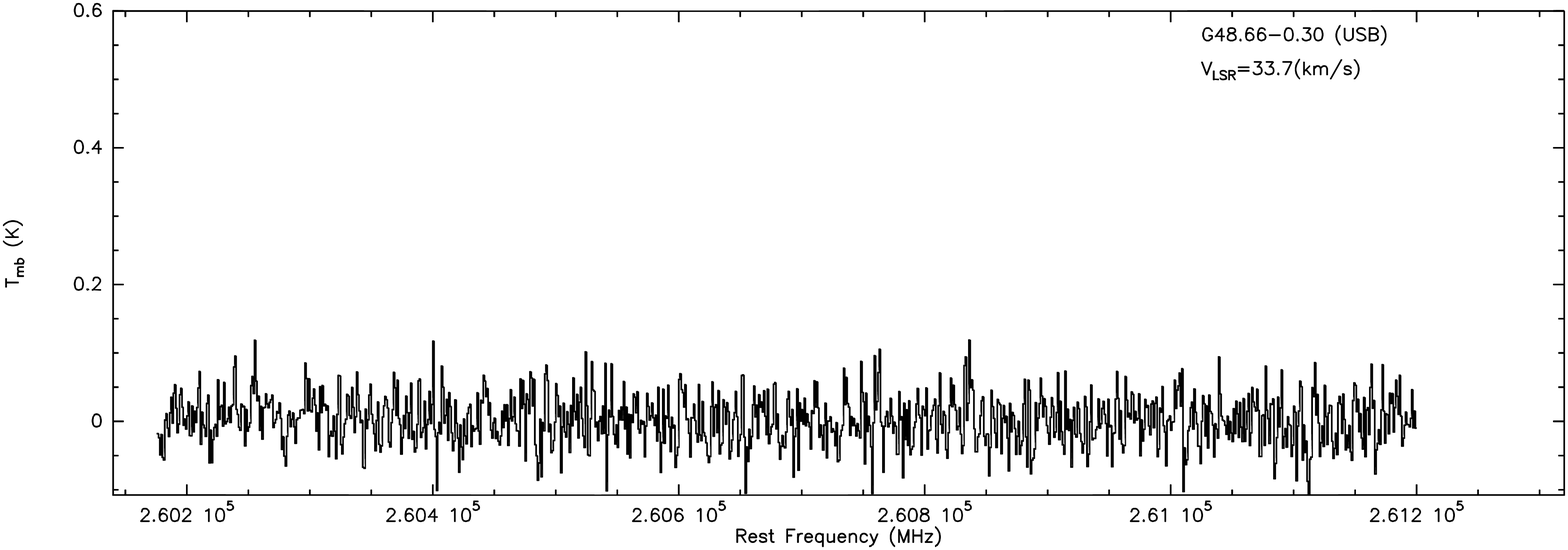}
\caption{(continued) For G48.66-0.30.}
\end{figure*}
 \addtocounter{figure}{-1}
\begin{figure*}
\centering
\includegraphics[scale=.30,angle=0]{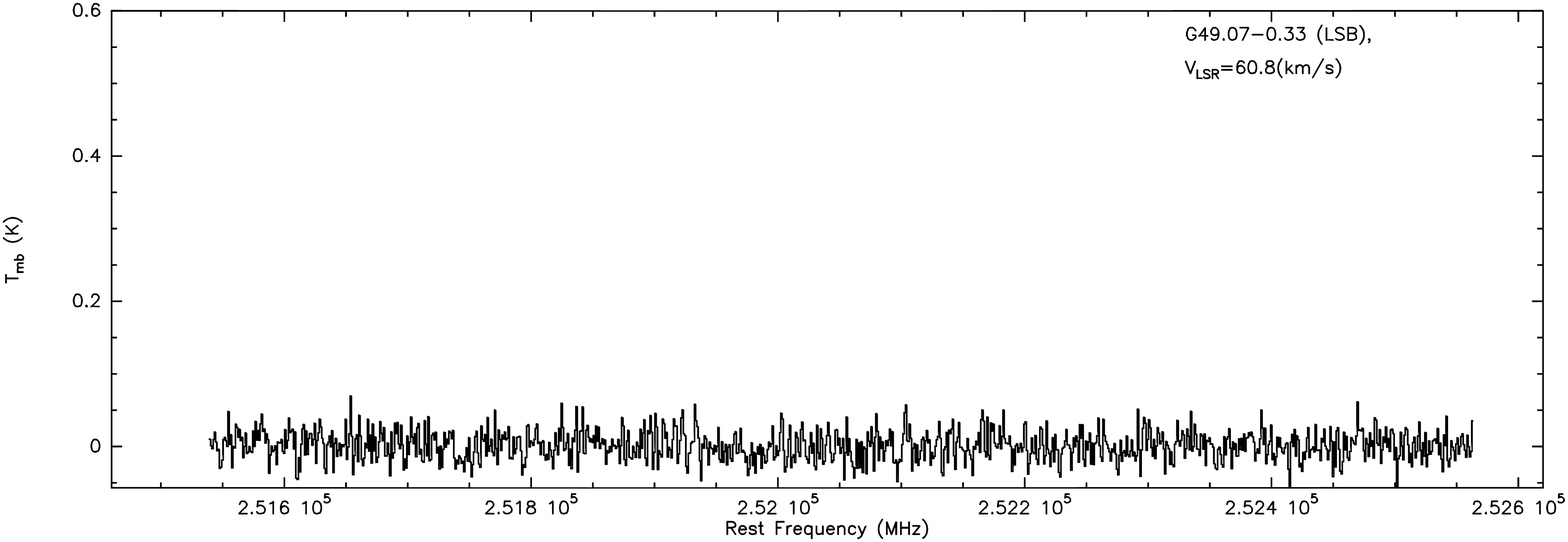}
\includegraphics[scale=.30,angle=0]{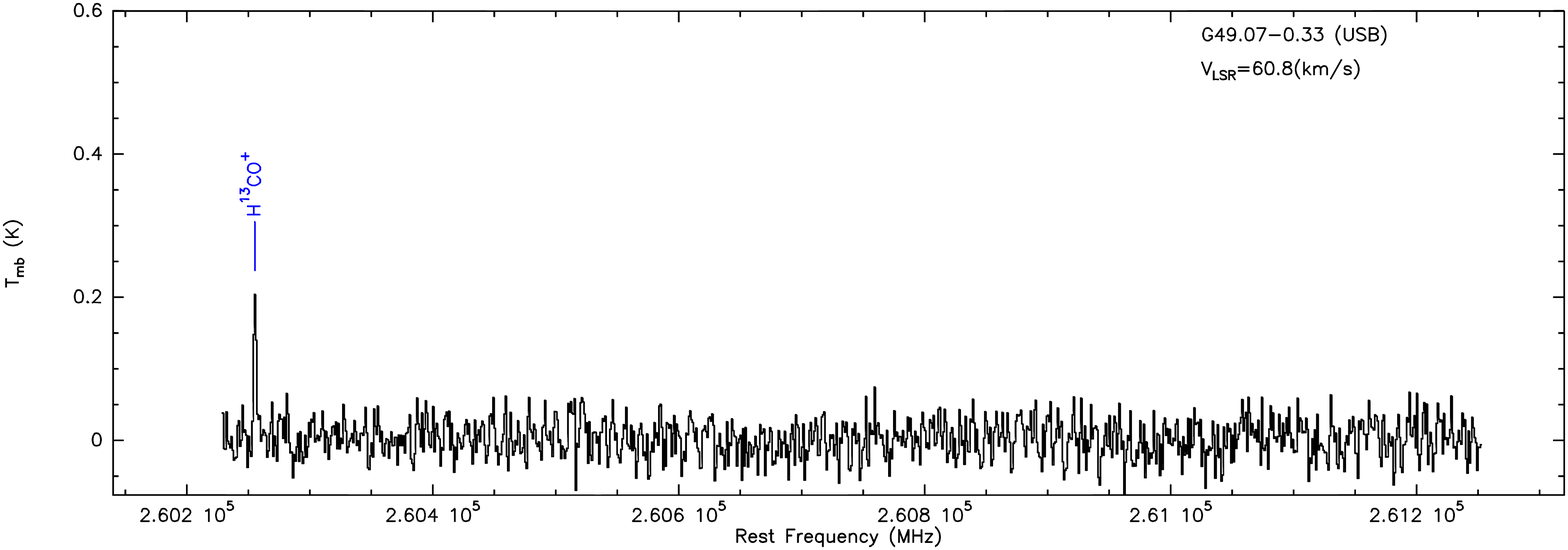}
\caption{(continued) For G49.07-0.33.}
\end{figure*}
\clearpage
 \addtocounter{figure}{-1}
\begin{figure*}
\centering
\includegraphics[scale=.30,angle=0]{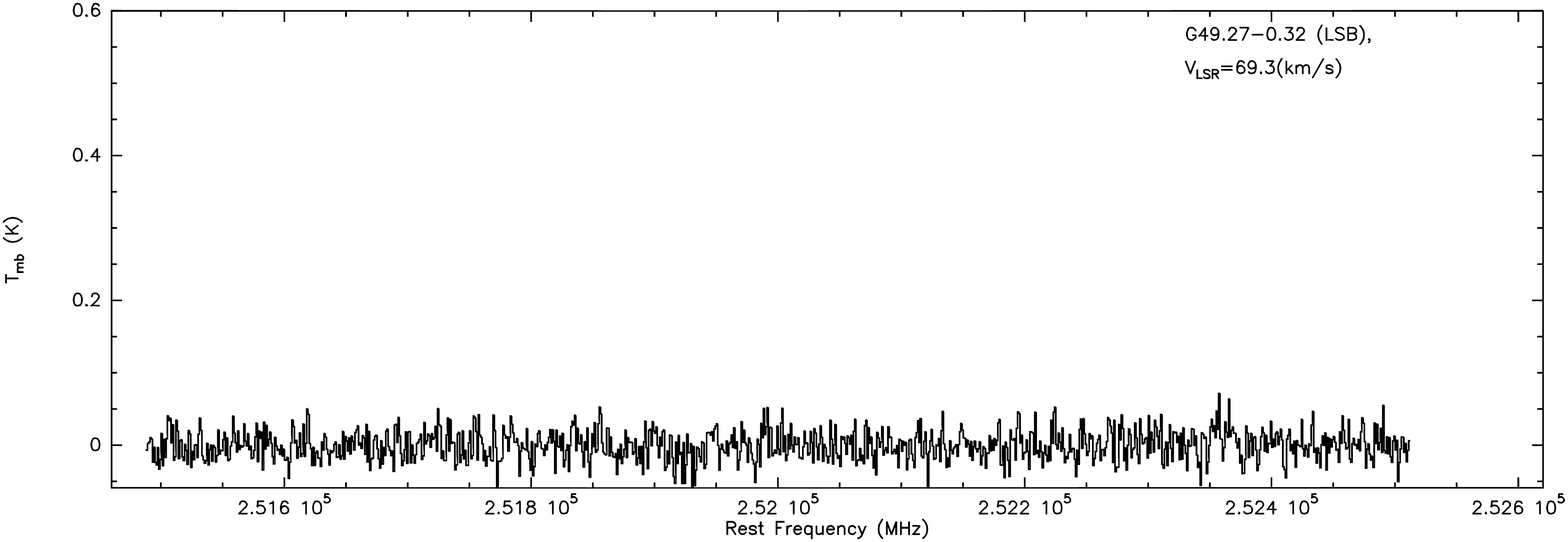}
\includegraphics[scale=.30,angle=0]{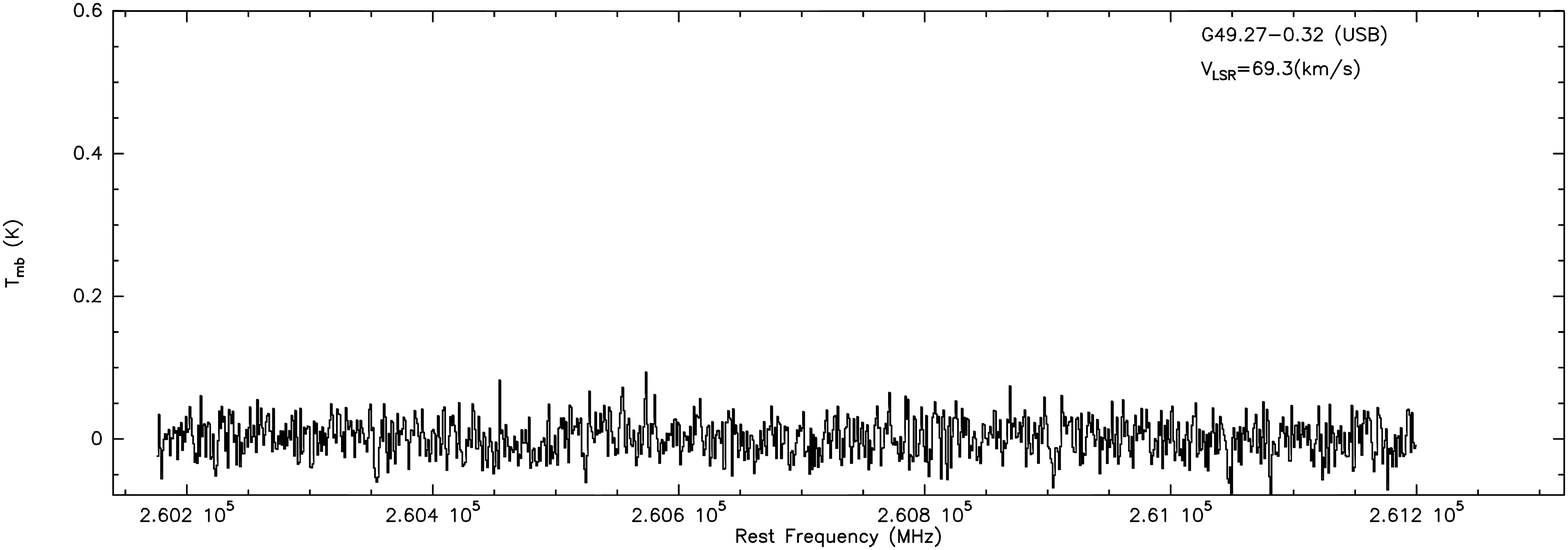}
\caption{(continued) For G49.27-0.32.}
\end{figure*}
 \addtocounter{figure}{-1}
\begin{figure*}
\centering
\includegraphics[scale=.30,angle=0]{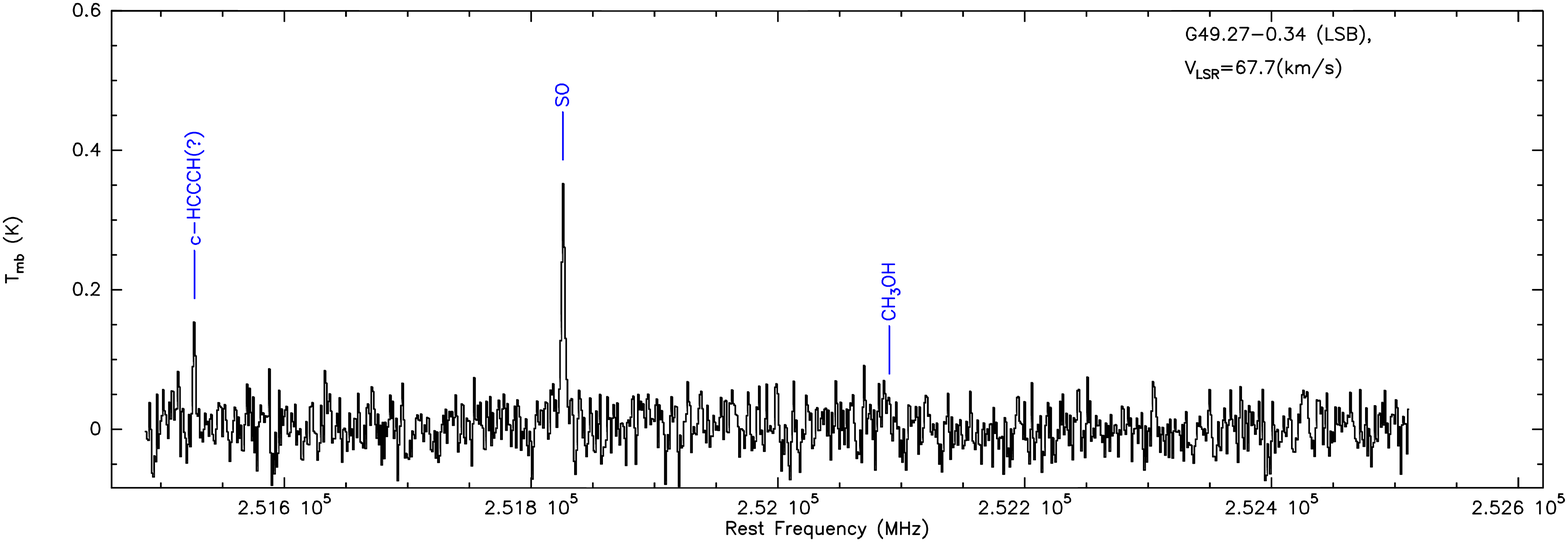}
\includegraphics[scale=.30,angle=0]{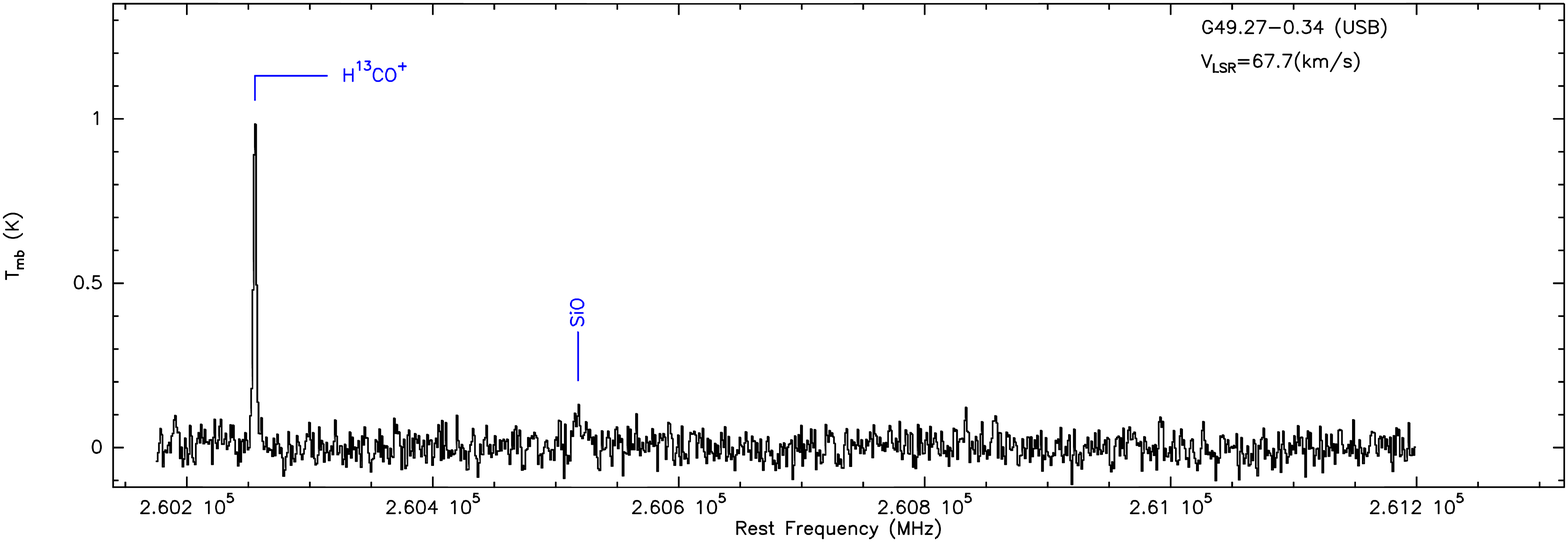}
\caption{(continued) For G49.27-0.34.}
\end{figure*}
\clearpage
 \addtocounter{figure}{-1}
\begin{figure*}
\centering
\includegraphics[scale=.30,angle=0]{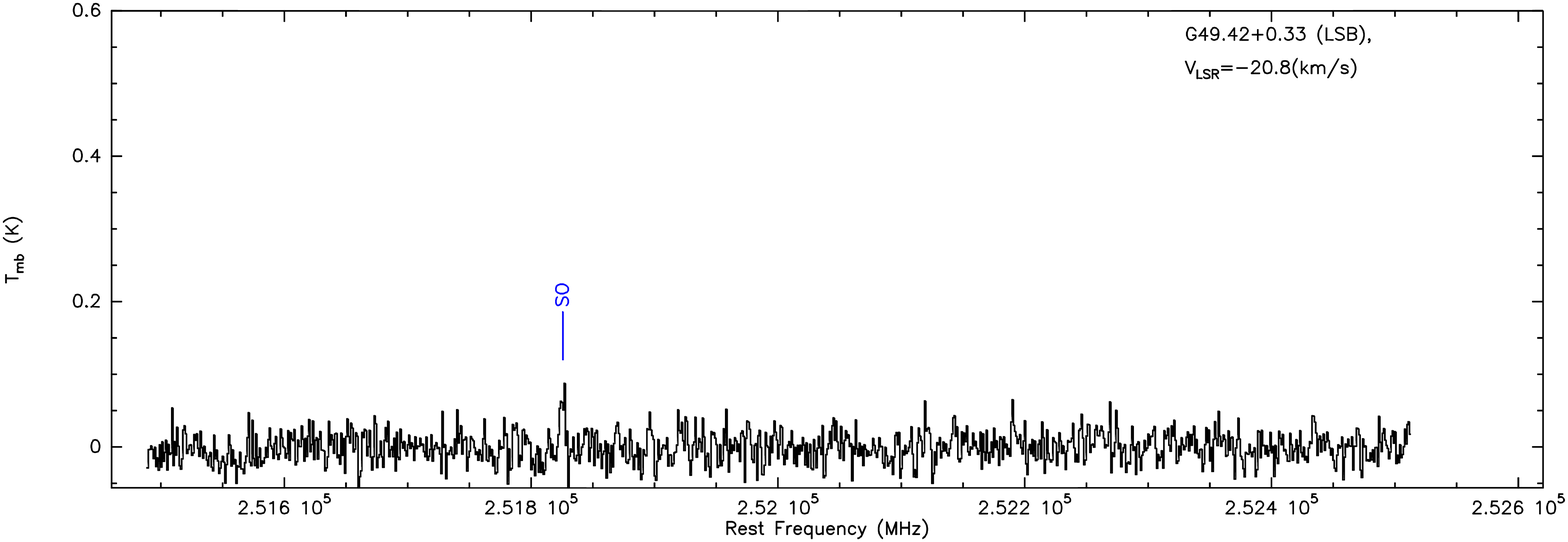}
\includegraphics[scale=.30,angle=0]{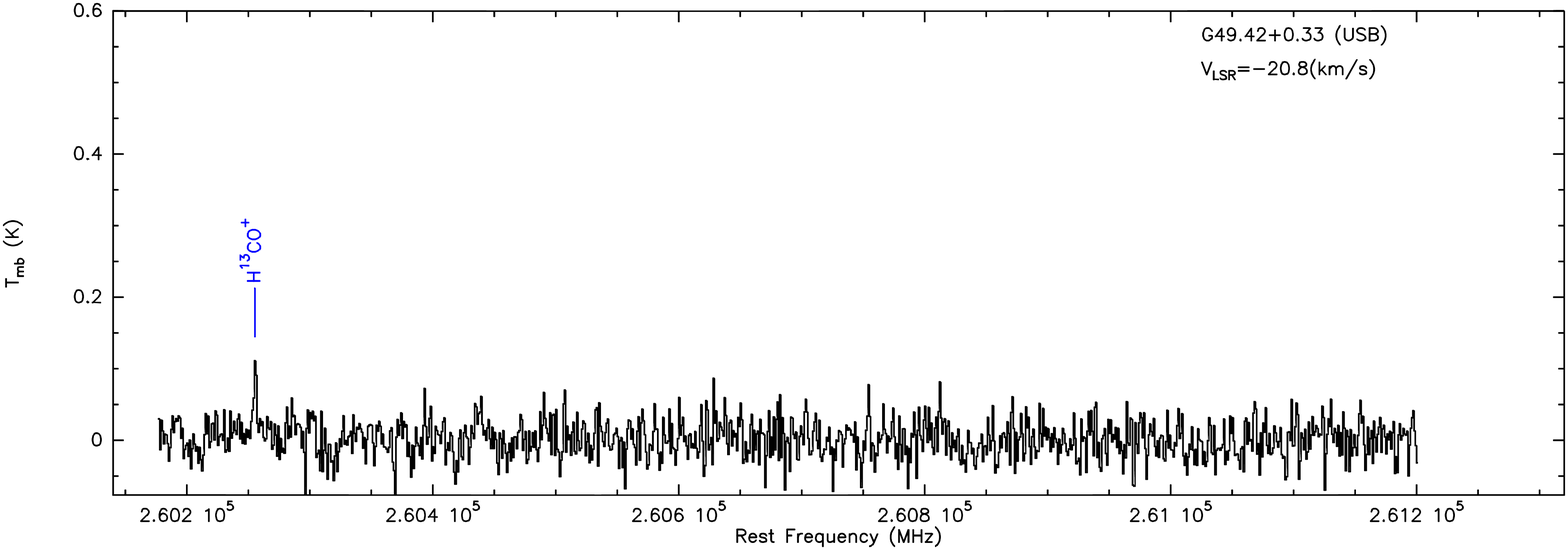}
\caption{(continued) For G49.42+0.33.}
\end{figure*}
 \addtocounter{figure}{-1}
\begin{figure*}
\centering
\includegraphics[scale=.30,angle=0]{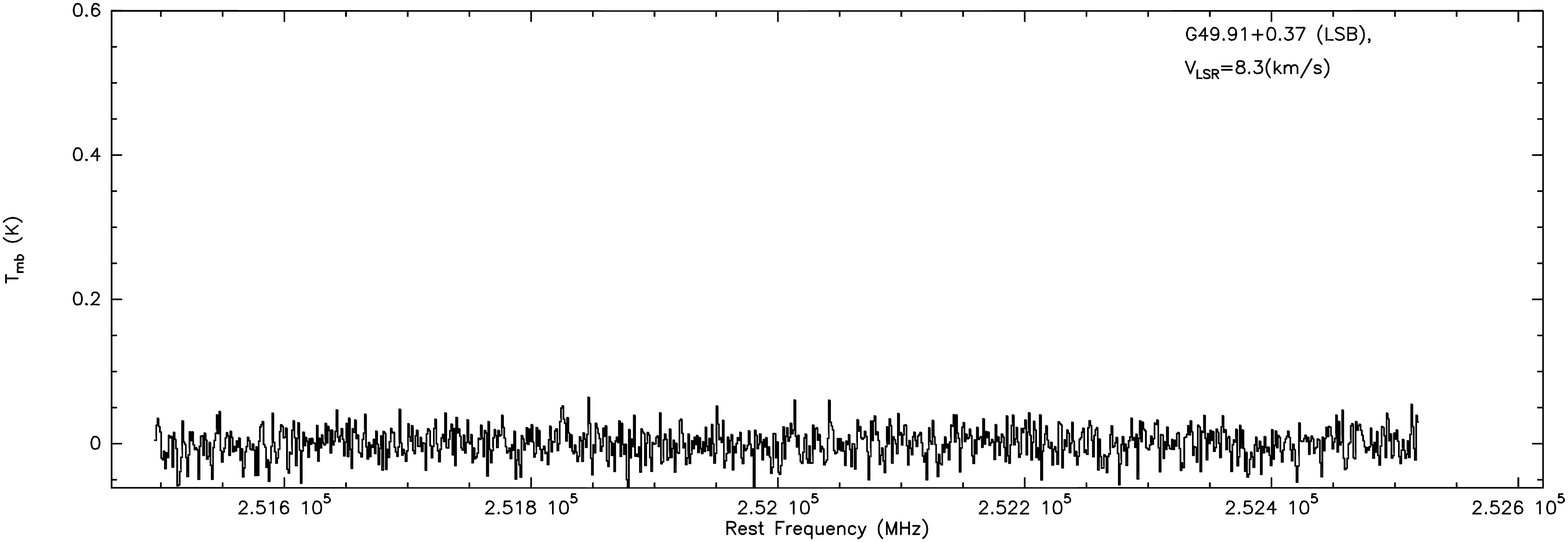}
\includegraphics[scale=.30,angle=0]{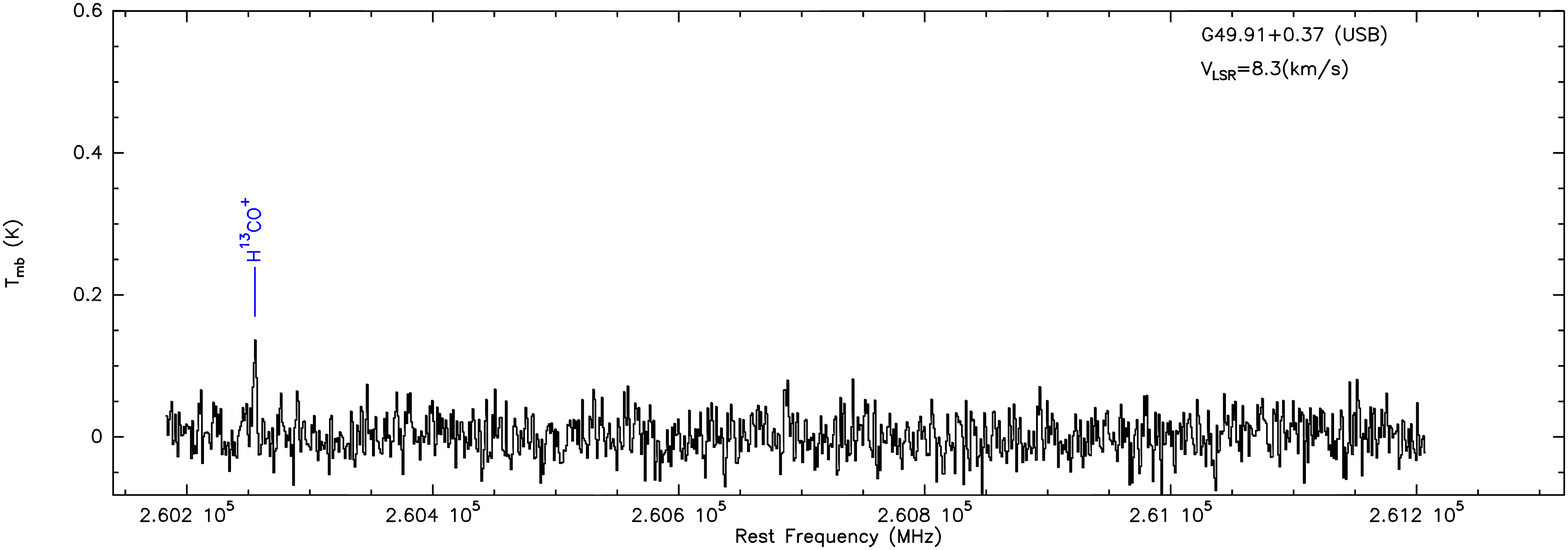}
\caption{(continued) For G49.91+0.37.}
\end{figure*}
\clearpage
 \addtocounter{figure}{-1}
\begin{figure*}
\centering
\includegraphics[scale=.30,angle=0]{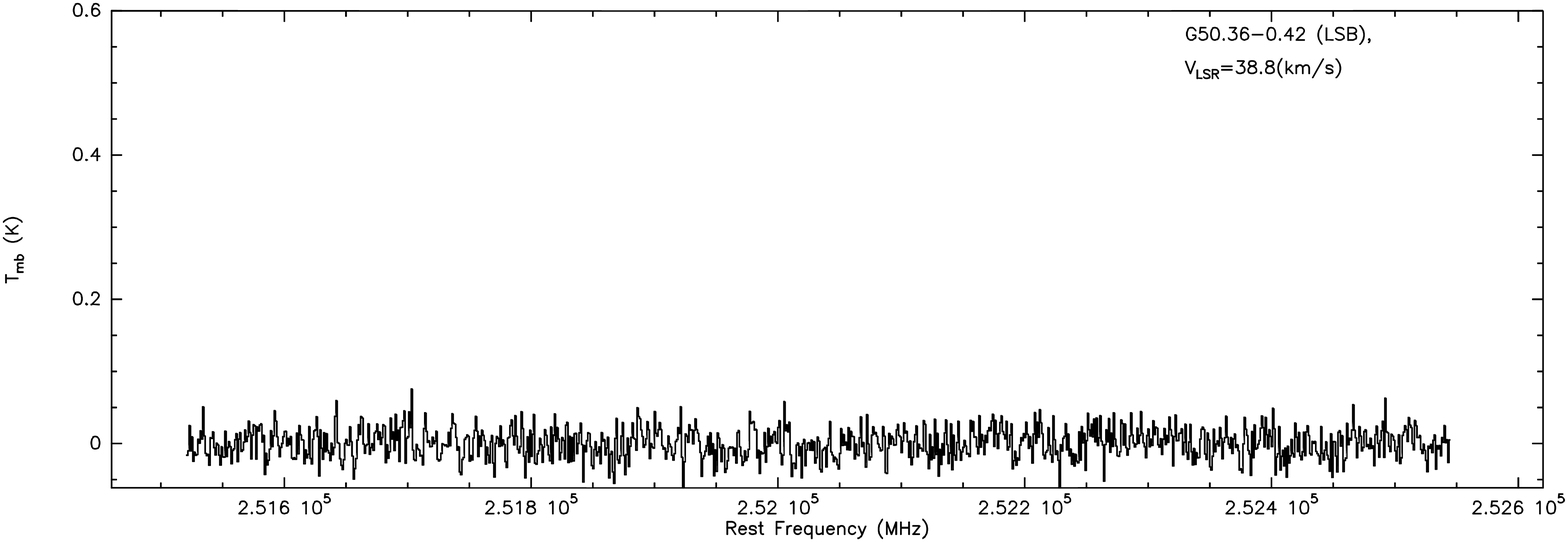}
\includegraphics[scale=.30,angle=0]{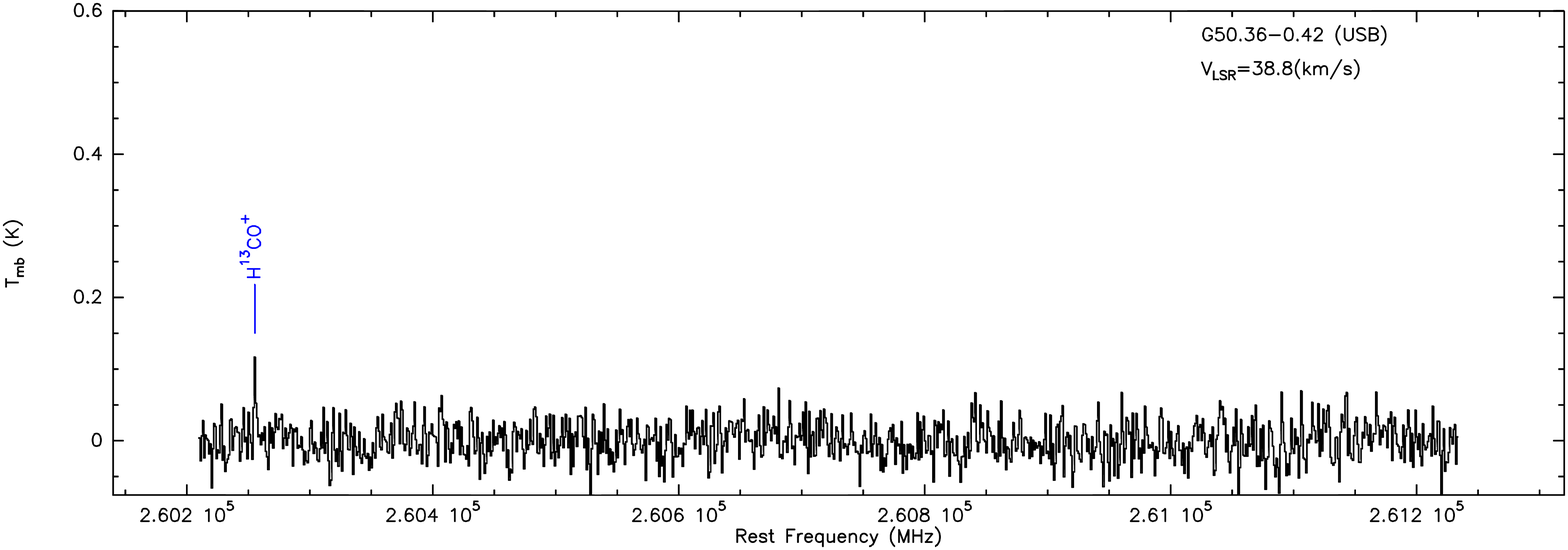}
\caption{(continued) For G50.36-0.42.}
\end{figure*}
 \addtocounter{figure}{-1}
\begin{figure*}
\centering
\includegraphics[scale=.30,angle=0]{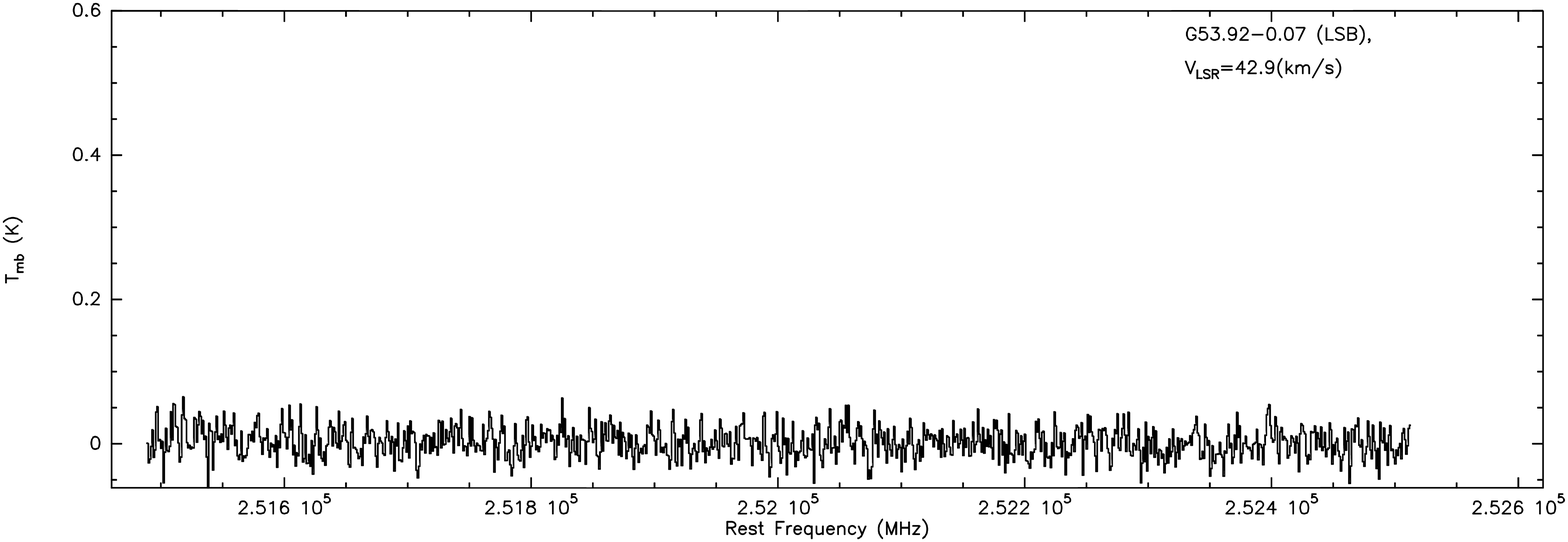}
\includegraphics[scale=.30,angle=0]{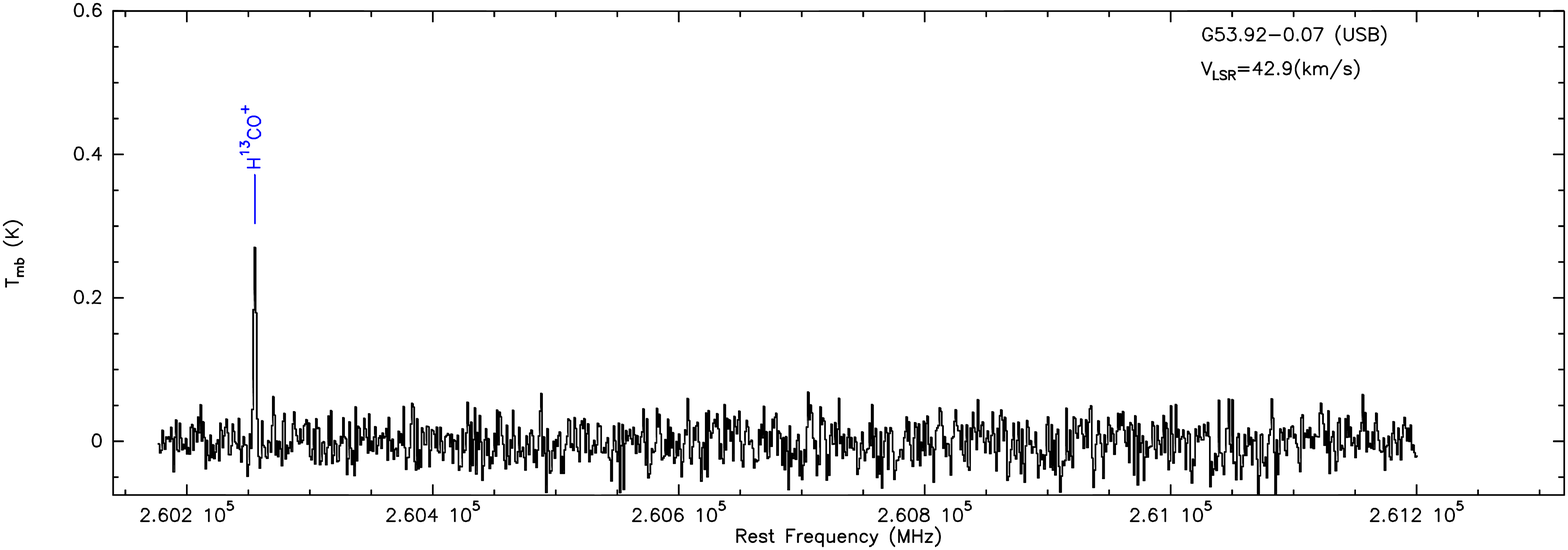}
\caption{(continued) For G53.92-0.07.}
\end{figure*}
\clearpage
 \addtocounter{figure}{-1}
\begin{figure*}
\centering
\includegraphics[scale=.30,angle=0]{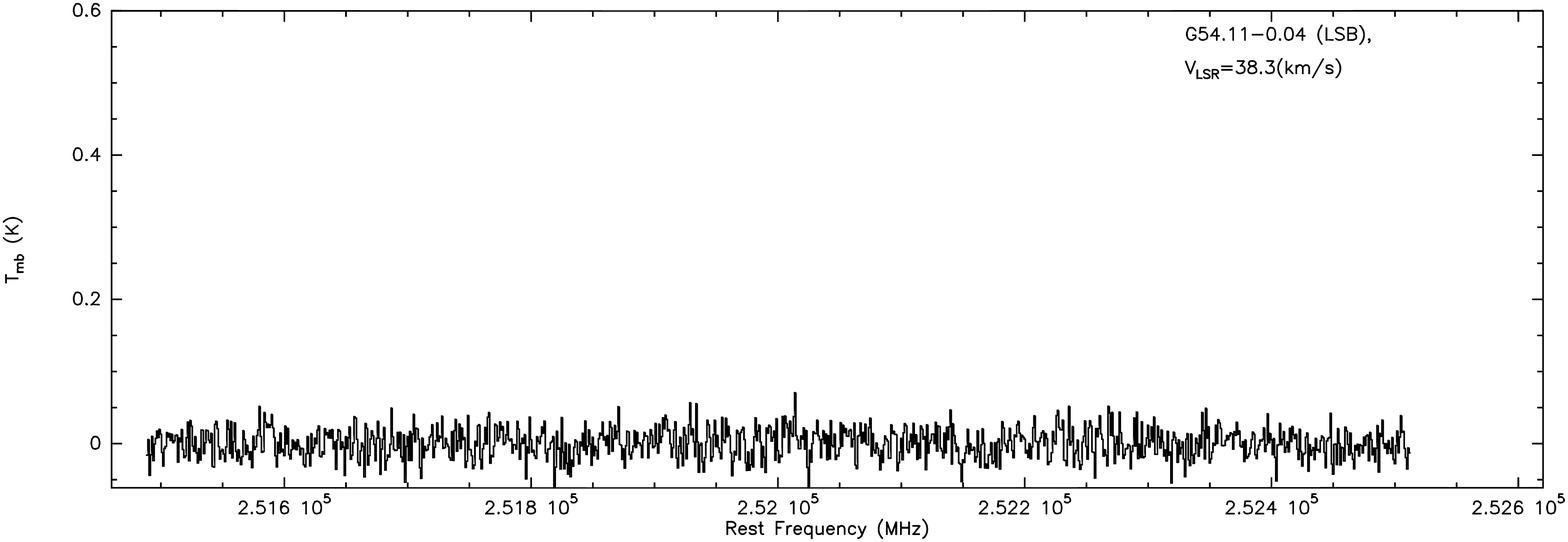}
\includegraphics[scale=.30,angle=0]{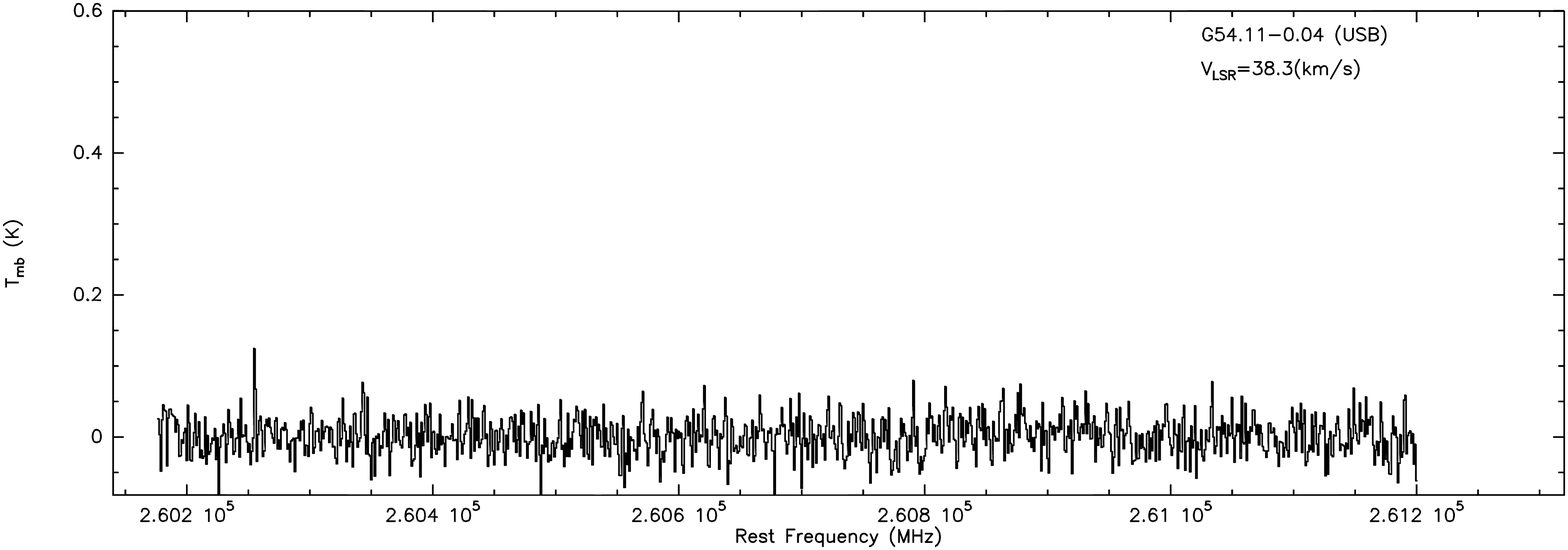}
\caption{(continued) For G54.11-0.04.}
\end{figure*}
 \addtocounter{figure}{-1}
\begin{figure*}
\centering
\includegraphics[scale=.30,angle=0]{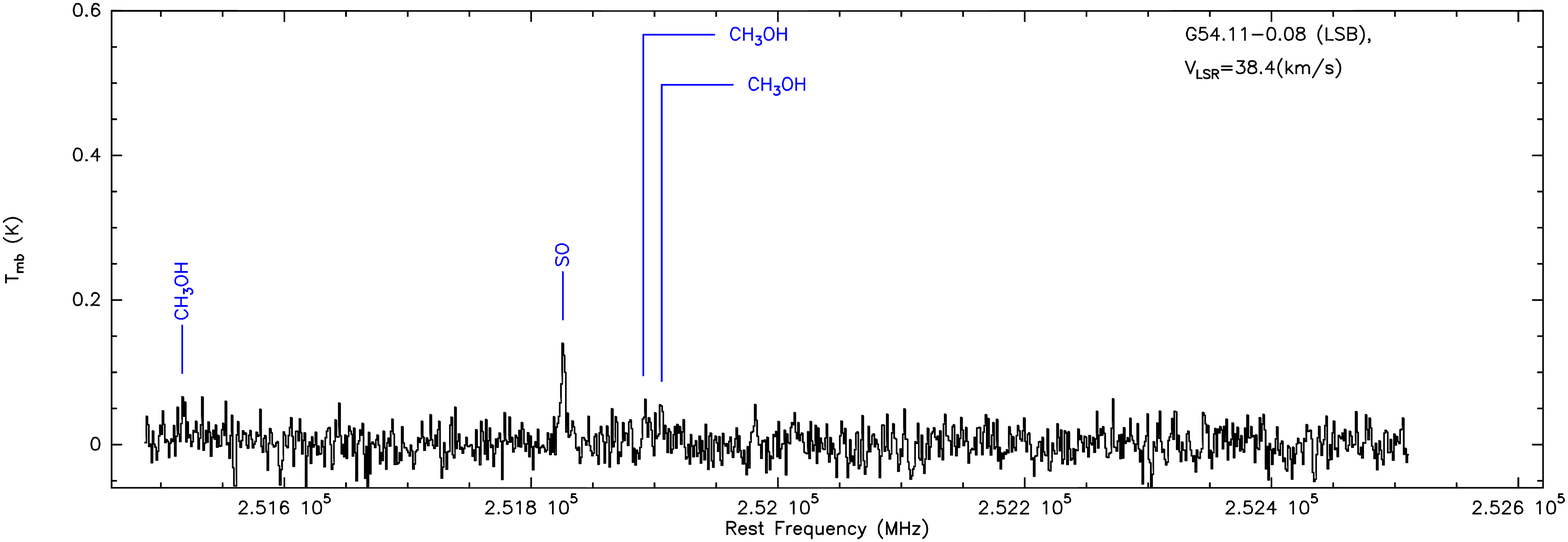}
\includegraphics[scale=.30,angle=0]{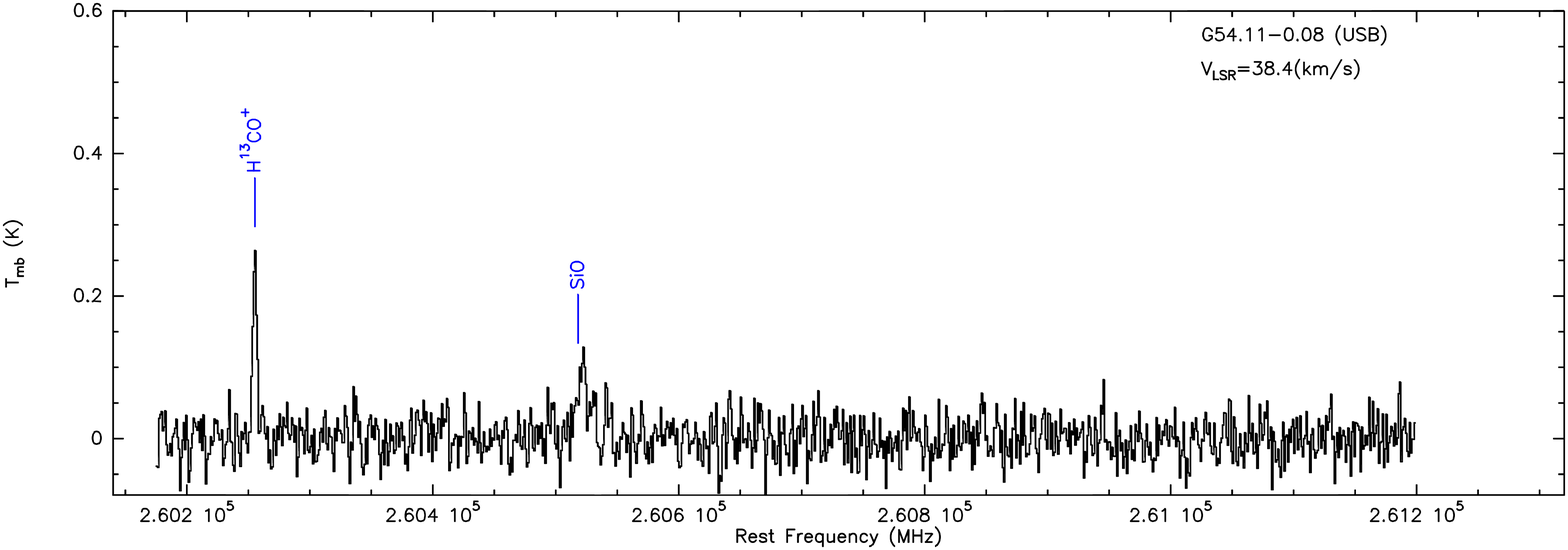}
\caption{(continued) For G54.11-0.08.}
\end{figure*}
\clearpage
 \addtocounter{figure}{-1}
\begin{figure*}
\centering
\includegraphics[scale=.30,angle=0]{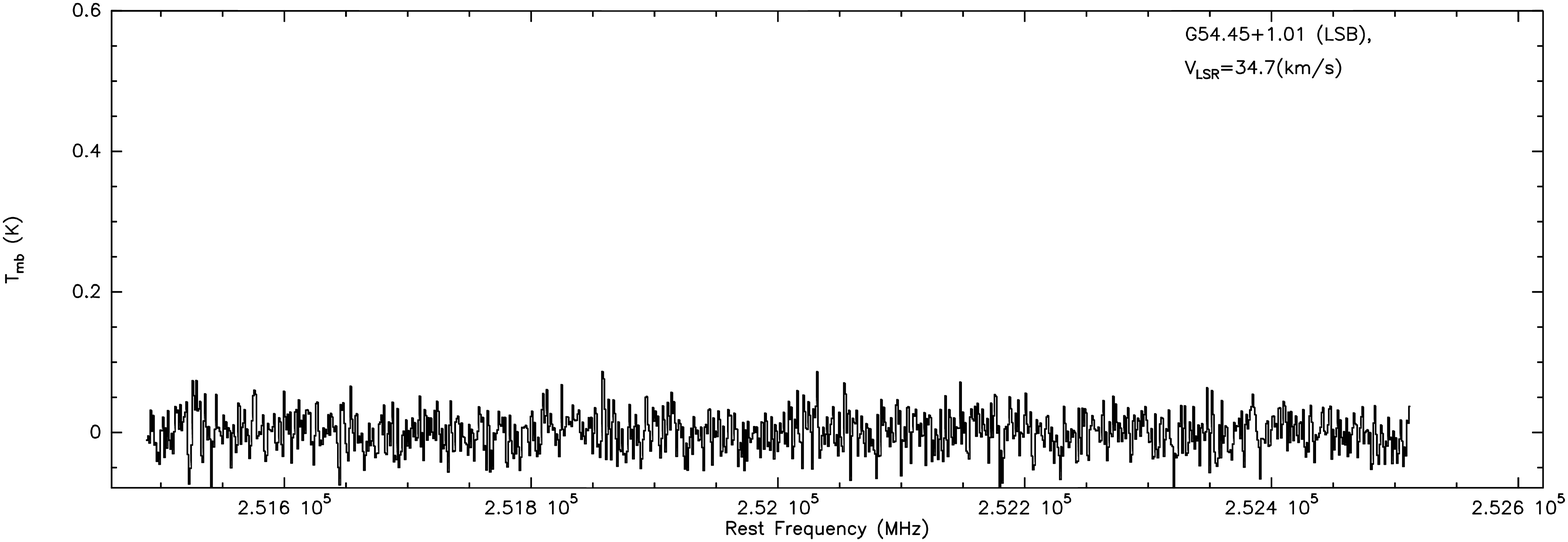}
\includegraphics[scale=.30,angle=0]{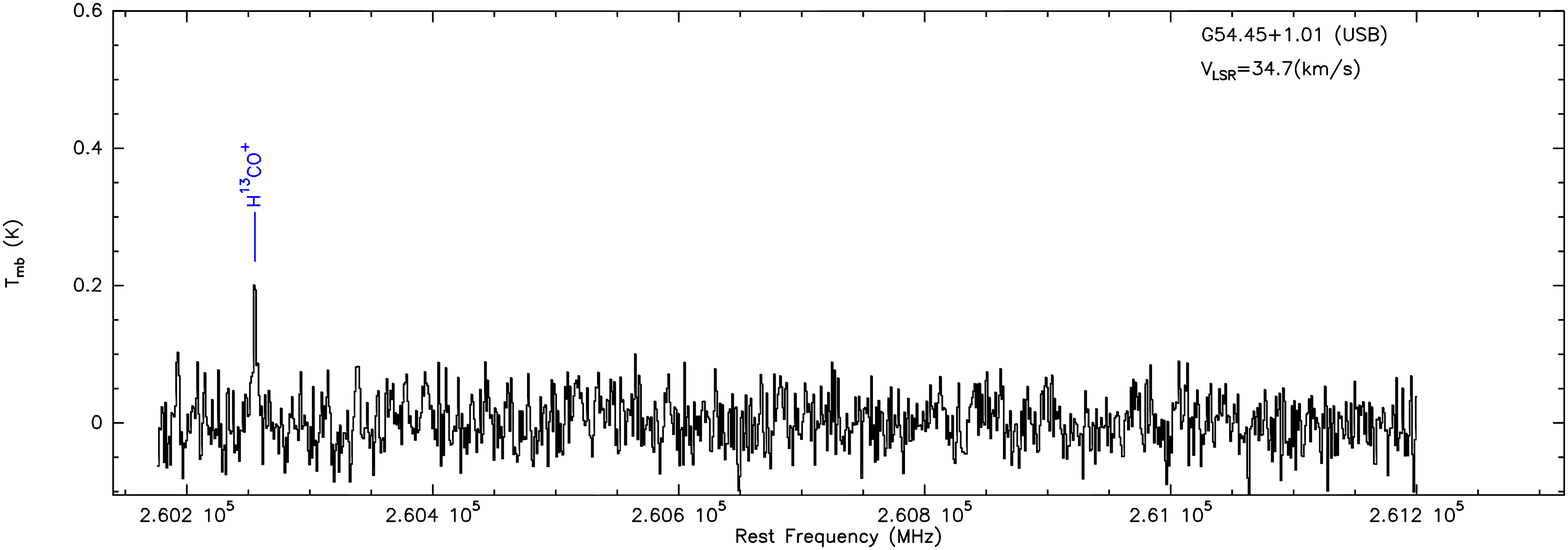}
\caption{(continued) For G54.45+1.01.}
\end{figure*}
 \addtocounter{figure}{-1}
\begin{figure*}
\centering
\includegraphics[scale=.30,angle=0]{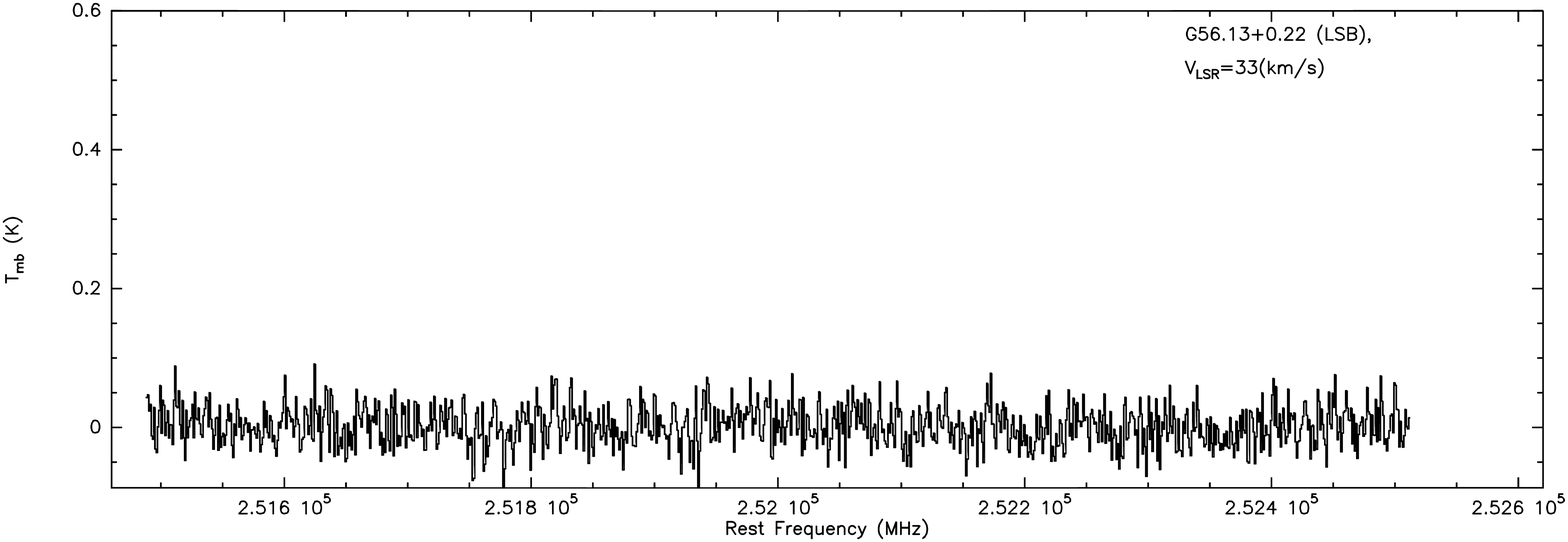}
\includegraphics[scale=.30,angle=0]{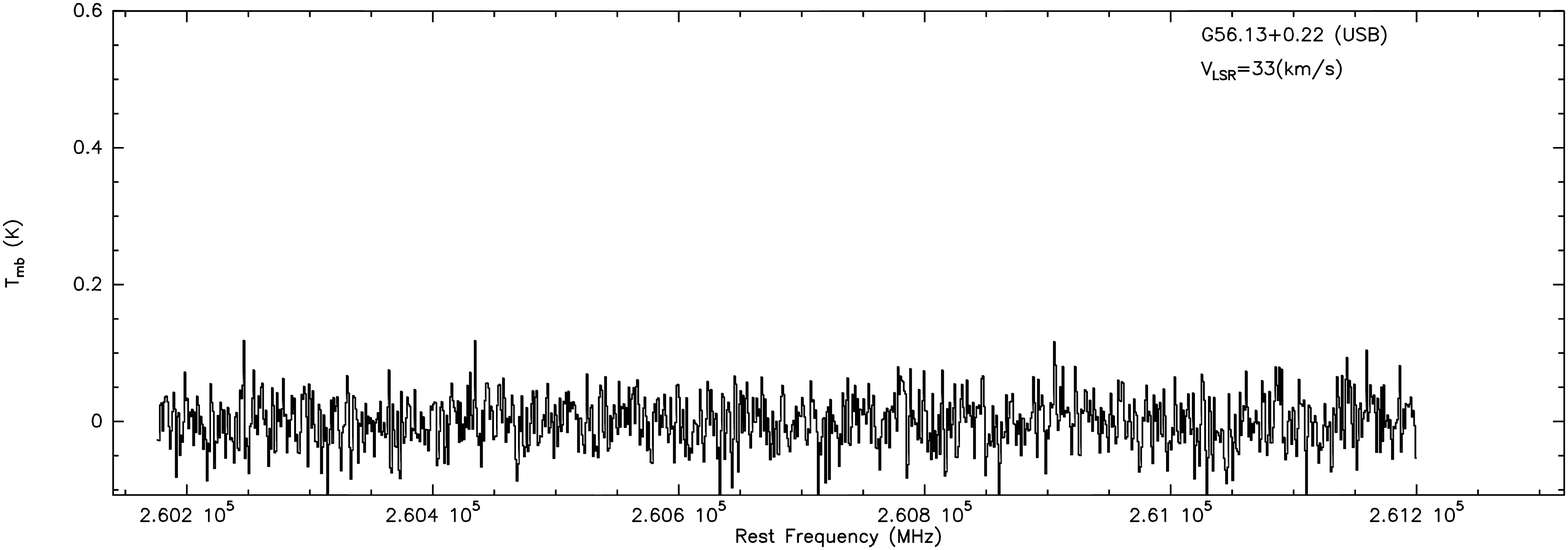}
\caption{(continued) For G56.13+0.22.}
\end{figure*}
\clearpage
 \addtocounter{figure}{-1}
\begin{figure*}
\centering
\includegraphics[scale=.30,angle=0]{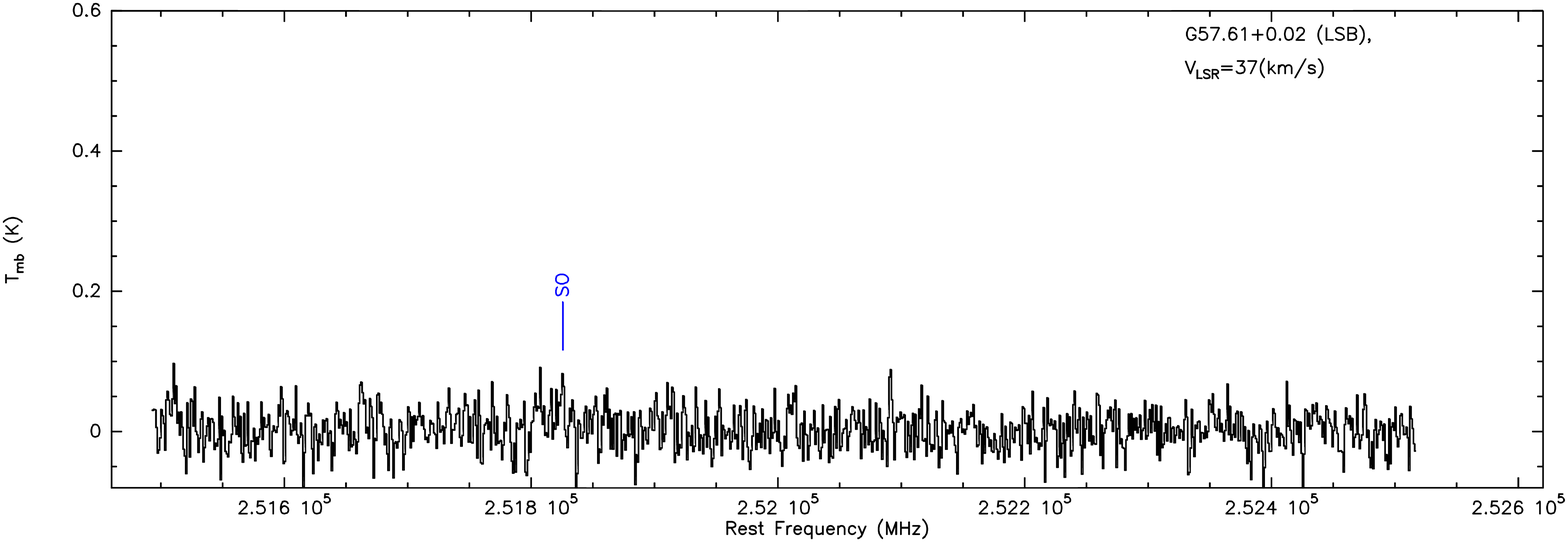}
\includegraphics[scale=.30,angle=0]{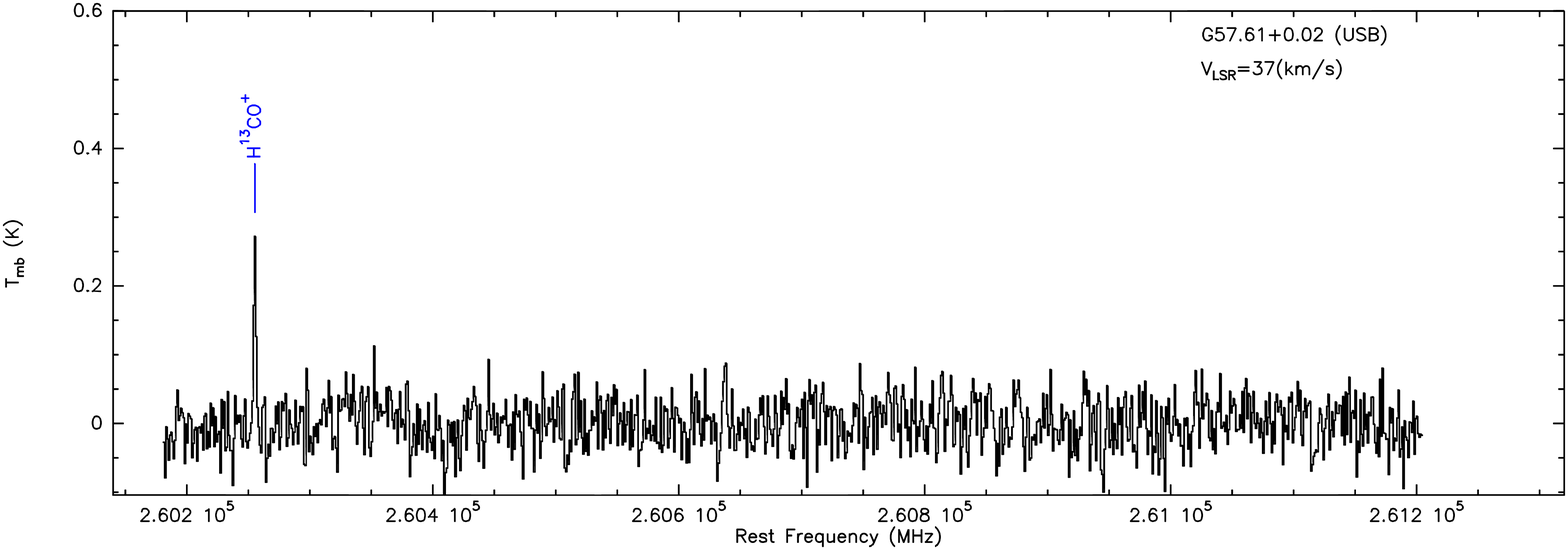}
\caption{(continued) For G57.61+0.02.}
\end{figure*}
 \addtocounter{figure}{-1}
\begin{figure*}
\centering
\includegraphics[scale=.30,angle=0]{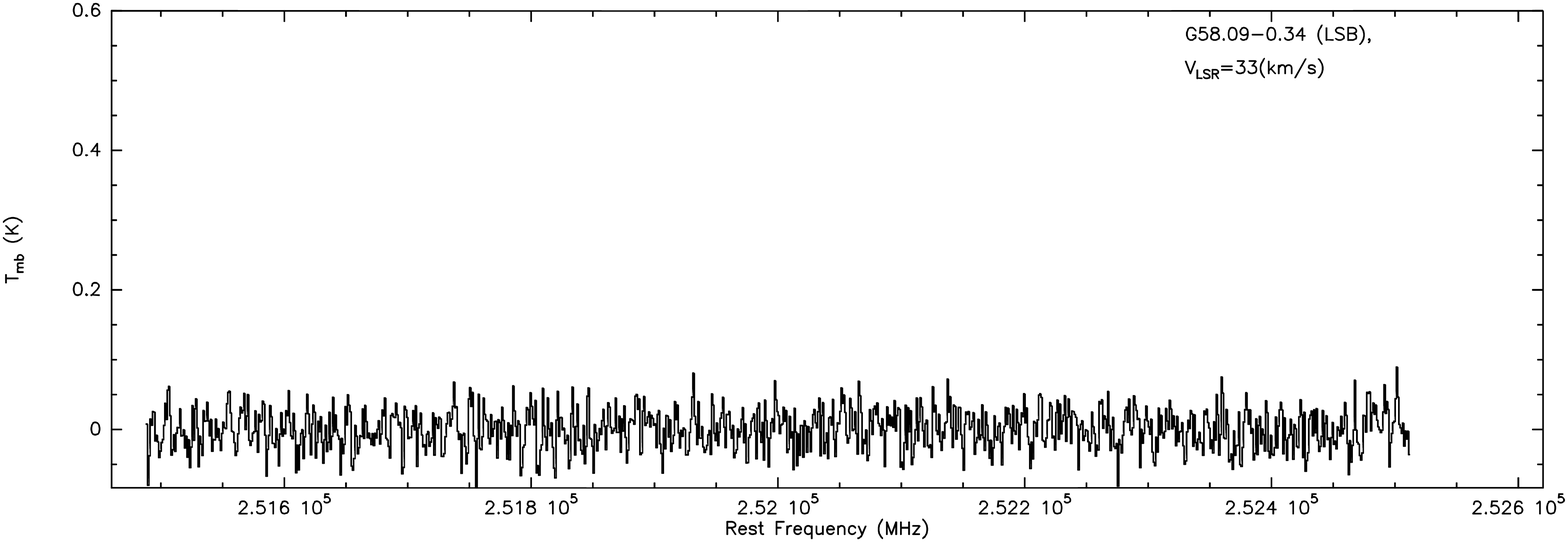}
\includegraphics[scale=.30,angle=0]{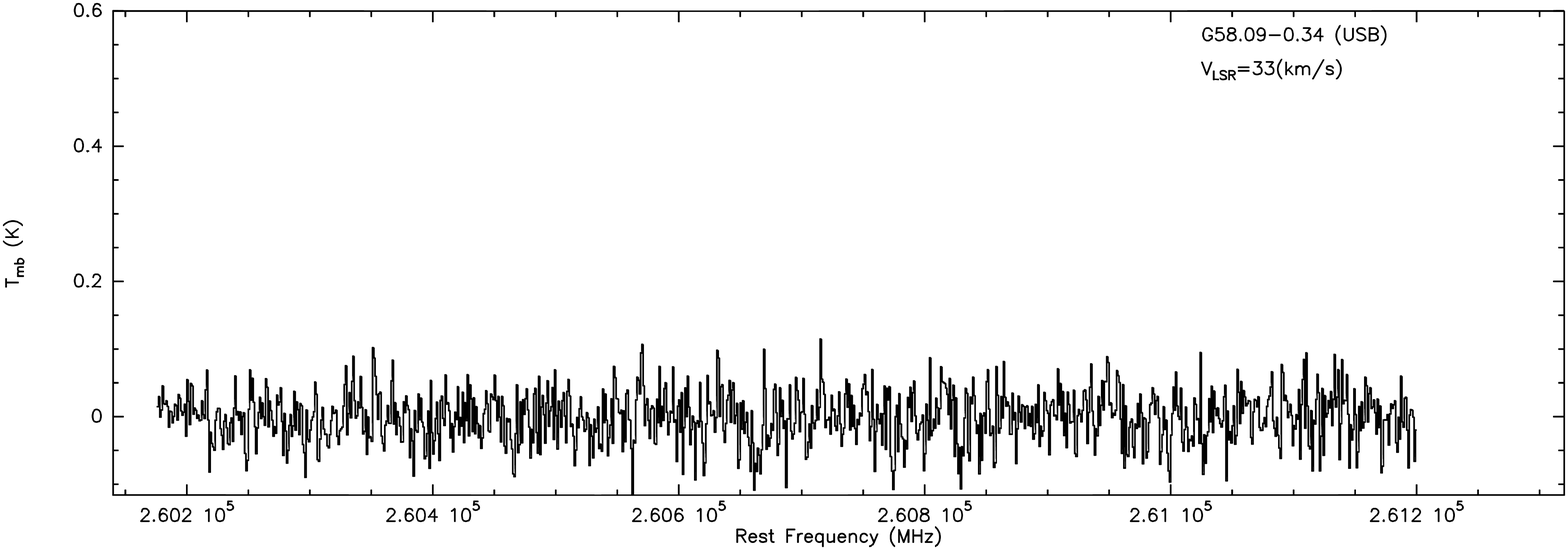}
\caption{(continued) For G58.09-0.34.}
\end{figure*}
\clearpage
 \addtocounter{figure}{-1}
\begin{figure*}
\centering
\includegraphics[scale=.30,angle=0]{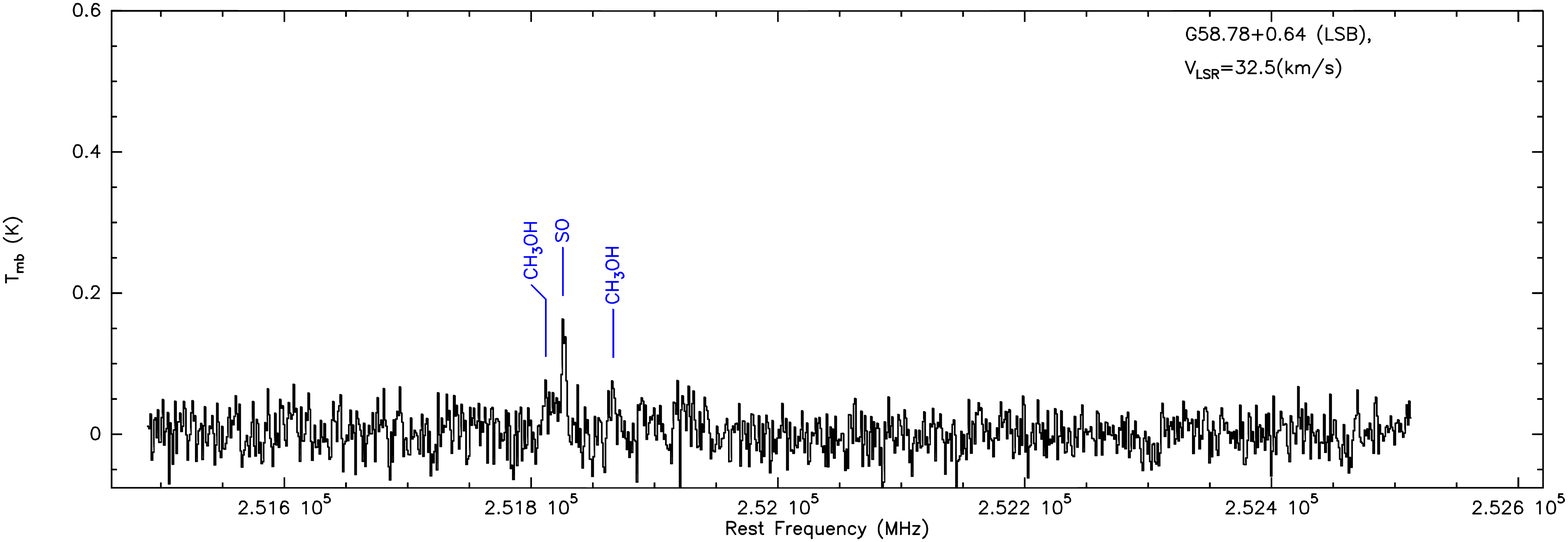}
\includegraphics[scale=.30,angle=0]{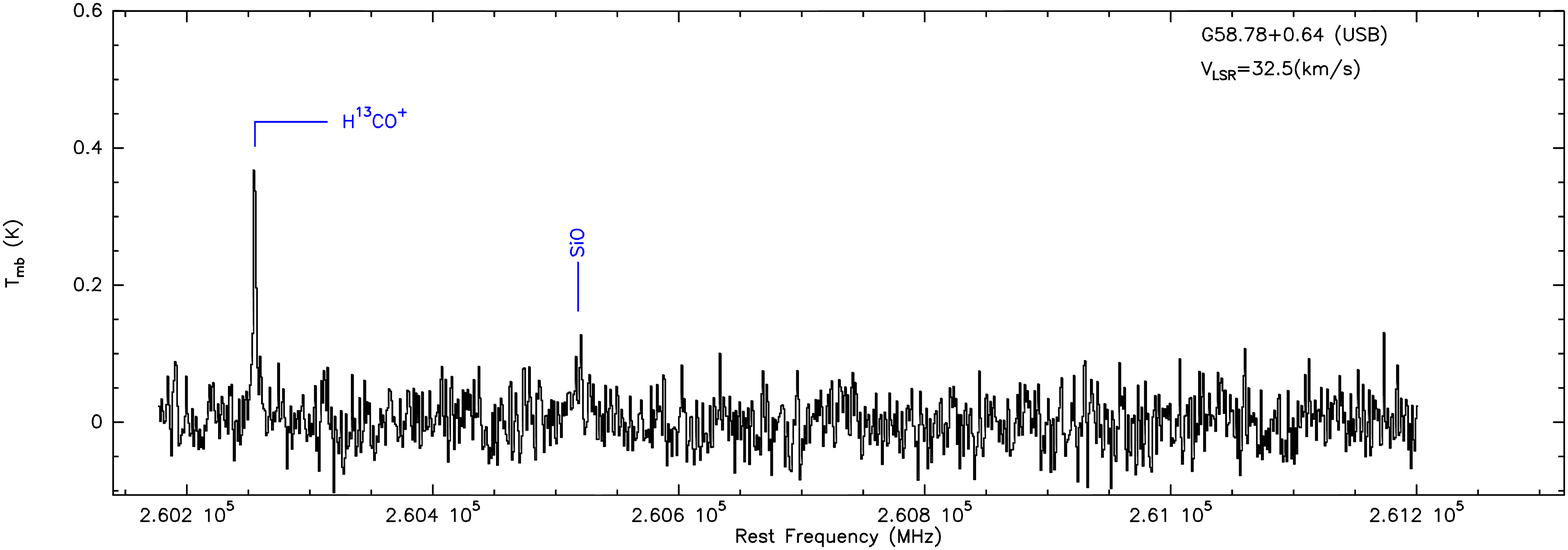}
\caption{(continued) For G58.78+0.64.}
\end{figure*}
 \addtocounter{figure}{-1}
\begin{figure*}
\centering
\includegraphics[scale=.30,angle=0]{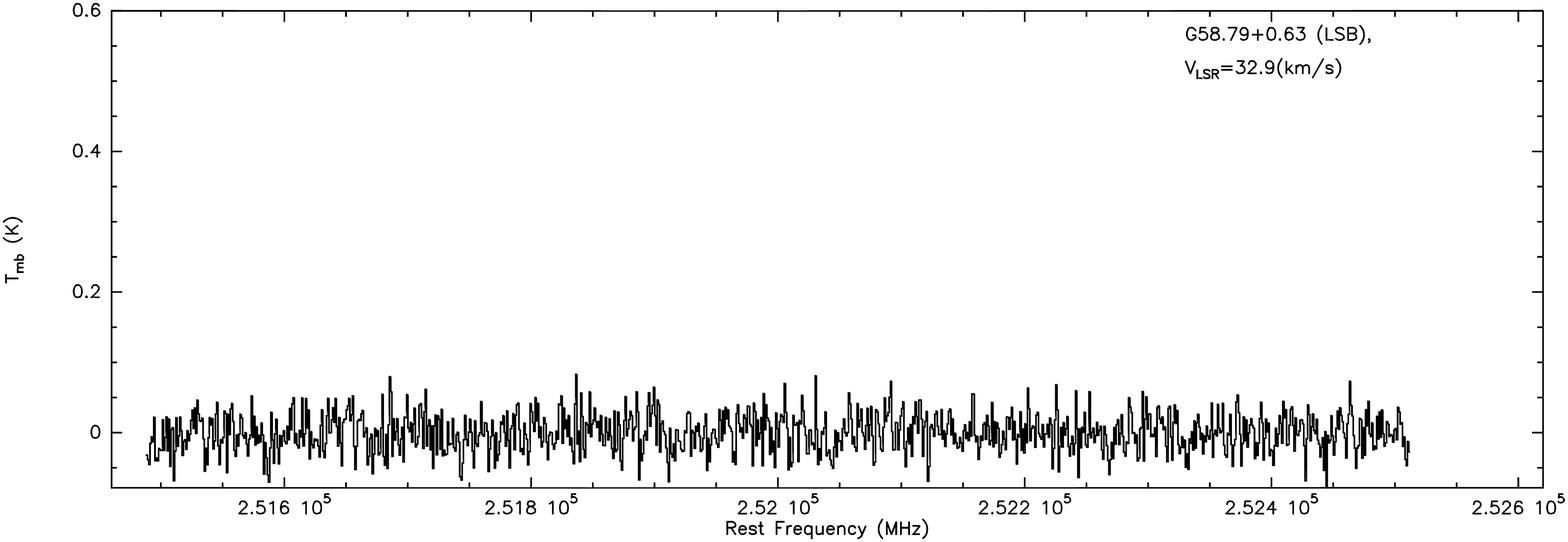}
\includegraphics[scale=.30,angle=0]{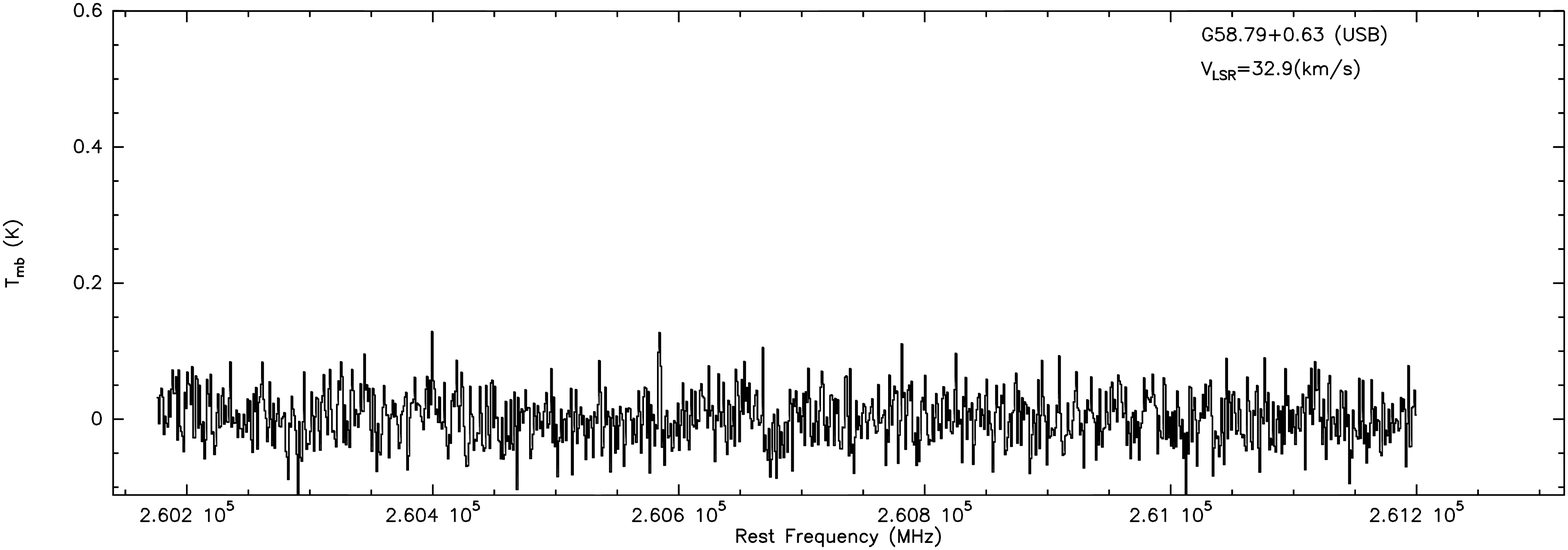}
\caption{(continued) For G58.79+0.63.}
\end{figure*}
\clearpage
 \addtocounter{figure}{-1}
\begin{figure*}
\centering
\includegraphics[scale=.30,angle=0]{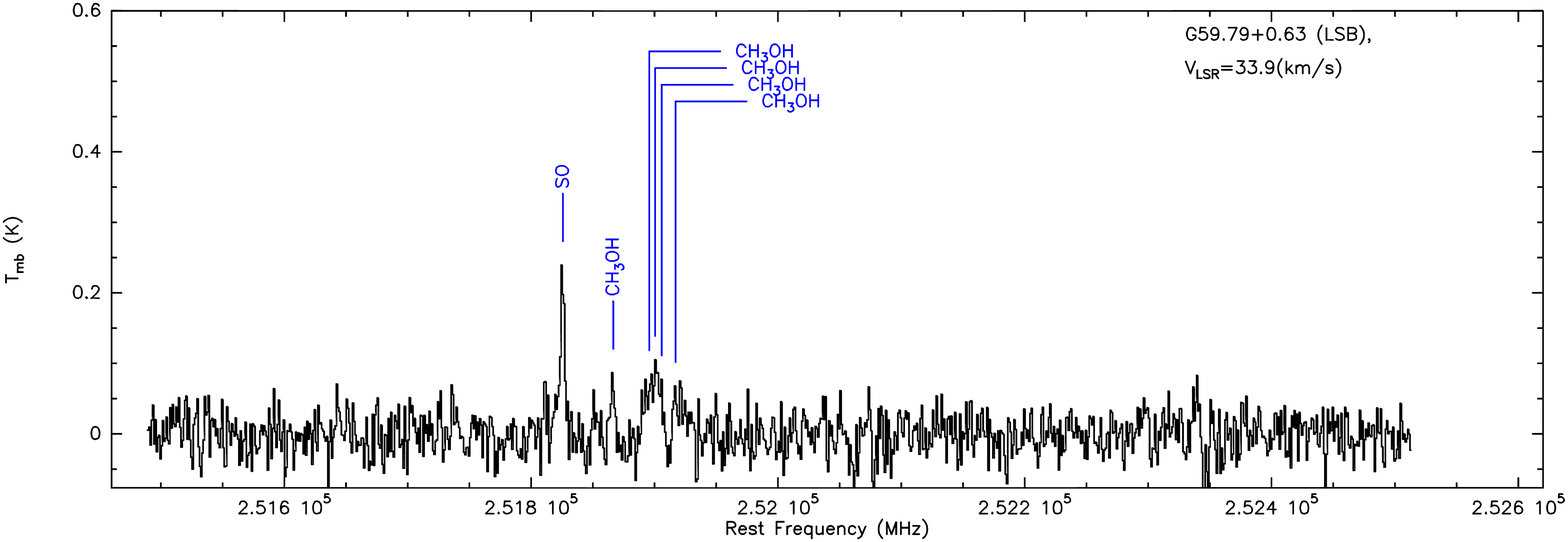}
\includegraphics[scale=.30,angle=0]{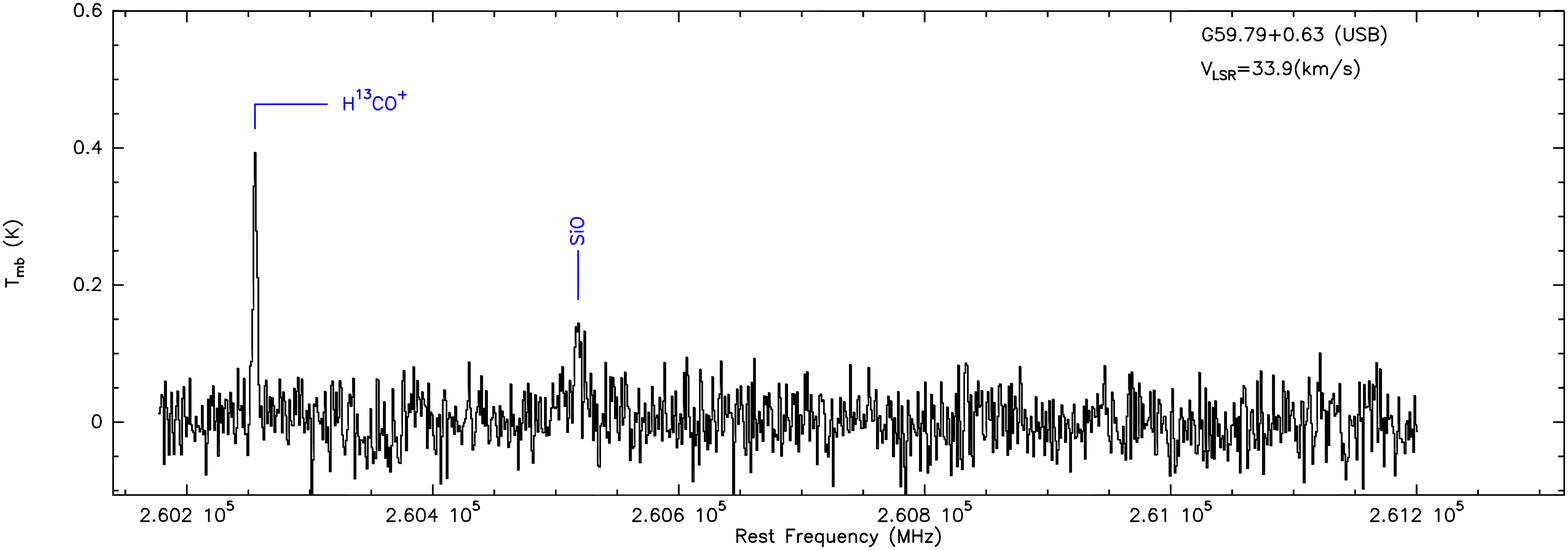}
\caption{(continued) For G59.79+0.63.}
\end{figure*}
 \addtocounter{figure}{-1}
\begin{figure*}
\centering
\includegraphics[scale=.30,angle=0]{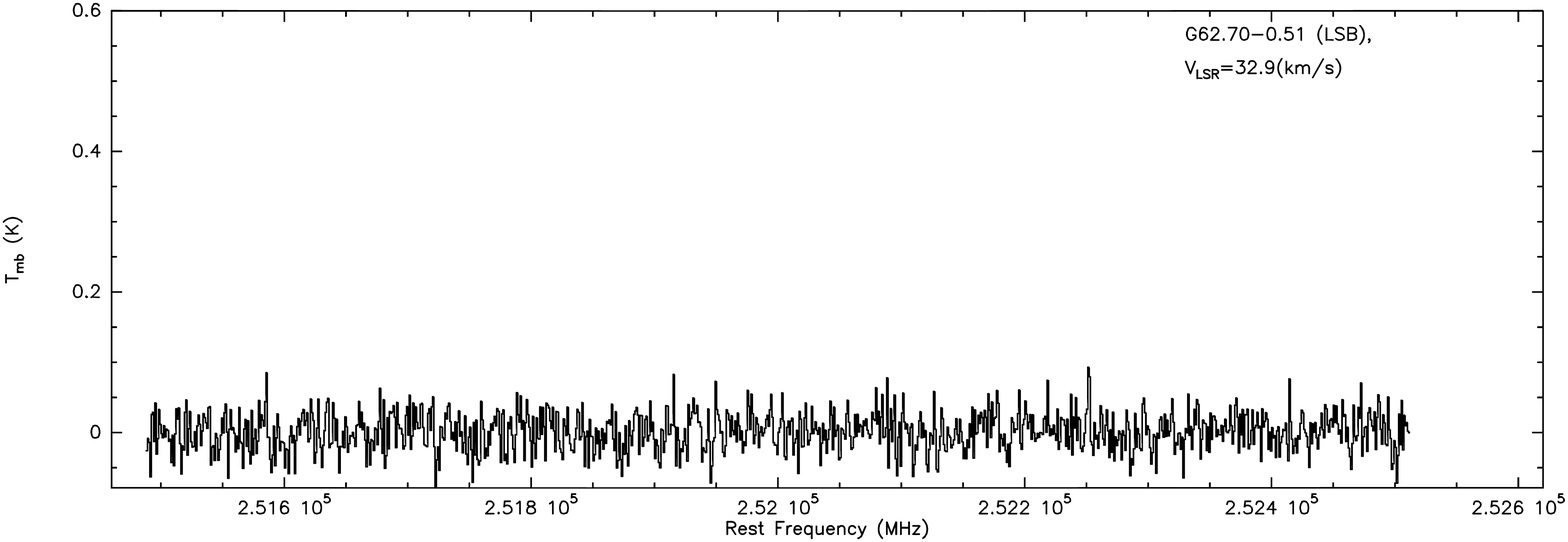}
\includegraphics[scale=.30,angle=0]{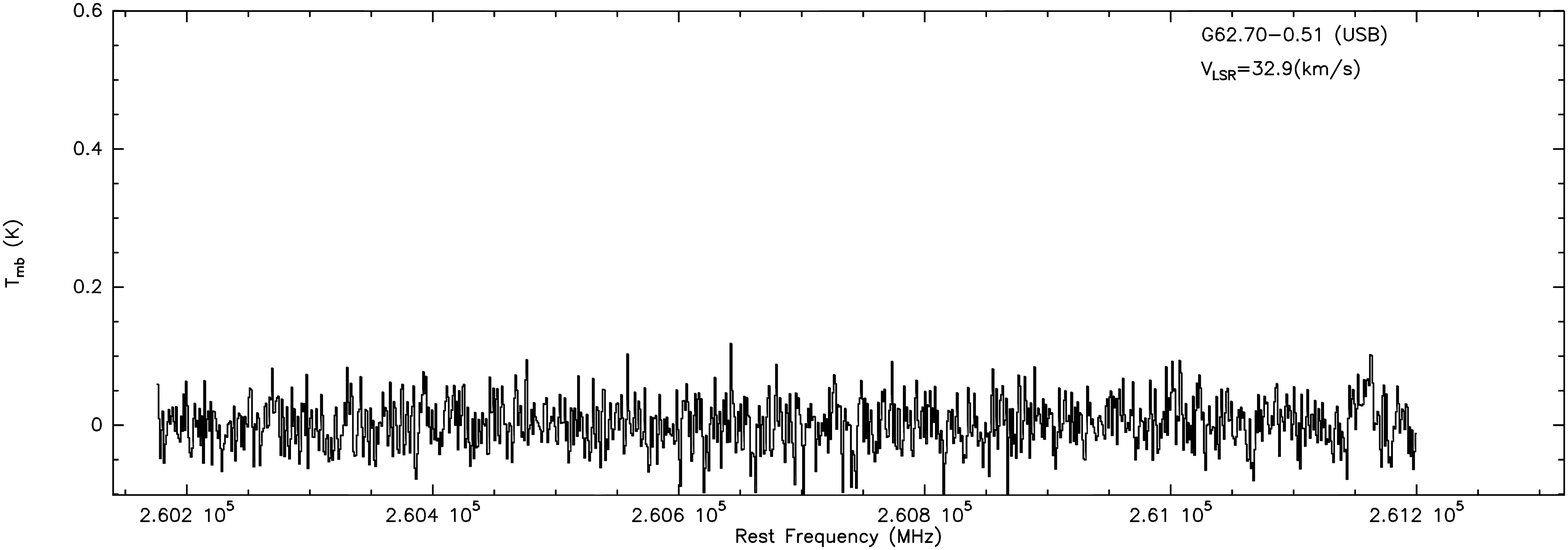}
\caption{(continued) For G62.70-0.51.}
\end{figure*}
\clearpage




\bibliography{ego-survey}{}
\bibliographystyle{apj}

\end{document}